\documentclass[a4paper]{article}

\usepackage{jheppub}
\usepackage[utf8]{inputenc}
\usepackage{amsmath,physics,slashed}
\usepackage{amsfonts}
\usepackage{mathtools}
\usepackage{amssymb,bbm}
\usepackage{dsfont}
\usepackage{subcaption}
\usepackage[svgnames]{xcolor}
\usepackage{tikz}
\usetikzlibrary{calc,arrows,cd,decorations.markings,snakes}
\tikzset{
->-/.style args={#1rotate#2}{decoration={markings, mark=at position #1 with {\arrow[scale=1.5,rotate = #2 ]{stealth}}}, postaction={decorate}}
}
\usepackage{braket}
\usepackage{amsthm}
\newtheorem{theorem}{Theorem}
\theoremstyle{remark}
\newtheorem*{example}{Example}

\numberwithin{equation}{section}

\newcommand{\nc}{\newcommand}
\nc{\rnc}{\renewcommand} 

\rnc{\a}{\alpha}
\rnc{\b}{\beta}
\nc{\g}{\gamma}
\rnc{\d}{\delta}
\nc{\e}{\epsilon}
\nc{\z}{\zeta}
\nc{\f}{\phi}
\nc{\m}{\mu}
\nc{\n}{\nu}
\rnc{\r}{\rho}
\rnc{\k}{\kappa}
\rnc{\l}{\lambda}
\nc{\s}{\sigma}
\rnc{\t}{\tau}
\nc{\w}{\omega}
\nc{\x}{\chi}
\nc{\F}{\Phi}

\nc{\ttb}{\mathrm{b}}
\nc{\rd}{\mathrm{d}}
\nc{\ttf}{\mathrm{f}}
\nc{\ttL}{\mathrm{L}}
\nc{\ttR}{\mathrm{R}}
\nc{\ttv}{\mathrm{v}}
\nc{\tts}{\mathrm{s}}
\nc{\ttc}{\mathrm{c}}
\DeclareMathOperator{\Spin}{\mathrm{Spin}}
\DeclareMathOperator{\SU}{\mathrm{SU}}
\DeclareMathOperator{\U}{\mathrm{U}}
\DeclareMathOperator{\SO}{\mathrm{SO}}
\DeclareMathOperator{\PO}{\mathrm{PO}}
\DeclareMathOperator{\Sp}{\mathrm{Sp}}
\DeclareMathOperator{\Rep}{\mathrm{Rep}}
\nc{\bZ}{\mathbb{Z}}
\nc{\cM}{\mathcal{M}}
\nc{\cC}{\mathcal{C}}
\nc{\cV}{\mathcal{V}}
\nc{\cN}{\mathcal{N}}
\nc{\cO}{\mathcal{O}}
\nc{\cA}{\mathcal{A}}

\nc{\module}{\mathrm{V}}
\nc{\Module}{\mathbf{V}} 
\nc{\mim}{\mathbf{Z}}  
\nc{\aca}{{{}_A\cC_A}} 
\nc{\bcb}{{{}_B\cC_B}}
\nc{\acb}{{{}_A\cC_B}}
\nc{\bca}{{{}_B\cC_A}}
\nc{\ca}{{\cC_A}}
\nc{\<}{\langle }
\rnc{\>}{\rangle }
\nc{\bm}{{\bf m}}
\nc{\bn}{{\bf n}}

\usepackage[status=draft,nomargin,inline]{fixme}
\FXRegisterAuthor{ko}{ako}{KO}
\fxusetheme{color}
\fxsetup{inlineface=\bfseries}
\definecolor{xred}{RGB}{207,38,89} 
\newcommand{\oneform}{Q}

\FXRegisterAuthor{kr}{akr}{KR}

\FXRegisterAuthor{ss}{ass}{SS}
\usepackage{array}
\usepackage{multirow}

\title{Symmetries and Strings of Adjoint QCD$_2$}
\author[1]{Zohar Komargodski,}
\author[1]{Kantaro Ohmori,}
\author[2]{Konstantinos Roumpedakis,}
\author[1]{Sahand Seifnashri}

\affiliation[1]{\it Simons Center for Geometry and Physics\\
Stony Brook University\\
Stony Brook, NY 11794-3636, USA}
\affiliation[2]{\it C.~N.~Yang Institute for Theoretical Physics\\
Stony Brook University\\
Stony Brook, NY 11794-3840, USA}

\emailAdd{zkomargodski@scgp.stonybrook.edu}
\emailAdd{komori@scgp.stonybrook.edu}
\emailAdd{kroumpedakis@gmail.com}
\emailAdd{sahand.seifnashri@stonybrook.edu}

\preprint{YITP-SB-20-28}

\date{}

\abstract{
We revisit the symmetries of massless two-dimensional adjoint QCD with gauge group $\SU(N)$. The dynamics is not sufficiently constrained by the ordinary symmetries and anomalies. Here we show that the theory in fact admits $\sim 2^{2N}$ non-invertible symmetries which severely constrain the possible infrared phases and massive excitations. We prove that for all $N$ these new symmetries enforce deconfinement of the fundamental quark. 
When the adjoint quark has a small mass, $m\ll g_\mathrm{YM}$, the theory confines and the non-invertible symmetries are softly broken. We use them to compute analytically the $k$-string tension for $N\leq 5$. Our results suggest that the $k$-string tension, $T_k$, is $T_k\sim |m| \sin(\pi k /N)$ for all $N$.
We also consider the dynamics of adjoint QCD deformed by symmetric quartic fermion interactions. These operators are not generated by the RG flow due to the non-invertible symmetries, thus violating the ordinary notion of naturalness. We conjecture partial confinement for the deformed theory by these four-fermion interactions, and prove it for $\SU(N\leq5)$ gauge theory. Comparing the topological phases at zero and large mass, we find that a massless particle ought to appear on the string for some intermediate nonzero mass, consistent with an emergent supersymmetry at nonzero mass.
We also study the possible infrared phases of adjoint QCD allowed by the non-invertible symmetries, which we are able to do exhaustively for small values of $N$. The paper contains detailed reviews of ideas from fusion category theory that are essential for the results we prove.
}

\begin{document}
\maketitle
\section{Introduction and Summary}
\label{sec:intro}
Confinement in gauge theories has been a central subject of research in quantum field theory for decades.
While the ultimate goal is to understand confinement in Yang-Mills (YM) theory and Quantum Chromodynamics (QCD) in 3+1-dimensions, these models so far forbid any direct analytical handle. It is therefore important to continue extracting lessons from similar but more tractable models.

In this spirit, in the current paper we analyze the 1+1-dimensional model of QCD with a Majorana quark in the adjoint representation of the gauge group. Pure YM theory in 1+1 dimensions is a solvable model without propagating local degrees of freedom~\cite{Migdal:1975zg}. Adding a Majorana fermion makes the model nontrivial and with rich dynamics. We only consider the case where the gauge group is $\SU(N)$ in this paper. There has been a lot of progress recently on some aspects of strongly coupled field theory dynamics, and, in particular, gauge theories with adjoint matter in 3 and 4 space-time dimensions. See for instance ~\cite{Gomis:2017ixy,Cordova:2017vab,Cordova:2018qvg,Anber:2018iof,Cordova:2018acb,Wan:2018djl,Poppitz:2019fnp,Bolognesi:2019fej,Cordova:2019bsd,Choi:2019eyl,Bi:2019gle,Cordova:2019jqi,Delmastro:2020dkz} and references therein. 

The adjoint Majorana fermion model in 1+1 dimensions was extensively studied in 90's, see for instance~\cite{Dalley:1992yy,Kutasov:1993gq,Boorstein:1993nd,Kutasov:1994xq,Gross:1995bp,Boorstein:1997hd,Gross:1997mx}. 
For the gauge group $\SU(2)$ it was rigorously concluded that when the adjoint quark is massless, the gauge theory is in the deconfined phase, i.e.\ the Wilson line in the fundamental representation obeys a perimeter law. This is a surprising claim as the quark is in the adjoint representation of the gauge group, so naively there are no particles that can screen the Wilson line in the fundamental representation. The physical mechanism at play here is that of a fractionalization of the adjoint quark. We will see this explicitly in an Abelian analog of this phenomenon. As soon as the quark is given a mass, the  Wilson line in the fundamental representation has an area law and thus confinement sets in.\footnote{In this paper we focus on the massless adjoint theory or the theory with a small mass. There are also many interesting questions about the theory at large mass which have been recently studied in \cite{Donahue:2019adv,Donahue:2019fgn}.}

From a modern point of view, this result for $\SU(2)$ gauge theory can be immediately understood from the symmetries of the theory and their anomalies. 
The theory has a $\mathbb{Z}_2^\chi$ chiral symmetry acting on the adjoint fermion as $(-1)^{F_L}$ with $F_L$ being the left fermion number. There is also a $\mathbb{Z}_{2}$ symmetry associated to the center of the group (i.e., a one-form symmetry in modern terminology \cite{Gaiotto:2014kfa}). 
There is a mixed anomaly between these two symmetries, implying that the topological $(-1)^{F_L}$ line is charged under the $\mathbb{Z}_{2}$ one-form symmetry. The topological line associated to $(-1)^{F_L}$, being an ordinary internal symmetry, cannot change the energy density. This means that, on the one hand, acting with $(-1)^{F_L}$ we must now reside in a state with a string, but on the other hand, it must have energy density (tension) equal to the original vacuum. Therefore, the fundamental Wilson line must have perimeter law. 
If the adjoint quark is massive, one cannot run the same argument since  $(-1)^{F_L}$ is no longer a symmetry and hence the energy density of the string could be nonzero, and indeed it is confined.\footnote{Many applications of anomalies involving one-form symmetries have been studied recently in diverse dimensions and situations. 
See ~\cite{Gaiotto:2014kfa,Gaiotto:2017yup,Tanizaki:2017bam,Komargodski:2017smk,Shimizu:2017asf,Kitano:2017jng,Gomis:2017ixy,Guo:2017xex,Aitken:2018kky,Anber:2018iof,Cordova:2018acb,Anber:2018xek,Hsin:2018vcg,Bolognesi:2019fej} for some examples and additional references. \label{one-form.refs}}

The situation for $\SU(N>2)$ gauge theory with an adjoint fermion is more complicated. With a massless quark, $(-1)^{F_L}$ is still a symmetry but now the anomaly with the center (one-form) $\mathbb{Z}_N$ symmetry is less constraining. 
Indeed, the topological line of $(-1)^{F_L}$ takes the string sector with $k\in \mathbb{Z}_N$ to $k+N/2\in \mathbb{Z}_N$ for even $N$ while for odd $N$ it does not shift $k$. So one can immediately conclude that the Wilson line for an external quark with $N$-ality $N/2$ must be deconfined for even $N$ and for odd $N$ one cannot draw any conclusion about deconfinement. This summarizes the nice observations of \cite{Cherman:2019hbq}. Therefore, from the ordinary symmetries and anomalies, one cannot conclude whether the fundamental Wilson line is confined or not for $\SU(N>2)$. 

Our main aim in this paper is to show that this gauge theory admits many  ``exotic'' symmetries, in fact, of the order of $2^{2N}$ ``exotic'' symmetries. These symmetries correspond to non-invertible topological lines. They constrain the massive particle spectrum as well as the long distance behavior. We analyze these new symmetries in detail and show that they lead to the deconfinement of the fundamental Wilson line for all $N$. As soon as we add a mass term for the quark, these new symmetries disappear and one finds that the Wilson lines become confined. However, some quartic fermion terms preserve a sufficiently interesting subset of these $\sim 2^{2N}$ symmetries and we therefore also analyze the dynamics of the theory with such quartic terms. 
Finally, if the mass of the adjoint quark is small, one can treat it perturbatively around the massless theory and we use such an expansion to compute the exact $k$-string tension for several low-rank cases. This is possibly the first such analytical computation without supersymmetry. We will summarize our findings below.

\paragraph{Non-Abelian bosonization}
A convenient tool for studying adjoint QCD is the non-Abelian bosonization which relates free fermions to a Wess-Zumino-Witten (WZW) model. The non-Abelian bosonization was first proposed in \cite{Witten:1983ar}, and was recently made more precise in \cite{Ji:2019ugf}. It states that 
\begin{equation}
    \text{$n$ Majorana fermions}/(-1)^F
    \leftrightarrow
    \text{$\mathrm{Spin}(n)_1$ WZW} ~,
    \label{eq:NABos}
\end{equation}
where the modding by $(-1)^F$ on left hand side denotes the spin-structure sum (a.k.a GSO projection) of the fermions. The $\mathrm{Spin}(n)$ symmetry on the right hand side is identified with the symmetry rotating the Majorana fermions. When $n = N^2-1$, the $\SU(N)$ symmetry acting on the fermions in the adjoint representation is embedded into the subgroup $\mathrm{PSU}(N)\subset \mathrm{Spin}(N^2-1)$ of the right hand side.
Gauging the $\SU(N)$ which acts on the $\mathrm{Spin}(N^2-1)_1$ WZW model results in the $\mathrm{Spin}(N^2-1)_1/\SU(N)_N$ coset model \cite{Witten:1983ar,Bardakci:1987ee} with a kinetic term for the $\SU(N)$ gauge fields. The Yang-Mills coupling $g_\mathrm{YM}$ for the $\SU(N)$ gauge fields provides a dimensionful scale in the problem. The model has nontrivial excitations at that scale and it is not known to be solvable. This coset model is exactly equivalent to the $\SU(N)$ gauge theory with Majorana fermions in the adjoint representation (and a sum over the spin structures).

If we ignore the kinetic term, the central charge of this coset turns out to be zero, and thus it is a topological quantum field theory (TQFT). From the bosonization in \eqref{eq:NABos}, we get the duality
\begin{equation}
    \text{($\SU(N)$ adj. QCD with $g_\mathrm{YM}\to\infty$)}/(-1)^F
    \leftrightarrow
    \text{$\mathrm{Spin}(N^2-1)_1/\SU(N)_N$ TQFT}~.
    \label{eq:QcdDualBos}
\end{equation}
Since removing the kinetic term  ($g_\mathrm{YM}\to\infty$) naively corresponds to taking the deep infrared limit, one expects to find the TQFT  $\mathrm{Spin}(N^2-1)_1/\SU(N)_N$ at long distances. We will see that this TQFT is indeed acted upon by our $\sim 2^{2N}$ symmetries. This large ground state degeneracy in massless adjoint QCD is forced by the $\sim 2^{2N}$ new symmetries. Note that it is not obvious that the infrared limit is the same as removing the kinetic term for the $\SU(N)$ gauge fields. Some of our results (about the symmetries, confinement vs deconfinement, and the effects of four-fermion interactions) actually do not depend on this assumption. We will also explore possible alternatives for the infrared limit.

A technical point is that the statement  in \eqref{eq:QcdDualBos} is correct for the GSO projected version of adjoint QCD. Of course, the gauging of  $(-1)^F$ does not change much the essential dynamical questions such as confinement vs deconfinement but it is important to make a distinction between the model with and without $(-1)^F$ gauged. They have different invertible and non-invertible symmetries. 
In the bulk of the paper we only discuss, for simplicity, the version of the model with gauged  $(-1)^F$ symmetry (i.e.\ the bosonic model). Some details about the fermionic theory are collected in the appendices.

The coset $\mathrm{Spin}(N^2-1)_1/\SU(N)_N$ can be considered as the precise version of the bosonization in \cite{Kutasov:1994xq}. 
The bosonic coset $\mathrm{Spin}(N^2-1)_1/\SU(N)_N$ has $3\times 2^{N-2}$ vacua. Therefore, in the large $N$ limit, it  leads to a Hagedorn behavior with zero Hagedorn temperature \cite{Kutasov:1993gq,Boorstein:1997hd}. 
This large number of degenerate vacua is enforced by the large number of new symmetries. (The number of vacua in the fermionic theory in the NS sector is $2^{N-1}$, which likewise leads to a zero temperature Hagedorn transition at large $N$.)

The existence of the $\sim 2^{2N}$ new symmetries is established in the full theory, and that they lead to deconfinement of the fundamental line, is likewise established without any assumptions about the infrared. To actually count the ground states one needs to, roughly speaking, understand which representation of the non-invertible symmetries is physically realized, and while the TQFT $\mathrm{Spin}(N^2-1)_1/\SU(N)_N$ is an appealing candidate, as we said, some of our conclusions do not depend on it being the right answer. (In some special cases such as $N=2,3,4$ we will be able to say much more about this question.)

\paragraph{Topological line operators in 1+1-d QFTs}
An important claim of this paper is that 1+1-dimensional massless adjoint QCD has \emph{non-invertible topological lines}. This serves as an ultraviolet reason for the complete deconfinement of the theory that we will show.
Recall that in QFT, given a symmetry group $G$ and an element $g\in G$ we have a topological operator $U(g)[\Sigma]$ defined on a codimension-1 subspace $\Sigma$ in spacetime.
The operator $U(g)$ induces the symmetry action on local operators as depicted in Fig.~\ref{fig:Ug}.
The topological operators $U(g)$ respects the multiplication law of the group $G$, and in particular have their inverse:
\begin{equation}
    U(g)U(g^{-1}) =\mathbbm{1}~,
\end{equation}
where $\mathbbm{1}$ stands for the trivial topological operator.
However, a generic topological operator in a QFT does not necessarily have an inverse, and we call such an operator a non-invertible topological operator.
In general, given topological operators $L_i$ of the same dimensions, we have a fusion rule:
\begin{equation}
    L_i \otimes L_j = \sum_{k} N_{i,j}^k L_k~,
    \label{eq:fusion}
\end{equation}
where the symbol $\otimes$ stands for the operation of putting the two operators on top of each other. In the co-dimension 1 case, this algebra is generally non-commutative. The information including the fusion rule~\eqref{eq:fusion} and other more subtle properties is packaged in the mathematical concept of fusion category, which is the main technical tool we employ in this paper.
\begin{figure}[t]
    \centering
    \begin{tikzpicture}[scale = .8, baseline = 0]
        \coordinate (o) at (0,0);
        \coordinate (d) at (.7,-1);
        \coordinate (m) at (.7,0);
        \coordinate (u) at (.7,1);
        \draw[fill] (o) circle [radius = .05];
        \node[anchor = south] at (o) {$\mathcal{O}$};
        \draw[->- = .5 rotate 0] (d) -- (u);
        \node[anchor = west] at (m) {$U(g)$};
    \end{tikzpicture}
    \hspace{1em}
    $\Rightarrow$
    \hspace{1em}
    \begin{tikzpicture}[scale = .8, baseline = 0]
        \coordinate (o) at (0,0);
        \coordinate (d) at (-1,-1);
        \coordinate (m) at (-1,0);
        \coordinate (u) at (-1,1);
        \draw[fill] (o) circle [radius = .05];
        \node[anchor = south] at (o) {$g \cdot \mathcal{O}$};
        \draw[->- = .5 rotate 0] (d) -- (u);
        \node[anchor = east] at ($(u)!.5!(d)$) {$U(g)$};
    \end{tikzpicture}
    \caption{The symmetry operator $U(g)$ for an element $g$ of the symmetry group $G$ causes the symmetry action on a local operator when the symmetry operator passes through the local operator.}
    \label{fig:Ug}
\end{figure}

The most basic example of a non-invertible topological line is the duality line $\mathcal{N}$ in the critical Ising CFT. The fusion algebra of the lines is 
\begin{equation}
    \eta \otimes \eta \simeq 1, \quad \mathcal{N} \otimes \eta \simeq \eta \otimes \mathcal{N} \simeq \mathcal{N}, \quad \mathcal{N} \otimes \mathcal{N} \simeq 1 + \eta,
\end{equation} 
where $\eta$ is the $\mathbb{Z}_2$ symmetry line. The line $\mathcal{N}$ implements the Kramers-Wannier duality, that is, it sends the spin operator $\sigma$ to the disorder operator $\mu$ (see Fig.~\ref{fig:KW}).
\begin{figure}[ht]
    \centering
    \begin{tikzpicture}[scale = .8, baseline = 0]
        \coordinate (o) at (0,0);
        \coordinate (d) at (.7,-1);
        \coordinate (m) at (.7,0);
        \coordinate (u) at (.7,1);
        \draw[fill] (o) circle [radius = .05];
        \node[anchor = south] at (o) {$\sigma$};
        \draw[->- = .5 rotate 0] (d) -- (u);
        \node[anchor = west] at (m) {$\mathcal{N}$};
    \end{tikzpicture}
    \hspace{1.5em}
    $=$
    \hspace{1.5em}
    \begin{tikzpicture}[scale = .8, baseline = 0]
        \coordinate (o) at (0,0);
        \coordinate (d) at (-1,-1);
        \coordinate (m) at (-1,0);
        \coordinate (u) at (-1,1);
        \draw[fill] (o) circle [radius = .05];
        \node[anchor = south] at (o) {$\mu$};
        \draw[->- = .5 rotate 0] (d) -- (m);
        \draw[->- = .5 rotate 0] (m) -- (u);
        \draw[->- = .5 rotate 0,dashed] (m) -- (o);
        \node[anchor = east] at ($(m)!.5!(d)$) {$\mathcal{N}$};
        \node[anchor = east] at ($(m)!.5!(u)$) {$\mathcal{N}$};
        \node[anchor = south] at ($(m)!.5!(o)$) {$\eta$};
    \end{tikzpicture}
    \caption{The non-invertible duality line $\mathcal{N}$ in the critical Ising CFT maps the spin operator $\sigma$ to the disorder operator $\mu$ which lives at the edge of the $\mathbb{Z}_2$ symmetry line $\eta$.}
    \label{fig:KW}
\end{figure}
In general, a non-invertible topological line maps a local operator to a sum of defect operators, which live at the edge of other topological lines.

In addition to the Ising CFT, the other most familiar construction of models with such non-invertible symmetries are the $G_k$ WZW models with simply-connected group $G$ at level $k$. These models have topological lines $L_\mu$ that preserve the left and right $\hat{\mathfrak{g}}_k$ affine algebras (and hence commute with the energy-momentum tensor and are thus topological),
and are labeled by the integrable representations $\mu$ of $\hat{\mathfrak{g}}_k$. They are called Verlinde lines \cite{Verlinde:1988sn}. 
In this case, the coefficients $N_{\mu\nu}^\rho$ in \eqref{eq:fusion} coincide with the fusion coefficients in the fusion rule of the primary operators $\mathcal{O}_\mu$, which are also labelled by the integrable representations.

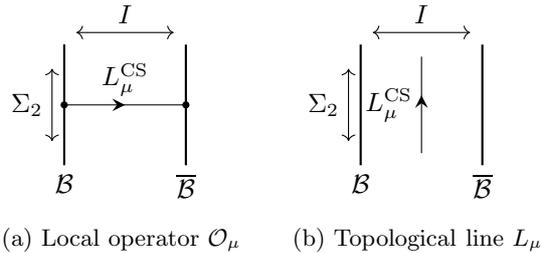
\begin{figure}[t]
    \centering
    \subcaptionbox{Local operator $\mathcal{O}_\mu$}[.25 \linewidth]{
    \begin{tikzpicture}[scale = .8]
        \draw[thick] (0,0) coordinate(ul) -- ++ (0,-2) coordinate(dl);
        \draw[thick] (ul) ++ (2,0) coordinate(ur) -- ++ (0,-2) coordinate (dr);
        \node[anchor = north] at (dl) {$\mathcal{B}$};
        \node[anchor = north] at (dr) {$\overline{\mathcal{B}}$};
        \draw (ul) -- ++ (0,-1) coordinate(ol);
        \draw (ol)  ++(2,0) coordinate (or);
        \draw[->-=.5 rotate 0] (ol) -- node[midway,anchor=south] {$L^\text{CS}_\mu$} ++(2,0) coordinate (or);
        \node[anchor = west] at (or) {\phantom{$\mathcal{O}_\mu$}};
        \draw[<->] (ul) ++ (-.2,-.4) -- node[midway,anchor = east] {$\Sigma_2$} ++ (0,-1.2);
        \draw[fill] (ol) circle [radius = .05];
        \draw[fill] (or) circle [radius = .05];
        \draw[<->] (ul) ++ (.2,.2) -- node[midway,anchor = south] {$I$} ++ (1.6,0);
    \end{tikzpicture}
    }
    \subcaptionbox{Topological line $L_\mu$}[.25\linewidth]{
    \begin{tikzpicture}[scale = .8]
        \draw[thick] (0,0) coordinate(ul) -- ++ (0,-2) coordinate(dl);
        \draw[thick] (ul) ++ (2,0) coordinate(ur) -- ++ (0,-2) coordinate (dr);
        \node[anchor = north] at (dl) {$\mathcal{B}$};
        \node[anchor = north] at (dr) {$\overline{\mathcal{B}}$};
        \draw (ul) ++ (1,-.2) coordinate(Lu);
        \draw[->- = .4 rotate 180] (Lu) -- node[midway,anchor=east] {$L^\text{CS}_\mu$} ++(0,-1.6) coordinate (Ld);
        \draw[<->] (ul) ++ (.2,.2) -- node[midway,anchor = south] {$I$} ++ (1.6,0);
        \draw[<->] (ul) ++ (-.2,-.4) -- node[midway,anchor = east] {$\Sigma_2$} ++ (0,-1.2);
        \node[anchor = west] at (or) {\phantom{$\mathcal{O}_\mu$}};
    \end{tikzpicture}
    }
    \caption{The $G_k$ WZW model can be obtained by considering the 2+1-d CS theory on $I\times \Sigma_2$, where $I$ is the interval and $\Sigma_2$ is the 2d spacetime of the WZW model.
    The boundary conditions $\mathcal{B}$ and $\overline{\mathcal{B}}$ are conjugate of each other.
    (a):  a local operator $\cO_\mu$ in the Verma module $\module_\mu\otimes \overline{\module}_\mu$ can be realized as the Wilson line $L^\text{CS}_\mu$ in the representation  $\mu$ bridging the two boundaries.
    (b):  a Verlinde line $L_\mu$ is identified with the bulk Wilson line $L_\mu^\text{CS}$ along the 2d spacetime $\Sigma_2$}
    \label{fig:CSWZW}
\end{figure}

A convenient way to understand the Verlinde lines is to realize the WZW model as a Chern-Simons (CS) theory on an interval \cite{Witten:1988hf} (See Fig.~\ref{fig:CSWZW}). 
In this picture, the primary operator $\mathcal{O}_{\mu}$ corresponds to the Wilson line $L^\text{CS}_\mu$ bridging the two boundaries, while the Verlinde line $L_\mu$ in the 2d theory can be interpreted as the Wilson line $L^\text{CS}_\mu$ in the CS theory along the 2d spacetime.
Therefore, the CS theory explains why the two objects, the Verlinde lines $L_\mu$ and the primary operators $\mathcal{O}_\mu$, obey the same fusion rule.
In this way, the fusion category of the Verlinde lines, which we denote by $\mathcal{C}=\Rep\hat{\mathfrak{g}}_k$, is given by the modular tensor category governing the Wilson lines of the CS theory \cite{Moore:1988qv}.\footnote{Modular tensor category is a fusion category equipped with a braiding (and a nondegenerate modular $S$-matrix). A general fusion category does not necessarily admit braiding. The Verlinde lines are therefore somewhat special non-invertible lines which are equipped with braiding.}

Recently, there has been a lot of interest in the study of topological line operators in 1+1-dimensional QFTs \cite{Carqueville:2012dk,Brunner:2013xna,Bhardwaj:2017xup,Chang:2018iay,Ji:2019ugf,Lin:2019hks, Thorngren:2019iar}.\footnote{See also \cite{Kapustin:2010if,Kitaev_2012,Fuchs:2012dt,Barkeshli:2014cna,Carqueville:2017aoe,Carqueville:2017ono,Johnson-Freyd:2020usu,kong2020algebraic,Rudelius:2020orz} for the study of topological extended operators in higher dimensions.} It has been noted that topological line operators can be thought of as a generalization of the notion of symmetry, which is invariant under renormalization group (RG) flow, and a ``non-anomalous part" of the lines can be gauged \cite{Carqueville:2012dk,Brunner:2013xna,Bhardwaj:2017xup,Chang:2018iay, Thorngren:2019iar}. 
In particular, \cite{Chang:2018iay} emphasized the RG flow invariance of topological lines and its ability to constrain the infrared (IR) physics;
for example, they found that the degenerate 2 vacua found in the integrable relevant flow from the tricritical Ising CFT can be explained by the non-invertible topological line preserved by the relevant operator.
We will find that adjoint QCD is another good example whose dynamics is tightly constrained by non-invertible topological line operators.

As a general comment, the reason that symmetries and their anomalies are RG invariants is that the space of symmetries and their anomalies is a discrete space. Similarly, the space of all possible non-symmetry lines and their fusion and other relevant structures is discrete. This is why such non-invertible symmetries place constraints on the full interacting theory. In fact, the non-invertible symmetries lead to amusing violations of naturalness. From Fig.~\ref{fig:KW} it is evident that RG transformations cannot generate local operators that do not freely pass through the non-invertible lines. This leads to examples of 1+1 dimensional quantum field theories where some operators that are invariant under all the standard symmetries are ``mysteriously'' never generated by the dynamics. We shall see this later concretely in adjoint QCD.

\paragraph{Topological line operators in adjoint QCD}
The main part of this paper is devoted to the study of topological line operators in $\SU(N)$ adjoint QCD.
Consider the topological lines in the theory of free massless $N^2-1$ Majorana fermions (with gauged fermion number symmetry). It is actually a difficult problem to enumerate them (for a single periodic scalar, i.e.\ two fermions, see \cite{Fuchs:2007tx}). Among all these lines, those that commute with the $\SU(N)$ currents $j$ are especially interesting, because they would survive the gauging of $\SU(N)$ and remain as (possibly non-invertible) symmetries of the full adjoint gauge theory.

Here, we have to be careful about the possible anomalies between the topological line operators and the $SU(N)$ symmetry that we would like to gauge. For example, in the 1+1d QED case, the axial symmetry has a mixed anomaly with the vector symmetry that we gauge and therefore the continuous axial symmetry eventually disappears. One can detect such an anomaly as follows.
Consider a (possibly-non-invertible) topological line $L$ on a closed curve $C_1$ in the spacetime and then move the line $L$ from the curve $C_1$ to the another curve $C_2$, sweeping the middle region $D$, $\partial D = C_1\sqcup C_2$. When $D$ is flat and we do not have any background gauge field, the expectation value does not change from the assumption that $L$ is topological.
\cite{Chang:2018iay} pointed out that on a curved space $L$ can fail to be truly topological and rather the correlation function gets the anomalous contribution
\begin{equation}
    \exp\left[ i \alpha_L\int_D \mathrm{d}^2\sigma\sqrt{g}R(g) \right],
\end{equation}
where $R(g)$ is the scalar curvature of the metric $g$ and $\alpha_L$ is a number depending on $L$.
Similarly, if we turn on a  background gauge field $A$, there might be an anomalous contribution
\begin{equation}
    \exp\left[ i \alpha_L'\int_D f_2(A) \right],
    \label{eq:line_anomaly}
\end{equation}
where $f_2(A)$ is a gauge-invariant quantity made of the background $A$ that can be integrated over a 2-dimensional surface.
For the axial symmetry in the QED case, $f_2(A)$ is simply proportional to the field strength $F$. Although we can remedy this anomaly by modifying $L$ by the counter term proportional to $A$ before gauging, after gauging the vector $\U(1)$ symmetry, the line becomes non-topological anyway. More generally, non-trivial $f_2$ exists only when the gauge group $G$ is not simply-connected.
Otherwise the anomalous phase \eqref{eq:line_anomaly} is always exact and does no harm.\footnote{Recall that for $\U(1)$ case $F$ is locally exact: $F = \mathrm{d}A$, but globally can carry a magnetic flux through the surface. This is precisely because of the nontrivial $\pi_1(\U(1))$. When $\pi_1(G)$ is discrete there is a generalized Stiefel-Whitney class $w_2$  valued in $\pi_1(G)$ which can play the same role as $F$ in $\U(1)$ case.} 
Therefore, for the case when $G=\SU(N)$ we do not have to worry about this type of anomaly.
On the other hand, we will see that indeed when $G=\mathrm{PSU}(N)$ some of the topological lines are broken. 
This anomaly in $\mathrm{PSU}$ gauge theory is also closely related to the mixed anomaly between some topological lines and the one-form symmetry. We will come back to this point soon.

In this paper, we employ the general theory of topological lines in RCFTs quite extensively to constrain the topological lines of adjoint QCD.\footnote{See  \cite{Petkova:2000ip,Fuchs:2002cm, Bachas:2004sy,Frohlich:2006ch,Alekseev:2007in,Bachas:2009mc} for some references on the subject of topological lines in RCFTs.}
Since adjoint QCD can be viewed as the coset $\mathrm{Spin}(N^2-1)_1/\SU(N)_N$ with a kinetic term for the $\SU(N)$ gauge fields, we can begin by studying the topological lines of $\mathrm{Spin}(N^2-1)_1/\SU(N)_N$. Those are guaranteed to exist also at finite $g_\mathrm{YM}$ since they are the same as the topological lines of the free fermion theory which commute with the $\SU(N)$ currents.  

Let us make some elementary remarks about the theory of $N^2-1$ free fermions. Using the conformal embedding $\hat{\mathfrak{su}}(N)_N\subset \hat{\mathfrak{so}}(N^2-1)_1$, we can regard the $N^2-1$ fermions as a non-diagonal RCFT of the $\hat{\mathfrak{su}}(N)_N$ affine algebra.
For example, when $N=3$, the partition function on the torus of 8 Majorana fermions in the adjoint representation summed over all spin structures can be written as
\begin{equation}\label{SU3f}
    \begin{split}
        Z_{\text{8 fermions}}[T^2]
        =&    \abs{\chi_{(0,0)}+\chi_{(3,0)}+\chi_{(0,3)}}^2 + 3 \abs{\chi_{(1,1)}}^2 \,,
    \end{split}
\end{equation}
where $\chi_{\lambda}$ are the $\hat{\mathfrak{su}}(3)_3$ characters corresponding to integrable representations whose finite Dynkin label is $\lambda$.
This decomposition can be obtained from the branching rules of the representations under the conformal embedding $\hat{\mathfrak{su}}(3)_3\subset\hat{\mathfrak{spin}}(8)_1$ as we will see in section \ref{sec:IR TQFT}. The theory of 8 fermions with a sum over the spin structure is therefore rational under the  $\hat{\mathfrak{su}}(3)_3$ chiral algebra since its partition function is a finite sum of irreducible characters. But importantly, it is a non-diagonal theory (unlike ordinary WZW models).

There are numerous works on the structure of non-diagonal RCFT (for reviews see \cite{Moore:1988ss,DiFrancesco:1997nk}). 
The main statement is that any non-diagonal RCFT $\mathcal{T}_A$ over the characters of $\hat{\mathfrak{g}}_k$  can be obtained by gauging, in the generalized sense, a ``subpart'' $A$ of the Verlinde lines $\mathcal{C}$ in the diagonal $G_k$ WZW model \cite{Kirillov:2001ti,Fuchs:2002cm,Frohlich:2009gb}.\footnote{
    The precise mathematical notion of ``non-anomalous subpart'' is called a symmetric Frobenius algebra object in $\Rep\hat{\mathfrak{g}}_k$. More precisely the Morita equivalence class of it is in one-to-one correspondence to an RCFT. We will explain this briefly in \ref{app:gauging}.
}
The $T^2$ partition function of the RCFT $\mathcal{T}_A$ looks like
\begin{equation}
    Z_{\mathcal{T}_A}[T^2] = \sum_{\mu,\bar{\mu}} \mim_{\mu\bar{\mu}}^A \chi_{\mu}\overline{\chi}_{\bar{\mu}}~,
\end{equation}
with some nonnegative integer coefficients $\mim^A_{\mu\bar{\mu}}$, known as the \emph{modular invariant matrix} of the theory.
Since we have gauged the subpart $A$ of the Verlinde lines $\mathcal{C}$ in the diagonal $G_k$ WZW model, the symmetries of the non-diagonal theory $\mathcal{T}_A$ need to be determined. As usual, some new symmetries appear due to the gauging of $A$. We denote the fusion category of the $\hat{\mathfrak{g}}_k$-preserving topological lines in $\mathcal{T}_A$ as $\aca$.
Many properties of the topological lines will be directly extracted from the coefficients $\mim^A_{\mu\bar{\mu}}$.  In this paper the case of interest is when $\mathcal{T}_A$ is the free fermion theory with gauged $(-1)^F$ where the fermion transforms in the adjoint representation of $G = \SU(N)$. We have seen in~\eqref{SU3f} how for the case of 8 fermions in the adjoint representation of $\SU(3)$ they can be viewed as a non-diagonal RCFT with current algebra $\hat{\mathfrak{su}}(3)_3$. In that particular case it is easy to understand which of the Verlinde lines of the diagonal $\SU(3)_3$ WZW model were gauged. Due to the equivalence $\mathrm{PSU}(3)_1\simeq \Spin(8)_1$, the non-diagonal model can be viewed as a quotient of the diagonal $\SU(3)_3$ model by the ordinary $\mathbb{Z}_3$ center symmetry acting on $\SU(3)$ group elements. So the case of $8$ fermions is special in that the symmetry we gauge is actually an ordinary, invertible, symmetry. This is not the case for higher $N$. 

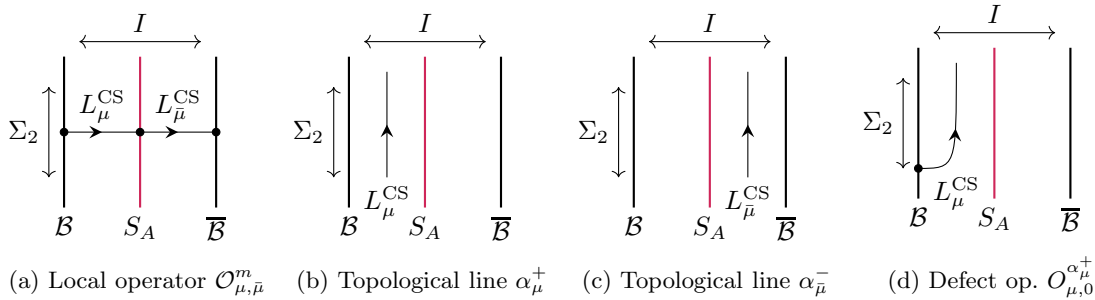
\begin{figure}[t]
    \centering
    \begin{minipage}[b]{.24\linewidth}
        \centering
        \begin{tikzpicture}
            \draw[thick] (0,0) coordinate(ul) -- ++ (0,-2) coordinate(dl);
            \draw[thick] (ul) ++ (2,0) coordinate(ur) -- ++ (0,-2) coordinate (dr);
            \draw[thick, xred] (ul) ++ (1,0) coordinate(um) -- ++ (0,-2) coordinate (dm);
            \node[anchor = north] at (dl) {$\mathcal{B}$};
            \node[anchor = north] at (dr) {$\overline{\mathcal{B}}$};
            \node[anchor = north] at (dm) {$S_A$};
            \draw (ul) -- ++ (0,-1) coordinate(ol);
            \draw (ol)  ++(2,0) coordinate (or);
            \draw[->-=.5 rotate 0] (ol) -- node[midway,anchor=south] {$L^\text{CS}_\mu$} ++(1,0) coordinate (om);
            \draw[->-=.5 rotate 0] (om) -- node[midway,anchor=south] {$L^\text{CS}_{\bar{\mu}}$} ++(1,0) coordinate (or);
            \node[anchor = west] at (or) {\phantom{$\mathcal{O}_{\mu,{\bar{\mu}}}$}};
            \draw[<->] (ul) ++ (-.2,-.4) -- node[midway,anchor = east] {$\Sigma_2$} ++ (0,-1.2);
            \draw[fill] (ol) circle [radius = .05];
            \draw[fill] (om) circle [radius = .05];
            \draw[fill] (or) circle [radius = .05];
            \draw[<->] (ul) ++ (.2,.2) -- node[midway,anchor = south] {$I$} ++ (1.6,0);
        \end{tikzpicture}
        \subcaption{Local operator $\mathcal{O}_{\mu,{\bar{\mu}}}^m$} \label{fig:KSOp}
    \end{minipage}
    \begin{minipage}[b]{.24\linewidth}
        \centering
        \begin{tikzpicture}
            \draw[thick] (0,0) coordinate(ul) -- ++ (0,-2) coordinate(dl);
            \draw[thick] (ul) ++ (2,0) coordinate(ur) -- ++ (0,-2) coordinate (dr);
            \draw[thick, xred] (ul) ++ (1,0) coordinate(um) -- ++ (0,-2) coordinate (dm);
            \node[anchor = north] at (dl) {$\mathcal{B}$};
            \node[anchor = north] at (dr) {$\overline{\mathcal{B}}$};
            \node[anchor = north] at (dm) {$S_A$};
            \draw (ul)  ++ (.5,-.2) coordinate(Lu);
            \draw (Lu)  ++(0,-1.4) coordinate (Ld);
            \draw[->- = .5 rotate 0] (Ld) -- (Lu);
            \node[anchor = north] at (Ld) {$L^\text{CS}_\mu$};
            \node[anchor = west] at (or) {\phantom{$\mathcal{O}_{\mu,{\bar{\mu}}}$}};
            \draw[<->] (ul) ++ (-.2,-.4) -- node[midway,anchor = east] {$\Sigma_2$} ++ (0,-1.2);
            \draw[<->] (ul) ++ (.2,.2) -- node[midway,anchor = south] {$I$} ++ (1.6,0);
        \end{tikzpicture}
        \subcaption{Topological line $\alpha^+_\mu$} \label{fig:KSAlphaP}
    \end{minipage}
    \begin{minipage}[b]{.24\linewidth}
        \centering
        \begin{tikzpicture}
            \draw[thick] (0,0) coordinate(ul) -- ++ (0,-2) coordinate(dl);
            \draw[thick] (ul) ++ (2,0) coordinate(ur) -- ++ (0,-2) coordinate (dr);
            \draw[thick, xred] (ul) ++ (1,0) coordinate(um) -- ++ (0,-2) coordinate (dm);
            \node[anchor = north] at (dl) {$\mathcal{B}$};
            \node[anchor = north] at (dr) {$\overline{\mathcal{B}}$};
            \node[anchor = north] at (dm) {$S_A$};
            \draw (ul)  ++ (1.5,-.2) coordinate(Lu);
            \draw (Lu)  ++(0,-1.4) coordinate (Ld);
            \draw[->- = .5 rotate 0] (Ld) -- (Lu);
            \node[anchor = north] at (Ld) {$L^\text{CS}_{\bar{\mu}}$};
            \node[anchor = west] at (or) {\phantom{$\mathcal{O}_{\mu,{\bar{\mu}}}$}};
            \draw[<->] (ul) ++ (-.2,-.4) -- node[midway,anchor = east] {$\Sigma_2$} ++ (0,-1.2);
            \draw[<->] (ul) ++ (.2,.2) -- node[midway,anchor = south] {$I$} ++ (1.6,0);
        \end{tikzpicture}
        \subcaption{Topological line $\alpha^-_{\bar{\mu}}$} \label{fig:KSAlphaM}
    \end{minipage}
    \begin{minipage}[b]{.24\linewidth}
        \centering
        \begin{tikzpicture}
            \draw[thick] (0,0) coordinate(ul) -- ++ (0,-2) coordinate(dl);
            \draw[thick] (ul) ++ (2,0) coordinate(ur) -- ++ (0,-2) coordinate (dr);
            \draw[thick, xred] (ul) ++ (1,0) coordinate(um) -- ++ (0,-2) coordinate (dm);
            \node[anchor = north] at (dl) {$\mathcal{B}$};
            \node[anchor = north] at (dr) {$\overline{\mathcal{B}}$};
            \node[anchor = north] at (dm) {$S_A$};
            \draw (ul)  ++ (.5,-.2) coordinate(Lu);
            \draw (Lu)  ++(-.5,-1.4) coordinate (Ld);
            \draw[->- = .5 rotate 0] (Ld) .. controls ++(.5,0) .. (Lu);
            \draw[fill] (Ld) circle[radius = .05];
            \node[anchor = north] at ($(Ld)+(.5,0)$) {$L^\text{CS}_\mu$};
            \node[anchor = west] at (or) {\phantom{$\mathcal{O}_{\mu,{\bar{\mu}}}$}};
            \draw[<->] (ul) ++ (-.2,-.4) -- node[midway,anchor = east] {$\Sigma_2$} ++ (0,-1.2);
            \draw[<->] (ul) ++ (.2,.2) -- node[midway,anchor = south] {$I$} ++ (1.6,0);
        \end{tikzpicture}
        \subcaption{Defect op.\ $O^{\alpha^+_\mu}_{\mu,0}$} \label{fig:KSOAlpha}
    \end{minipage}
    \caption{
    The RCFT $\mathcal{T}_A$ can be obtained by considering the 2+1-d CS theory on $I\times \Sigma_2$ as in Fig.~\ref{fig:CSWZW} with the surface operator $S_A$ inserted in the middle of the interval $I$.
    (a): a local operator $\mathcal{O}^m_{\mu,{\bar{\mu}}}$ in the Verma module $\module_\mu\otimes \overline{\module}_{\bar{\mu}}$ can be realized as the Wilson lines $L^\text{CS}_\mu$ and $L^\text{CS}_{\bar{\mu}}$ connected at $S_A$. The number of possible ways of such connections determines the multiplicity of the Verma module in the total Hilbert space of $\mathcal{T}_A$.
    (b): The bulk Wilson line $L^\text{CS}_\mu$ along the 2d spacetime $\Sigma_2$ and to the left of $S_A$ defines a 2d topological line $\alpha^+(L_\mu)$ in $\mathcal{T}_A$.
    (c): The bulk Wilson line $L^\text{CS}_\mu$ along the 2d spacetime $\Sigma_2$ and to the right of $S_A$ defines another 2d topological line $\alpha^-(L_{\bar{\mu}})$ in $\mathcal{T}_A$.
    (d): A defect operator at the edge of the line $\alpha^+_\mu$ can be constructed by ending the bulk Wilson line $L^\mathrm{CS}_\mu$ on the left boundary $\mathcal{B}$. The resulting defect operator $O^{\alpha^+_\mu}_{\mu,0}$ is in the Verma module $\module_\mu \otimes \overline{\module}_0$.}
    \label{fig:KS}
\end{figure}

We can regard $\mathcal{T}_A$ as a CS theory compactified on an interval, with a (2-dimensional) surface operator $S_A$ inserted in the middle of the interval (see Fig.~\ref{fig:KS}) \cite{Kapustin:2010if,Fuchs:2012dt,Carqueville:2017ono}.
The construction of the surface operator $S_A$, given a ``non-anomalous subpart" $A$ of $\Rep\hat{\mathfrak{g}}_k$, is obtained in \cite{Carqueville:2017ono}.
An operator $\mathcal{O}_{\mu, \bar{\mu}}^m$ in the Verma module $\module_\mu \otimes \overline{\module}_{\bar{\mu}}$ (where $m=1,\dots,\mim^A_{\mu\bar{\mu}}$) can be constructed as in Fig.~\ref{fig:KSOp}.
Here, the bulk Wilson lines $L^\text{CS}_\mu$ and $L^\text{CS}_{\bar{\mu}}$ are connected at the surface operator $S_A$.
We can further construct a subset of the topological lines $\aca$ in $\mathcal{T}_A$ by placing the bulk Wilson lines on the left, or on the right of $S_A$, as depicted in Fig.~\ref{fig:KSAlphaP} and Fig.~\ref{fig:KSAlphaM}.
These operations define maps, or tensor functors, between fusion categories:
\begin{equation}
    \alpha^\pm: \mathcal{C} \to \aca ~,
\end{equation}
which are called $\alpha$-inductions \cite{Longo:1994xe,Bockenhauer:1998ca, Bockenhauer:1998in, Bockenhauer:1998ef,Bockenhauer:1999wt, ostrik2003module}.
These functors will be one of the main tools to analyse the topological lines in adjoint QCD.
We abbreviate the image $\alpha^\pm(L_\mu)$ of the Verlinde line $L_\mu$ under these functors as $\alpha^\pm_\mu$.

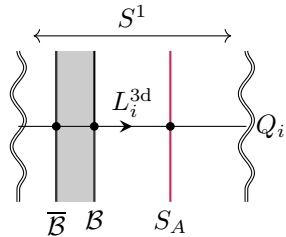
\begin{figure}[t]
    \centering
    \begin{tikzpicture}
        \draw[thick] (0,0) coordinate(ub) -- ++ (0,-2) coordinate(db);
        \draw[thick] (ub) ++ (-.5,0) coordinate(ubb) -- ++ (0,-2) coordinate (dbb);
        \draw[draw = none, fill = gray, opacity = 0.4] (ub) -- (ubb) -- (dbb) -- (db) -- cycle;
        \draw[thick, xred] (ub) ++ (1,0) coordinate(us) -- ++ (0,-2) coordinate (ds);
        \node[anchor = north] at (db) {$\mathcal{B}$};
        \node[anchor = north] at (dbb) {$\overline{\mathcal{B}}$};
        \node[anchor = north] at (ds) {$S_A$};
        \draw (ubb) ++ (-.5,0) coordinate (uls);
        \draw (us) ++ (1,0) coordinate (urs);
        \draw[double, snake=snake, segment length=7mm] (uls) -- ++(0,-2) coordinate (dls);
        \draw[double, snake=snake, segment length=7mm] (urs) -- ++(0,-2) coordinate (drs);
        \draw (ul) -- ++ (0,-1) coordinate(ol);
        \draw (ol)  ++(2,0) coordinate (or);
        \draw[<->] (uls) ++ (.2,.2) -- node[midway,anchor = south] {$S^1$} ++ (2.6,0);
        \draw[->-=.5 rotate 0] (uls) ++ (0,-1) coordinate (Ol)  -- ++ (.5,0) coordinate(Ob) --++ (.5,0) coordinate(Obb)  -- node[midway,anchor=south] {$L^\mathrm{3d}_i$} ++(1,0)  coordinate(Os) -- ++(1,0) coordinate(Or);
        \node[anchor = west] at (Or) {$\oneform_i$};
        \draw[fill] (Ob) circle[radius = .05];
        \draw[fill] (Obb) circle[radius = .05];
        \draw[fill] (Os) circle[radius = .05];
    \end{tikzpicture}
    \caption{1+1d adjoint QCD from coupled 2+1d systems on $S^1$.
    The left wiggly line and the right wiggly line are identified.
    The white region is filled with the $\SU(N)_N$ CS theory, while the gray region is filled with the $\SU(N)_0$ Yang-Mills theory with finite YM coupling without CS term.
    In the CS theory region we have the topological surface operator $S_A$. On the interfaces $\mathcal{B}$ and $\overline{\mathcal{B}}$ between these theories, left- and right-moving fermions reside.
    To obtain the 1+1d adjoint QCD, we take the limit where the width of the gray region becomes zero.
    The one-form symmetry generator $\oneform_1$ in 1+1d adjoint QCD is realized by the topological line operator $L^\mathrm{3d}_i$ realizing 2+1d one-form symmetry wrapping $S^1$.
    In the CS region, $L^\mathrm{3d}_1$ coincide with $L^\text{CS}_\mu$ with highest weight $\mu = (0,\dots,0,N)$.}
    \label{fig:AdjQCDCS}
\end{figure}

Adjoint QCD in 1+1 dimensions can be thus engineered by further gluing the CS boundaries $\mathcal{B}$ and $\overline{\mathcal{B}}$
with $\SU(N)_0$ gauge fields, as illustrated in Fig.~\ref{fig:AdjQCDCS}.
From the picture, it is clear that the topological lines $\aca$ also exists in adjoint QCD.
Since our results are derived using the properties of the symmetries of the massless adjoint theory, they are in fact valid not just for the theory with the canonical kinetic term for the gauge field, but we can take any function of $F^2$. Additionally, we will see that some quartic fermion interactions preserves an interesting subset of the symmetries $\aca$.

\paragraph{Topological lines and deconfinement}
Now we can explain the relationship between the topological lines in massless adjoint QCD and the IR behavior of the theory.
Because the $\aca$ symmetry is preserved along the RG flow, the IR physics should also admit the same set of lines. Therefore we hope to extract information on the IR physics from the topological lines, as we often do from a group symmetry and its anomaly.
Here, we are interested in the asymptotic behavior -- perimeter law or area law -- of the Wilson line $W_\mu$ of the dynamical gauge field in the representation $\mu$.\footnote{
    In 1+1d, one-form symmetry is always preserved because its symmetry operator is a topological local operator, but a Wilson line still can have either perimeter law or area law.
    Therefore, it is appropriate to define confinement/deconfinement by the behavior of Wilson lines.}\footnote{
Here, the Wilson line $W_\mu$ denotes the one composed of the 1+1d gauge field, and should be distinguished from the Wilson line $L^\text{CS}_\mu$ in the 2+1d CS theory, or its projection $\alpha^\pm_\mu$ onto the 2d theory. In other words $W_\mu$ lives in the gray region in Fig.~\ref{fig:AdjQCDCS} while $L^\text{CS}_\mu$ lives in the white region. In particular, $W_\mu$ is not topological while $L^\text{CS}_\mu$ and $\alpha^\pm_\mu$ are.}

A naive argument for confinement in adjoint QCD is that there is nothing to screen the Wilson line $W_\mathrm{fund}$ in the fundamental representation because any composite of adjoint quarks cannot be in the fundamental representation.
To understand better how the new non-invertible symmetries help settle this question, one has to understand in what way the non-invertible symmetries ``talk'' to the $\mathbb{Z}_N$ one-form symmetry of the model. 
The one-form symmetry is nothing but a collection of local topological operators $\oneform_i$, $i=1, \cdots N$, which generate the $\mathbb{Z}_N$ one-form symmetry, corresponding to the center of the gauge group acting trivially on the quarks.
They can thus transform nontrivially as they pass through the topological lines in $\aca$. From here the argument proceeds very much like in the $\SU(2)$ case. The fundamental Wilson line is always charged under the $\mathbb{Z}_N$ symmetry, \begin{equation}
    \oneform_i W_\mathrm{fund} = e^{\frac{2\pi\mathrm{i}}{N}i} W_\mathrm{fund} \oneform_i~,
\end{equation}
but we do not know whether $W_\mathrm{fund} $ is confined or not and $W_\mathrm{fund} $ is not in general a topological line. As in the $\SU(2)$ case, the idea is to find a topological line which is {\it also} charged under $\oneform_i$. 
The key observation is that the topological line $\alpha^+_{\mathrm{fund}}$ is charged under the $\mathbb{Z}_N$ one-form symmetry
 \begin{equation}
    \oneform_i \alpha_\mathrm{fund}^\pm = e^{\frac{2\pi\mathrm{i}}{N}i} \alpha_\mathrm{fund}^\pm \oneform_i~.
    \label{eq:alphaCharge}
\end{equation}

\begin{figure}[ht]
    \centering
    \begin{tikzpicture}[baseline = -15]
        \draw[->- = .4 rotate 180] (ul) -- ++(2.4,0) coordinate (ur) -- ++(0,-1) coordinate (dr) -- node[midway,anchor=north] {$W_\mathrm{fund}$} ++(-2.4,0) coordinate (dl) -- cycle;
    \end{tikzpicture}
    \hspace{3mm}
    $\Rightarrow$
    \hspace{3mm}
    \begin{tikzpicture}[baseline = -15]
        \draw[->- = .28 rotate 180] (ul) -- ++(.5,0) coordinate (uml) ++(1.4,0) coordinate(umr) -- ++(.5,0) coordinate(ur) -- ++(0,-1) coordinate (dr) -- node[midway,anchor=north] {$W_\mathrm{fund}$} ++(-2.4,0) coordinate (dl) -- (ul);
        \draw[fill] (uml) circle[radius = .05];
        \draw[fill] (umr) circle[radius = .05];
        \draw[->- = .5 rotate 0, dashed] (umr) -- node[midway,anchor=south] {$\alpha^+_{\mathrm{fund}}$} (uml);
        \node[anchor = north] at (uml) {$\mathcal{O}_\text{LC}^*$};
        \node[anchor = north] at (umr) {$\mathcal{O}_\text{LC}$};
    \end{tikzpicture}
    \hspace{3mm}
    $=$
    \hspace{3mm}
    \begin{tikzpicture}[baseline = -15]
        \draw[->- = .28 rotate 180] (ul) -- ++(.5,0) coordinate (uml) ++(1.4,0) coordinate(umr) -- ++(.5,0) coordinate(ur) -- ++(0,-1) coordinate (dr) -- node[midway,anchor=north] {$W_\mathrm{fund}$} ++(-2.4,0) coordinate (dl) -- (ul);
        \draw[fill] (uml) circle[radius = .05];
        \draw[fill] (umr) circle[radius = .05];
        \draw[->- = .5 rotate 0, dashed] (umr) .. controls ++(-.7,0) and ++(.7,0) .. ++(-.7,.5) node[anchor=south] {$\alpha_\mathrm{fund}^+$} .. controls ++(-.7,0) and ++(.7,0) .. (uml);
      \node[anchor = north] at (uml) {$\mathcal{O}_\text{LC}^*$};
     \node[anchor = north] at (umr) {$\mathcal{O}_\text{LC}$};
    \end{tikzpicture}
    \caption{A segment in a Wilson loop $W_\mathrm{fund}$ can be replaced by the topological line $\alpha^+_\mathrm{fund}$ using the line-changing operator $\mathcal{O}_\text{LC} = \mathcal{O}^{\alpha^+_{\mathrm{fund}}}_{\mathrm{fund},0}$ and its conjugate operator $\mathcal{O}_\text{LC}^*$.
    The replacement only causes an effect proportional to the length of the replacement.
    After the replacement, we can arbitrarily deform the topological line $\alpha^+_\mathrm{fund}$ without changing its value. This is why the Wilson line has perimeter law -- it is the existence of the generally non-invertible topological line $\alpha^+_{\mathrm{fund}}$. }
    \label{fig:WDeform}
\end{figure}
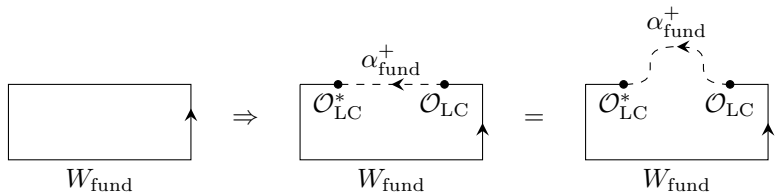

This can be deduced from the 2+1d viewpoint in Fig.~\ref{fig:AdjQCDCS}.
The 2+1d set up contains two theories, the $\SU(N)_N$ CS theory and $\SU(N)$ YM theory, both having $\mathbb{Z}_N$ one-form symmetry. Furthermore, through the interfaces $\mathcal{B}$ and $\overline{\mathcal{B}}$ these two one-form symmetries are identified.
The one-form symmetry in 1+1d just comes from this one-form symmetry in 2+1d. 
The one-form symmetry (local) operator $\oneform_i$ in 1+1d is obtained by wrapping the one-form symmetry (line) operator $L^\mathrm{3d}_i$ on $S^1$, going through the interfaces $\mathcal{B}$, $\overline{\mathcal{B}}$ and the surface $S_A$.
The equation \eqref{eq:alphaCharge} comes from the fact that the one-form symmetry charge of the topological line $L^\text{CS}_\mu$ in the CS theory is the $N$-ality of the representation $\mu$.
Hence by acting with $\alpha^+_{\mathrm{fund}}$ on the vacuum we create a string state. But since $\alpha^+_{\mathrm{fund}}$ is topological,\footnote{Once again, the subscript fund $\alpha^+_{\mathrm{fund}}$ should not be confused with the fundamental Wilson line in the 1+1 dimensional gauge theory -- rather it stands for the representation of the Verlinde line in $\SU(N)_N$ which is mapped by $\alpha$-induction to our non-diagonal coset.} the string tension has to vanish and hence we have a perimeter law. This picture allows us to also identify a little more concretely the object that is capable of screening the fundamental Wilson line. Indeed, the defect operator which lives at the edge of topological line $\alpha^+_{\mathrm{fund}}$, $\mathcal{O}^\mathcal{\alpha^+_\mathrm{fund}}_{\mathrm{fund},0}$, is in the fundamental representation of the gauge group $\SU(N)$, as depicted in Fig.~\ref{fig:KSOAlpha}.
Therefore, roughly speaking this defect operator can screen the fundamental Wilson line. Since it is in the fundamental representation of $\SU(N)$ we can attach it to the non-topological Wilson line and then on the other side to $\alpha^+_{\mathrm{fund}}$, which is a topological line. This allows to break the string into ``topological matter'' and hence one finds perimeter law. See Fig.~\ref{fig:WDeform}. In the case of $\SU(2)$ gauge theory this argument is identical to the one we have already mentioned, where $(-1)^{F_L}$ (or its dual symmetry upon gauging fermion number) creates a string from the vacuum, but on the other hand, it is a symmetry. 

Later we will give a more formal argument for the perimeter law of the fundamental Wilson line. There is also a less formal argument that explains all of the above in simple classical terms. While this argument is inaccurate and somewhat heuristic, it is worth mentioning it as it gives invaluable intuition to why many massless 2d models deconfine. To demonstrate the idea, start with a WZW model where the basic variable is the group element $g \in \SU(N)$. This transforms under $\SU(N)_\mathrm{L}\times \SU(N)_\mathrm{R}$ global symmetry as usual. Consider the current operator $J \sim g^{-1} \partial g$. If we study how it transforms under local diagonal $\SU(N)$ transformations we find that $J$ transforms like a one-form: 
\begin{equation}
J\to \Omega J \Omega^{-1} +i\;\Omega^{-1}\partial \Omega~.
\end{equation}
Therefore, the WZW model admits some extended operators which are invariant under local diagonal $\SU(N)$ transformations \cite{Bachas:2004sy,Alekseev:2007in,Bachas:2009mc}: 
\begin{equation}\label{Vclass} \Tr_\mathcal{I} P e^{i\int J}~,\end{equation}
where $\mathcal{I}$ is an arbitrary representation. 
Clearly these are just the classical analogs of Verlinde lines.

The main observation is that by the equations of motion $J$ is a flat connection and hence these extended operators are all topological. The operators~\eqref{Vclass} would survive gauging the diagonal $\SU(N)$, though $J$ would no longer be flat anymore. However, ignoring this for a moment, we see clearly that we can insert segments with the operators~\eqref{Vclass} inside ordinary Wilson lines made out of the dynamical gauge field. So our Wilson line, in any representation, can morph into the topological operators. This can be also seen from the equations of motion -- integrating out the gauge field we obtain $A\sim g^{-1} \partial g +\cdots$.  One thing we have not explained is why the operators~\eqref{Vclass} can be fixed in the theory with gauged $\SU(N)$ such that they still remain topological. Another thing we glossed over is that the operators could receive quantum corrections due to the need to normal order these extended operators.
Both issues are rigorously addressed using the fusion category tools we use in this paper. But the argument showing that the ordinary Wilson line (in any representation) can morph into the composite topological operator~\eqref{Vclass} is nevertheless appealing and explains why deconfinement takes place in many massless two-dimensional theories. It essentially boils down to the fact that there is left-right decoupling before the diagonal symmetry is gauged.

\paragraph{More on one-form symmetry in 1+1 dimensions}
We have thus far given a fairly extensive introduction to the theory of non-invertible lines and explained how we are going to use them to establish deconfinement.
Here we would like to talk more about one-form symmetry in 1+1 dimensions, which leads to some peculiarities that will be important later. 

\medskip 

\begin{itemize}
\item One-form symmetry (such as the $\mathbb{Z}_N$ one-form symmetry in Abelian gauge theory with a fermion of charge $N$ or $\SU(N)$ gauge theory with an adjoint fermion) implies the existence of local operators which are topological: $\oneform_i(x)$, with $i=1,..,N$. Since they generate a $\mathbb{Z}_N$ symmetry we can take $\oneform_N=1$ and $\oneform_1^k=\oneform_k$.
It may not be easy to write them using the fundamental fields of the theory, but they nevertheless exist. There must exist a line operator in the theory which obeys the following algebra \begin{equation}\label{alg}\oneform_1 W = e^{2\pi i / N }W \oneform_1~. \end{equation}
Of course, in general, there could be many such lines, and $W$ is not required to be topological. In gauge theories, we can choose $W$ to be the fundamental Wilson line. In the context of fusion categories, which is the simplest formalism that allows to discuss non-invertible topological lines, one  assumes that local topological operators are absent. However, including them is not that difficult and we summarize the main modifications that this leads to below. Mathematically, one obtains a multi-fusion category in this way.

\item It will be convenient to define the notion of a ``universe'': Sometimes in the infinite volume limit of QFT we encounter degenerate super-selection sectors. These could be due to a spontaneously broken symmetry, for instance.
For a usual super-selection sectors, we can separate them by domain walls which have finite tension. We will use the notion of universes  for superselection sectors which are not separated by finite tension domain walls. Such superselection sectors are peculiar because they would not mix in compact space. This notion of ``universes'' as  superselection sectors that are not separated by finite tension domain walls has also been recently emphasized in~\cite{Aminov:2019hwg,Sharpe:2019ddn,Tanizaki:2019rbk} (see also an older exposition \cite{Hellerman:2006zs}).

\item A lemma that is rather easy to prove is that two Poincar\'e invariant vacua with different expectation values of $\oneform_1$ must belong to different universes -- i.e.\ there cannot be finite mass domain walls between these vacua (regardless of whether the energy density in these vacua is the same or not).
The proof is to imagine a finite energy configuration which interpolates between the two states. Then $\langle \oneform_1(x) \rangle$ must be a nontrivial function of position contradicting it being a topological operator. Therefore such a finite energy domain wall cannot exist. This is why if we have a $\mathbb{Z}_N$ one-form symmetry it is generic to develop universes. But is it necessary that at least $N$ universes exist?

\item It is in fact true that (at least) $N$ universe must exist. Take the infinite space-like line $W$ and act on the vacuum. This must be a new universe due to~\eqref{alg}. Indeed the expectation values of $\oneform_1$ must abruptly jump. (It cannot be that $\oneform_1=0$ either before or after the line since $\oneform_1^N=1$.) These universes can be thought of as the QFTs that live on the flux tubes which $W$ creates from the vacuum. Of course, the energy density of these flux tubes is not determined from these general arguments and we will see that it may have a complicated behavior as a function of the parameters. Can there be more than $N$ universes? More precisely, can there be more universes than the number of elements in the one-form symmetry group? We expect the answer is negative except in theories without propagating degrees of freedom.

\item Given line operators that interpolate between different universes, the question of whether such lines are confined or deconfined should be interpreted as the question of whether a large rectangular such line has area or perimeter law. This is the same as comparing the energy densities of the two universes.\footnote{In principle, the breaking of one-form symmetry can have two facets. One is that Wilson lines have perimeter law and another is that the theory compactified on $S^1$ admits a broken ordinary Abelian zero-form symmetry. These two definitions do not necessarily coincide. Indeed, in 1+1 dimensions the second definition does not make sense because zero-form symmetries cannot be broken in quantum mechanics (instead, the ground state could be degenerate in a linear representation of the zero-form symmetry).
A similar situation can be found in systems with a discrete $d$-form symmetry in $d+1$-dimensions, see~\cite{Tanizaki:2019rbk}.
}

\item Consider a theory with $\mathbb{Z}_N$ one-form symmetry compactified on the torus. We can tinker with how the universes are summed up in the partition function of the theory by adding background fields for the one-form symmetry. This is done by coupling the theory to a discrete $\mathbb{Z}_N$ two-form gauge field 
\begin{equation}
    B\in H^2(X,\mathbb{Z}_N)~,
\end{equation}
and imposing $\int_X B = k \mod N$. Such background fields preserve two-dimensional Poincar\'e invariance.
The way this affects the partition function is as follows. Inserting the $B$ flux is equivalent to inserting a co-dimension 2 defect topological operator, i.e.\ some power of $\oneform_1$. That means that different universes are now weighted with different phases 
\begin{equation}\label{Usum} Z=\sum_{p=0}^{N-1} e^{2\pi i pk / N} Z_p~,\end{equation}
where $Z_p$ is the partition function in the universe $p$.

\item If we were to gauge the one-form symmetry, naively, we would need to sum over $k$. The partition function of the theory with the gauged one form symmetry is therefore $Z_0$. In other words, all the universes except one are projected out. 
This is however imprecise. There is not a unique way to gauge the one-form symmetry. Indeed, we can modify the action by the counter-term $-\frac{2\pi p_0}{N} \int B$, where $p_0$ is an arbitrary integer in $\mathbb{Z}_N$ which we are free to choose. 
If we do not sum over the $B$ fluxes this would be tantamount to multiplying the partition function by a phase. So~\eqref{Usum} becomes
\begin{equation}Z=\sum_{p=0}^{N-1} e^{2\pi i (p-p_0)k / N} Z_p~.\end{equation}
Now gauging the one form symmetry means that we need to sum over $k$ again and hence we project to the universe $Z_{p_0}$.

\item To summarize: when we gauge the one form symmetry, a new discrete theta angle emerges. Computing the partition function with some fixed discrete theta angle is tantamount to projecting on one of the $N$ universes.\footnote{In fact, the theta angle of the Abelian theory~\eqref{pure} which we analyze in the next section can be viewed in precisely this way. We could have started with the gauge group $\mathbb{R}$ which has one-form symmetry $\mathbb{R}$ and no theta angle (since $\int F_{01}$ vanishes on any compact manifold). Then gauging the subgroup $\mathbb{Z}$ of the one-form symmetry we arrive at a $\U(1)$ gauge theory which has a theta angle valued in $[0,2\pi)$.  The theta angle can be thought of as putting projective representations of the gauge group at infinity. In our $\U(1)$ example~\eqref{pure}, this corresponds to putting charge $\theta/2\pi$ at infinity.  In $\SU(N)/ \mathbb{Z}_N$ gauge theory the discrete theta angle corresponds to a projective representation of $\SU(N)/ \mathbb{Z}_N$~\cite{Witten:1978ka}.} This allows us a new point of view on the question of confinement or deconfinement. Instead of talking about large rectangular line operators, we can compare the partition functions directly: $Z_p/Z_0$. This can be viewed as the ratio between the partition function of the theory with gauged one form symmetry with a discrete theta angle $p$ and the one without a discrete theta angle. If this ratio is exponentially small in the volume of space then the corresponding line is confined.
\end{itemize}

\medskip

In the context of massless adjoint QCD, we have argued above that one has exponentially many degenerate ground states as a result of the non-invertible symmetries. We will have to classify these ground states into different universes according to the expectation value of the one-form symmetry local operator $\oneform_1$. 
From this we will learn which vacua are separated by finite mass kinks and which are not. 

\paragraph{Quartic deformation}

As soon as we add a mass for the adjoint quark, all the non-invertible topological lines disappear and we have confinement of the fundamental Wilson line. It is however noteworthy that the theory has three quartic fermion interactions. All of them preserve the chiral $\mathbb{Z}_2$ symmetry of the massless theory and of course also the one-form symmetry: 
\begin{equation}
    \mathcal{O}_1= \Tr (\psi_+\psi_+ \psi_-\psi_-)~,\quad \mathcal{O}_2= \Tr (\psi_+ \psi_-)\Tr (\psi_+ \psi_-)~,\quad \mathcal{O}_3=\Tr(\psi_+ \psi_- \psi_+ \psi_-)~.
\end{equation}
The operator $\mathcal{O}_3$ is odd under charge conjugation so it would not be generated by RG transformations. $\mathcal{O}_1$ can be thought of as a $j\bar j$ in terms of the $\hat{\mathfrak{su}}(N)$ currents and hence it preserves all the non-invertible lines of $\aca$ and will be generated by the RG flow. Adding it to the Lagrangian would not change our discussion about the symmetries or the results about confinement vs deconfinement.
The most interesting case is that of $\mathcal{O}_2$, which is invariant under all the ordinary symmetries but in general breaks some of the non-invertible lines. (A more precise statement is that $ \mathcal{O}_2$ is invariant under all the ordinary symmetries of the fermionic theory.)

It is natural to expect that adding $ \mathcal{O}_2$ to the action we would restore confinement of the fundamental line. In addition, there must be deconfinement of the line $W_\mathrm{fund}^{N/2}$ for even $N$ as follows from the standard symmetries of the system. For $N\leq5$ we find that indeed, as expected, the fundamental line now confines while $W_F^{N/2}$ (for even $N$) does not. It is tempting to conjecture that for all $N$ the fundamental line would confine upon deforming the theory with $\mathcal{O}_2$.

\paragraph{Some concrete results for $\SU(3)$, $\SU(4)$ and $\SU(5)$ adjoint QCD}

In the main text we give a very concrete treatment of the fusion categories of symmetries in $\SU(3)$ and $\SU(4)$ gauge theory and we also analyze the detailed action on various local operators such as the quartic operators above. We also study the vacua of these theories, the universes, and the superselection sectors. 
An explicit analysis becomes harder for larger $N$. But we will still make some conjectures about the large $N$ limit and various other properties of the theory. 

One noteworthy result that we have in fact derived for all $N$ is that there must be some finite nonzero mass for the adjoint quark where massless fermions appear in the string state. We have done this by comparing the fermionic SPT phases at large mass and at vanishing mass using our non-invertible symmetries. This is in agreement with the result of~\cite{Kutasov:1993gq} that for some nonzero finite mass the theory develops $\mathcal{N}=1$ supersymmetry which is spontaneously broken in the string state and hence there is a massless Majorana fermion there. The existence of this massless fermion was recently pointed out in \cite{Dubovsky:2018dlk}.

\paragraph{Exact results about $k$-string tensions}

The theory with a massless adjoint fermion deconfines for the reasons we have briefly explained above.
We can add a mass for the adjoint quarks and ask about the $k$-string tension.
Namely, we ask what is the smallest possible energy density in the universe with $k$ fundamental strings.
Typically there could be many super-selection sectors with $k$ strings, so what we do is to minimize over them.
Physically that is equivalent to asking about the behaviour of a Wilson loop with $k$ boxes in the Young diagram. The minimization over super-selection sectors corresponds to allowing the worldline of the probe quark to be dressed with finite mass kinks, which would happen dynamically anyhow in order to minimize the string tension.
For quark mass much bigger than $g_\mathrm{YM}$ the answer can be derived from pure Yang-Mills theory in 2d, minimizing over the quadratic Casimir.
One finds that the $k$-string tension behaves as 
\begin{equation}
    T_k\sim g_\mathrm{YM}^2 k (N-k)
\end{equation}
up to a $k$-independent constant.
It is always minimized by the fully anti-symmetric representation with $k$ boxes.
At small mass compared to $g_\mathrm{YM}$ the problem is much more difficult seeing as the theory is in the strongly coupled regime.
We use the non-invertible symmetries to compute the tension for small rank gauge groups explicitly.
This involves computing the NIM-rep matrices which we do not know how to do efficiently for all gauge groups $\SU(N)$ so we present explicit results for $N\leq 5$.
Yet, quite miraculously, in all cases the string tension appears to obey the relation  \begin{equation}\label{sten}T_k \sim g_\mathrm{YM}|m| \sin(\pi k /N)~,\end{equation}
where $m$ is the mass of the adjoint quark. The fact that the string tension is proportional to $|m|$ is not surprising, it follows from the fact that the $m=0$ theory is gapped and deconfined. It is tempting to conjecture that~\eqref{sten} holds for all $N$, but at present we do not have a proof. Our computation of the $k$-string tension is possibly the first such exact computation in a strongly coupled theory.

The techniques in this paper can be extended to many other gauge theories in 1+1 dimensions. In particular it would be nice to revisit the model with two adjoint fermions, considered in \cite{Gopakumar:2012gd}.

\subsection{Organization of the Paper}

In section \ref{sec:Vacua vs Universes} we consider the Schwinger model with a dynamical charge $q$. While this model has no non-invertible symmetries, it exhibits the mechanism of complete deconfinement at zero mass through charge fractionalization. This can be understood due to an anomaly involving the chiral and one-form symmetry, hence, the idea is essentially the same as in the adjoint $\SU(2)$ gauge theory. 
In section \ref{sec:review.adjQCD} we review in a little more detail previous literature on adjoint QCD. In section \ref{sec:bos} we review non-Abelian bosonization, focusing on some global aspects of the problem. In section \ref{sec:top.lines} we discuss the notion of topological lines, the gauging of non-invertible lines, and the classification of gapped theories with topological lines. Then, we discuss the topological lines of adjoint QCD and prove deconfinement. In section \ref{sec:deformation} we study the mass and quartic deformations, and compute the $k$-string tensions in the small mass limit. In section \ref{sec:IR TQFT} we discuss the possible infrared phases and vacua of the theory. Our analysis is very explicit for $\SU(2),\SU(3),\SU(4),\SU(5)$ and increasingly less explicit afterwards.

Several appendices cover essential technical computations as well as an extensive review of various results. Appendix \ref{app:FusionCat} contains detailed discussion about fusion categories and theorems stated in section \ref{sec:top.lines}. In appendix \ref{app:ActionOfLines} we discuss the action of topological lines on local operators in RCFTs. We include known explicit expressions for $N=3$ and $4$, and new results for $N=5$. In appendix \ref{app:TQFT} we collect some facts about 2d $G/G$ TQFTs and 3d Chern-Simons theories. In appendix \ref{app:branching} we have worked out the branching rules for the conformal embedding $\hat{\mathfrak{su}}(N)_N \subset \hat{\mathfrak{so}}(N^2-1)_1$. The branching rules can be used to determine the number of vacua in the $\Spin(N^2-1)_1 / \SU(N)_N$ TQFT. In appendix \ref{app:modular.invariants} we list modular invariant partition functions of the known (and some new) non-diagonal $\hat{\mathfrak{su}}(N)_N$ WZW models (the list is complete for $N\leq 4$). These modular invariant partition functions correspond to alternative IR TQFTs that adjoint QCD can flow to, apart from the $\Spin(N^2-1)_1 / \SU(N)_N$ coset. Finally, in appendices \ref{app:fermionic} and \ref{app:fermioniclines} we discuss the fermionic adjoint QCD. In the former, we count the number of vacua in the possible fermionic symmetric TQFT in the IR and in the latter we study the topological lines of the theory, in particular (fermionic) $\SU(3)$ adjoint QCD.

\textit{When this work was being completed, we learned of \cite{Gaiotto:2020iye} which has overlap with section \ref{sec:IR.TQFT}.}
\section{Vacua vs Universes in the Schwinger Model}
\label{sec:Vacua vs Universes}

We will first review the physics of the free photon in two spacetime dimensions. Since the photon in two spacetime dimensions has no propagating degrees of freedom, the theory of the free photon is a little peculiar. We begin with the action
\begin{equation}
\label{pure}
S =\int d^2x \left( -\frac1{4e^2}F_{\mu\nu}^2+\frac{\theta}{2\pi} F_{01}\right)~.
\end{equation}
Let us quantize it on the circle of radius $R$. In the gauge $A_0=0$ we have the constraint $\partial_1F_{01}=0$ which means that the electric field is constant in space. 
This is solved by $A_1(x,t)=G(t)+F(x)$ with $F$ and $G$ arbitrary functions. By a space dependent gauge transformation we can set $F=0$ and we remain with $A_1(x,t)=G(t)$. Even after setting $A_0=0$ and $F=0$ we still have one important remaining gauge transformation with gauge parameter $\Omega=e^{ix/R}$ which is a well defined map from $S^1$ to $\U(1)$ since $\Omega$ does not change if we shift $x\to x+2\pi R \mathbb{Z}$. This implies that $G\equiv G+\frac1{R}\mathbb{Z}$, i.e.\ $G$ is a periodic variable with periodicity $\frac{1}{R}$.
Plugging $A_1(x,t)=G(t)$ into the action and using $F_{01}=\dot G$ (and integrating over the circle) we find a quantum mechanical model for $G(t)$ with action 
\begin{equation}
S =2\pi R \int dt \left( \frac{1}{2e^2}\dot G^2+\frac{\theta}{2\pi} \dot G\right)~.
\end{equation}
This describes an Aharonov-Bohm particle on the circle. The physics depends only on $\theta \mod 2\pi$.
The Hamiltonian is 
\begin{equation}
H = \frac{1}{2} \frac{e^2}{2\pi R} (\Pi_G-\theta R)^2~.
\end{equation}
The eigenstates are $\Psi_n(G) = e^{2\pi inGR}$, with $n\in\mathbb{Z}$. The energy levels are
\begin{equation}\label{Elev}
E_n=\frac{1}{2}\frac{e^2R}{2\pi}\left(2\pi n - \theta\right)^2~.\end{equation}
These states of the QFT~\eqref{pure} on the circle are all translationally invariant (i.e.\ they have no momentum on the circle). The expectation value of the electric field in the state $\Psi_n$ is 
\begin{equation}\label{Efield}\langle n | F_{01} |n \rangle=\langle n | \dot G |n \rangle= e^2\left( n -\frac{ \theta}{2\pi}\right)~. 
\end{equation}
We see that the states of the theory on the circle are simply states with constant electric field (both in time and space) and constant energy density. Indeed from~\eqref{Elev} 
we see that the total energy of the state is proportional to $R$, which is the volume of space. We can ask which state has the lowest energy density.
We can restrict ourselves to $\theta\in[0,2\pi)$ since the spectrum is periodic in $\theta$. For $\theta\in[0,\pi)$ the lowest state is $|0\rangle$, For $\theta\in(\pi,2\pi)$ the lowest state is $|1\rangle$, and for $\theta=\pi$ we have a two-fold degenerate ground state on the circle. 

Under normal circumstances, pure states with constant energy density are expected to evolve in time and dissipate their energy to infinity. For instance, imagine we had propagating massive particles of charge $\pm1$ and set $\theta=0$ for simplicity. Then if we prepared the system in the state with constant electric field $\langle   F_{01}  \rangle= e^2$, the system would create two massive particles from the vacuum with charges $\pm1$, such that the charge -1 would be created to the left of the charge +1, and they would repel each other. In this way, between the two charges the expectation value of the electric field vanishes and the energy is instead converted to kinetic energy of the massive particles that proceed to infinity.   

Therefore it is somewhat of a peculiarity of the theory~\eqref{pure} that it admits such states. Of course, this is due to the absence of charged propagating degrees of freedom. Another related fact is that the two ground states at $\theta=\pi$ are exactly degenerate. Normally, on a compact space, we expect instantons which lead to energy differences suppressed by the volume of space. But such instantons are absent due to the lack of charged particles. 
  
Much of the same comments could be made about the free $\U(1)$ gauge theory in any number of dimensions. However, a point that is absolutely crucial in two dimensions is that the above states with constant energy density do not break two-dimensional Poincar\'e invariance in the large $R$ limit!
Indeed, while in higher dimensions states with constant electric field necessarily break Poincar\'e invariance, in two dimensions they do not because the order parameter $\langle F_{01}\rangle$ is  Poincar\'e invariant. Therefore, we should think about the states $|n\rangle$ as superselection sectors. However unlike ordinary superselection sectors which are typically separated by finite potential barriers (such that we can create bubbles of superselection sectors of lower energy inside wrong vacua) here the potential barriers are infinite since the states do not mix on $S^1$. We will refer to such superselection sectors which are separated by infinite potential barriers as ``universes''. 

The free photon theory in two spacetime dimensions therefore admits infinitely many universes. As a function of $\theta$ we have a first order transition at $\theta=\pi$ with two universes simultaneously having the lowest energy density.\footnote{We may compactify Euclidean time $\tau\sim\tau+\beta$ and compute the partition function as a function of $R,\beta$. We find 
$$Z=\sum_{n}e^{-\frac{e^2\beta R}{4\pi}\left(2\pi n - \theta\right)^2}~.$$
The partition function only depends on the area of the torus, i.e.\ $\beta R$, which is a reflection of the familiar statement that the theory is invariant under area-preserving diffeomorphisms~\cite{Witten:1992xu}.}

Let us consider the Wilson lines in the free photon theory. In dimensions higher than $1+1$ as soon as there is a free massless photon we are in the Coulomb phase and all Wilson lines are deconfined (perimeter law). 
In two spacetime dimensions this is not so since there is no propagating photon degree of freedom. The Wilson line $P e^{i \int A }$ describes the world-line of a unit charge particle. As a result the electric field must jump by $e^2$ across the line. So the Wilson line can be said to separate the universes $|n\rangle$ and $|n+1\rangle$. Similarly the Wilson line $P e^{ik \int A }$  separates the universes $|n\rangle$ and $|n+k\rangle$.
Now consider a large rectangular Wilson line of unit charge. If outside of the line the universe is $|n\rangle$ then inside the rectangle we have the universe $|n\pm1\rangle$ depending on the orientation of $P e^{i \int A }$. For instance, at $\theta=0$, if the vacuum outside corresponds to $|0\rangle$, which is the lowest ground state, then inside the loop we have either of $|\pm1\rangle$, which have energy density $E_{\pm1} = \frac12 e^2$. This excited state inside the loop should be thought of as a confined string and hence the Wilson line is confined: 
\begin{equation}
    \langle W_{\square} \rangle\sim e^{-\frac12 e^2 LT}~,
\end{equation}
where $LT$ is the area of the Wilson loop ($L$ is the length and $T$ is the height).
We see that the question of confinement or deconfinement is determined by the energy density of the universe that is created inside the loop.
An interesting case to consider is $\theta=\pi$, in which case the universes $|0\rangle$, $|1\rangle$ are exactly degenerate. If we prepare the state $|0\rangle$ at infinity, then with an appropriately oriented rectangular Wilson loop we can create the state $|1\rangle$ or the state $|-1\rangle$ inside. In the former case the Wilson line will have perimeter law and in the latter case it will have area law.

Thinking more broadly about two-dimensional theories, there could be multiple super-selection sectors (for instance due to symmetries) and also multiple universes. Let us see how this comes about when we add a fermion particle.

We consider the $\U(1)$ gauge theory with a fermion of (nonzero) charge $q\in\mathbb{Z}$. This is a variation of the more familiar Schwinger model, in which there is a fermion of charge $q=1$~\cite{Coleman:1975pw,Coleman:1976uz}. This variation was recently considered in several papers, e.g.~\cite{Seiberg:2010qd,Komargodski:2017dmc,Armoni:2018bga,Misumi:2019dwq}. Our presentation here is self contained.

The absence of a dynamical particle of the minimal possible charge would lead to some interesting differences from the more familiar $q=1$ case. The Lagrangian is
\begin{equation}\label{qQED} - \frac{1}{ 4e^2}F^2+\bar \Psi \gamma^\mu D_\mu \Psi+m\bar \Psi \Psi~,\end{equation}
with $D_\mu=\partial_\mu-iq A_\mu$. 
We do not include a $\theta$ term because there is an axial transformation 
\begin{equation}\label{mainL}m\to m e^{-i\alpha}~,\quad \theta\to\theta+q \alpha~.\end{equation}
Therefore, if we study the model as a function of complex $m$ we will cover the full parameter space. 
In the meanwhile we do not include quartic interactions of $\Psi$. If we imagine that the theory~\eqref{mainL} arises from a lattice system we would expect quartic fermion interactions suppressed by the cutoff. For now we take this cutoff to infinity.  The physics at large $|m|$ (i.e.\ $|m|^2\gg e^2$) is rather simple since the fermion fluctuations are suppressed. The long-distance limit (at distances much longer than $m^{-1}$) is the free Abelian theory with $\theta = Arg(m^{q})$. The free Abelian theory by itself has infinitely many universes, which means that there are infinitely many Poincar\'e invariant states with infinite energy barriers between them.
It may be that now, with a massive fermion, the barriers are of order $|m|$ or that some of the universes are identified altogether. It may also be that some of the universes remain. All of these scenarios are consistent with the decoupling limit $|m|\to\infty$. 
The infinitely many universes of the free Abelian theory are labeled by the expectation values of the electric field~\eqref{Efield}. But since we now have dynamical excitations of charge $q$ (and the conjugate representation with charge $-q$), the states $|n\rangle$ and $|n\pm q\rangle$ can communicate and thus now are resides in the same universe.
Therefore the number of universes is reduced from infinity to $q$, labeled by an integer modulo $q$.

Next, we will find the lowest energy density states in each universe.
For $q\in 2\mathbb{Z}+1$ and $\theta\in[0,\pi)$ these are $|0\rangle$, $|\pm 1\rangle$,..., $|\pm {{|q|-1}\over 2}\rangle$. The lowest energy density universe is $|0\rangle$. For $\theta\in(\pi,2\pi)$ the $|q|$ lowest energy density states in each universe are $|-{{|q|}\over 2}+{3\over 2}\rangle$, $|-{{|q|}\over 2}+{5\over 2}\rangle$,...,$| {{|q|+1}\over 2}\rangle$. The lowest energy density universe in this range is always $|1\rangle$. The case of $\theta=\pi$ is quite more interesting; the states with $|q|+1$ lowest energy densities are $|0\rangle$, $|\pm 1\rangle$,..., $|\pm {{|q|-1}\over 2}\rangle$, $| {{|q|+1}\over 2}\rangle$, and now the states $|-{{|q|-1}\over 2}\rangle$, $| {{|q|+1}\over 2}\rangle$ can communicate by creating massive charged particles from the vacuum. These states are exactly degenerate in infinite volume but on the circle they will not be because there are finite action instantons which would lead to energy splitting of order $e^{-mR}$ (where $R$ is the circle radius as usual). In the infinite volume limit the worldline of the dynamical particle $\Psi$ is a domain wall between $|-{{|q|-1}\over 2}\rangle$ and $| {{|q|+1}\over 2}\rangle$. The tension of the domain wall is $m$. So at $\theta=\pi$ we have $|q|$ universes but one of the universes contains two ordinary super-selection sectors. 

Another very useful way to think about this is that for (nonzero) $q\in 2\mathbb{Z}+1$ there are always $|q|$ universes which do not communicate (and their energies depend on $\theta$ in some way) but inside one of the universes there is an additional ordinary first order transition  (as a function of $\theta$) with a finite energy barrier at $\theta=\pi$. It is only for $q=1$ that the two superselection sectors $|-{{|q|-1}\over 2}\rangle$ and $| {{|q|+1}\over 2}\rangle$ are also the true vacua (with the smallest energy density) at $\theta=\pi$. For $q\in 2\mathbb{Z}+1\neq 1$ this phase transition happens on the world-volume of some flux tube (instead of in the vacuum). Either way, this degeneracy between the two superselection sectors can be understood as being due to spontaneously broken charge conjugation symmetry. Sometimes charge conjugation symmetry breaking takes place in a state with a flux tube and sometimes in the true vacuum.

For (nonzero) $q\in 2\mathbb{Z}$ the story is a tad different. If $\theta\in(0,\pi)$ the lowest energy density states are $|-{|q|\over 2} +1\rangle$, $|-{|q|\over 2} +2\rangle$,..., $|{|q|\over 2} \rangle$. In fact the same states remain as the lowest energy density states at $\theta=\pi$ and also $\theta\in(\pi,2\pi)$. None of these are separated by a finite mass domain wall and hence these are $|q|$ universes. There is no first order phase transition inside any of these universes at $\theta=\pi$ (only which universe becomes the lowest energy density one changes). However something interesting happens at $\theta=0$ (and therefore for all $\theta = 0 \ {\rm mod} \ 2\pi \mathbb{Z}$). The states $|{|q|\over 2} \rangle$ and $|-{|q|\over 2} \rangle$ are degenerate at $\theta=0$ but they can communicate due to the fact that the domain wall between them (the particle of charge $q$) has finite mass. They are therefore degenerate at infinite volume but split at finite volume. This should be simply interpreted as spontaneous breaking of charge conjugation symmetry in the flux tube $|{|q|\over 2} \rangle$. There are therefore two superselection sectors in this universe. As $\theta$ goes through $0 \ {\rm mod} \ 2\pi \mathbb{Z}$ there is always a first order phase transition in one of the universes which is associated to a flux tube. 

This almost finishes the analysis of the model at large $|m|$. It only remains to discuss the question of confinement and deconfinement (in the sense of whether the rectangular Wilson loops have area or perimeter law). The Wilson line $P e^{iq \int A }$  is always screened since there is a particle of charge $q$. The rectangular Wilson line $P e^{i \int A }$ is always confined except, again, at $\theta=\pi$ where with one possible orientation it is deconfined and with the other orientation it is confined. In summary, $W_{\square}$ is always confined except for $\theta=\pi$ and $W_{\square}^q$ is always deconfined.

Now we turn to the physics of the model at small $|m|$, i.e.\ $|m|^2\ll e^2$. We start from $m=0$.
For $m=0$ we have a $\mathbb{Z}_{2q}$ axial symmetry acting as \begin{equation}\label{axial}\Psi\to e^{\gamma_3{2\pi i k \over 2q}}\Psi~,\quad k=0,...,2q-1~.\end{equation}
Note that the $\mathbb{Z}_{2}$ subgroup corresponding to $k=q$ acts as fermion number. The $\gamma_3$ matrix ensures that $\Psi_+$ and $\Psi_-$ are acted upon with the opposite phases. Since the gauge symmetry acts as $\Psi\to e^{i	q\alpha}\Psi$ (without a $\gamma_3$ matrix), we see that the transformation~\eqref{axial} with $k=q$ is in fact a gauge symmetry. But since the transformation $k=q$ is fermion number, this means that this model is non-spin. In other words, although there are fermions in the underlying Lagrangian~\eqref{qQED} the path integral includes a sum over all spin structures automatically. In particular, there are no fermionic local operators and the axial symmetry is really $\mathbb{Z}_q$ and not $\mathbb{Z}_{2q}$.

The massless model is most easily solved with bosonization. Since the subject of spin structure will be crucial later it is paramount to review in detail the Abelian bosonization case. Our boson will be always $2\pi$ periodic: $\phi\simeq \phi+2\pi$. Consider the action 
\begin{equation}
S={f^2\over 2}\int d^2x (\partial \phi)^2~,
\end{equation}
with $f$ a free parameter. An important subset of the local operators in the theory is given by $e^{i \bn \phi+i\bm\tilde \phi}$, where $\tilde\phi$ is the dual scalar, which is likewise $2\pi$ periodic. 
The scaling dimension and spin are 
\begin{equation}
\Delta = {1\over 4\pi f^2}\bn^2+\pi f^2 \bm^2~,\quad S = \bn \bm~.
\end{equation}
Clearly, this is a bosonic theory and it has no operators of half-integer spin (T-duality takes $f^2\to {1\over 4\pi^2f^2} $). For $f^2=1/4\pi$ this model is often said to be the bosonic dual to the free fermion. But clearly that is not quite right. For instance, the ``would be'' free fermion corresponds to $\bn=1/2$, $\bm=1$ but this is not in the spectrum since the corresponding operator is nonlocal. In addition, $\bn=0$, $\bm=1$ corresponds to an operator of dimension $1/4$ which does not exist in the free fermion theory. 
The precise statement of bosonization is that the model with $f^2={1\over 4\pi}$ is dual to a free complex fermion with gauged fermion number. This has been recently emphasized in~\cite{Thorngren:2018bhj,Karch:2019lnn}.

Since above we have  argued that the model~\eqref{qQED} indeed includes a sum over the spin structures of the fermion we can thus apply the usual bosonization dictionary.  The massless model can be thus rewritten as
\begin{equation}\label{boson}{1\over 8\pi} (\partial \phi)^2-{1\over 4e^2}F^2+{q\over 2\pi } \phi F_{01}~,\end{equation}
with $\phi\simeq \phi+2\pi$.
The $\mathbb{Z}_q$ axial symmetry acts by $\phi\to \phi+2\pi k / q$ with $k=0,...,q-1$. 
There are two ways to think about this model. One is to drop the term $-{1\over 4e^2}F^2$ at long distances (since from the infrared point of view it looks irrelevant) and then remain with ${1\over 8\pi} (\partial \phi)^2+{q\over 2\pi } \phi F_{01}$ which can be identified with the $\U(1)_q/\U(1)_q$ topological field theory. This is a simple Abelian TQFT with $q$ exactly degenerate vacua on the circle. One can think about it as $BF$ theory or a discrete $\mathbb{Z}_q$ gauge theory (or $\U(1)$ Chern-Simons theory at level $q$ on a circle). It is a trivial special case of the more general $G/G$ construction reviewed in~\cite{Blau:1993hj}. The second way is to integrate out $F$ more rigorously. Indeed, we can first treat $\phi$ as a classical fixed source. Then integrating out $F$ we would get the energy density of the true ground state with $\theta=q\phi$. (The contributions from other universes in the free $\U(1)$ theory are exponentially suppressed.) If there are two such ground states which simultaneously minimize the energy then we have to include both. In this way, from~\eqref{Elev} we find the following potential for $\phi$:
\begin{equation}
   V(\phi)= \pi e^2 R\  \min_{n} \left(n-{q\phi\over 2\pi}\right)^2~. 
\end{equation}
This potential is nicely consistent with the global symmetry $\phi\to \phi+2\pi k / q$. Therefore the effective theory for $\phi$ is \begin{equation}\label{EFTi}{1\over 8\pi} (\partial \phi)^2-\pi e^2 R\  \min_{n} \left(n-{q\phi\over 2\pi}\right)^2~.\end{equation}
The potential has $q$ degenerate critical points at $\phi=2\pi k /q$ with $k=0,..,q-1$. The  $\mathbb{Z}_q$ symmetry is spontaneously broken!
However, unlike in the usual situation of spontaneous breaking of a discrete symmetry in two spacetime dimensions, the $q$ vacua remain exactly degenerate on the circle. This is obvious from the fact that there are no charged particles in the system that could furnish finite mass domain walls. But it is less obvious from~\eqref{EFTi} since if we plot the potential, the energy barrier that separates the minima naively seems finite. This confusion is resolved by noting that the potential in~\eqref{EFTi} is non-differentiable and hence we cannot compute the tension of the kink reliably from the effective theory. And indeed, the correct answer is that the mass of domain walls is infinite. 

Let us now discuss the Wilson lines of the massless theory. Clearly, since we have $q$ degenerate universes, $W_{\square}$ is deconfined in the sense that a large rectangular loop admits a perimeter law. 
Therefore, the massless theory with a charge $q$ particle leads to deconfinement of the elementary charge! This is rather surprising and has received some attention in the literature. See~\cite{Gross:1995bp} for a review and references.
(Often it is phrased as the statement that fractional charges are deconfined in the Schwinger model. Of course, as long as the gauge group is $\U(1)$ we are not allowed to consider the world-lines of fractional charges, strictly speaking.)

This story changes rather drastically as soon as we add a small nonzero mass. Adding a mass $m=|m|e^{i \arg(m)}$,  the potential becomes
\begin{equation}\label{smallmass}V=\pi e^2 \  \min_{n} \left(n-{q\phi\over 2\pi}\right)^2-|m|\Lambda\cos(\phi+ \arg m)~.\end{equation} 
$\Lambda$ stands for some mass scale, which will not be important for us (except that it is positive).
For generic $\arg m$ only one universe has the truly lowest energy density. Therefore, confinement is restored for any nonzero $m$. The tension of the confining string is linearly proportional to $|m|$. If $\arg m ={\pi\over q}\ {\rm mod}\ \frac{2\pi}{q}  $ then we find that two of the $q$ universes have an exactly degenerate energy density. Therefore the fundamental Wilson loop with one orientation is deconfined  and confined for the other orientation. Now we are prepared to compare these results to those of the large mass theory. 

For $q\in 2\mathbb{Z}$ this almost coincides with the large mass limit. At small mass we have $q$ universes, and the only thing that happens as we change the $\theta$ angle is that the identity of the universe that is the lowest one changes. Therefore there is no phase transition neither in the ground state nor in any of the string (flux tube) states.  But at large mass there was a phase transition inside one of the universes which describes a flux tube at $\theta=0 \ {\rm mod} \ 2\pi \mathbb{Z}$. This means that there is some finite mass $m_*$ (corresponding to a vanishing $\theta$ angle) where a second order transition (most likely in the Ising universality class) takes place on this flux tube at $\theta=0$. This is an example of a flux tube theory that becomes massless while the bulk is gapped and has a unique vacuum. The flux tube theory where this transition takes place consists of $q/2$ confining strings on top of each other. 
For $q\in 2\mathbb{Z}+1$  there is a similar disagreement between the small and large mass limits. The disagreement is that, for large mass, inside the universe $|{|q|+1\over 2}\rangle$ there is an additional first order transition at $\theta=\pi$. All we see at small mass is that there is a first order transition between the universes themselves but we do not see additional superselection sectors inside the universe $|{|q|+1\over 2}\rangle$. Therefore, there must be an Ising-type (second order) transition inside this universe for any odd $q$ and for some non-zero value of the mass (such that the effective theta angle is $\pi$). In particular, for $q=1$, which is the ordinary two-dimensional QED, there must be an Ising type fixed point at  $m_*e^{i\pi}$ with some positive $m_*$. For odd $q>1$ this phase transition is still second order, but it does not happen in the true ground state, instead, it happens in a state with a higher energy density. One could say that it happens when one puts $(q+1)/2$ confining strings on top of each other. These phase transitions at finite mass, where two-dimensional QED flows to the Ising critical point, were recently discussed also in~\cite{Komargodski:2017dmc}. 

It is not unfamiliar that new massless particles appear on the string's world-volume. Here we see a situation (both for even and odd $q$) where this can be proven to take place. Only for $q=1$ this transition takes place in the true ground state and hence corresponds to a standard quantum critical point.

A useful additional exercise to carry out before discussing the non-Abelian theory is to add small quartic interactions to~\eqref{qQED}.\footnote{We thank Yuya Tanizaki for discussions about this model.} By ``small' what we mean is that it is suppressed by an energy scale much higher than $e$ or $m$. 
A possible quartic interaction to contemplate is  $|\alpha|e^{i\arg\alpha} \bar \Psi_+\bar \Psi_+\Psi_-\Psi_-+c.c.$. 
Another term we can add is $\beta \bar \Psi_+\bar \Psi_-\Psi_+\Psi_-$.
The term proportional to $\beta$ is invariant under both the shift symmetry and the dual shift symmetry, which means that in terms of the variable $\phi$ it cannot lead to an interesting potential for $\phi$ (it may lead to derivative terms). 

The operator proportional to $|\alpha|$ on the other hand leads to 
\begin{equation}
\label{quartic}V\sim |\alpha|\cos(2\phi+\arg\alpha)~.
\end{equation}
Such a quartic interaction preserves a chiral $\mathbb{Z}_2$ symmetry for even $q$. For odd $q$ it does not preserve any chiral symmetry so we assume that $q$ is even in the following few paragraphs.  
Let us set $m=0$ and include the quartic interaction~\eqref{quartic}.
Such a potential always leaves two degenerate universes (at least) since $\phi$ and $\phi+\pi$ give the exact same energy density and hence the universes $k$ and $k+q/2$ are degenerate.
This means that charge $q/2$ particles are deconfined! More precisely, while $W_{\square}$ is confined due to the quartic interaction (even in the massless theory), $W_{\square}^{q/2}$ is deconfined. In fact $W_{\square}^{q/2}$ is deconfined with either of its two possible orientations. 

For even $q$ and $m=0$ this model is a little similar to the massless nonAbelian adjoint model.
The similarity is that both models have a chiral $\mathbb{Z}_2$ 0-form symmetry and  a one-form $\mathbb{Z}_q$ symmetry.\footnote{One can think of $\oneform_1$, the generator of the one-form symmetry, as, roughly speaking, $e^{{2\pi i\over e^2q} F_{01}}$. It is essentially topological because particle creation only changes the electric fields by integer multiples of $e^2 j$.}
Furthermore, both models have a mixed anomaly between the $\mathbb{Z}_2$ 0-form symmetry and a one-form $\mathbb{Z}_q$ symmetry.
The argument for this in the Abelian theory is the standard argument for a mixed anomaly between a chiral symmetry and a one-form symmetry (see footnote \ref{one-form.refs} for references). 
Namely, if we were to gauge the one-form $\mathbb{Z}_q$ symmetry then the periodicity of the $\theta$ angle would be extended to $2\pi q$. Alternatively we could introduce an additional discrete theta angle ranging over $1,...,q$. Then under a chiral $\mathbb{Z}_2$ transformation the discrete theta parameter jumps by $q/2$ in accord with the deconfinement of these representations.
 
In the context of the Abelian theory it is possible to demonstrate explicitly how inside a rectangular Wilson line a different universe appears. Let us insert the worldlines of a charge 1 and charge -1 particles $P e^{\pm i\int A}$ separated by distance $L$. 
The equations of motion in the presence of these sources in the bosonic variables for static configurations are given by
\begin{equation}\label{forPhi}{1\over 4\pi}\partial_1^2 \phi-|m|\Lambda \sin(\phi+\arg m) + {q\over 2\pi} F_{01}=0~, \end{equation} 
\begin{equation}\label{ForF}{1\over e^2}\partial_1F_{01} + {q\over 2\pi}\partial_1 \phi=J_0^{EXT} ~, \end{equation}
where $J_0^{EXT} = \delta(x+L/2)-\delta(x-L/2)$. 
The formula for the total energy of the system is 
\begin{equation}
    E= \int dx\left[ {1\over 2e^2}F_{01}^2+{1\over 8\pi}(\partial_1\phi)^2-m|\Lambda| \cos(\phi+\arg m)\right]~.
\end{equation}
If $m$ is the largest parameter in the problem (and $\arg m=0$) then to a good approximation, between the sources (i.e.\ $x\in [-L/2,L/2]$), $F_{01} =e^2$ and outside of the sources $F_{01}=0$. The field $\phi$ is not quite at the origin between the sources, rather, $\phi\sim qe^2/2\pi|m|\Lambda$. So the field $\phi$ is slightly displaced from its vacuum while $F_{01}$ is displaced significantly. Inside the loop we are in a different universe which has energy density $e^2/2$. The energy density mostly comes from $F_{01}^2$. 

In the massless limit it is not beneficial for $F_{01}$ to be activated between the sources, rather, it is energetically favorable for $\phi$ to attain (approximately) the value ${2\pi \over q}$ between the sources. The reason is simply that a constant value of $\phi$ does not cost energy in the massless limit. 
The field configuration is exactly solvable since if we set $m=0$ in~\eqref{forPhi} the equations become linear. 

The equations are solved by $F_{01}=Ae^{qex/\sqrt \pi}$ for $x<-L/2$ and $Ae^{-qex/\sqrt\pi}$ for $x>L/2$ and $B(e^{qex/\sqrt \pi}+e^{-qex/\sqrt \pi})$ for $-L/2<x<L/2$.
The solution for $\phi$ is  $\phi=-{2\pi\over qe^2}Ae^{qex/\sqrt \pi}$ for $x<-L/2$ and $-{2\pi\over qe^2}Ae^{-qex/\sqrt\pi}$ for $x>L/2$ and $-{2\pi\over qe^2}B(e^{qex/\sqrt\pi}+e^{-qex/\sqrt\pi})+\phi_0$ for $-L/2<x<L/2$. The constant $\phi_0$ must be adjusted so that the jump across the source is consistent with~\eqref{ForF}. This is achieved by setting $\phi_0=2\pi /q$. 

To determine the constants $A,B$ we note that from the massless version of equation~\eqref{forPhi} it follows that $\phi$ must be continuous and its first derivative must be likewise continuous at $x=\pm L/2$. This leads to two relations which are solved by $B=e^2/2$ and $A={e^2\over 2}(e^{qeL/\sqrt \pi}-1)$. We thus see that if $L$ is very large the electric field quickly decays far away from the sources, as expected, while $\phi$ in between the sources is exponentially close to $2\pi/q$, as expected from the universe corresponding to $n=1\ {\rm mod}\ q$. Qualitatively, we see that near the source it is $F_{01}$ that jumps according to the Gauss law (while $\phi$ and $\partial_1\phi$ are continuous) but then this jump in the electric field, which is energetically costly, is quickly screened by $\phi$. 

While we have phrased the results in terms of the bosonic variables, it is worth noting that this was not necessary. In terms of the current $J_\mu = \bar\Psi\gamma_\mu \Psi $, the equations~\eqref{forPhi} and~\eqref{ForF} take the form ${1\over e^2} \partial_1 F_{01}+q J_0 = J_0^{EXT}$ and $\partial_1 J_0 +{q\over \pi} F_{01}=0$. The second equation follows readily from the fermionic degrees of freedom, being nothing but the axial anomaly equation for static configurations. We see that the charge density $J_0$ is a continuous function of space. Most interestingly, if we integrate $J_0$ over a region much bigger than $e^{-1}$ and smaller than $L$ around the external sources, we find that $\int dx J_0 =\pm 1/q$ -- namely, even though the dynamical particle has charge $q$, the charge accumulated around the sources is, respectively, $\pm 1$, i.e.\ in some sense the dynamical charge has fractionalized, which is what allows for screening. Presumably the same mechanism is responsible for deconfinement in the massless adjoint theory but it is hard to demonstrate it as explicitly. It may be possible in the limit of many matter fields~\cite{Armoni:1997bu}. 

Let us now give an argument for deconfinement in the massless Abelian theory which is more in the spirit of this paper and relies only on the symmetries and anomalies. We are dealing with a bosonic system with a $\mathbb{Z}_q$ zero-form symmetry and a $\mathbb{Z}_q$ one-form symmetry. Most importantly there is an anomaly between the two symmetries which means that if we take the topological local operator $\oneform_1$ that generates the one-form symmetry and pass it through the  topological line $U$ that generates the $\mathbb{Z}_q$ zero-form symmetry we obtain 
\begin{equation}
    \oneform_1 U = e^{2\pi i / q} U \oneform_1~.
\end{equation}
This can be concluded by noticing that $U$ can end on charge 1 defect operator $e^{i\frac{1}{q}\tilde{\phi}}$ in the bosonized theory, therefore connected to $W_\square$ at the defect operator. A line-changing operator exists only when the connected lines have the same one-form symmetry charge.

The anomaly means that if we insert $U$ in a time-like fashion this will separate space into the universe without the string and a universe with the string. But since this time-like line is topological we can move it freely and this implies that the energy density in both universes is the same, namely, the string tension vanishes. 
We see that the deconfinement of the Abelian model really follows straightforwardly from the ordinary symmetries and anomalies of the theory.  
Once we add the quartic fermion interaction proportional to $\alpha$ above, the $\mathbb{Z}_q$ zero-form symmetry is broken to $\mathbb{Z}_2$ for even $q$ and to nothing otherwise. Then we no longer have a topological line that can create the universe with one string and confinement sets in (for even $q$, $q/2$ strings can be created with a topological line and hence that string is still deconfined). 

From this point of view the adjoint theory is actually simpler than the Abelian theory. Even though it only has $\mathbb{Z}_2$ and $\mathbb{Z}_N$ zero-form and one-form symmetries respectively, it has (exponentially) many other non-invertible symmetries which allow to construct the universe with one string via a topological line operator. So deconfinement can be proven without having to solve the full theory.

To close this section let us present some results about the $k$-string tensions in this model (without quartic interactions and at $\theta=0$ -- both of which can be easily relaxed). First, in the large mass limit, \begin{equation}\label{AbelianS} T_k=e^2 R\pi k^2~,\end{equation}  where for $q\in 2\mathbb{Z}+1$ $k$ ranges over the integers $k=0,\pm1, \pm2,...,\pm{|q|-1\over 2}$, and for $q\in 2\mathbb{Z}$ $k$ ranges over the integers $0,\pm1,\pm2,...,\pm |q|/2-1,|q|/2$.
This does not describe physical $k$-strings as bound states of the fundamental string since the binding energy has the wrong sign. Hence, the conclusion is that in the large mass limit $k$-strings do not exist. This is in contrast with the non-Abelian theory, as we will see. That fact that $k$-strings do not exist in the large mass limit is hardly surprising -- it simply boils down to the fact that same sign charges repel. 
The small mass limit is a little less trivial. We have computed the potential in~\eqref{smallmass}. For simplicity consider nonzero positive mass, such that $\arg m=0$. For small mass $m\ll e$ the energy density in the vacua $j=0,...,q-1$ is given by $E_j = -m \Lambda \cos(2\pi j / q)$. The $k$-string tension is therefore 
\begin{equation}
    T_k = m\Lambda(1 -\cos(2\pi k / q))=2m\Lambda  \sin^2(\pi k / q)~.
\end{equation}
This again does not describe genuine $k$-strings as the binding energy has the wrong sign. In conclusion, the Abelian model has no genuine $k$-strings, unlike the non-Abelian theory. This also makes sense from the following point of view: the least energy $k$-strings in the non-Abelian case turn out to correspond to completely anti-symmetric representations, which have no analog in the Abelian case.

\section{Review of Previous Proposals \label{sec:review.adjQCD}}

In the previous section we studied in depth the Abelian gauge theory with and without matter. 
Let us recall the properties of the pure non-Abelian theory now (the non-Abelian model was first solved in~\cite{Migdal:1975zg}). The Lagrangian of the theory without matter is given by \begin{equation}\label{na}{-1\over 4 g_\mathrm{YM}^2 }\Tr F^2~.\end{equation}
We will consider the gauge group $\SU(N)$. There is no theta angle in this case and hence this theory has no free parameters (since $g_\mathrm{YM}$ has dimensions of mass, it is not a parameter).
The theory has no propagating degrees of freedom and it is solvable. Quantizing the theory on a circle, we find that it essentially reduces to a quantum mechanical model with eigenstates labeled by irreducible representations of the group $\SU(N)$. These eigenstates have energy  $E\sim g_\mathrm{YM}^2RC_2(\mathcal{I})$ with $C_2(\mathcal{I})$ the quadratic Casimir of the irreducible representation $\mathcal{I}$ and $R$ is the radius of the circle. Therefore, these eigenstates have finite energy density in the infinite volume limit. There are therefore infinitely many universes labeled by irreducible representations of $\SU(N)$. However, unlike in the $\mathrm{U}(1)$ case, the universes are not associated to a standard one-form symmetry. Indeed, the set of representations of a group is not itself a group. This is therefore an example of a situation where the universes are not determined by a standard one-form symmetry. (The one-form symmetry of the model~\eqref{na} is simply $\mathbb{Z}_N$.)
It might seem perplexing that the adjoint representation universe cannot communicate with the universe corresponding to the trivial representation. This is of course because the pure gauge theory has no propagating gluons. This means that all Wilson lines are confined and the string tension is proportional to $C_2(\mathcal{I})$.

Just to be concrete, let us quote the $\SU(2)$ gauge theory partition function on a circle of size $R$ and time direction compactified on a circle of size $L$: 
\begin{equation}
    Z_{\SU(2)} = \sum_{j=0,\frac{1}{2},1,\frac{3}{2}...} e^{-\frac{1}{2}g_\mathrm{YM}^2LRj(j+1)}~.
\end{equation}
The dimension of the representation corresponding to $j$ is as usual $2j+1$.

Now let us consider the gauge theory based on $\mathrm{PSU}(N)=\SU(N)/\mathbb{Z}_N$. This is obtained from the gauge theory $\SU(N)$ by gauging the one-form symmetry $\mathbb{Z}_N$. The $\mathrm{PSU}(N)$ gauge theory does not have one-form symmetry, but as our general discussion showed, there is now a discrete theta angle $p\in \mathbb{Z}_N$~\cite{Witten:1978ka}. We can thus ask about the partition functions $Z_{\mathrm{PSU}(N)}^p$. As argued before, choosing $p$ is tantamount to choosing those universes in $\SU(N)$ gauge theory where the one-form symmetry generator takes the same expectation values. 
Therefore, the partition function $Z_{\mathrm{PSU}(N)}^p$ picks up a contribution from those universes where the corresponding representation has $p$ boxes mod ${N}$~\cite{Aminov:2019hwg,Sharpe:2019ddn}. Note that those are generally not $\mathrm{PSU}(N)$ (linear) representations.
This is in accord with the idea that the theta angle corresponds to a projective representation in two dimensions. For instance, in $\SU(2)$ gauge theory we have $\mathrm{PSU}(2)=\SO(3)$ and we have the two partition functions: 
\begin{align}
    Z_{\SO(3)}^0 & =\sum_{j=0,1,2...} e^{-\frac{1}{2}g_\mathrm{YM}^2LRj(j+1)}~, \\
    Z_{\SO(3)}^1 & =\sum_{j=\frac12,\frac32...} e^{-\frac{1}{2}g_\mathrm{YM}^2LRj(j+1)}~.
\end{align}
Note that $Z_{\SO(3)}^1/Z_{\SO(3)}^0$ is exponentially small on a large torus, which is the same as the statement that the fundamental Wilson line in $\SU(2)$ gauge theory is confined.

Let us now include an adjoint Majorana fermion. Namely, we add to the theory the adjoint fermions $\psi_+$ and $\psi_-$ such that they are themselves real. The Lagrangian takes the form 
\begin{equation}\label{BasicL}
S_{\ttf}=\int \rd^2 x \, \left[-\frac{1}{4g_\mathrm{YM}^2} \Tr F^2 + \Tr \psi^Ti \slashed{D}_\r \psi +im\psi^T\psi\right]~. 
\end{equation}
By virtue of the reality of $\psi$, $m$ must be real. There is no $\theta$ angle in $\SU(N)$ gauge theory (neither continuous or discrete)  and hence the change of variables $\psi_+\to -\psi_+$ does not change the measure. The theory with mass $m$ is therefore equivalent to the theory with mass $-m$. As a result, we can take $m$ to be nonnegative without loss of generality. For the same reason the massless theory $m=0$ has such a chiral $\mathbb{Z}_2^\chi$ symmetry. This means that the massless point is distinguished. 

While the $\mathrm{U}(1)$ model with a charge $q$ fermion from the previous section is a non-spin theory, the adjoint model~\eqref{BasicL} is a spin theory because the fermion number symmetry $(-1)^F:\psi_{\pm}\to -\psi_{\pm}$ is not part of the gauge group.\footnote{The same is true also for $\SU(2)$ gauge theory; while there is a gauge transformation implementing $\psi_{\pm}\to -\psi_{\pm}$, it also acts non-trivially on the gauge field and it is hence not the same as fermion number. To see this explicitly consider the fermionic gauge invariant operator $Tr (F_{01} \psi)$. It does not vanish in $\SU(2)$ gauge theory.}

The dynamics for  $m^2\gg g_\mathrm{YM}^2$ is governed by the pure $\SU(N)$ gauge theory, but now the existence of massive adjoint fermions reduces the universes only to the $N$ universes with the smallest values of $C_2(\mathcal{I})$. This of course always includes the empty universes and then $N-1$ additional universes with nonzero energy density representing the flux tubes ending on the representation $\mathcal{I}$. These flux tube theories are all a priori different (i.e.\ they have a different spectrum of excitations). For instance, in $\SU(2)$ gauge theory we have the trivial and doublet representations furnishing the two universes. In the $\SU(3)$ gauge theory case, there are three universes corresponding to the representations $1,3,\bar3$. The latter two have the same energy density due to charge conjugation symmetry.  In the universe corresponding to the trivial representation, $\psi$ itself is not an excitation -- only the two particle state $\psi\psi$ exists. So the mass gap above the vacuum in the trivial universe should be around $2m$. In the universe corresponding to the fundamental representation, since in the product ${\rm fund}\otimes {\rm adjoint}$ the fundamental representation itself can be found (this is true for all $N$), the state $\psi$ exists. This means that the mass gap above the flux tube should be approximately $m$ rather than $2m$. (Another way to think about it is that the universe corresponding to the fundamental representation is described by the excitations of the fundamental Wilson line, and we can insert a single $\psi$ on such a Wilson line in a gauge invariant fashion.)

In the $\mathrm{PSU}(N)$ gauge theory language the same story is told in a different language. It is interesting to repeat it because of the different role played by the chiral $\mathbb{Z}_2^\chi$ symmetry. We now have theories labeled with $p\in \mathbb{Z}_N$ and mass $m$. Now the chiral symmetry flips the sign of $m$ but at the same time it also takes~\cite{Cherman:2019hbq}
\begin{equation}
\label{mshift}p\to p+N/2~.
\end{equation} 
(For odd $N$, the index $p$ is unchanged by the $\mathbb{Z}_2^\chi$ symmetry. For even $N$, the measure of the $\mathrm{PSU}(N)$ theory is not invariant under $\mathbb{Z}_2^\chi$ which leads to the above shift of the index $p$.) We can therefore again restrict $m$ to be positive and study the theories with all possible values of $p$. Different values of $p$ are just the different universes we saw above. Either way, we see that for $m^2\gg g_\mathrm{YM}^2$ the fundamental Wilson line is confined $W_{F}\sim e^{-g_\mathrm{YM}^2 Area}$.

The small mass limit is much less obvious. Some facts about the massless limit directly follow from the symmetries and anomalies.  For the $\SU(N)$ gauge theory, at $m=0$ there is an axial symmetry $\mathbb{Z}_2^\chi$ which has a mixed anomaly with the one-form $\mathbb{Z}_N$ symmetry for even $N$. This is reflected in equation~\eqref{mshift}. This means that, for even $N$, the universes $p$ and $p+N/2$ have the same energy density and hence the line $W_F^{N/2}$ is deconfined.  Another way of saying the same thing is that in the $\mathrm{PSU}(N)$ gauge theory language the theories with $p=0$ and $p=N/2$  are related by a $\mathbb{Z}_2$ symmetry and hence 
\begin{equation}
    {Z_{\mathrm{PSU}(N)}^{N/2}/Z_{\mathrm{PSU}(N)}^{0} =1}~.
\end{equation}
For $N=2$ which is the simplest nontrivial case this means that the fundamental line is deconfined. But for $N>2$ ordinary symmetries and anomalies alone do not imply that the fundamental line is deconfined. We will later see that there are theories with the same symmetries and anomalies where the fundamental line for $N>2$ is confined or deconfined, depending on the parameters.

In~\cite{Gross:1995bp} it was argued (see after equation (5.28)) that there are $2(N-1)$ zero modes in each universe. This would mean that all the Wilson lines are deconfined in the massless theory. However, it is not true that such zero modes exist~\cite{Cherman:2019hbq}. The only zero modes which exist are those that allow to connect the universes $p$ and $p+N/2$ which imply deconfinement of $W_F^{N/2}$. Identifying this error, the authors of~\cite{Cherman:2019hbq} were therefore led to suggest that the fundamental line is confined and only $W_F^{N/2}$ is deconfined. In fact, this is exactly what happened in the Schwinger model with quartic interactions and $\mathbb{Z}_2$ axial symmetry~\eqref{quartic}! So the suggestion of~\cite{Cherman:2019hbq} is certainly attractive from the point of view of anomaly matching.

What we will show in this note is that one can harness the power of certain topological defects which are unrelated to ordinary zero-form or one-form symmetries. These topological defects are non-invertible in general and hence do not correspond to a symmetry group. We will show that these topological defects exist only in the massless theory and they allow to prove some very concrete results about the dynamics. Renormalization-group invariants that are associated with non-invertible topological defects should be viewed as a generalization of the idea of anomaly matching \cite{Chang:2018iay}.
The adjoint gauge theory is a striking demonstration that these renormalization group invariants are really powerful and can shed light on the physics of strongly coupled gauge theories.

We will show that the massless adjoint $\SU(N)$ gauge theory admits a large number of such  topological non-symmetry defects. We use them to prove that the theory has a large ground state degeneracy. In fact, as $N$ is taken to be large, the degeneracy is exponential in $N$. In particular, all the Wilson lines are deconfined.  Furthermore, the Hagedorn temperature of the planar massless theory vanishes. 
We also analyze the effects of some quartic interactions. We show that sometimes they break the non-symmetry defects. This means that sometimes the model deformed by quartic interactions is confined. This is in accord with our expectations from the usual  symmetries and anomalies. Finally, we study the theory at small nonzero mass, where these non-invertible defects cease to be topological. For small enough mass the breaking of topological invariance is soft and we use that in order to obtain exact predictions for the tension of $k$-strings at small mass.

\section{Adjoint QCD and Non-Abelian Bosonization} \label{sec:bos}

Let us begin by studying 2d adjoint QCD through non-Abelian bosonization. The action of adjoint QCD i.e.\ $\SU(N)$ gauge theory coupled to Majorana fermions in the adjoint representation, in two Eucledian dimensions is given by
\begin{equation}
S_\ttf=\int \rd^2 x \, \left[-\frac{1}{4g_\mathrm{YM}^2} \Tr F^2 + \Tr \psi^Ti \slashed{D}_\r \psi \right]~, \label{Spsi}
\end{equation}
where the subscript $\r$ in the covariant derivative denotes the spin structure chosen to define the fermions. Summing over spin structures, i.e.\ gauging the $(-1)^F$ symmetry, the theory is dual to a gauged WZW model. To explain this duality, first we review the non-Abelian bosonization \cite{Witten:1983ar}.

The duality states that $n$ free Majorana fermions with gauged $(-1)^F$ symmetry can be bosonized to the $\Spin(n)_1$ WZW model \cite{Ji:2019ugf}. Since the global aspects of the theory are crucial for us here, we check that the duality is exact and all the global symmetries of the two sides of the duality match. The global symmetry group of $n$ free Majorana fermions before gauging $(-1)^F$ is $G_\ttf=\mathrm{O}(n)_\ttL \times \mathrm{O}(n)_\ttR$. This group can be decomposed into its connected and disconnected parts as
\begin{equation}
    G_\ttf = \begin{cases} 
    {\SO(n)}_\ttL \times {\SO(n)}_\ttR \times \bZ_2^{F} \times \bZ_2^{\x} ~~~& n\text{ odd}~,\\
    \left( {\SO(n)}_\ttL \times {\SO(n)}_\ttR \right) \rtimes \left( \bZ_2^{C} \times \bZ_2^{C_\ttL} \right) ~~~& n\text{ even}~,
    \end{cases} \label{global.sym}
\end{equation}
as will be described in detail in the following.

\paragraph{Symmetries: odd $n$}  
$\bZ_2^F$ is generated by the fermion number symmetry $(-1)^F$ and $\bZ_2^\x$ is generated by $(-1)^{F_\ttL}$ which is the left-moving fermion number symmetry. Alternatively we can write $\bZ_2^{F} \times \bZ_2^{\x} = \bZ_2^{F_\ttL} \times \bZ_2^{F_\ttR}$. There is a mixed anomaly between $\bZ_2^F$ and $\bZ_2^{\x}$. An easy way to derive this anomaly is to add a mass term and then reverse the sign of the mass by acting with  $\bZ_2^{\x}$. This reverses the sign of the mass for an odd number of Majorana fermions and hence induces the Arf contact term for the spin structure. Therefore, naively, after gauging $\bZ_2^F$ -- summing over spin structures -- the axial symmetry should have disappeared. Instead $\bZ_2^{\x}$ becomes a non-invertible line\cite{Ji:2019ugf}.  At the same time, as usual, gauging fermion number leads to a dual ``quantum symmetry'', which in this case acts as $+1$ and $-1$ on the untwisted and twisted sectors respectively. Hence, after bosonization, the global symmetry of the model becomes a non-trivial $\bZ_2$ extension of ${\SO(n)}_\ttL \times {\SO(n)}_\ttR$, namely 
\begin{equation}
G_\ttb = \frac{\Spin(n)_\ttL \times \Spin(n)_\ttR}{Z\left(\Spin(n)_\mathrm{diag}\right)} \quad \text{(odd $n$)}~. \label{sym.n.odd}
\end{equation}
The group $Z\left(\Spin(n)_\mathrm{diag}\right)$, is the center of the diagonal $\Spin(n)$ which for odd $n$ is $\bZ_2$. To see that the extension is a Spin group, notice that the spin field in the twisted sector transforms in the spinor representation of the $\mathfrak{so}(n)$ algebra. The  $\bZ_2$ associated with the center of the group $\Spin(n)$ is precisely the symmetry dual to the gauged fermion number symmetry.

\paragraph{Symmetries: even $n$}   
Now let us discuss the bosonic symmetry group for even $n$~\eqref{global.sym}. The disconnected part of the symmetry group $G_\ttf$, is generated by determinant $-1$ orthogonal matrices. These matrices do not commute with the connected component of the symmetry group that includes the identity, hence we have the semi-direct product in \eqref{global.sym}. This semi-direct product is defined by the $\bZ_2$ outer automorphism group of $\mathfrak{so}(n)$ which is known as the charge conjugation. More precisely, $\bZ_2^C$ is the diagonal charge conjugation, and $\bZ_2^{C_\ttL}$ acts as charge conjugation only on left-moving fermions. We can now gauge $\bZ_2^F$. The $\bZ_2^F$ symmetry becomes a dual invertible $\bZ_2$ symmetry which we will soon identify as before to be in the center of the $\Spin(n)$ symmetry of the bosonic theory.\footnote{Similarly to the fate of $\bZ_2^{\x}$ for odd $n$, after gauging fermion number symmetry, $\bZ_2^{C_\ttL}$  becomes a non-invertible line. Only that now it does not commute with the current algebra and hence we do not follow it further.}
So we conclude that the bosonic symmetry group is a $\bZ_2$ extension of
\begin{equation}
    \frac{{\SO(n)}_\ttL \times {\SO(n)}_\ttR}{\bZ_2^F}  \rtimes  \bZ_2^{C}~,
\end{equation}
namely
\begin{equation}
    G_\ttb = \frac{\Spin(n)_\ttL \times \Spin(n)_\ttR}{Z\left(\Spin(n)_\mathrm{diag}\right)}\rtimes \bZ_2^C \quad \text{(even $n$)}~, \label{sym.n.even}
\end{equation}
where $Z\left(\Spin(n)_\mathrm{diag}\right)$ is $\bZ_2 \times \bZ_2$ for $n=4k$ and $\bZ_4$ for $n=4k+2$. 
The dual symmetry to $\bZ_2^F$ is now a $\bZ_2$ subgroup of $Z\left(\Spin(n)_\mathrm{diag}\right)$. We refer to it as $\bZ_2^\ttv$. It is the subgroup of the center for which $\Spin(n)_\ttL/\bZ_2^\ttv = \SO(n)_\ttL$.\footnote{Here we have conventionally chosen the $\bZ_2$ in $\Spin(n)_\ttL$; one can equally well choose it to be inside $\Spin(n)_\ttR$. These choices are equivalent since the diagonal center acts completely trivially.}
A natural question concerns the other elements of the center $Z\left(\Spin(n)_\mathrm{diag}\right)$. For even $n$ one can define an axial transformation on the fermions $\psi_+\to -\psi_+$. This $\bZ_2$ symmetry resides inside ${\SO(n)}_\ttL$ and it commutes with it. For $n\equiv0 \pmod{4}$ this symmetry has no mixed anomaly with $\bZ_2^F$ and this is why $Z\left(\Spin(n)_\mathrm{diag}\right)$ is $\bZ_2 \times \bZ_2$ for $n=4k$. For $n=4k+2$ there is actually a mixed anomaly between the axial symmetry $\psi_+\to -\psi_+$ and $\bZ_2^F$ but this anomaly is such that upon gauging $\bZ_2^F$ the axial does not disappear, but instead, it squares to the generator of $\bZ_2^\ttv$, which is why the center becomes
$\bZ_4$. A similar phenomenon in a different context was discussed in~\cite{Ang:2019txy}.

\paragraph{Is bosonization unique?}
So far we have listed the symmetries of the theory which is obtained from $n$ free fermions upon gauging $\bZ_2^F$. But in fact the procedure of bosonization is not yet uniquely specified since 
one has a freedom to insert spin-structure dependent phases when summing over different partition functions. The choice of such phases correspond to invertible spin-TQFTs (a.k.a fermionic SPT phases)~\cite{Gaiotto:2015zta,Seiberg:2016rsg,Bhardwaj:2016clt,Kapustin:2017jrc}.\footnote{More precisely, such choices correspond to fermionic SPT phases modulo bosonic SPT phases.} In 1+1d there are two invertible spin-TQFTs, the trivial theory and the Kitaev chain whose partition function is given by the Arf invariant of the spin structure \cite{Moore:2006dw,Kapustin:2014dxa,Gaiotto:2015zta, Karch:2019lnn, Ji:2019ugf, Hsieh:2020uwb}. Therefore these two choices correspond to adding the Arf invariant to the action of the fermionic theory, and in the bosonic theory are related by gauging the quantum $\bZ_2^\ttv$ symmetry (without an Arf term).
We fix the scheme unambiguously as follows. We give the fermions large {\it positive} mass and require that at low energies there is no Arf term. In this scheme the bosonized theory at large positive mass has two vacua with the quantum $\bZ_2^\ttv$ symmetry spontaneously broken.
Flipping the sign of the mass for all the fermions introduces the Arf contact term for odd $n$ because then the total number of fermions is odd. So for odd $n$ one can go between the two possible choices of the invertible spin TQFT by flipping the mass of the fermions. Since flipping the sign of the mass term is the same as acting with the axial fermion number symmetry, the two choices are related by the Verlinde line that emerges after we gauge fermion number. In other words, for odd $n$ the two choices of how we may bosonize the theory are related by a duality line and so there is essentially only one way to bosonize modulo relabeling the operators. We will see that this is realized by a Kramers-Wannier type duality in the dual bosonic theory. 

For even $n$ there are genuinely two distinct bosonized theories since the Arf term is not  generated by flipping the sign of the mass term.
By requiring that at large mass the Arf term is absent, we are choosing one of these two possible ways to bosonize the fermions. We will stick to this choice of the bosonized theory for the rest of the paper.

\paragraph{The dual description \label{par:sigma.model}} 
Both for even and odd $n$, the dual description satisfying all the requirements above is $\Spin(n)_1$ WZW model \cite{Witten:1983ar, Ji:2019ugf}, defined in terms of group valued fields $g(x)\in \Spin(n)$. The model has a global symmetry that acts as $g(x)\rightarrow U_\ttL^{-1} g(x) \; U_\ttR $ with $U_\ttL,U_\ttR\in \Spin(n)$. However, the diagonal center acts trivially on $g(x)$. Moreover, for even $n$ there is also the $\bZ_2^C$ charge conjugation symmetry which acts by the outer automorphism action on $g(x)$. Therefore, the global symmetries of the WZW model match exactly with those of the free fermion theory after bosonization -- see \eqref{sym.n.odd} and \eqref{sym.n.even}. In fact we have an exact duality between the $\Spin(n)_1$ WZW model, and $n$ Majorana fermions with gauged $(-1)^F$ symmetry, as long as the $(-1)^F$ symmetry is gauged with the choice of Arf contact term specified above. 
The center of the symmetry group of the WZW model also precisely matches what we have anticipated above, with $\bZ_2^\ttv$ being part of the center always:
\begin{equation}
    Z\left( \frac{\Spin(n)_\ttL \times \Spin(n)_\ttR}{Z\left(\Spin(n)_\mathrm{diag}\right)} \right) = Z\left(\Spin(n)_\ttL\right) = \begin{cases} \bZ_2^\ttv ~~~& n=2k+1\\
    \bZ_2^\tts \times \bZ_2^\ttc ~~~& n=4k\\
    \bZ_4 ~~~& n=4k+2
    \end{cases}~,
\end{equation}

By gauging this $\bZ^\ttv_2$ symmetry with the Arf twist, we can fermionize the theory and go back to the fermionic theory with the $(-1)^F$ symmetry \cite{AlvarezGaume:1987vm, Gaiotto:2015zta, Kapustin:2017jrc,Thorngren:2018bhj,Karch:2019lnn}.
If we gauge the $\bZ^\ttv_2$ symmetry without the Arf twist, then we are choosing among the two possible bosonized theories we discussed above.
When $n$ is odd, as anticipated, the $\Spin(n)_1$ WZW model is self-dual in the sense of Kramers-Wannier, i.e.\ $\Spin(n)_1$ and $\SO(n)_1$ are equivalent conformal field theories. The actual isomorphism requires re-labeling some operators, and the  corresponding duality defect line is given by a non-invertible Verlinde line of the $\Spin(n)_1$ WZW model \cite{Chang:2018iay}. When $n$ is even, gauging $\bZ_2^\ttv$ results again in the $\SO(n)_1$ WZW model which is now not dual to the $\Spin(n)_1$ WZW model. It is now clear why we choose to bosonize with the choice of the Arf contact term as we did: the bosonic theory can be then chosen to be $\Spin(n)_1$ WZW model both for even and odd $n$, which greatly simplifies the notation.

\paragraph{The coset construction}
Now when $n=N^2-1$, we can gauge the $\SU(N)\subset\Spin(N^2-1)$\footnote{More precisely $\mathrm{PSU}(N)\subset\Spin(N^2-1)$, and the center of $\SU(N)$ does not act faithfully on the theory; see \eqref{embeddings}.} symmetry on both sides of the duality to get a duality between
\begin{equation}\label{WZWlan}
    \int \mathcal{D}g \,\mathcal{D}A \; \exp \left( -S_\mathrm{WZW}[g,A] +\int_\Sigma \rd^2 x \, \frac{1}{4g_\mathrm{YM}^2} \Tr F^2 \right)~, 
\end{equation}
and
\begin{equation}
    \sum_\r \int \mathcal{D}\psi \,\mathcal{D}A \; \exp \left(-\int_\Sigma \rd^2 x \, \left[-\frac{1}{4g_\mathrm{YM}^2} \Tr F^2 + \Tr \psi^Ti \slashed{D}_\r \psi \right]  \right) ~,
\end{equation}
where $S_\mathrm{WZW}[g,A]$ is the action of the $\Spin(N^2-1)_1/\SU(N)_N$ gauged WZW model.

Let us for a second concentrate on the piece  $S_\mathrm{WZW}[g,A]$, formally sending $g_\mathrm{YM}\to \infty$ and ignoring the kinetic term. An observation that will be useful later, is that the $\Spin(N^2-1)_1$ WZW model has the same central charge as the $\SU(N)_N$ WZW model. It means that the coset $\Spin(N^2-1)_1/\SU(N)_N$ has zero central charge and is a trivial CFT. Trivial CFT here means that all the Virasoro generators act trivially on the states of the theory, therefore the theory is a TQFT. Hence, the two theories $\Spin(N^2-1)_1$ and $\SU(N)_N$ must be the same up to a discrete generalized gauging (orbifolding) \cite{Frohlich:2009gb,Bhardwaj:2017xup}; see section \ref{sec:modular.invariants} for details. It turns out that for $N=3$ this is an ordinary gauging and $\mathrm{PSU}(3)_3 = \Spin(8)_1$ \cite{Gepner:1986wi}, while for $N\geq 4$ it is a generalized gauging of some non-invertible lines.

\section{Topological Lines in WZW Model and Adjoint QCD \label{sec:top.lines}}
In this section, first we review some aspects of topological line operators in two dimensions, which are going to be used later to study the infrared behaviour of adjoint QCD. We then show that the existence of non-invertible topological lines is responsible for the deconfinement in the massless theory, and we study the $N=2$ and $3$ cases in detail. For simplicity we focus on the bosonic theory, with the $(-1)^F$ symmetry gauged, since it turns out that the qualitative IR behaviour of the theory is the same. We elaborate on the fermionic theory in appendices \ref{app:fermionic} and \ref{app:fermioniclines}.

\subsection{Topological Lines in 2d}

\subsubsection{Fusion Categories}
As reviewed in the introduction, when a quantum field theory have a symmetry with group $G$, there is the codimension-one topological operator $U(g)$ for each element $g\in G$.
One can fuse them by bringing them close and parallel to each other.
The operation, called fusion, is represented by the symbol $\otimes$, and respects the group multiplication:
\begin{equation}
    U(g_1)\otimes U(g_2) \simeq U(g_1 g_2) \quad g_1,g_2\in G.
\end{equation}
In particular, a symmetry operator $U(g)$ always has its inverse operator $U(g)^{-1}=U(g^{-1})$:
\begin{equation}
    U(g)\otimes U(g)^{-1} \simeq 1 ~,
\end{equation}
here 1 is the invisible identity codimension-one operator.
In the rest of this paper, we write the line operator $U(g)$ (for a 0-from symmetry) as just $g$. 

A generalization of the notion of symmetry is the $q$-form symmetry, which corresponds to topological codimension-$(q+1)$ operators that are invertible and can be thought of as the exponentiation of a conserved Noether charge when the group is continuous~\cite{Gaiotto:2014kfa}.
Such $q$-form symmetries act on $q$-dimensional operator by linking, and the topological nature of them leads to the conservation of the corresponding charges assigned to the $q$-dimensional operators.
In two dimensions, $0$-form and $1$-form symmetries are associated with topological line and local operators. They act on local operators and on lines, respectively. 

In this section and most of the rest of the paper, we consider another generalization of symmetry, which we call ``category symmetry'', since it is governed by the mathematical notion of category with additional structures, instead of a group.
It is composed of all topological lines operators that the 1+1d quantum field theory has. They are not required to have inverses.
A basic example is the duality defect line in the Ising CFT implementing the Kramers-Wannier duality, which is not invertible and hence does not correspond to any group-like symmetry.
Other familiar examples of topological lines include the Verlinde lines in WZW models.

Topological lines (invertible or not) in a bosonic quantum field theory form a mathematical structure known as tensor category.
If one imposes a certain finiteness, one gets a rigid structure known as \emph{fusion category} \cite{etingof2005fusion,etingof2016tensor, Bhardwaj:2017xup}, and \emph{unitary fusion category} in a unitary theory.
This type of category symmetry should be regarded as a generalization of finite group symmetry.
The rigidity of fusion categories, known as Ocneanu rigidity \cite{etingof2005fusion}, implies that such structures do not admit continuous deformations and are invariant under the RG flow \cite{Chang:2018iay}.
Hence they can lead to very concrete and powerful constraints on the dynamics.
The fusion category includes not only the algebra of the (potentially non-invertible) lines but also the the proper generalization of 't Hooft anomaly matching conditions, as we review below.

In the remainder of the section, we give a brief introduction to fusion categories by stating some of their axioms which we use later on. For a complete list of the axioms see appendix \ref{app:axioms} (see \cite{Bhardwaj:2017xup} for a review accessible to physicists which we follow closely).
\begin{enumerate}
	\item \textbf{Lines (Objects):} The objects in a fusion category $\mathcal{C}$ correspond to oriented topological line operators that implement the symmetry. More precisely, for an oriented path $C$ and an object $a\in \mathcal{C}$, there exist the topological line operator $a(C)$ whose dependence on $C$ is topological.
	\item \textbf{Defect Operators (Morphisms):} The morphisms in $\mathcal{C}$, correspond to local topological defect operators which turn a line $a$ into another line $b$. Such defect operators\footnote{By using the folding of axiom \ref{folding}, $m\in\mathrm{Hom}(a,b)$ can be thought as a defect operator living at the end of the line $a\otimes b^\ast$. Instead one might call it a line-changing operator.} form a complex vector space denoted by $\mathrm{Hom}(a,b)$
	\begin{equation*}
    \begin{tikzpicture}[scale = 1, baseline = 0]
        \coordinate (d) at (-1.2,0);
        \coordinate (m) at (0,0);
        \coordinate (u) at (1.2,0);
        \draw[fill] (m) circle [radius = .05];
        \draw[thick, ->- = .5 rotate 0] (d) -- (m);
        \draw[thick, ->- = .6 rotate 0] (m) -- (u);
        \node[anchor = south] at (m) {$m$};
        \node[anchor = south] at ($(m)!.5!(d)$) {$a$};
        \node[anchor = south] at ($(m)!.5!(u)$) {$b$};
    \end{tikzpicture} \qquad m\in \mathrm{Hom}(a,b)~.
\end{equation*}
 	\item \textbf{Additive Structure:} Given two lines $a$ and $b$, there exist a new line given by their sum $a \oplus b$, such that $\langle \cdots (a \oplus b) (C) \cdots \rangle = \langle \cdots a(C) \cdots \rangle + \langle \cdots b(C) \cdots \rangle$.
	\item \textbf{Fusion:} Given two parallel lines $a$ and $b$, one can bring them close and consider them as a single line denoted by $a \otimes b$. The invisible identity line correspond to $\mathbbm{1} \in \mathcal{C}$, and acts as the identity element under fusion, i.e.\ $a\otimes \mathbbm{1} = \mathbbm{1} \otimes a =a$.
    \item \textbf{Simplicity, Semisimplicity, and Finiteness:} Simple lines are defined to be irreducible meaning that they cannot be decomposed as a sum of two other lines.
    Equivalently, for simple lines $a\in\mathcal{C}$,  the defect spaces $\mathrm{Hom}(a,a)$ are one-dimensional and thus isomorphic to $\mathbb{C}$.
    Fusion categories are finite and semisimple in the sense that there is a finite number of simple lines and every other line is isomorphic to a direct sum of simple lines.
    Furthermore, in fusion categories as opposed to multifusion categories, the identity line $\mathbbm{1}$ is simple.\footnote{
        The category symmetry of the $\SU(N)$ adjoint QCD actually does not satisfy this condition, because it has the topological local operators corresponding to the one-form symmetry.
        In this paper, however, we first ignore these topological local operators and analyze the topological line operators only, then, afterwards, we study the algebra between topological lines and topological local operators. In particular, one has to ignore the topological local operators in order to make sense of equation~\eqref{QD}, as in general the loop could contract to a topological local operator. We will see that it is indeed consistent to first ignore that we are dealing with a multifusion category and then add the topological local operators. The point is that we apply the tools of fusion categories to analyze the topological lines of a WZW model, where the simplicity of the trivial line is satisfied, and then we will gauge the WZW model appropriately. }
	\item \textbf{Associativity Structure:} The fusion operation is associative $(a\otimes b) \otimes c \simeq a \otimes (b \otimes c)$, and the associativity structure is a particular isomorphism
	\begin{equation}
		\alpha_{a,b,c} \in \mathrm{Hom}((a\otimes b) \otimes c, a \otimes (b \otimes c))~,
	\end{equation}
    which is called the \textit{associator}. The associators capture the data of crossing relations of lines under joining and splitting, and satisfy a consistency condition known as the pentagon identity, see \cite{Moore:1988qv,Bhardwaj:2017xup, Chang:2018iay}.
    When the symmetry is a group, the associator encodes the 't Hooft anomaly.
	
	\item \label{folding} \textbf{Dual Structure (Folding):} For any line $a$, there exist the dual line $a^\ast$ which is the orientation reversal of $a$, such that $(a^\ast)^\ast \simeq a$ and $(a\otimes b)^\ast = b^\ast \otimes a^\ast$. If $a$ is invertible, $a^* \simeq a^{-1}$. A line $a$ can be folded to form the line $a^\ast\otimes a$, and this data is captured by defect operators at the ends of the line $a^\ast\otimes a$ which are known as the evaluation and co-evaluation morphisms $\epsilon_a \in \mathrm{Hom}(a^\ast\otimes a,\mathbbm{1})$ and $\epsilon^a \in \mathrm{Hom}(\mathbbm{1},a^\ast\otimes a)$ respectively. Using these folding defect operators one can calculate the expectation value of an empty loop of $a$ by
	\begin{equation}\label{QD}
		\begin{tikzpicture}[scale = .7, baseline = -2.5]
			\draw[ ->- = .07 rotate -25] (0,0) arc (0:360:.5);
			\node[anchor = west] at (0,0) {$a$};
			\node[anchor = south] at (-0.5,0.5) {$\epsilon_a$};
			\draw[fill] (-0.5,0.5) circle [radius = .05];
			\node[anchor = south] at (-0.5,-0.5) {$\epsilon^a$};
			\draw[fill] (-0.5,-0.5) circle [radius = .05];
		\end{tikzpicture} = 
		\epsilon_a \circ \epsilon^a = (\mathrm{dim}\,a) \mathbb{I} \in \mathrm{Hom}(\mathbbm{1},\mathbbm{1})\,,
	\end{equation}
	where $\langle\,\begin{tikzpicture}[scale = 1, baseline = -2.5,>=stealth]
	\draw[->- = .055 rotate -21] (0,0) arc (0:360:.19);
	\node[anchor = west] at (0,0) {$a$};
	\end{tikzpicture}\rangle = \mathrm{dim}\,a $ is known as the quantum dimension of $a$.
	\item \textbf{Unitary Structure:} For any defect operator $m \in \mathrm{Hom}(a,b)$ which takes the line $a$ to $b$, there exist the Hermitian conjugate defect operator $m^\dagger \in \mathrm{Hom}(b,a)$ form $b$ to $a$.
\end{enumerate}
\paragraph{Action of line operators on local operators}
The above axioms describe the properties of topological line operators in a two dimensional theory. These line operators can also act on the local operators of the theory. The action of a topological line $L$ on a local operator $\mathcal{O}$ is defined by shrinking a loop of $L$ that encircles the local operator. Since the line operator is topological, shrinking the line does not change the correlation functions and the resulting configuration is equivalent as another local operator that we denote by $L \cdot \mathcal{O}$
\begin{equation}
	\begin{tikzpicture} [scale = 1, baseline = -2.5]
        \draw[thick, ->- = .02 rotate -10] (0,0) arc (0:360:0.6);
        \draw[fill] (-0.6,0) circle [radius = .05];
        \node[anchor = south] at (-0.6,0) {$\mathcal{O}$};
        \node[anchor = west] at (0,0) {$L$};
	\end{tikzpicture} =
    \begin{tikzpicture} [scale = 1, baseline = -2.5]
        \draw[fill] (0,0) circle [radius = .05];
        \node[anchor = south] at (0,0) {$L \cdot \mathcal{O}$};
	\end{tikzpicture}~. \label{lines.on.ops}
\end{equation}
A line operator $L$ is said to commute with a local operator $\mathcal{O}$, if the loop of $L$ in \eqref{lines.on.ops} can be moved through $\mathcal{O}$ without changing the correlation functions -- meaning that an empty loop of $L$ commutes with $\mathcal{O}$. Equivalently, $L$ commutes with $\mathcal{O}$ if
\begin{equation}
    L \cdot \mathcal{O} = (\mathrm{dim}\,L) \, \mathcal{O}~. \label{l.commutes.o}
\end{equation}
For instance when $L$ is invertible, the quantum dimension of $L$ is one and \eqref{l.commutes.o} reduces to $L \cdot \mathcal{O}=\mathcal{O}$. As explained in \cite{Chang:2018iay}, in a unitary theory if equation \eqref{l.commutes.o} holds, then $\mathcal{O}$ commutes with any line of $L$, and not just with an empty loop of $L$.

Generally when a non-invertible topological line $L$ passes through a local operator $\mathcal{O}$, it transforms the local operator into $L \cdot \mathcal{O}/(\mathrm{dim}\,L)$ plus a defect operator attached to the original line operator
\begin{equation}
    \begin{tikzpicture}[scale = 1, baseline = 0]
        \coordinate (o) at (0,0);
        \coordinate (d) at (0.5,-1);
        \coordinate (m) at (0.5,0);
        \coordinate (u) at (0.5,1);
        \draw[fill] (o) circle [radius = .05];
        \node[anchor = south] at (o) {$\mathcal{O}$};
        \draw[thick, ->- = .55 rotate 0] (d) -- (u);
        \node[anchor = west] at (m) {$L$};
    \end{tikzpicture} = \frac{1}{\mathrm{dim}\,L} \left(
    \begin{tikzpicture}[scale = 1, baseline = 0]
        \coordinate (o) at (0.7,0);
        \coordinate (d) at (0,-1);
        \coordinate (m) at (0,0);
        \coordinate (u) at (0,1);
        \draw[fill] (o) circle [radius = .05];
        \node[anchor = south] at (o) {${L \cdot \mathcal{O}}$};
        \draw[thick, ->- = .55 rotate 0] (d) -- (u);
        \node[anchor = east] at (m) {$L$};
    \end{tikzpicture} \right) +
    \begin{tikzpicture}[scale = 1, baseline = 0]
        \coordinate (o) at (1,0);
        \coordinate (d) at (0,-1);
        \coordinate (m) at (0,0);
        \coordinate (u) at (0,1);
        \draw[fill] (o) circle [radius = .05];
        \node[anchor = south] at (o) {$\mathcal{O}_{\mathrm{defect}}$};
        \draw[thick, ->- = .55 rotate 0] (d) -- (m);
        \draw[thick, ->- = .55 rotate 0] (m) -- (u);
        \draw[->- = .5 rotate 0,dashed] (m) -- (o);
        \node[anchor = east] at ($(m)!.5!(d)$) {$L$};
        \node[anchor = east] at ($(m)!.5!(u)$) {$L$};
    \end{tikzpicture}~. \label{non.inv.lines}
\end{equation}
Note that the dashed line in \eqref{non.inv.lines} that attaches the defect operator $\mathcal{O}_{\mathrm{defect}}$ to $L$, can be a non-simple (composite) line but it does not include the identity line in its decomposition into simple lines.\footnote{Note that when $L$ is invertible, the second term in \eqref{non.inv.lines} vanishes.} Therefore, when the line $L$ is a closed loop, the last term in \eqref{non.inv.lines} becomes a tadpole which vanishes and we recover equation \eqref{lines.on.ops}. The vanishing of tadpole was proved in \cite{Chang:2018iay} by assuming that all the topological lines act on the local operators faithfully. Furthermore, assuming the vanishing tadpole property and unitarity, it was shown in \cite{Chang:2018iay} that when \eqref{l.commutes.o} holds and $L$ commutes with $\mathcal{O}$, then the second term on the RHS of \eqref{non.inv.lines} vanishes. Therefore, if an empty loop of $L$ commutes with $\mathcal{O}$, then any line of $L$ has to commute with $\mathcal{O}$. This property will be important when we study the deformations of adjoint QCD in section \ref{sec:deformation}.

\paragraph{Selection rules and naturalness} Category symmetries like the ordinary group symmetries lead to Ward identities and selection rules on the amplitudes. These new selection rules violate the ordinary notion of naturalness.

Starting with a QFT in the UV with category symmetry $\cC$, symmetric deformation cannot break the category symmetry. In other words, an operator $\cO$ transforming non-trivially under a topological line $L\in\cC$ cannot be generated by radiative quantum corrections along the RG flow. To show this, we consider the simple situation where
\begin{equation}
    L\cdot \cO = \l \cO \neq \braket{L}\cO~.
\end{equation}
All the amplitudes in the theory with an insertion of a single operator $\cO$ (and no other non-invariant operators) must vanish. Consider the sphere amplitude $\braket{\cO(x_0) \phi_1(x_1) \cdots \phi_n(x_n)}_{S^2}$
where all the operators $\phi_i$ commute with $L$, i.e.\ $L\cdot \phi_i = \braket{L} \phi_i$. Now contracting a loop of $L$ enclosing the points $x_0,x_2, \dots, x_n\in S^2$, we can shrink the loop in two ways. If we shrink the loop on the operators we only get a factor of $\l$ when shrinking it on $\cO$ (note that this only works when there is a single insertion of $\cO$), but shrinking it on the other side of the sphere we get the quantum dimension $\braket{L}$. Hence
\begin{equation}
    \left( \l - \braket{L} \right) \braket{\cO(x_0) \phi_1(x_1) \cdots \phi_n(x_n)}_{S^2} = 0~.
\end{equation}
This shows that all the amplitudes involving $\cO$ must vanish unless $\cO$ commutes with $L$. Hence, the operator $\cO$ breaking any topological line will not be generated by the radiative corrections, even if it is preserved by all the ordinary invertible symmetries. This leads to a violation of the ordinary notion of naturalness. We conclude this subsection with two examples of fusion categories, namely the Verlinde lines in WZW models and group-like symmetries with 't Hooft anomalies.
\paragraph{Examples}
The Verlinde lines of a $G_k$ WZW model with a simply connected Lie group $G$, are by definition the topological lines that preserve the affine Lie algebra $\hat{\mathfrak{g}}_k$ of $G$ at level $k$.
These lines correspond to the integrable highest-weight representations of $\hat{\mathfrak{g}}_k$.
Although the group $G$ is not a finite group, since it has only finitely many integrable irreducible highest-weight representations at fixed $k$, it satisfies all the above axioms and forms a fusion category, which we denote by $\mathrm{Rep}\,\hat{\mathfrak{g}}_k$.\footnote{The WZW model contains other topological lines that are the ordinary symmetry lines of left- and right-$G$ symmetry, which are not included in the Verlinde lines.} The example of $\mathrm{Rep}\,\hat{\mathfrak{g}}_k$ is special in the sense that it admits braiding, which makes the fusion ring of the category commutative.
Braiding can be thought of as a commutative structure on the fusion category.
In a braided fusion category one can define modular $S$ and $T$ matrices similar to that of a CFT, and when the $S$-matrix is nondegenerate -- as is the physically relevant situations -- the resulting category is known as a \textit{modular tensor category}, or a \textit{modular category} for short \cite{Moore:1988qv,Kirillov:2001ti,Kitaev:2006lla,etingof2016tensor}. A useful way of thinking about such categories, as emphasized in the introduction, is in the framework of 3d TQFTs, in which the Wilson lines form such a modular category with a non-degenerate $S$-matrix.
For instance, $\mathrm{Rep} \,\hat{\mathfrak{g}}_k$ describes the Wilson lines of 3d $G_k$ Chern-Simons theory.
By the correspondence between 3d TQFTs and 2d RCFTs~\cite{Witten:1988hf,Moore:1989yh,Fuchs:2002cm,Kapustin:2010if} (see section \ref{sec:intro}), such a category describes the category of representations in a chiral RCFT.

Group symmetries are generated by topological lines which are invertible.
In such a fusion category, simple lines correspond to the group elements, and their fusion is given by the group multiplication law.
For three simple lines $g_{1,2,3}$, we have the isomorphism associator $\alpha_{g_1,g_2,g_3} \in \mathrm{Hom}(g_1g_2g_3, g_1g_2g_3)$, and since the line $g_1g_2g_3$ is simple its associator is just a $\U(1)$ phase.
The pentagon identity implies that the associators define a 3-cocycle $\alpha \in H^3(G,U(1))$, which turns out to be equivalent to the 't Hooft anomaly of the symmetry \cite{Bhardwaj:2017xup}.\footnote{In the mathematical literature this is often denoted by $\mathrm{Vec}_G^\alpha$ -- the category of $G$-graded vector spaces with associator $\alpha$. \label{VecG}} (The connection between anomalies and cocycles for ordinary symmetries goes back to~\cite{Faddeev:1985iz}.)
In the other direction, every fusion category whose simple lines are invertible is equivalent to such a group symmetry with 't Hooft anomaly.

For the case of invertible lines the corresponding fusion category includes both the information of the group and its 't Hooft anomaly.
For a group symmetry, we can gauge a non-anomalous subgroup of the group.
The notion of gauging is generalized into category symmetry case in \cite{Frohlich:2009gb, Carqueville:2012dk}, and reviewed in \cite{Brunner:2013xna,Bhardwaj:2017xup}.
The precise generalization of ``non-anomalous subpart" in category symmetry case is called symmetric Frobenius algebra.
It often happens that the whole category symmetry cannot be gauged. In such a case, analogous to the anomalous group symmetry, the theory cannot be gapped with single vacuum as we will see.

In the following subsections (\ref{sec:gauging}, \ref{sec:modular.invariants} and \ref{sec:3d}) we will discuss gauging of category symmetries, TQFTs with category symmetries, and modular invariants of $\SU(N)_N$. As we will see all these concepts are very closely related and there is a one-to-one correspondence between them. These correspondences are all summarized in table \ref{tab:correspondences}, and the corresponding category theory notions in table \ref{tab:notations}.
\begin{table}[t]
\centering
\begin{tabular}{|c|l|c|}
\hline
$\cC$ & \multicolumn{1}{c|}{Physical concepts}                                     & Notation      \\ \hline
\multirow{3}{*}{\begin{tabular}[c]{@{}c@{}}For a\\ fusion category $\cC$\end{tabular}} &
  Ways of gauging a non-anomalous subpart of $\cC$ &
  $A$ \\ \cline{2-3} 
      & 2d TQFTs with symmetry category $\cC$                                      & $T_\ca$       \\ \cline{2-3} 
      & 2d TQFTs with symmetry category $_B\cC_B$ & $T_{_B\cC_A}$ \\ \hline
\multirow{2}{*}{\begin{tabular}[c]{@{}c@{}}If $\cC$ is also\\ modular\end{tabular}} &
  2d RCFTs with chiral algebra $\mathcal{V}$ such that $\mathrm{Rep} \mathcal{V}=\cC$ &
  $\mathcal{T}_A$ \\ \cline{2-3} 
      & Surface operators of the 3d TQFT associated with $\cC$                     & $S_A$         \\ \hline
\end{tabular}
\caption{The physical concepts that are in one-to-one correspondence. Given a modular tensor category $\cC$, every physical concept in the above table is uniquely determined by the Morita equivalence class of a symmetric Frobenius algebra $A$ in $\cC$. When $\cC$ is only a fusion category, only the first two rows are in one-to-one correspondence. In the third row, $B$ denotes a way of gauging a subpart of $\cC$.}
\label{tab:correspondences}
\end{table}
\begin{table}[t]
\centering
\begin{tabular}{|c|c|c|}
\hline
Physical concepts & Categorical notions & Notation \\ \hline
Topological lines of the ungauged theory & Fusion Category & $\cC$ \\
Ways of gauging a subpart of $\cC$ & Frobenius algebras in $\cC$ & $A$ \\
Topological boundary conditions & $\cC$-module categories & $\ca$ \\
Topological lines of the gauged theory & Dual fusion category & $\aca$ \\ \hline
\end{tabular}
\caption{Fusion category theory notions and the corresponding physical concepts.}
\label{tab:notations}
\end{table}
\subsubsection{Gauging Symmetries and Symmetric TQFTs \label{sec:gauging}}
The purpose of this section is to discuss the generalized notion of gauging in a category symmetry $\cC$, and its relation with gapped phases with that category symmetry which we call $\cC$-symmetric TQFTs. This turns out to be crucial for studying the adjoint QCD. This is because, as we will see, the topological lines of adjoint QCD can be obtained as a gauging of the Verlinde lines of the $\SU(N)_N$ WZW model. Thus to understand the topological lines of adjoint QCD, one has to understand the gauging procedure. Furthermore, the possible gapped phases of adjoint QCD are restricted by the existence of these topological lines, and such possibilities are in one-to-one correspondence with different ways of gauging a subpart of these lines. We begin by first reviewing the gauging for ordinary group symmetries.

\paragraph{Gauging.}
For a 2d theory $T$ with a discrete group symmetry $G$, a non-anomalous subgroup of it such as $H\subset G$ can be gauged. To gauge the symmetry, one has couple the theory to a background discrete $H$ gauge fields. Turning on these gauge fields is equivalent to inserting topological lines around the non-trivial cycles of the 2d spacetime manifold $M$. Then, the gauging operation is done by summing over all the gauge field configurations, or equivalently summing over all the gauge-inequivalent insertions of different topological lines in $H$. Alternatively, one could consider the line $A = \bigoplus_{h\in H} h$ given by the sum of the topological lines in $H$. Then summing over all the insertions is equivalent to inserting a fine-enough trivalent mesh of $A$ into the path integral \cite{Carqueville:2012dk,Brunner:2013xna,Brunner:2014lua,Bhardwaj:2017xup}. Note that, one also has to fix a choice of defect operator in the three-way junctions of $A$ in the mesh, which is equivalent to a choice of phase coefficients when summing over different insertions. Thus the gauging is determined by $A$ and the choice of defect operator, that together form an algebra as discussed in appendix \ref{app:gauging}. This procedure can be easily generalized to category symmetries. 

More precisely, for a category symmetry $\cC$, gauging a non-anomalous subpart of $\cC$ is done by inserting a fine mesh of a \emph{symmetric Frobenius algebra} object $A$ in $\cC$ into the path integral. For a theory $T$ with a category symmetry $\cC$, we denote the gauged theory by $T/A$ where $A$ denotes the gauging. In the rest, we do not need to know the details of $A$ as an algebra, and we simply refer to $A$ as a way of gauging. All we need to know is the following theorem whose proof is given in appendix \ref{app:module.categories}.
\begin{theorem}\label{theorem1}
    For a category symmetry $\cC$, different ways of gauging a non-anomalous subpart of $\cC$ are in one-to-one correspondence with different $\cC$-symmetric TQFTs. For a way of gauging $\cC$ such as $A$, the corresponding $\cC$-symmetric TQFT is denoted by $T_{\cC_A}$.
\end{theorem}

The physical interpretation of this statement in our context is  as follows: one begins with the standard diagonal $\SU(N)_N$ WZW model. It has a collection of Verlinde lines $\cC$. If we gauge a non-anomalous subpart $A$ of $\cC$ we obtain a non-diagonal WZW model based on the same affine Lie-algebra $\hat{\mathfrak{su}}(N)_N$~\cite{Frohlich:2009gb}. This will be further reviewed in \ref{sec:modular.invariants}. These two models have the same central charge and hence the quotient of these two models is given by a 1+1 dimensional TQFT. The crucial point is that after we have gauged the subpart $A$, the topological lines of the resulting non-diagonal theory are not the same as the original Verlinde lines. Thus we have to discuss the lines of the theory after gauging to understand the symmetries of QCD.

\paragraph{Topological lines of the gauged theory.} A natural question to ask is the description of the topological lines of a theory after gauging. This is well known for the case of finite Abelian group symmetries. For instance, starting with a theory with a $\mathbb{Z}_n$ symmetry and gauging it, the gauged theory is known to have a new $\mathbb{Z}_n$ symmetry \cite{Vafa:1989ih}, which is also known as the quantum symmetry.

As explained in \cite{Bhardwaj:2017xup}, starting with a theory with category symmetry $\mathcal{C}$ and gauging a subpart of it associated with $A$, the gauged theory will have a new quantum symmetry that we denote by $\aca$, and call it the gauging of $\mathcal{C}$ with respect to $A$.\footnote{More precisely, $\aca$ is the category of $(A,A)$-bimodules in $\cC$.} This finite gauging is an invertible and associative operation.\footnote{Fusion categories $\mathcal{C}$ and $\aca$ that are related by gauging are called \textit{categorically Morita equivalent}.} More precisely, if $\mathcal{C}'=\aca$ is a gauging of $\mathcal{C}$ with respect to $A$ and $\mathcal{C}''={}_{A'}{\mathcal{C}'}_{A'}$ is a gauging of $\mathcal{C}'$ with $A'$, then $\mathcal{C}''$ is also a gauging of $\mathcal{C}$ with respect to the combined gauging operation that we denote by $A' \circ A$. Furthermore, if $\mathcal{C}'=\aca$ is a gauging of $\mathcal{C}$, then $\mathcal{C}$ is also a gauging of $\mathcal{C}'$ by a symmetric Frobenius algebra object that can be called $A^{-1}$, i.e.\ $A^{-1} \circ A = A \circ A^{-1} = 1$.

Now we can give a construction for the $\cC$-symmetric TQFT $T_{\cC_A}$ of theorem \ref{theorem1}. This TQFTs can be constructed by starting with a canonical $\aca$-symmetric TQFT, and then gauging a non-anomalous subpart of its $\cC$ symmetry by $A^{-1}$. For details and a definition of $T_{\cC}$ see appendix \ref{app:module.categories}. In particular, the number of vacua in $T_{\cC}$ is the same as the number of simple lines in $\cC$. But here we only need to know that for the case of $\SU(N)_N$ Verlinde lines, this TQFT is the $\SU(N)_N/\SU(N)_N$ coset.

For category symmetries $\cC$ and $\aca$ that are related by gauging, it is easy to see that different choices of gauging $\aca$ are in one-to-one correspondence with choices of gauging $\cC$. In particular, given $B$ a choice of gauging $\cC$, there exist a choice of gauging $\aca$ by $B \circ A^{-1}$. We first gauge $\aca$ back to $\cC$ by $A^{-1}$, and then gauge $\cC$ with $B$ to get the combined gauging operation $B \circ A^{-1}$
\begin{equation*}
\begin{tikzcd}
&\cC  \ar[rd,"B"] \\
\aca \ar[,ru,"A^{-1}"]  \ar[rr,"B \circ A^{-1}"] && _B\cC_B
\end{tikzcd}
\end{equation*}
According to theorem \ref{theorem1}, there must exist an $\aca$-symmetric TQFT associated with the gauging $B \circ A^{-1}$. This TQFT is given by starting with the canonical $\bcb$-symmetric TQFT $T_{\bcb}$, and then gauging a non-anomalous subpart of its $\bcb$ symmetry by $A\circ B^{-1}$. We denote the resulting TQFT by $T_{\acb}$.\footnote{The theory $T_{\acb}$ also has a right $\bcb$ symmetry; see appendix \ref{app:module.categories} for detail.} In section \ref{sec:modular.invariants} we describe a practical way of analyzing the theory $T_{\acb}$, for when $\mathcal{C}$ is the Verlinde lines of $\SU(N)_N$ or more generally when it is a modular tensor category.
\subsubsection{Topological lines in RCFTs and Modular Invariants \label{sec:modular.invariants}}
Here we review the topological lines in RCFTs. An RCFT contains a chiral vertex algebra $\mathcal{V}$, and has a modular invariant for this chiral algebra. The chiral aspects of the theory are captured by the representations of the chiral algebra which form a modular category $\mathrm{Rep}\mathcal{V}$, also known as the Moore-Seiberg data \cite{Moore:1988ss,Moore:1988qv}. To describe the full theory one needs a choice of a modular invariant matrix $\mim_{\mu\nu}$ that gives the torus partition function 
\begin{equation}
	Z (\tau,\bar{\tau}) = \sum_{\mu,\nu} \mim_{\mu\nu} \chi_\mu(\tau) \bar{\chi}_\nu(\bar{\tau})~, \label{partition function}
\end{equation}
where $\chi_\mu(\tau)$ is the character of the irreducible representation $\module_\mu$ of the chiral algebra. As emphasized above, there could be multiple choices of the $\mim_{\mu\nu}$ that lead to consistent theories.

In the diagonal modular invariant theory we have $\mim_{\mu\nu}=\delta_{\mu\nu}$, and $\mathrm{Rep}\mathcal{V}$ describes many important aspects of the theory, e.g. its local operators, OPE structure constants, and line operators~\cite{Fuchs:2002cm}. For instance, in the $G_k$ WZW model -- that is the diagonal modular invariant RCFT associated with chiral algebra $\hat{\mathfrak{g}}_k$ -- the representations of the chiral algebra form the Verlinde modular category $\mathrm{Rep}\,\hat{\mathfrak{g}}_k$. The Verlinde modular category describes both the primary operators of this theory, and the Verlinde lines -- which are the topological line operators that commute with the chiral algebra $\hat{\mathfrak{g}}_k$ \cite{Verlinde:1988sn}.

A relevant question to us is the study of non-diagonal modular invariant RCFTs, and their topological line operators. The essential observation made in \cite{Kirillov:2001ti,Fuchs:2002cm,Frohlich:2009gb} is that physical modular invariant RCFT with chiral algebra $\mathcal{V}$, are in one-to-one correspondence with distinct gauging (generalized orbifolding) of $\mathrm{Rep}\mathcal{V}$ with some algebra (what we previously called ``subpart'') $A$. The topological lines of the theory that commute with $\mathcal{V}$ are not the same before and after gauging $A$. Before gauging these are just the Verlinde lines $\mathrm{Rep}\mathcal{V}$ and after gauging we denote them by  ${}_A\left(\mathrm{Rep}\mathcal{V}\right)_A$\cite{Fuchs:2002cm}.

Therefore instead of finding distinct ways of gauging the category symmetry $\mathrm{Rep}\mathcal{V}$ directly, one can first try to classify physical modular invariant matrices $\mim_{\mu\nu}$ associated with a non-diagonal RCFT with chiral algebra $\mathcal{V}$.\footnote{Note that a non-diagonal RCFT cannot necessarily be uniquely determined by just the modular invariant matrix $\mim$. In particular, different theories can share the same modular invariant matrix (same bulk states) but different boundary states (see \ref{app:su3}). Nonetheless $\min$ packages many of the necessary information and is easier to handle hand the data of Frobenius algebra itself.}
The Hilbert space on a circle of such a theory decomposes as
\begin{equation}
	\mathcal{H} = \bigoplus_{\mu,\nu} \mim_{\mu\nu} \, \module_\mu \otimes \bar{\module}_\nu~,
\end{equation}
where $\mim_{\mu\nu}$ is the degeneracy of the states in the holomorphic $\otimes$ antiholomorphic module $\module_\mu\otimes\bar{\module}_\nu$. In particular $\mim_{\mu\nu}$ must have nonnegative integer entries, commute with the modular $S$ and $T$ matrices of $\mathcal{C}$, and $\mim_{00}=1$ reflecting a non-degenerate single vacuum. More precisely, a physical modular invariant RCFT corresponds to left and right extensions of the chiral algebra and an isomorphism between their modular category of representations \cite{Moore:1988ss}. In the general case the torus partition function takes the form
\begin{equation}
Z = \sum_{a} \big( b_{a\mu} \, \chi_\mu \big) \big( \bar{b}_{a\nu} \, \bar{\chi}_\nu \big)~,
\end{equation}
so that $\Module_a = \bigoplus_\mu b_{a\mu} \,\module_\mu$ and $\bar{\Module}_a = \bigoplus_\nu \bar{b}_{a\nu}\,\bar{\module}_\nu$ are the irreps of the extended chiral algebras. Note when $\bar{b}_{a\mu}=b_{a\mu}$, such modular invariants are known as type I.\footnote{For type I invariants, the vacuum block corresponds to the algebra object $A$ that we gauge, i.e.\ $\module_A=\bigoplus_\mu b_{0\mu}\, \module_\mu$ \cite{Fuchs:2002cm}.} But in general $\bar{b}_{a\mu}$ and $b_{a\mu}$ are related by a permutation $\omega$, i.e.\ $\bar{b}_{a\mu}=b_{\omega(a)\mu}$ \cite{Bockenhauer:1999vv}. Finding modular invariant matrices, although is an open problem in the general case, is an easier and more tractable task than finding Frobenius algebras in $\mathcal{C}$.

Finally, we describe a practical way of finding TQFTs with the category symmetry $\aca$ where $\mathcal{C}=\mathrm{Rep}\mathcal{V}$ for some chiral algebra $\mathcal{V}$. Since $\aca$ is a gauging of $\cC$ by $A$, $\aca$-symmetric TQFTs are in one-to-one correspondence with $\cC$-symmetric TQFTs. More precisely, given a $\cC$-symmetric TQFT such as $T_{\cC_B}$, we can gauge its $\cC$ symmetry by $A$ to get the $\aca$-symmetric TQFT $T_{\acb}$, and vice versa. Since $\mathcal{C}$ is braided, there exist a map form the lines of $\cC$ into the lines of $\aca$ that preserves their fusions and crossing relations.\footnote{This map, or more precisely tensor functor, is known as $\alpha$-induction \cite{Bockenhauer:1999wt, ostrik2003module}.} Thus any $\aca$-symmetric theory is also $\cC$-symmetric.
In particular, the theory $T_{{}_A\cC_B}$ is also a $\mathcal{C}$-symmetric TQFT denoted by $T_{\cC_{A^{\mathrm{op}}\otimes B}}$. Note that ${A^{\mathrm{op}}\otimes B}$ corresponds to a gauging of $\cC$, or equivalently a modular invariant matrix of $\mathcal{V}$ that we denote by $\mim(A^{\mathrm{op}}\otimes B)$.
The operation $\otimes$ is different from $\circ$ and, as explained in appendix \ref{app:modular.invariants}, defines a fusion between algebras that is given by taking the product of the corresponding modular invariant matrices
\begin{align}
	\mim\left(A \otimes B\right) &= \mim\left(A\right) \mim\left(B\right)~, \label{product.of.algebras} \\
	\mim\left(A \oplus B\right) &= \mim\left(A\right)+ \mim\left(B\right)~,\\
	\mim\left(A^\mathrm{op}\right) &= \mim\left(A\right)^\mathsf{T}~.
\end{align}
Note that gauging a theory $T$ with $A\oplus B$ results in a sum of two decoupled theory given by $T/(A\otimes B) = T/A \oplus T/B$. With the formulae given above we can count the number of vacua in $\cC$-symmetric TQFTs and also the number of simple lines in $\aca$ as
\begin{align}
	\text{Number of vacua in $T_{\cC_A}$} &= \Tr \left[ \mim\left(A\right) \right]~, 
	\label{dim.of.ca}\\
	\text{Number of simple lines in $\aca$} &= \Tr \left[ \mim\left(A\right)^\mathsf{T} \mim\left(A\right) \right]~. 	\label{dim.of.aca}\\
	\text{Number of vacua in $T_{\acb}$} &= \Tr \left[ \mim\left(A\right)^\mathsf{T} \mim\left(B\right) \right]~. \label{dim.of.acb}
\end{align}

\subsubsection{3d TQFT and \texorpdfstring{$\a$}{Alpha}-Induction}
\label{sec:3d}
In the introduction we briefly reviewed the correspondence between 2d modular invariant RCFTs and 3d CS theories on an interval. In the previous section, we also saw that modular invariant RCFTs are in one-to-one correspondence with algebra objects $A$. In 3d, these RCFTs correspond to the insertion of different surface operators. The diagonal theory (trivial $A$) corresponds to inserting a trivial surface, i.e.\ no surface at all. Moreover, surface operators in the 3d TQFT form a 2-category \cite{Kapustin:2010if, Fuchs:2012dt, Carqueville:2017ono, Douglas:2018fusion2cat} and the fusion between algebras defined in \eqref{product.of.algebras} can also be understood from the 3d point of view. For instance, the product \eqref{product.of.algebras}, can be understood as the fusion of surface operators in the 3d TQFT.

The objects of the 2-category are surface operators whose 1-morphisms are line operators sitting at the junction of two surface operators, mapping one surface to another. The 2-morphisms (morphisms between 1-morphisms) are local operators sitting at the junction of two line operators. The bulk local operators in the 3d TQFT are the 2-morphisms of the trivial line. Similarly the bulk line operators are the 1-morphisms of the trivial surface. 

The correspondence between 2d RCFTs and 3d TQFTs provides an intuitive way to understand the $\a$-induction map which is one of the main tools for analysing the topological lines of adjoint QCD. The purpose of the current section is to use the 3d-2d correspondence and give an intuitive proof of some properties of the $\a$-induction used in the rest of this note.    

Consider a CS on $\Sigma_2 \times I$ theory with gauge group $G$ at level $k$, where $I$ is the interval $\left[0,1\right]$. At the boundary we impose boundary conditions $\mathcal{B}$ and $\overline{\mathcal{B}}$ where the latter is the conjugate of the former. We denote the surface that corresponds to the RCFT with modular invariant matrix $\mim(A)$ by $S_A$.\footnote{In the language of \cite{Gaiotto:2019xmp,Johnson-Freyd:2020usu}, $S_A$ is the categorical condensate of $A$.} The lines $L^\text{CS}_\mu$ of the 3d CS theory form the modular category $\cC=\Rep \hat{\mathfrak{g}}_k$ which we briefly review in appendix \ref{app:CS}. In the case where $S_A$ is trivial the corresponding 2d RCFT is the diagonal theory. Starting from a diagonally modular invariant RCFT whose category symmetry is $\cC$, we can gauge an algebra object $A$ and construct a new modular invariant RCFT whose category symmetry is $\aca$. 

In the 3d TQFT this corresponds to inserting a surface operator $S_A$. We can then insert bulk line operators on either sides of the surface $S_A$ leading to two different maps $ \alpha^\pm: \mathcal{C} \to \aca$ (see figure \ref{fig:KS}).
For simplicity, we denote the image of a line $L^\text{CS}_\mu$ by just $\a^\pm_\m$. \footnote{The lines $\alpha^\pm_\mu$ are possibly equivalent to each other -- for instance if the surface operator is trivial, the lines coincide as one can move the Wilson line freely in the bulk.}
We can combine the two maps and consider the more general lines $\a^+_\m \otimes \a^-_{\bar{\n}}$. In general, the lines $\a^+_\m \otimes \a^-_{\bar{\n}}$ are not simple but rather a sum of simple lines. It was claimed in \cite{ostrik2003module} and proved in \cite{Frohlich:2006ch} that the image of the $\a$-induction is the whole $\aca$. More precisely, any simple line of $\aca$ appears in the decomposition of the line $\a^+_\m \otimes \a^-_{\bar{\n}}$ into simple lines, for some $\m,\bar{\n}\in\cC$.

The main goal of this section is to understand how to decompose the lines $\a^+_\m \otimes \a^-_{\bar{\n}}$ into simple lines of  $\aca$. For this purpose we define the ``inner product'':
\begin{equation}
    \< \a^+_\m \otimes \a^-_{\bar{\m} }, \a^+_{\n} \otimes \a^-_{\bar{\n}}\>= \text{Dim Hom}\left( \a^+_\m \otimes \a^-_{\bar{\m}}, \a^+_{\n} \otimes \a^-_{\bar{\n}}\right)~.
\end{equation}
When the self-product $\< \a^+_\m \otimes \a^-_{\bar{\m} }, \a^+_{\m} \otimes \a^-_{\bar{\m} }\>$ is 1 it means that line $\a^+_\m \otimes \a^-_{\bar{\m} }$ is a single simple line. When it is bigger that 1, it means the line $\a^+_\m \otimes \a^-_{\bar{\m} }$ contains several simple lines. In general, $\a^+_\m \otimes \a^-_{\bar{\m} }$ and $\a^+_{\n}  \otimes \a^-_{\bar{\n} }$ might have common simple lines, which is exactly what the inner product measures.

\begin{figure}[t]
\centering
\hspace{-3cm}
\begin{subfigure}[b]{0.3\textwidth}
    \centering
         \begin{tikzpicture}
            \draw[thick] (-1,-1.5) -- (-1,1.5);
            \draw[thick] (1,-1.5) -- (1,1.5);
            \draw[thick, xred] (0,-1.5)  -- (0,1.5);
            \node[anchor = north] at (-1,-1.5) {$\mathcal{B}$};
            \node[anchor = north] at (1,-1.5) {$\overline{\mathcal{B}}$};
            \node[anchor = north] at (0,-1.5) {$S_A$};
            \draw[->- = .5 rotate 0] (-0.5,-1) -- (-0.5,0);
            \draw[->- = .65 rotate 0,dashed] (-0.5,0) -- (-0.5,1);
            \draw[->- = .65 rotate 0] (-0.5,0) -- (0,0);
            \draw[->- = .5 rotate 0,dashed] (0.5,-1) -- (0.5,0);
            \draw[->- = .65 rotate 0] (0.5,0) -- (0.5,1);
            \draw[->- = .65 rotate 0] (0,0) -- (0.5,0);
            \draw[fill] (0,0) circle [radius = .05];
            \draw[fill] (0.5,0) circle [radius = .05];
            \draw[fill] (-0.5,0) circle [radius = .05];
            \node[anchor = north] at (-0.5,-1) {$\m$};
            \node[anchor = north] at (0.5,-1) {$\mathbbm{1}$};
            \node[anchor = north] at (-0.25,0) {$\m$};
            \node[anchor = north] at (0.25,0) {$\bar{\m}$};
            \node[anchor = north] at (-0.5,1.5) {$\mathbbm{1}$};
            \node[anchor = north] at (0.5,1.5) {$\bar{\m}$};
        \end{tikzpicture}
\caption{Property 1}
\label{fig:prop1}
\end{subfigure}
\hspace{1cm}
\begin{subfigure}[b]{0.3\textwidth}
    \centering
         \begin{tikzpicture}
            \draw[thick] (-1,-1.5) -- (-1,1.5);
            \draw[thick] (1,-1.5) -- (1,1.5);
            \draw[thick, xred] (0,-1.5)  -- (0,1.5);
            \node[anchor = north] at (-1,-1.5) {$\mathcal{B}$};
            \node[anchor = north] at (1,-1.5) {$\overline{\mathcal{B}}$};
            \node[anchor = north] at (0,-1.5) {$S_A$};
            \draw[->- = .5 rotate 0] (-0.5,-1) -- (-0.5,0);
            \draw[->- = .65 rotate 0] (-0.5,0) -- (-0.5,1);
            \draw[fill] (-0.5,0) circle [radius = .05];
            \node[anchor = north] at (-0.5,-1) {$\m$};
            \node[anchor = north] at (-0.5,1.5) {$\n$};

            \node[anchor = north] at (2,0) {$=$};
            
            \draw[thick] (3,-1.5) -- (3,1.5);
            \draw[thick] (5,-1.5) -- (5,1.5);
            \draw[thick, xred] (4,-1.5)  -- (4,1.5);
            \node[anchor = north] at (3,-1.5) {$\mathcal{B}$};
            \node[anchor = north] at (5,-1.5) {$\overline{\mathcal{B}}$};
            \node[anchor = north] at (4,-1.5) {$S_A$};
            \draw[->- = .5 rotate 0] (3.2,-1) .. controls (3.20,-0.45) .. (3.5,0);
            \draw[->- = 0.7 rotate 0] (3.5,0) .. controls (3.80,-0.45) .. (3.80,-1) ;
            \draw[->- = .65 rotate 0,dashed] (3.5,0) -- (3.5,1);
            \draw[fill] (3.5,0) circle [radius = .05];
            \node[anchor = north] at (3.2,-1) {$\m$};
            \node[anchor = north] at (3.80,-1) {$\n$};
             \node[anchor = south] at (3.5,1) {$\mathbbm{1}$};
        \end{tikzpicture}
\caption{Property 2}
\label{fig:prop2}
\end{subfigure}
\caption{Pictorial proofs of the properties of the $\a$-induction.}
\end{figure}
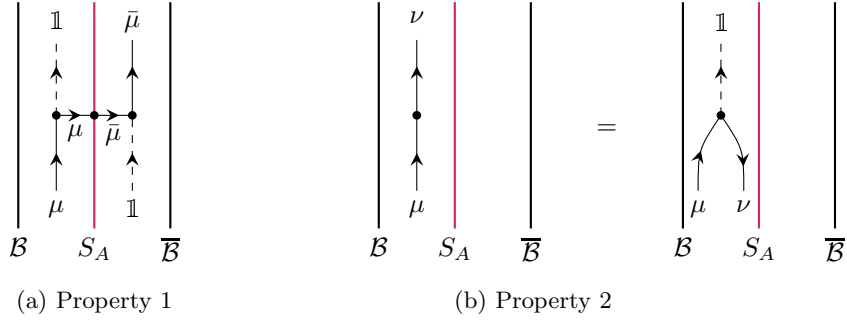

This product has two very important properties \cite{Bockenhauer:1999wt, ostrik2003module}
\begin{enumerate}
    \item  $\< \a^+_\m ,\a^-_{\bar{\m} }\>=\mim_{\m {\bar{\m} }}$~, \label{prop1}
    \item $\< \a^+_\m \otimes\a^-_{\bar{\m} }, \a^+_{\n } \otimes \a^-_{\bar{\n} }\>=\< \a^+_\m \otimes \a^+_{\n^*}, \a^-_{{\bar{\m} }^*} \otimes \a^-_{{\bar{\n} }}\>$~, \label{prop2}
\end{enumerate}
where $\mim_{\m{\bar{\m} }}$ is the modular invariant matrix \eqref{partition function} of the corresponding RCFT, and $\m^*$ is the conjugate representation of $\m$. Given these two properties one can easily calculate any product from
\begin{align}
     \< \a^+_\m \otimes \a^-_{\bar{\m} }, \a^+_{\n } \otimes \a^-_{\bar{\n} }\>  = \sum_{\r {\bar{\r} }} N_{\m \r} ^{\n} N_{\bar{\n} \bar{\r}} ^{\bar{\m} } \mim_{\r \bar{\r}}~.
     \label{eq:alphainner}
\end{align}
 
The above properties can be given an intuitive 3d proof. For property \ref{prop1}, as shown in figure \ref{fig:prop1} the number of morphisms between $\a^+_\m$ (line on the left of $S_A$) and $\a^-_{\bar{\m}}$ (line on the right of $S_A$) is the equal to the number of morphisms between lines $\m$ and $\bar{\m}$ meeting at $S_A$. This is exactly the number of primary operators of the RCFT transforming in the $(\m,\bar{\m})$ representation (see section \ref{sec:intro}) which is equal to $\mim_{\m\bar{\m}}$. For property \ref{prop2}, consider for simplicity $\< \a^+_\m , \a^+_{\bar{\m}} \>$. As shown in figure \ref{fig:prop2}, we can deform the picture on the left to the picture on the right, which implies that the number of morphisms from $\m$ to $\bar{\m}$ is equal to the number of morphisms from $\m \otimes \bar{\m}^* $ (it is $\bar{\m}^*$ because the arrow is outgoing) to the identity. This translates into the analogous property for the images of the $\a$-induction.

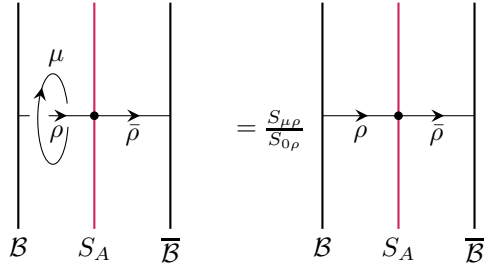
\begin{figure}[ht]
     \centering
         \begin{tikzpicture}
            \node[anchor = north] at (-1,-1.5) {$\mathcal{B}$};
            \node[anchor = north] at (1,-1.5) {$\overline{\mathcal{B}}$};
            \node[anchor = north] at (0,-1.5) {$S_A$};
             \draw[thick] (-1,-1.5) -- (-1,1.5);
            \draw[thick] (1,-1.5) -- (1,1.5);
            \draw[thick, xred] (0,-1.5)  -- (0,1.5);
            \draw[->- = .7 rotate 7] (0,0) ++ (100:2 and -.15) arc (350:20:.2 and .6);
            \draw (-1,0) -- (-0.85,0);
            \draw[->- = .4 rotate 0] (-0.6,0) -- (0,0);
            \draw[->- = .6 rotate 0] (0,0) -- (1,0);
            \node[anchor = north] at (-0.5,1) {$\m$};
            \node[anchor = north] at (-0.5,0) {$\r$};
            \node[anchor = north] at (0.50,0) {$\bar{\r}$};
            \draw[fill] (0,0) circle [radius = .05];
            
             \node[anchor = north] at (2,0) {$=$};
             \node[anchor = north] at (2.5,0.2) {$\frac{S_{\m \r} }{S_{0\r}}$};
            
            \node[anchor = north] at (3,-1.5) {$\mathcal{B}$};
            \node[anchor = north] at (5,-1.5) {$\overline{\mathcal{B}}$};
            \node[anchor = north] at (4,-1.5) {$S_A$};
            \draw[thick] (3,-1.5) -- (3,1.5);
            \draw[thick] (5,-1.5) -- (5,1.5);
            \draw[thick, xred] (4,-1.5)  -- (4,1.5);
            \draw[->- = .6 rotate 0] (3,0) --  (4,0);
            \draw[->- = .6 rotate 0] (4,0) -- (5,0);
            \node[anchor = north] at (3.5,0) {$\r$};
            \node[anchor = north] at (4.5,0) {$\bar{\r}$};
            \draw[fill] (4,0) circle [radius = .05];
        \end{tikzpicture}
    \caption{Action of the line $\a^+_\m$ on a local operator in the $(\rho,\bar{\rho})$ representation.}
    \label{fig:action on op}
\end{figure}

We end this section by analyzing the action of $\a^+_\m \otimes \a^-_{\bar{\m}}$ lines on the local operators of the RCFT. Recall from section \ref{sec:intro} that a 2d local operator $\mathcal{O}_{\r,\bar{\r}}^m$ corresponds to two horizontal lines meeting at the surface $S_A$, namely a $\r$ line ($L_\r^\mathrm{3d}$) from the left and a $\bar{\r}$ line from the right. The the multiplicity label $m=1,\dots,\mim_{\r\bar{\r}}$ determines the morphism at the intersection of these lines on the surface. In 2d the action of a topological line is represented by a circle surrounding the operator, which in the bulk is equivalent to a circular line linking a straight line as shown in figure \ref{fig:action on op}.
Unliking the two lines we get an S-matrix factor (see appendix \ref{app:CS}) and we arrive at
\begin{equation}
    \a^+_\m \cdot \mathcal{O}_{\r,\bar{\r}}^m  = \frac{S_{\m \r} }{S_{0\r}} \, \mathcal{O}_{\r,\bar{\r}}^m  ~.
\end{equation}
This should not be confused with a topological line in 2d passing a local operator, since that produces more terms (see \eqref{non.inv.lines}). However, when the $\m$ line is a loop these extra terms give tadpole diagrams and  therefore they vanish  \cite{Chang:2018iay}.  For more general lines we get 
\begin{equation}
    (\a^+_\m \otimes \a^-_{\bar{\m}}) \cdot \mathcal{O}_{\r,\bar{\r}}^m  = \frac{S_{\m \r}}{S_{0\r} } \frac{S_{ \bar{\m} \bar{\r}} }{S_{0\bar{\r}}}\, \mathcal{O}_{\r,\bar{\r}}^m  ~. \label{action of alpha ind}
\end{equation}
Note that the index $m$ is untouched by the action~\eqref{action of alpha ind} because intuitively the index $m$ lives near the topological surface $S_A$, away from the lines.

\subsection{Topological Lines in Adjoint QCD \label{adjQCD.lines.bos}}
As we discussed in section \ref{sec:bos}, adjoint QCD with gauged $(-1)^F$ can be bosonized to the $\Spin(N^2-1)_1/\SU(N)_N$ gauged WZW model with a kinetic term for the $\SU(N)$ gauge fields. Here we discuss the topological lines of the bosonic theory. We summarize the fermionic theory in appendix \ref{app:fermioniclines} and only discuss the theory with gauged $(-1)^F$ here.

Since $\hat{\mathfrak{su}}(N)_N \subset \hat{\mathfrak{so}}(N^2-1)_1$ is a conformal embedding, the diagonal $\Spin(N^2-1)_1$ WZW model can be regarded as a non-diagonal modular invariant $\SU(N)_N$ WZW model \cite{Moore:1988ss}.
Therefore, the topological lines of the $\Spin(N^2-1)_1/\SU(N)_N$ coset include the topological lines of the diagonal $\Spin(N^2-1)_1$ WZW that commute with its $\hat{\mathfrak{su}}(N)_N$ chiral algebra, i.e.\ those that survives the $\SU(N)_N$ gauging.\footnote{Although we do not have a proof, we believe that these lines are all the topological lines of the theory, except for the line that generates the charge conjugation symmetry.}
In the rest of this section, we use the formalism reviewed in \ref{sec:gauging} to describes these lines more systematically.

As we reviewed in \ref{sec:modular.invariants}, the $\Spin(N^2-1)_1$ WZW is a generalized gauging of the $\SU(N)_N$ WZW model. 
It turns out that for $N=3$ we have $\Spin(8)_1 \cong \mathrm{PSU}(3)_3=\SU(3)_3/\bZ_3$ -- which is the usual $\bZ_3$ gauging of $\SU(3)_3$-- while for higher values of $N$ it is a generalized gauging.
Let $\cC=\Rep \hat{\mathfrak{su}}(N)_N$ be the Verlinde lines of the $\SU(N)_N$ WZW model.
There exist an object $A \in \mathcal{C}$ which can be gauged so that the chiral algebra of $\SU(N)_N$ is extended to that of $\Spin(N^2-1)_1$. 
That is,
\begin{equation}
	\frac{\SU(N)_N \; \mathrm{WZW}}{A} = \Spin(N^2-1)_1 \; \mathrm{WZW}~,
\end{equation} 
where $A$ denotes a way of gauging $\cC$.
This gauging changes the category symmetry from $\cC$ to $\aca$.
The topological lines in $\aca$ describe all the lines that act on the theory but commute with $\hat{\mathfrak{su}}(N)_N$ affine algebra \cite{Fuchs:2002cm,Petkova:2000ip}.\footnote{This can also be deduced from an alternative definition of $\aca$ given in appendix \ref{app:lines.of.gauged.th}.}
These are the topological lines of the $\Spin(N^2-1)_1$ WZW that survive after gauging to the $\Spin(N^2-1)_1/\SU(N)_N$ gauged WZW model.
Note that a $\Spin(N^2-1)_1$ WZW model apart from the Verlinde lines that preserve the whole $\hat{\mathfrak{so}}(N^2-1)_1$ affine algebra, has more topological lines that preserve only the smaller $\hat{\mathfrak{su}}(N)_N$ subalgebra.
Adding further the kinetic term for the gauge field preserves these lines.
This is because the kinetic term is written in terms of the $\SU(N)$ gauge fields which couple to $\SU(N)$ currents.
These are invariant under $\aca$ and so are the gauge fields.
The existence of these topological lines in $\aca$ -- which are mostly non-invertible, restricts the possible gapped phases of adjoint QCD which we study in section \ref{sec:IR.TQFT}.
\subsection{Deconfinement in Adjoint QCD}
\label{sec:deconf}
So far we have considered the category symmetry $\aca$ in the UV. Since this is preserved by the RG flow, the vacuum Hilbert space has to form a representation of $\aca$. Different vacua of the theory correspond to different boundary conditions at infinity. If the IR TQFT is just $\Spin(N^2-1)_1/\SU(N)_N$, these boundary conditions correspond to different elements of $\cC_A$.\footnote{$\cC_A$ is a module category over $\aca$ whose action on it can be determined from the data given in section \ref{sec:ocv graphs}} These vacua are distributed between the $N$ universes of the theory which are labeled by the $\bZ_N$ one-form symmetry and separated by the Wilson lines (see section \ref{sec:intro}). The purpose of this section is to explain how the existence of topological lines requires that all these universes are degenerate, and therefore all Wilson loops have a perimeter law which proves that the theory is in the deconfined phase.  

The deconfinement can actually be shown without assuming what the IR TQFT is. 
That is, given a vacuum $\ket{0}$ in some universe, we can obtain a superposition of vacua in other universes as $L^n\ket{0}$ with topological line $L$ with the unit one-form charge.
Because $L$ is topological, $\ket{0}$ and $L^n\ket{0}$ necessarily have the same energy, and therefore every universe has the same lowest energy, meaning complete deconfinement.

To complete this argument, we must show that $L\ket{0}\neq0$ for a vacuum $\ket{0}$ (in flat space $\mathbb{R}$). This is achieved by invoking reflection positivity.
Instead of working on the non-compact space $\mathbb{R}$, we work on a large circle $S^1$.  
 The vacuum $\ket{0}$ is approximated (up to terms exponentially small in the circle size) by a boundary state (satisfying the generalized Cardy condition: \cite{Moore:2006dw}), which is also denoted by $\ket{0}$. This is a particular state on the circle which approximates the vacuum in the infinite volume limit.
The norm of the state $\ket{1} = L\ket{0}$ is
\begin{equation}
    \braket{1|1} =\braket{0|L^{\dag}L|0} = \braket{0|0} + \sum_{L_a\neq \mathbbm{1}} N_{L^\dag L}^a \braket{0|L_a|0}~,
\end{equation}
where $L^\dag$ is the dual of the line $L$, and $L^{\dag}\otimes L = \mathbbm{1}+\sum_{L_a\neq \mathbbm{1}}N_{L^\dag L}^a L_a$ with nonnegative integers $N_{L^\dag L}^a$.
The term $\braket{0|L_a|0}$ is reflection-symmetric when $L_a$ is put along a time slice, where the reflection is taken to be the timelike line perpendicular to $L_a$. To be more precise, $\braket{0|L_a|0}$ can be regarded, by 90-degree rotation, as the norm $\braket{L_a|L_a}$, where the state $\ket{L_a}$ is the lowest energy state in the Hilbert space $\mathcal{H}_{0,L_a,0}$ on $\mathbb{R}$ whose boundary condition at infinity is fixed by the flat space vacuum $\ket{0}$ and timelike $L_a$ is inserted in the middle. Therefore $\braket{0|L_a|0} = \braket{L_a|L_a}\ge 0$ (it is $0$ when $\mathcal{H}_{0,L_a,0} = \{0\}$).\footnote{In fact $\braket{0|L_a|0}$ is identified with the matrix element $n_{a,0}^0$ of the NIM-rep discussed below when the radius of the circle is very large.}
Thus we have 
\begin{equation}
    \braket{1|1} \ge \braket{0|0} > 0~.
\end{equation}
Note that to show $L\ket{0}\neq 0$ we used the fact that $\ket{0}$ is the vacuum, or a boundary state.
On the other hand, for a generic state $\ket{\psi}$ and a non-invertible line $L$, $L\ket{\psi}$ can be 0, which we will see explicitly in the following when $N=3$.
To illustrate how the lines acts on vacua and universes more explicitly, in the rest of the section we look at the examples of $\SU(2)$ and $\SU(3)$.

Consider the bosonic adjoint QCD based on the $\SU(2)$ group. In this case $\Spin(3)_1$ actually means $\SU(2)_2$ and the theory at the $g_\mathrm{YM}\rightarrow \infty$ is dual to the $\SU(2)_2/\SU(2)_2$ TQFT. In this case the category of lines $\aca$ is $\cC$ itself. Hence, the lines of adjoint QCD are just the Verlinde lines of the $\SU(2)_2$ WZW model. These correspond to the integrable representations of  $\SU(2)_2$, namely the trivial line $\mathbbm{1}$, the fundamental line $L_{\mathrm{fund}}$ and the adjoint line $L_{\mathrm{adj}}$. They obey the fusion algebra
\begin{equation}
    L_\mathrm{adj} \otimes L_\mathrm{adj}=\mathbbm{1}, \quad L_\mathrm{fund} \otimes L_\mathrm{adj}=L_\mathrm{fund}, \quad L_\mathrm{fund}\otimes L_\mathrm{fund}=\mathbbm{1}+ L_\mathrm{adj} ~. \label{su(2) alg}
\end{equation}
and they carry one-form symmetry charge equal to the $N$-ality of the corresponding representation (see section \ref{sec:intro}), namely $q_\mathbbm{1}=0=q_\mathrm{adj}$ and $q_\mathrm{fund}=1$. This adjoint QCD has two universes labeled by their one form-charge. It is now obvious that $L_\mathrm{fund}$ creates a universe whose one-form charge is 1 out of the universe with charge 0, and the two universes are degenerate. In other words the fundamental Wilson line has a perimeter law and thus the theory is in the deconfining phase. (Note that before gauging $(-1)^F$, the fundamental string universe could be created by the invertible $\bZ_2^\chi$ axial symmetry, which leads to a quick proof of deconfinement. But after turning the theory to a bosonic theory, the axial symmetry disappears and is replaced by the non-invertible symmetry $L_\mathrm{fund}$.)

It is possible that each universe has several vacua in it. At this point, we have only showed that there are two degenerate universes but what about the exact number of vacua? As we have argued in appendix \ref{app:module.categories} the vacuum Hilbert space has to form a representation of the topological lines. It is described by a TQFT where these lines act. The possible TQFTs in the IR is the subject of section \ref{sec:IR TQFT}. Here we just show that the smallest allowed representation of the above algebra \eqref{su(2) alg} is three dimensional. To see this, first note that in a TQFT the topological lines form a nonnegative integer valued matrix representation, or \emph{NIM-rep} for short.\footnote{More precisely, such a NIM-rep describes the action of topological lines on the boundary conditions (the true vacua of the gapped theory) forming the module category $\cC_A$ -- see also \cite{Cardy:1989ir,Behrend:1999bn,Gannon:2001ki,Gaberdiel:2002qa}.} Let us now assume that there is a one-dimensional NIM-rep. If we represent the action of the above lines by matrices $n_\mathrm{fund}$ and $n_\mathrm{adj}$, the above algebra implies that
\begin{equation}
    n_\mathrm{adj}n_\mathrm{adj}=1, \quad n_\mathrm{fund} n_\mathrm{adj}=n_\mathrm{fund}, \quad n_\mathrm{fund} n_\mathrm{fund}=1+ n_\mathrm{adj} ~. 
\end{equation}
which cannot be satisfied in integers. A two dimensional representation is also not allowed but it is more involved to see why.
Although one can find $2\times2$ matrices that obey the algebra it turns out that the crossing kernel equations \cite{Chang:2018iay} are not satisfied.\footnote{We thank Yifan Wang for pointing it out to us.}
On the other hand, a three dimensional NIM-rep, namely the ``regular'' representation, exists and it is given by the fusion coefficients of the algebra.
In fact since there is no non-diagonal $\SU(2)_2$ modular invariant, every allowed NIM-rep is decomposed into copies of the regular NIM-rep.
The regular NIM-rep is the smallest allowed representation of \eqref{su(2) alg}.
If the vacuum Hilbert space belongs to the smallest representation of \eqref{su(2) alg}, we conclude that $\SU(2)$ adjoint QCD has two degenerate universes, the first with two vacua and the second with a single vacuum.
Regardless of the symmetry realization in the infrared, the above analysis shows that there are two degenerate universes with one-form charge 0 and 1 respectively, leading to deconfinement. 

Before moving on to the case of $\SU(3)$ gauge theory it is worth contemplating the consequences of our result thus far,
i.e.\ that there are two vacua in the first universe (the one without a string) and one vacuum in the universe with a string.
It is interesting to add back the mass term for the adjoint quark and compare this to the situation at large mass for the quark. In the fermionic language the mass term of course preserves $(-1)^F$, which in the bosonized language means that the mass term operator ought to be invariant under $g\to -g$. 
Therefore, in the bosonized $\SU(2)_2/\SU(2)_2$ language the mass term corresponds to deforming the action by 
\begin{equation}
    \delta S = \frac12c\int \mathrm{d}^2 x \, \Tr(g^2)~,
\end{equation}
with some real coefficient $c$. 
We can diagonalize $g=\mathrm{diag}(e^{ix},e^{-ix})$ with $x\simeq x+2\pi$ but in addition also $x\simeq -x$ due to the Weyl group. The potential is $c \cos(2x)$ and hence there are two cases: for negative $c$ there are two vacua $x=0$ and $x=\pi$ while for positive $c$ there is only one vacuum $x=\pi/2$. This tells us the number of vacua in one of the two universes. But at large quark mass the number of vacua will be the same in all universes as the string is not important and hence we have two possibilities, corresponding to negative and positive $c$ respectively: two vacua in each universe or one vacuum in each universe. Either way there is a phase transition compared to the vanishing mass case. Indeed, for zero mass or very small mass the number of vacua in the two universes ought to remain (2,1) as the two-fold degeneracy is protected by the $g\to -g$ symmetry. This phase transition is clearly of the Ising type as two vacua merge into one or one vacuum splits into two, depending on the sign of the mass. This is beautifully consistent with the emergence of a massless Majorana Goldstino particle on the string at the supersymmetric point~\cite{Kutasov:1993gq}. After bosonization, the massless Majorana fixed point becomes precisely the Ising fixed point.\footnote{From our discussion it seems that for negative and positive masses the transition occurs in different universes. Actually, it always occurs on the confining string due to the fact that when we flip the sign of the mass, due to the mixed anomaly with one-form symmetry, the universes are exchanged.} \footnote{We can translate this discussion to the fermionic theory. After gauging $\bZ_2^\mathrm{v}$ with $\mathrm{Arf}$ twist,  $\bZ_2^\mathrm{v}$ doublet states in the bosonic theory are identified and give a single state. The phase of this vacuum differs from that of the $\bZ_2^\mathrm{v}$ singlet state in the bosonic theory by the $\mathrm{Arf}$ invertible phase which depends on the spin structure. In our convention, at large mass all the vacua have the same invertible phase as the state coming from a doublet. Thus, a singlet vacuum should experience a phase transition while the mass is cranked up towards infinity.} This argument for an Ising transition on the string at finite mass can be carried out for all $N$ in principle, though in practice it is very difficult (as we will see) to count the number of vacua in each universe in the TQFT. We will check it for $N=3,4,5$ explicitly in the following.

Let us now consider the bosonic adjoint QCD based on $\SU(3)$. In the UV the category symmetry is the same as in the $\Spin(8)_1/\SU(3)_3$ gauged WZW model.\footnote{As noted in Sec.~\ref{sec:bos}, there are two bosonization of the theory. Here we are concerning the bosonization whose category symmetry is that of $\Spin(8)_1/\SU(3)_3$ coset. The other one has a different category symmetry isomorphic to that of $\SO(8)_1/\SU(3)_3$ coset model. Note that although $\Spin(8)_1$ and $\SO(8)_1$ shares the same torus partition function, the theories are different as $\SU(3)_3$ RCFT, so as the category symmetries of them does.} Working out the $\a$-induction map (see section \ref{sec:3d}) from  $\Rep \hat{\mathfrak{su}}(3)_3$ to the topological lines of the gauged WZW model we get
\begin{equation}
\begin{aligned}
&\a^+_{(0,0)}=\mathbbm{1}~,
\qquad&&\a^+_{(0,1)}\a^-_{(0,1)}=2L_5 ~,\\
&\a^+_{(1,0)}=L_1~,
&&\a^+_{(1,0)}\a^-_{(1,0)}=2L_6~,\\
&\a^+_{(0,1)}=L_2~,
&&\a^+_{(1,0)}\a^-_{(0,1)}=s_1+s_2+s_3+s_4~,\\
&\a^-_{(1,0)}=L_3~,
&&\a^+_{(0,1)}\a^-_{(1,0)}=s_5+s_6+s_7+s_8~,\\
&\a^-_{(0,1)}=L_4~,
&&\a^+_{(1,1)}=a+b+ab ~.
\end{aligned}
\label{a ind in su3}
\end{equation}
The lines $s_i$, $a,b$ and $ab$ are invertible and generate a $(\mathbb{Z}_2 \times \mathbb{Z}_2)\rtimes \mathbb{Z}_3$ symmetry. The $\mathbb{Z}_2 \times \mathbb{Z}_2$ factor is the center of $\Spin(8)$, while the second factor is the $\mathbb{Z}_3$ subgroup of the $S_3$ triality in $\Spin(8)_1$. The $\mathbb{Z}_2$ part of the triality acts as charge conjugation on  $\SU(3)_3$ and therefore is not an element of $\aca$ since it does not preserve the $\hat{\mathfrak{su}}(N)_N$ affine algebra.\footnote{The charge conjugation is still a symmetry of both the $\Spin(8)_1$ WZW model and the theory after $\SU(3)$ gauging.}
One can further determine the fusion algebra of these lines but for simplicity we do not include it here. The conclusion is that the $L_i$ lines listed above are non-invertible.
In addition, $L_1,L_3$ and $L_5$ have one-form symmetry charge equal to 1 while $L_2,L_4$ and $L_6$ have one-form symmetry charge equal to 2. It follows that the former create a universe with one-form charge equal to 1, while the latter create a universe with one-form charge equal to 2, leading to deconfinement of all Wilson lines. 

If we assume that the IR theory is captured by the $g_\mathrm{YM} \rightarrow \infty$ limit, we can say more about how the vacua of the theory are distributed in the 3 universes. In this case, the vacua correspond to the 6 elements of $C_A$. It is sufficient to study the action of the $\a^+_{(1,0)}=\a^+(L_{(1,0)})$ line on these vacua from 
\begin{equation}
    \a^+_{(1,0)} \ket{\mathsf{v}_a}= \sum_b n_{(1,0)a} \! ^b \ket{\mathsf{v}_b}~,
\end{equation}
where $L_{(1,0)}\in \mathcal{C}$ is the fundamental Verlinde line of $\SU(3)_3$, and the NIM-rep matrices $n_\mu$ are discussed in appendix \ref{sec:ocv graphs}. Using equation \eqref{su3.nimrep} we get
\begin{align}
    \a^+_{(1,0)} \ket{\mathsf{v}_1}&=\ket{\mathsf{v}_5}~, \\
    \a^+_{(1,0)} \ket{\mathsf{v}_2}&=\ket{\mathsf{v}_5}~, \\
    \a^+_{(1,0)} \ket{\mathsf{v}_3}&=\ket{\mathsf{v}_5}~, \\
    \a^+_{(1,0)} \ket{\mathsf{v}_4}&=\ket{\mathsf{v}_5}~, \\
    \a^+_{(1,0)} \ket{\mathsf{v}_5}&=2\ket{\mathsf{v}_6}~, \\
    \a^+_{(1,0)} \ket{\mathsf{v}_6}&=\ket{\mathsf{v}_1}+\ket{\mathsf{v}_2}+\ket{\mathsf{v}_3}+\ket{\mathsf{v}_4}~.
\end{align}
If we assign one-form charge $0$ to the first vacuum, since the charge of the $\a^+_{(1,0)}$ line is $1$, we conclude that\footnote{To justify the assignment of charge 0 to the universe with four vacua, one should use the relation between the one-form symmetry and the outer automorphism of the chiral algebra explained in section \ref{sec:coset.TQFT}.}
\begin{equation}
    q_1=q_2=q_3=q_4=0, \quad q_5=1, \quad q_6=2~.
\end{equation}
This implies that there are $(4,1,1)$ vacua in the universes with one-form symmetry charges $0$, $1$, and $2$ respectively.
Further studying the action of the $\bZ_2^\mathrm{v}$ line on these vacua, we find that there are $(\underline{2}+\underline{2},\underline{1},\underline{1})$ vacua where $\underline{1}$ means a $\bZ_2^\mathrm{v}$ invariant vacuum and $\underline{2}$ a pair of vacua related by the $\bZ_2^\mathrm{v}$ symmetry.
Thus, in the fermionic theory, there are $(2,1,1)$ vacua in the universes $0,1,2$.
The pattern of $\bZ_2^\mathrm{v}$ symmetry breaking/preserving in each vacuum predicts Ising phase transitions in these universes as we increase the quark mass. This is because in the large mass limit, the $\bZ_2^\mathrm{v}$ symmetry is completely broken in all universes from the analysis analogous to what we have done for $N=2$ case. 
Thus we predict the existences of massless fermions in the string universes for some positive value of mass in the fermionic theory, consistent with the emergence of supersymmetry.

Let us consider what happens if we compactify the theory with $N=3$ on a circle with radius $R$. For simplicity, let us consider the fermionic theory (and the anti-periodic boundary condition on $S^1$), where we have two vacua $\ket{\mathsf{v}_1}$ (identified with $\ket{\mathsf{v}_2}$ by $\bZ_2^\mathrm{v}$ symmetry) and $\ket{\mathsf{v}_3}$ (identified with $\ket{\mathsf{v}_4}$) in universe 0, $\ket{\mathsf{v}_5}$ in universe 1, and $\ket{\mathsf{v}_6}$ in universe 2. 
On the vacua $\ket{\mathsf{v}_1}$ and $\ket{\mathsf{v}_3}$ the $\bZ_2^\chi$ chiral symmetry is broken. With a finite YM coupling $g_\text{YM}$ (equivalently with a finite scale), there will be a kink solution between $\ket{\mathsf{v}_1}$ and $\ket{\mathsf{v}_3}$ and thus the lowest energy state on $S^1$ is 
$\ket{+} = \ket{\mathsf{v}_1}+\ket{\mathsf{v}_3}$, and the energy gap to the second low-lying state $\ket{-}=\ket{\mathsf{v}_1}-\ket{\mathsf{v}_3}$ behaves like $e^{-g_{\mathrm{YM}}R}$.
The states $\ket{\mathsf{v}_5}$ and $\ket{\mathsf{v}_6}$ have the same energy with $\ket{+}$, because of $\alpha^+_{(1,0)}\ket{+} = 2 \ket{\mathsf{v}_5}$ and $(\alpha^+_{(1,0)})^2\ket{+} = 4 \ket{\mathsf{v}_6}$.
On the other hand, acting with topological lines on $\ket{-}$ gives $0$, which is consistent with the absence of low-lying states with exponentially suppressed energy in other universes.
Therefore, on $S^1$ we have 3 vacua $\ket{+},\ket{\mathsf{v}_5},\ket{\mathsf{v}_6}$, one in each universe. 
One can define a different basis $\ket{\tilde{\ell}_j} = \ket{+} + e^{\frac{2\pi \mathrm{i}}{3}j}\ket{\mathsf{v}_5} + e^{\frac{4\pi \mathrm{i}}{3}j}\ket{\mathsf{v}_6}$ for $j=1,2,3$, which diagonalizes the topological line $\alpha^+_{(1,0)}$.
This basis is close to what is considered in \cite{Cherman:2019hbq}, which diagonalizes the fundamental Wilson line.
However, since $\ket{\tilde{\ell}_j}$ diagonalizes the \emph{topological} line, there cannot be any tunneling event between $\ket{\tilde{\ell}_j}$.

For $\SU(4)$ using the NIM-rep matrix given in equation \eqref{su4.nimrep}, we study the action of the $\a^+_{(1,0,0)}$ line on the $12$ elements of $\cC_A$ and find $(4,3,2,3)$ vacua in universes $0$, $1$, $2$, and $3$ respectively.
Furthermore, by studying the action of the $\bZ_2^\ttv$ line which in this case is given by the $\a^+_{(4,0,0)}$ line (see equation \eqref{su4.lines}), we find $(4,3,2,3)=(\underline{2}+\underline{2},\underline{1}+\underline{2},\underline{1}+\underline{1},\underline{1}+\underline{2})$. 
Moreover, in the next section, we will see that with small mass deformation the $\underline{1}$ states in the universes $1,2,3$ will remain the lowest energy states in each universe.
Thus there are $(0,1,2,1)$ $\bZ_2^\ttv$ invariant vacua in universes $0$, $1$, $2$, and $3$ respectively.
This again predicts the existences of massless fermion transitions in universes $1$, $2$, and $3$ in the fermionic theory.

Finally, for $\SU(5)$ using the NIM-rep matrix \eqref{su5.nimrep} we find $(8,4,4,4,4)=(4\times\underline{2},2\times\underline{1}+\underline{2},2\times\underline{1}+\underline{2},2\times\underline{1}+\underline{2},2\times\underline{1}+\underline{2})$ vacua in universes $(0,1,2,3,4)$. Again, there are $\bZ_2^\mathrm{v}$ invariant vacua only in the non-trivial universes and these remain the lowest energy states in each universe with small mass. Thus we predict massless fermions on the string as we increase the mass, compatible with the spontaneous breaking of supersymmetry in the string state.
\section{Mass and Quartic Deformations \label{sec:deformation}}
In this section we study some interesting deformations of adjoint QCD. One of them is a mass term for the adjoint quark and the others are quartic fermion operators. As we saw above adjoint QCD has generally many degenerate vacua, even small deformations which are quadratic or quartic in the fermions could change the long distance physics dramatically.  The purpose of this section is to understand how these deformations affect the IR structure of the theory.  The way we proceed is by analysing which topological lines are broken by these deformations.  

It is sufficient to look at the free theory to analyze which lines are broken. The free theory is a diagonal $\Spin(N^2-1)_1$ WZW or equivalently a non-diagonal $\SU(N)_N$ WZW model. Since we are interested in topological lines that preserve only an $\SU(N)_N$ subalgebra, the second view point is more suitable. Hence, the strategy will be to consider the non-diagonal $\SU(N)_N$ RCFT and use the machinery of RCFT to determine the action of the topological lines on local operators. Adding an operator to the UV action that transforms non-trivially under a topological line will break such a line, leading to a smaller  symmetry category in the IR. Consequently, some of the vacua might get lifted affecting the phases of the theory. 

The procedure outlined in the above paragraphs is independent of which TQFT admitting the symmetry category is realised in IR of the (bosonized) adjoint QCD. Assuming that the theory flows to the  $\Spin(N^2-1)_1/\SU(N)_N$ TQFT enables us to extract even more information. Under the assumption, one can compute the vacuum expectation value (condensate) of various operators and determine that first order correction to the vacuum energies induced by a deformation. With this computation we verify that the mass deformation restores confinement, and calculate the tension of the confining strings.

In the rest of the section, we begin by studying the various relevant or classically marginal deformations of adjoint QCD. For $N=2,3$ and $4$ we then study the action of symmetry lines on those deformations and we determine which lines remain unbroken. This allows us to determine whether the deformed theory is confined or not. We also study the first order correction to the vacuum energies in the $\Spin(N^2-1)_1/\SU(N)_N$ TQFT for $N\leq 5$ and we calculate the tension of the confining strings. For $N=5$ even if we do not know the action of symmetry lines on operators, our TQFT calculation shows that in the deformed theory all the non-trivial universes are lifted. We conclude that certain deformations break all the lines with non-zero one-form charge. This shows that the theory confines even without assuming that the IR TQFT is given by $\Spin(N^2-1)_1/\SU(N)_N$. In this way we have studied confinement for $N\leq 5$.

\paragraph{Relevant and classically marginal operators}
In the rest of the section we use CFT notation. We denote left and right moving fermions by $\psi^i_{k+}$ and $\psi^i_{k-}$, where upper and lower indices are fundamental and antifundamental respectively (and $\pm$ denote the chirality). The mass deformation is just 
\begin{equation}
    \cO_\mathrm{m} = \psi^i_{k+} \psi^k_{i-}~, \label{mass.operator}
\end{equation}
and it belongs to the adjoint representation with highest weight $(1,0,\dots,0,1)$ in both left and right sectors. (Of course the operator is in the singlet with respect to the diagonal of the left and right current algebras, after its indices contracted.) 
The various quartic couplings are built out of the bilinears $\psi^i_k \psi^j_l$ and their right-moving counterparts.
We define the current
\begin{equation}
J^i_l=\psi^i_{k+} \psi^k_{l+}~, \label{currents}
\end{equation}
which generates the $\SU(N)$ transformations on left-movers. We can similarly define $\bar J^l_i$ for the right movers. In addition, there are quadratic primary operators in the left-moving sector:
\begin{equation}
O^{ij}_{kl}=\psi^i_{k+} \psi^j_{l+}-\frac{1}{N} \d_l^i \psi^n_{k+} \psi^j_{n+}-\frac{1}{N} \d_k^j \psi^i_{n+} \psi^n_{l+}~,
\end{equation} 
This set of operators contains two irreducible components that are
\begin{equation}
O^{(ij)}_{[kl]}, \quad O^{[ij]}_{(kl)}~, \label{traceless bilinear}
\end{equation}
where $(ij)$ and $[ij]$ denotes the symmetrization and anti-symmetrization of the indices.
We can likewise define the operators $\bar O$ that are made out of the right movers.
The currents in \eqref{currents} are descendants in the vacuum module $(0,0,\dots,0,0)$, while $O^{(ij)}_{[kl]}$ and $ O^{[ij]}_{(kl)}$ correspond to the primaries in the representations $(2,0,\dots,0,1,0)$ and $(0,1,0,\dots,0,2)$ respectively and are related by charge conjugation.  The various quartic interactions can be built out of the quadratics as 
\begin{align}
\mathcal{O}_1= J^i_l \bar{J^l_i} &=\Tr (\psi_+ \psi_+ \psi_- \psi_-) ~, \\
\mathcal{O}_2=O^{(ij)}_{[kl]} \bar{O}^{[lk]}_{(ji)}+O^{[ij]}_{(kl)} \bar{O}^{(lk)}_{[ji]} &= \Tr(\psi_+ \psi_-)\Tr(\psi_+ \psi_-)-\frac{2}{N} \Tr (\psi_+ \psi_+ \psi_- \psi_-) \label{double trace} ~,\\
\mathcal{O}_3=O^{(ij)}_{[kl]} \bar{O}^{[lk]}_{(ji)}-O^{[ij]}_{(kl)} \bar{O}^{(lk)}_{[ji]} &= \Tr (\psi_+ \psi_-\psi_+  \psi_-)~.
\end{align}
We see that $\mathcal{O}_1$ and $\mathcal{O}_2$ are even under charge conjugation, while $\mathcal{O}_3$ is odd. Moreover, since $\mathcal{O}_1$ is in the vacuum module, it does not break any topological line and therefore can be generated by the RG flow. On the other hand $\mathcal{O}_3$ cannot be generated since it breaks charge conjugation. In the following we analyze the deformation $\mathcal{O}_2$. The question in two-fold. On one hand, whether it can be generated by the RG flow, and on the other hand, the consequences of turning it on in the UV action. The standard considerations of naturalness would suggest that $\mathcal{O}_2$ can be generated as it is invariant under all the ordinary symmetries. Here we will see that the non-invertible symmetries impose surprising constraints on it.

To answer these questions we need to determine how topological lines act on local operators.
In the RCFT literature there is a whole machinery for this purpose known as the Ocneanu graphs which we review in appendix \ref{sec:ocv graphs}.
In the $\Spin(N^2-1)_1$ RCFT, a local operator $\cO_{\mu,\bar{\mu}}^m$ is labeled by a representations $\mu$ and $\bar\mu$ of the $\SU(N)_N$ algebra, and the multiplicity label $m$.
However after gauging $\SU(N)_N$, only the diagonal operators of the form $\cO_{\mu,\mu}^m$ survives, as they contain the gauge invariant operators.
This is also evident from the 3d point of view, since gauging $\SU(N)_N$ correspond to identifying the left and right chiral algebras and compactifying the CS theory on the circle, hence the only operators that survives are the diagonal ones -- see Fig.~\ref{fig:KSOp} and Fig~\ref{fig:AdjQCDCS}.
The free fermion theory has $\SU(N)_\mathrm{L}\times\SU(N)_\mathrm{R}$ left and right symmetries.
But the gauged $\SU(N)$ subgroup corresponds to the anti-diagonal elements $(g,g^\ast) \in \SU(N)_\mathrm{L}\times \SU(N)_\mathrm{R}$. Therefore, since $O^{(ij)}_{[kl]}$ transforms in $(2,0,\dots,0,1,0)$ and $\bar{O}^{(ij)}_{[kl]}$ transforms in $(0,1,\dots,0,2)$, the above quartic couplings are diagonal. Thus when we discuss the gauged theory, we often use the notation $\cO_{\mu}^m:=\cO_{\mu,\mu}^m$. In this notation
\begin{align}
    \cO_\mathrm{m} &= \cO^\mathrm{v}_{(1,0,\dots,0,1)}~,\\
    \cO_2 &= \cO_{(2,0,\dots,0,1,0)}+\cO_{(0,1,0,\dots,0,2)}~.
\end{align}
Note that the representations that appear above do not have multiplicities, except for the mass operator in the $N=3$ case -- see appendix \ref{app:branching}. Because the mass operator corresponds to the vector representation of $\Spin(N^2-1)_1$, we have denote the multiplicity label in this case by `$\mathrm{v}$'.

\paragraph{Action of line operator on local operators}
In general, the action of a  topological line $L_\a$ on a primary local operator/state $\cO_{\m,\bar{\m}}^m$, in the sense of equation \eqref{lines.on.ops}, would be written as 
\begin{equation}
 L_\a \cdot \cO_{\m,\bar{\m}}^m = \sum_{m'} \frac{\F^{(\m,\bar{\m};m,m')}_\a}{\sqrt{S_{0\m} S_{0 \bar{\m}}}}  \, \cO_{\m,\bar{\m}}^{m'}   \label{action of lines} ~, 
\end{equation}
where $S_{\m\n}$ is the modular $S$-matrix of $\SU(N)_N$, and $\Phi$ is the data to be determined.
Knowing the modular invariant matrix of the theory there is a way to determine these coefficients (though not uniquely) which can be encoded in the so called Ocneanu graphs.
We review how the Ocneanu graphs encode this information in appendix \ref{sec:ocv graphs}.
An alternate way to determine the action of topological line is the $\a$-induction map in section \ref{sec:3d}.
The two approaches might seem independent but they should actually be thought of as being complementary for our purposes.
For instance, the $\a$-induction map from the Verlinde lines of $\SU(N)_N$ to the lines adjoint QCD determines the one-form symmetry charge from the $N$-ality of the $\SU(N)_N$ lines.
However, the map is not one-to-one and therefore in general cannot completely determine the action of a general topological lines on local operators.
We use both approaches to study the cases of $N=2,3,4,5$.  

The topological lines are generally non-invertible, hence their quantum dimension
\begin{equation}
\mathrm{dim}(L_\a) = \frac{\F^{00}_\a}{\sqrt{S_{00} S_{0 0}}}~,
\end{equation}
is generally nontrivial ($\neq1$).
If the action of a line $L_\a$ on a local operators is equal to the quantum dimension of the line, it means that the local operators is invariant under $L_\a$ -- see equation \eqref{l.commutes.o}.

\paragraph{Condensate and the string tension}
If we assume that the IR TQFT is the $\Spin(N^2-1)/\SU(N)_N$ coset, we can further explicitly compute the effect of the deformation.
In the $\Spin(N^2-1)_1/\SU(N)_N$ TQFT there are two useful basis for the local operators/states.
After gauging $\SU(N)_N$, only the diagonal primary operators of the CFT survive which correspond to the boundary states of the CFT.
The first basis corresponds to the diagonal primary operators (Ishibashi states) of the CFT, that we denoted by $\cO_{\mu}^m = \cO_{\mu,\mu}^m$, where $m=1,\dots,\mim_{\mu\mu}$ labels the multiplicities.
The second basis (Cardy states) forms the vacua (where cluster decomposition holds in the flat space limit) of the gapped theory and corresponds to the \emph{normalized} boundary conditions which we denote by
\begin{equation}
    \mathsf{v}_a = \sum_{\mu,m} \phi_a^{(\mu,m)} \cO_\mu^m~.
\end{equation}
The matrix $\phi_a^{(\mu,m)}$, reviewed in more detail in Appendix~\ref{sec:ocv graphs}, is \emph{unitary} and contains the information about the NIM-reps $(n_\m)_a^b$ and the OPE coefficients $M_{(\mu,m),(\nu,n)}^{(\rho,r)}$ of local operators as
\cite{Petkova:1995fw,Behrend:1999bn,Gaberdiel:2002qa}
\begin{align}
(n_\m)_a^b & =\sum_{\n,m} \f^{(\n,m)}_a  \frac{S_{\m \n}}{S_{0\n}}\f^{(\n,m)*}_b ~, \label{nim.rep}\\
M_{(\mu,m),(\nu,n)}^{(\rho,r)} &=\sum_{a} \frac{\phi_a^{(\mu,m)} \phi_a^{(\nu,n)} \phi_a^{(\rho,r)\ast}}{\phi_a^0} ~.
\end{align}
Here, $n_\m$ describes the action of Verlinde lines ($\cC$) on the vacua (boundary conditions $\ca$) as
\begin{equation}
L_\mu \cdot \mathsf{v}_a = \sum_b n_{\mu a}^b \mathsf{v}_b~,
\end{equation}
and the OPEs among the TQFT operators are given by
\begin{align}
    \mathsf{v}_a \mathsf{v}_b &= \frac{\delta_{ab}}{\phi_a^0} \mathsf{v}_a~,\\
    \cO_\mu^m \cO_\nu^n &= \sum_{\rho,r} M_{(\mu,m),(\nu,n)}^{(\rho,r)} \cO_\rho^r~.
\end{align}
Note that for the diagonal WZW model, or equivalently the $G_k/G_k$ TQFT, $\phi$ is the same as the $S$-matrix, i.e.\ $\phi_a^{\mu} = S_{a \mu}$ (see appendix \ref{app:G/G}). 
By using the above formulae we find the condensates of primary operators as
\begin{equation}
    \bra{\mathsf{v}_a} \cO_\mu^m \ket{\mathsf{v}_a} = \frac{\phi_a^{(\mu,m)\ast}}{\phi_a^0}~.
\end{equation}
Furthermore inverting \eqref{nim.rep} we find
\begin{equation}
    \sum_m \phi_a^{(\mu,m)} \phi_b^{(\mu,m)\ast} = \sum_{\nu} S_{0\mu} S_{\mu\nu}^\ast n_{\nu a}^b~.
\end{equation}
Thus when there are no multiplicity in $\mu$ ($\mim_{\mu\mu}=1$), one can calculate the condensates using the NIM-rep matrices.

Using these formalism, we can compute the condensates of various deformation operators discussed above. In particular, we study the condensate of the mass operator in different universes and compute the tension of the confining strings. Moreover, we study the $\mathcal{O}_2$ condensate in different universes and verify the confinement of the fundamental Wilson line in the theory deformed by $\cO_2$.

\subsection{SU(2)}
For $N=2$, the quartic deformations $\mathcal{O}_2$ and $\mathcal{O}_3$ do not exist. As we already mentioned,  $\mathcal{O}_1$ does not break any topological lines. Hence it is generated by the RG flow and turning it on in the UV action does not lift any universe and the theory is still in the deconfining phase. Let us now consider the mass term. Since the theory in this case is given by the $\SU(2)$ gauging of the \emph{diagonal} $\SU(2)_2$ WZW model, the topological lines are just the Verlinde lines of $\SU(2)_2$ and  there are no multiplicities. In this case equation \eqref{action of lines} simplifies to 
\begin{equation}
  L_\n  \cdot \cO_\m =\frac{S_{\n \m}}{ S_{0 \m }} \,\cO_\m  ~, \label{action of Ver}
\end{equation}
where we have omitted the antiholomorphic index since $\bar{\m}=\m$. From the known expressions for the $S$-matrix one can easily calculate the action of topological lines on the mass operator $\cO_\mathrm{m}=\cO_\mathrm{adj}$:
\begin{equation}
 L_\mathrm{adj} \cdot \cO_\mathrm{m}=\cO_\mathrm{m}~, \qquad   L_\mathrm{fund}\cdot \cO_\mathrm{m}=- \sqrt{2} \; \cO_\mathrm{m} ,
\end{equation}
while the quantum dimensions are
\begin{equation}
   \mathrm{dim}( L_\mathrm{adj})=1~, \qquad \mathrm{dim}( L_\mathrm{fund})=\sqrt{2}~.
\end{equation}
Since $L_\mathrm{fund}$ is not invertible we see that its quantum dimension is not one.
All in all, we see that the mass term breaks the only non-invertible line $L_\mathrm{fund}$ which carries non-zero one-form symmetry charge.
The fact that the mass term breaks the $L_\mathrm{fund}$ line, can also be explained in the fermionic theory by the observation that the $\bZ_2^\mathrm{v}$ chiral symmetry flips the sign of the mass term.
Hence, the universe of the undeformed theory with one-form symmetry equal to $1$ is lifted after turning on a mass for the fermions.
This shows that Wilson loops separating the two universes acquire an area law and the theory confines.
We note that the line  $L_\mathrm{adj}$ is preserved by the mass term as it should be;  this line should be thought of as the bosonic avatar of fermion number (Gauging it with an Arf term we arrive at the massive fermionic theory).

In this case the only (indecomposable) candidate for the IR TQFT is the $\SU(2)_2/\SU(2)_2$ TQFT. Since we started with a diagonal modular invariant theory, the value of the condensates are given by the $S$-matrix:
\begin{equation}
\begin{array}{c|ccc}
\bra{\mathsf{v}} \cO \ket{\mathsf{v}} & \mathsf{v}_{0} &\mathsf{v}_\mathrm{adj} &\mathsf{v}_\mathrm{fund} \\ \hline
\cO_0 & 1 & 1 & 1 \\
\cO_\mathrm{adj} & 1 & 1 & -1 \\
\cO_\mathrm{fund} & \sqrt{2} & -\sqrt{2} & 0 
\end{array}
\end{equation}
This computation verifies that the mass deformation $m\cO_\mathrm{m}=m\cO_\mathrm{adj}$ lifts one of the universes -- depending on the sign of the mass $m$ -- and restores confinement.
Note that for one sign of the mass ($m>0$ in our conventions), $\mathsf{v}_\mathrm{fund}$ is lifted and we are left with the vacua $\mathsf{v}_0$ and $\mathsf{v}_\mathrm{adj}$.
These two vacua are related by the $\bZ_2^\mathrm{v}$ symmetry line $L_\mathrm{adj}$, thus the $\bZ_2^\mathrm{v}$ symmetry is spontaneously broken.
For the other sign of the mass parameter ($m<0$), we get the unique vacuum $\mathsf{v}_\mathrm{fund}$ and the $\bZ_2^\mathrm{v}$ symmetry is preserved.  This reshuffling of the vacua depending on the sign of the mass in the bosonized version of adjoint QCD simply reflects the mixed anomaly between the chiral symmetry and the one-form symmetry (which is responsible for changing the $N$-ality of the representations on the two sides) and the mixed anomaly between the chiral symmetry and fermion number symmetry (which is responsible for changing the number of vacua on the two sides). This theory of course has no interesting $k$-strings to discuss.

\subsection{SU(3)}
Now we study the deformations in the $\SU(3)$ adjoint QCD. This is the simplest non-diagonal case where the formalism of $\a$-induction in section \ref{sec:3d} and the Ocneanu graphs reviewed in \ref{sec:ocv graphs} are being particularly useful.  In this case the $\a$-induction completely determines the action of topological lines on local operators. One can also check the two approaches agree given the known the Ocneanu graphs (see appendix \ref{sec:ocn graphs for 3 and 4}), but for simplicity here we only use the $\a$-induction. The map is given in \eqref{a ind in su3}. Let's start with the quartic deformation 
\begin{equation}
  \mathcal{O}_2= \cO_{(3,0)}+\cO_{(0,3)}~.
\end{equation}
The topological lines do not act diagonally on $\mathcal{O}_2$. Let us focus on the first part only.  From \eqref{action of alpha ind} we have
\begin{equation}
    (\a^+_{\m} \otimes \a^-_{\bar{\m}})  \cdot \cO_{(3,0)} = \frac{S_{\m(3,0)}}{ S_{(0,0)(3,0)} } \frac{  S_{\bar{\m} (3,0)}}{ S_{(0,0) (3,0)}} \, \cO_{(3,0)} ~.  \label{action of a}
\end{equation}
Using this formula one can determine the action of the topological lines $L_i$ in \eqref{a ind in su3}. One can further determine the action of the invertible lines. For instance, consider the image of the Verlinde line $L_{(1,1)}$ which maps to $\a^+_{(1,1)}=a+b+ab$. From \eqref{action of a} we have that 
\begin{equation}
    \a^+_{(1,1)} \cdot  \cO_{(3,0)}  =3 \;\cO_{(3,0)}  ~.
\end{equation}

Since both the $a$ and $b$ lines generate invertible $\mathbb{Z}_2$ symmetries their action on states is either $1$ or $-1$. From the above, it follows that both $a$ and $b$ act trivially and therefore are not broken by $\mathcal{O}_2$. As a side remark, to arrive at the fermionic theory we need to gauge the diagonal $\mathbb{Z}_2^\ttv$ symmetry and the remaining axial $\mathbb{Z}_2$ will become the chiral $\bZ_2^\chi$ symmetry of the fermionic theory. The same line of arguments can be applied for the $s_i$ lines and shows that all of them are preserved. In fact, $\mathcal{O}_2$ breaks all the non-invertible lines and preserves the invertible ones.
Note that the operator $\mathcal{O}_2$ is not protected by any invertible symmetries neither in the fermionic nor in the bosonic description and yet it is not generated by the RG flow.
The unbroken lines generate the $(\mathbb{Z}_2 \times \mathbb{Z}_2)\rtimes \mathbb{Z}_3$ symmetry, and they all have trivial one-form symmetry charges. Therefore, $\cO_2$ breaks the topological lines with non-trivial one-form symmetry charge, and restores confinement.

Similarly we can study the action of topological lines on the mass term $\cO_\mathrm{m}=\cO^\mathrm{v}_{(1,1)}$. We find that the mass term breaks all the topological lines, except for the $\bZ_2^\mathrm{v}$ symmetry line which has a trivial one-form symmetry charge. Hence adding the mass term also restores confinement. 

Assuming that the theory flows to the $\Spin(8)_1/\SU(3)_3$ TQFT we can compute the condensate of the mass operators and further calculate the first order correction to the energy of each vacua in the deformed theory. By determining the lowest energy state in each universe, we find the tension of the confining string that creates that universe. We denote the tension of the string with $N$-ality $k$ by $T_k$. Then
\begin{equation}
    T_k \sim \min_{\mathsf{v}_a\in \text{universe }k} \bra{\mathsf{v}_a} m\cO_\mathrm{m} \ket{\mathsf{v}_a} - \min_{\mathsf{v}_a\in \text{universe }0} \bra{\mathsf{v}_a} m\cO_\mathrm{m} \ket{\mathsf{v}_a}~,
\end{equation}
where $m$ is the mass parameter. The value of the condensates can be computed using the $\phi$ matrix given in equation \eqref{su3.phi.matrix} as
\begin{equation}
\begin{array}{c|cccccc}
\bra{\mathsf{v}} \cO \ket{\mathsf{v}} & \mathsf{v}_1 &\mathsf{v}_2 &\mathsf{v}_3 &\mathsf{v}_4 &\mathsf{v}_5 &\mathsf{v}_6 \\ \hline
\cO_\cA & 1 & 1 & 1 & 1 & e^{2\pi i/3} & e^{4\pi i/3} \\
\cO_2 & -2 & -2 & -2 & -2 & 1 & 1 \\
\cO_\mathrm{m} & -\sqrt{3} & -\sqrt{3} & \sqrt{3} & \sqrt{3} & 0 & 0
\end{array}~
\end{equation}
Note that $\cO_\cA=\cO_{(0,3)}$ is the one-form symmetry operator (denoted by $\oneform_1$ in the introduction) whose condensate determines the universe in which a vacuum lives -- see section \ref{sec:coset.TQFT}.
Note that in vacua $\ket{\mathsf{v}_5}$ and $\ket{\mathsf{v}_6}$ although the fermion bilinear $\cO_\mathrm{m}$ does not condense, the $\mathcal{O}_2$, composed of four fermions, condenses.
Let us remark on the dynamics of the massless theory deformed by $\mathcal{O}_2$. From the table above we see that with one sign for this deformation the universes with the strings are lifted. Therefore we have confinement as expected. The other sign for the deformation behaves in a rather exotic fashion. The string-less universe is lifted while the universes with the $3,\bar 3$ strings remain degenerate. This degeneracy is protected by charge conjugation symmetry. This is quite strange seeing as this means that the strings develop negative tension and therefore dip below the ordinary vacuum. We do not see why such behaviour is disallowed.

Given the value of the mass operator condensates we find the string tensions (independent of the sign of $m$)
\begin{equation}
    \left(
\begin{array}{c}
 T_0 \\
 T_1 \\
 T_2
\end{array}
\right) \sim \abs{m} \left(
\begin{array}{c}
 0 \\
 \sqrt{3} \\
 \sqrt{3}
\end{array}
\right)~.
\end{equation}
This shows that the tension of the two strings must be the same, which is expected from the charge conjugation symmetry.
\subsection{SU(4)}
We now turn to the $\SU(4)$ adjoint QCD. In this case
\begin{equation}
  \mathcal{O}_2= \cO_{(2,1,0)}+\cO_{(0,1,2)}~,
\end{equation}
and we only need $\F^{(2,1,0),(2,1,0)}_\a$ and $\F^{(0,0,0),(0,0,0)}_\a$ coefficients in \eqref{action of lines} to study the action of topological lines on $\mathcal{O}_2$. Using the known Ocneanu graphs (see appendix \ref{sec:ocn graphs for 3 and 4}) we find that only three topological lines remain unbroken with quantum dimensions $1,1$ and $\sqrt{2}$. This shows that the unbroken lines satisfy the Ising fusion rules as in \eqref{su(2) alg} (and form the $\mathrm{TY}_+$ fusion category). These lines are the image of the following $\SU(4)_4$ Verlinde lines 
\begin{equation}
L_0=\a^+_{(4,0,0)}, \quad  L_2=\frac12 \a^+_{(1,1,1)}- \a^+_{(0,1,0)}~. \label{su4.lines}
\end{equation}
with 
\begin{equation}
    L_0 \otimes L_0=\mathbbm{1}, \quad L_2 \otimes L_0=L_2, \quad L_2 \otimes L_2=\mathbbm{1}+ L_0 ~. 
\end{equation}
These lines are precisely the Verlinde lines of the $\Spin(15)_1$ WZW model. Furthermore, this shows that the $\bZ_2^\ttv$ line $L_0$ has $0$ one-form symmetry charge, while the duality line $L_2$ has one-form symmetry charge equal to $2$. Similar to the $\SU(2)$ case, this algebra leads to at least three degenerate vacua distributed in two different universes. Hence, we have shown that the rest of the universes are lifted (those with one-form charge 1 and 3). We conclude that the Wilson lines with one-form symmetry charge equal to 2 are deconfined while those with one-form symmetry charge equal to 1 or 3 are confined. This is actually the scenario proposed in \cite{Cherman:2019hbq}.

Adding the mass term, we find that all the topological lines except for the $\bZ_2^\mathrm{v}$ symmetry lines are broken. Since the $\bZ_2^\mathrm{v}$ symmetry line is not charged under the one-form symmetry, the mass deformation, again restores the confinement of all the Wilson lines with non-trivial $N$-ality. 

Again assuming that the theory flows to the $\Spin(15)_1/\SU(4)_4$ TQFT we can calculate the condensate for the mass operator  $\cO_\mathrm{m}=\cO_{(1,0,1)}$ as well as $\cO_2$, using the NIM-rep matrix in \eqref{su4.nimrep}. We arrive at
\begin{equation}
\begin{array}{c|cccccccccccc}
\bra{\mathsf{v}} \cO \ket{\mathsf{v}} & \mathsf{v}_1 &\mathsf{v}_2 &\mathsf{v}_3 &\mathsf{v}_4 &\mathsf{v}_5 &\mathsf{v}_6 & \mathsf{v}_7 &\mathsf{v}_8 &\mathsf{v}_9 &\mathsf{v}_{10} &\mathsf{v}_{11} &\mathsf{v}_{12} \\ \hline
\cO_\cA & 1 & 1 & 1 & 1 & i & i & i & -1 & -1 & -i & -i & -i \\
\cO_2 & 0 & 0 & 0 & 0 & -2 & 2 & 2 & 0 & 0 & 2 & -2 & -2 \\
\cO_\mathrm{m} & -\sqrt{2}-1 & -\sqrt{2}-1 & \sqrt{2}-1 & \sqrt{2}-1 & -1 & 1 & 1 & 1-\sqrt{2} & \sqrt{2}+1 & -1 & 1 & 1
\end{array}
\end{equation}
The expectation value of $\cO_\cA = \cO_{(0,0,4)}$ distinguishes different universes. We observe that $\cO_2$ lifts universes $0$ and $2$, while leaving universes $1$ and $3$, related by the Wilson line with $N$-ality $2$, unlifted. This verifies that the fundamental Wilson line confines while the Wilson line with $N$-ality $N/2$ remains deconfined. Adding the mass deformation $m\cO_\mathrm{m}$ with $m>0$, only the vacua $\mathsf{v}_1$ and $\mathsf{v}_2$ remain unlifted which are related by the spontaneously broken $\bZ_2^\mathrm{v}$ symmetry. On the other hand, for $m<0$ the vacuum $\mathsf{v}_9$ remains unlifted and the $\bZ_2^\mathrm{v}$ symmetry is unbroken. By determining the lowest energy state in different universes we find the string tensions -- independent of the sign of $m$ -- as
\begin{equation}
    \left(
\begin{array}{c}
 T_0 \\
 T_1 \\
 T_2 \\
 T_3
\end{array}
\right) \sim \abs{m} \left(
\begin{array}{c}
 0 \\
 \sqrt{2} \\
 2 \\
 \sqrt{2}
\end{array}
\right)~.
\end{equation}
Note that one can summarize the $k$-string tension succinctly with the formula 
\begin{equation}
    T_k\sim |m|\sin(\pi k/4)~.
\end{equation}
up to a $k$-independent coefficient.

\subsection{SU(5)}
Finally, for our last example, we study the $\SU(5)$ adjoint QCD. In this case, the Ocneanu graphs have not appeared in the literature to our knowledge. Therefore, we do not have a complete list of the topological lines and their action on the deformation operators. However, the NIM-rep matrices in this case are derived in appendix \ref{app:su5.nimrep} and we use them to study the action of lines on the operators indirectly.

Dropping the kinetic term, we focus on the $\Spin(24)_1/\SU(5)_5$ gauged WZW model and calculate the condensates of the deformation operators $\cO_2=\cO_{(2,0,1,0)}+\cO_{(0,1,0,2)}$ and $\cO_\mathrm{m}=\cO_{(1,0,0,1)}$. We find that both of these operators lift all the universes with non-trivial $N$-ality, and the deformed theories by these operators confine. Hence we conclude that all the topological lines with non-trivial $N$-ality must be broken by these operators. Note that this conclusion is independent of whether the undeformed theory flows to the $\Spin(24)_1/\SU(5)_5$ TQFT in the IR.

Using the NIM-rep matrix $n_{(1,0,0,0)}$ given in equation \eqref{su5.nimrep}, we find the condensates of the deformation operators as
\begin{align}
&\begin{array}{c|cccccccc}
\bra{\mathsf{v}} \cO \ket{\mathsf{v}} & \mathsf{v}_1 &\mathsf{v}_2 &\mathsf{v}_3 &\mathsf{v}_4 &\mathsf{v}_5 &\mathsf{v}_6 & \mathsf{v}_7 &\mathsf{v}_8 \\ \hline
\cO_\cA & 1 & \omega & \omega^2 & \omega^3 & \omega^4 & \omega^4 & \omega^4 & \omega \\
\cO_2 & -4 & 1-\sqrt{5} & 11-5 \sqrt{5} & 11-5 \sqrt{5} & 1-\sqrt{5} & 2\sqrt{5}-4 & 2\sqrt{5}-4 & 2\sqrt{5}-4 \\
\cO_\mathrm{m} & -2 & 1-\sqrt{5} & \sqrt{5}-3 & \sqrt{5}-3 & 1-\sqrt{5} & 0 & 0 & 0
\end{array} \notag\\
&\begin{array}{c|cccccccc}
\bra{\mathsf{v}} \cO \ket{\mathsf{v}} & \mathsf{v}_9 &\mathsf{v}_{10} &\mathsf{v}_{11} &\mathsf{v}_{12} &\mathsf{v}_{13} &\mathsf{v}_{14} & \mathsf{v}_{15} &\mathsf{v}_{16} \\ \hline
\cO_\cA & \omega & \omega^3 & \omega^3 & 1 & 1 & 1 & 1 & 1 \\
\cO_2 &  2\sqrt{5}-4 & 2\sqrt{5}-4 & 2\sqrt{5}-4 & 36-16\sqrt{5} & 36-16\sqrt{5} & -4 & 36-16\sqrt{5} & 36-16\sqrt{5} \\
\cO_\mathrm{m} & 0 & 0 & 0 & 4-2 \sqrt{5} & 4-2 \sqrt{5} & -2 & 2\sqrt{5}-4 & 2\sqrt{5}-4
\end{array} \notag
\\
&\begin{array}{c|cccccccc}
\bra{\mathsf{v}} \cO \ket{\mathsf{v}} & \mathsf{v}_{17} &\mathsf{v}_{18} &\mathsf{v}_{19} &\mathsf{v}_{20} &\mathsf{v}_{21} &\mathsf{v}_{22} & \mathsf{v}_{23} &\mathsf{v}_{24} \\ \hline
\cO_\cA & \omega^2 & \omega^2 & \omega^2 & \omega^3 & \omega & \omega^4 & 1 & 1 \\
\cO_2 & 2\sqrt{5}-4 & 2\sqrt{5}-4 & 11-5 \sqrt{5} & 11-5 \sqrt{5} & 1-\sqrt{5} & 1-\sqrt{5} & -4 & -4 \\
\cO_\mathrm{m} & 0 & 0 & 3-\sqrt{5} & 3-\sqrt{5} & \sqrt{5}-1 & \sqrt{5}-1 & 2 & 2
\end{array} \label{su5.condensates}
\end{align}
where $\omega=e^{2\pi i/5}$. Again the $\cO_\cA=\cO_{(0,0,0,5)}$ condensate determines the universes. We see that the $\cO_2$ condensate lifts most of the vacua except for four of them ($\mathsf{v}_1,\mathsf{v}_{14},\mathsf{v}_{23},\mathsf{v}_{24}$) in universe 0. These four vacua correspond to the spontaneous breaking of the $\bZ_2^\mathrm{s}\times\bZ_2^\mathrm{c}$ center of $\Spin(24)$. Therefore, this is again consistent with the scenario that $\cO_2$ breaks all the non-invertible lines and preserves the invertible ones. The $\cO_\mathrm{m}$ condensate also lifts most of the vacua except for two of them, $\mathsf{v}_1$ and $\mathsf{v}_{14}$, in universe 0. Again, this suggest that these two vacua correspond to the spontaneous breaking of the $\bZ_2^\mathrm{v}$ symmetry, consistent with the scenario that the mass term breaks all the topological lines, except for the unbreakable $\bZ_2^\mathrm{v}$ symmetry line. By minimizing the $\cO_\mathrm{m}$ condensate in different universes we find the string tensions as
\begin{equation}
    \left(
\begin{array}{c}
 T_1 \\
 T_2 \\
 T_3 \\
 T_4
\end{array}
\right) \sim \abs{m} \left(
\begin{array}{c}
 1 \\
 \frac{1}{2} \left(\sqrt{5}+1\right) \\
 \frac{1}{2} \left(\sqrt{5}+1\right) \\
 1 
\end{array}
\right)~.
\end{equation}

The string tensions we have computed so far are only valid for the very small mass limit $\abs{m} \ll g_\mathrm{YM}$. This is because they are the first order correction to the energy of string states due to the perturbation $m\Tr(\psi_+ \psi_-) = m\cO_\mathrm{m}$. But we can also do a computation for the very large mass limit. In this limit, we can integrate out all the fermions and are left with the pure $\SU(N)$ gauge theory reviewed in section \ref{sec:review.adjQCD}. In this case, the energy of the state created by the Wilson line in a given representation is proportional to the quadratic Casimir of that representation. Minimizing the quadratic Casimir of representations with $N$-ality $k \pmod{N}$ we find
\begin{equation}
    T_k \sim g_\mathrm{YM}^2 \frac{k(N-k)}{N}~.
\end{equation}
The representation that minimizes the tension with $N$-ality $k$, is the $k$-th fundamental representation with $k$ antisymmetric indices and highest weight $\omega_k=(0,\dots,0,1,0,\dots,0)$. If we look at the condensates for the case of $N=5$ given in \eqref{su5.condensates}, we see that the tensions are minimized by the states $\mathsf{v}_1,\mathsf{v}_2,\mathsf{v}_3,\mathsf{v}_4,\mathsf{v}_5$ in universes $0,1,2,3,4$ respectively. Curiously, these vacua are related to the representations $\omega_0,\omega_1,\omega_2,\omega_3,\omega_4$ by the $\a$-inudction given in equation \eqref{su5.alpha.induction}. This suggests the possibility that the string tensions are minimized by the same representation for all values of the mass parameter and there is no first order phase transitions in different universes as we change the mass parameter away from the massless point $m=0$. Finally, we want to point out that our string tensions for the small mass limit and small $N$ agrees with the simple formula (also found in  different contexts in \cite{Douglas:1995nw,Hanany:1997hr,Herzog:2001fq,Armoni:2011dw})
\begin{equation}
    T_k \sim \abs{m} \sin(\pi k /N)~. \label{string.tension}
\end{equation}

It will be interesting to understand whether this formula is correct for all $k$ and $N$. We observe that the string tensions we have computed for small $N$, all coincide with the string tensions of the gauged $\SU(N)_N/\SU(N)_N$ WZW model. Note that adjoint QCD is related to the $\SU(N)_N/\SU(N)_N$ WZW model by discrete gauging as we have argued in previous sections. It is a tempting conjecture that, at least to this order in the quark mass, string tensions are not changed by discrete gauging and thus can be computed in the $\SU(N)_N/\SU(N)_N$ WZW model for all $N$. Calculating the string tension in the $\SU(N)_N/\SU(N)_N$ WZW model is easy and reduces to the computation of the ratios $\bra{\mathsf{v}_\mu} \cO_\mathrm{m} \ket{\mathsf{v}_\mu}=\frac{S_{\mu (1,0,\dots,0,1)}}{S_{\mu0}}$. Assuming this conjecture, we were able to extend the string tension formula \eqref{string.tension} to all $k$ and $N$.

We close this section by noting that we gauge the $\bZ_2^\mathrm{v}$ symmetry with an Arf twist to arrive at the fermionic theory. The conclusions about confinement remain true in the fermionic theory of course. It is tempting to conjecture that for all $N$, adding $\mathcal{O}_2$, one arrives at a confined phase where all the Wilson lines except the one with $N$-ality $N/2$ (for even $N$) are confined.
\section{IR TQFT and Adjoint QCD}
\label{sec:IR TQFT}
The aim of this section is to study the TQFT that characterizes the infrared (gapped) phase of adjoint QCD. We first study the bosonic TQFT by bosonization, and then fermionize this theory to get the fermionic IR TQFT.

As we discussed in section \ref{sec:bos}, the bosonized adjoint QCD\footnote{As discussed in Section~\ref{sec:bos}, there are two inequivalent bosonizations of the adjoint QCD when $N$ is odd and consider the specific one described in the section.} is dual to the $\Spin(N^2-1)_1/\SU(N)_N$ gauged WZW model plus the kinetic term
\begin{equation}
    \int \rd^2 x \, \frac{-1}{4g_\mathrm{YM}^2} \Tr F^2 ~,
\end{equation}
for the $\SU(N)$ gauge fields.
Since the coupling constant $g_\mathrm{YM}$ is super-renormalizable, one could imagine that the $g_\mathrm{YM}\rightarrow \infty$ limit should correspond to the deep IR limit.
Under this assumption, the kinetic term vanishes under the RG flow and the IR theory becomes the $\Spin(N^2-1)_1/\SU(N)_N$ coset with zero central charge, i.e.\ a TQFT.
Note that since $g_\mathrm{YM}$ is the only scale in the theory, the $g_\mathrm{YM}\rightarrow \infty$ limit is not meaningful and one cannot rigorously show that the IR TQFT is the same as the coset.
Nevertheless, the $\Spin(N^2-1)_1/\SU(N)_N$ coset is the most natural candidate for the IR TQFT.
In the following we study this coset in details, and then study other possibilities in section \ref{sec:IR.TQFT}.
Let us emphasize again that our previous conclusions about confinement vs deconfinement (in the theory not deformed by the mass or the quartic coupling) are independent of the assumption that the low energy theory is given by $\Spin(N^2-1)_1/\SU(N)_N$.
\subsection{\texorpdfstring{$\Spin(N^2-1)_1/\SU(N)_N$}{Spin/SU} TQFT \label{sec:coset.TQFT}}
Here we remove the kinetic term (i.e.\ $g_\mathrm{YM}=\infty$) and study the pure gauged WZW model $\Spin(N^2-1)_1/\SU(N)_N$.
This is one possible realization of ground states consistent with the category symmetries of the theory.
Since the central charge of the $\Spin(N^2-1)_1/\SU(N)_N$ gauged WZW model is zero, the coset is described by the conformal embedding of $\SU(N)_N$ inside $\Spin(N^2-1)_1$.
The vacua of the theory are determined by the branching rules of this embedding.
Each Kac-Moody primary of $\Spin(N^2-1)_1$ decompose into primaries of $\SU(N)_N$.
We use Latin letters for the former and Greek letters for the latter.
We write the branching rules as
\begin{equation}
\Module_a= \bigoplus_\m b_{a, \m } \, \module_\m ~, \label{branching}
\end{equation}
where $\Module_a$ and $\module_\m$ correspond to $\Spin(N^2-1)_1$ and $\SU(N)_N$ primaries respectively. The vacua of the $\Spin(N^2-1)_1/\SU(N)_N$ model are in one-to-one correspondence with the branching spaces labeled by $(a,\mu)$ with multiplicity $b^2_{a, \m}$. By the state/operator correspondence, we denote the local operators of the TQFT by $\cO_{a,\mu}^m$ with $m = 1,\dots,b^2_{a, \m}$.

\paragraph{Branching Rules}
The branching rules of the conformal embedding $\hat{\mathfrak{su}}(N)_N \subset \hat{\mathfrak{so}}(N^2-1)_1$, are studied in detail in appendix \ref{app:branching}. The spinor representations of $\hat{\mathfrak{so}}(N^2-1)_1$ decompose into $2^{\lfloor N/2 \rfloor-1}$-multiple of $\module_{[1,\dots,1]}$ \cite{Kac:1988tf}, while the vacuum $\Omega_0$ and vector $\Omega_1$ representations decompose into\footnote{We are following the notation of \cite{DiFrancesco:1997nk}, where the affine Dynkin labels of $\l$ are denoted by $[\l_0, \dots, \l_{N-1}]$, and its finite part by $(\l_1, \dots, \l_{N-1})$ omitting the zeroth Dynkin label given by $\l_0=N-\sum_{i=1}^{N-1} \l_i$ for $\hat{\mathfrak{su}}(N)_N$.}
\begin{equation}
    \Module_{\Omega_0}\oplus \Module_{\Omega_1} = \bigoplus_{\lambda} \module_{[\lambda_0,\dots,\lambda_{N-1}]}~, \label{branching.rule}
\end{equation}
where
\begin{align}
    \sum_{i=0}^{N-1} (\lambda_i+1) = 2N \quad &\mathrm{and} \quad (\lambda_i+1)\geq1~, \nonumber\\
    \sum_{i\leq k \leq j}(\lambda_k+1) \neq N \quad &\mathrm{for} \quad 0\leq i < j \leq N-1~,
\end{align}
and all such weights appear with multiplicity one. As shown in the appendix, $2^{N-1}$ representations appear on the RHS of \eqref{branching.rule} and along with the spinor representations they form a $3\times2^{N-2}$ dimensional vacuum Hilbert space for the bosonic theory.

\paragraph{Vacua vs. Universes}
The vacua of the coset TQFT can live in different universes, which  are labeled by the $\mathbb{Z}_N^{[1]}$ one-form symmetry. More precisely, the true vacua of the theory are eigenstates of the topological local operator that generates the one-form symmetry. To study this local operator and its eigenstates, first we take a detour to review the $\SU(N)_N/\SU(N)_N$ model or more generally $G/G$ models.

\paragraph{G/G TQFT}
As reviewed in appendix \ref{app:G/G}, the $G_k/G_k$ TQFT has topological lines $L_\mu$ -- Verlinde lines in $G_k$ WZW that survive after gauging -- and local topological primary operators $\mathcal{O}_\mu$ (we call them ``primary'' operators). These operators are both labeled by irreducible representations of the $\hat{\mathfrak{g}}_k$ affine Lie algebra and satisfy the same fusion algebra
\begin{align}
    L_\mu  L_\nu &= \sum_\l N_{\mu\nu}^\l \, L_\l~, \label{L fusion alg}\\
    \mathcal{O}_\mu \mathcal{O}_\nu &= \sum_\l N_{\mu\nu}^\l \, \mathcal{O}_\l~, \label{OPE}\\
    L_\mu \cdot \cO_\nu &= \frac{S_{\mu\nu}}{S_{0\nu}} \, {\cO_\nu}~,\label{line.on.states}
\end{align}
where $N_{\mu\nu}^\l$ and $S_{\mu\nu}$ are the fusion coefficients and the modular $S$-matrix of the affine Lie algebra.\footnote{$ \module_\mu \otimes \module_\nu= \bigoplus_\l N_{\mu\nu}^\l \module_\l $} Note that equation \eqref{line.on.states} is interpreted as the action of the topological line $L_\mu$ on the local operator $\cO_\nu$ by shrinking the line around $\cO_\nu$ -- see equation \eqref{lines.on.ops}. 
When $\cO_\nu$ is an invertible element of the fusion algebra \eqref{OPE}, we have the commutation relation
\begin{equation}
    \mathcal{O}_\nu L_\mu = \frac{S_{\mu\nu}^\ast S_{00}}{S_{0\mu}S_{0\nu}} \, L_\mu \mathcal{O}_\nu~, \label{ops.on.lines}
\end{equation}
that describes the action of $\cO_\nu$ on $L_\mu$ by passing $\cO_\nu$ through $L_\mu$ -- see equation \eqref{non.inv.lines}.

The true normalized vacua of the theory correspond to the boundary conditions $\ket{\mathsf{v}_\mu}$ imposed at infinity of $\mathbb{R}^2$ which generate superselection sectors on $\mathbb{R}^2$. These normalized vacua, written as linear combinations of the primaries, are
\begin{equation}
    \mathsf{v}_\mu = \sum_\nu S_{\mu\nu} \mathcal{O}_\nu ~.
\end{equation}
The $\mathsf{v}_\mu$ are analogous to boundary (Cardy) states in conformal field theory. They satisfy the generalized Cardy condition \cite{Moore:2006dw} and are related to the more conventional basis $\mathcal{O}_\nu$ (Ishibashi states) via the transformation above. 
The action of lines and primaries on these vacua is given by 
\begin{align}
    L_\mu \ket{\mathsf{v}_\nu} &= \sum_\l N_{\mu\nu}^\l \ket{\mathsf{v}_\l}~, \label{line.on.vacua} \\
    \cO_\mu \ket{\mathsf{v}_\nu} &= \frac{S^\ast_{\mu\nu}}{S_{0\nu}} \ket{\mathsf{v}_\nu}~. \label{ops.on.vacua}
\end{align}
Thus the Verlinde lines act on the primary operators diagonally, and on the boundary conditions by NIM-reps.

The invertible elements of the fusion algebras \eqref{L fusion alg} and \eqref{OPE} generate global zero-form and one-form symmetries respectively. For the case of $G_k=\SU(N)_N$, these invertible elements are generated by the representation $\module_{(0,\dots,0,N)}$ whose fusion with other representations generates the $\bZ_N$ outer automorphism\footnote{The outer automorphism group of an affine Lie algebra $\hat{\mathfrak{g}}$ is isomorphic to the center of the simply connected gauge group $\tilde{G}$.} of $\hat{\mathfrak{su}}(N)_N$ algebra \cite{DiFrancesco:1997nk}. To see this, note that $\module_{(0,\dots,0,N)}=\module_{\mathcal{A}0}$ where $\module_0=\module_{(0,\dots,0)}$ is the vacuum representation, and $\mathcal{A}$ denotes the action of the outer automorphism that acts on the affine weights by permuting the Dynkin labels cyclically. Using $N_{\mathcal{A}\mu,\mathcal{A}^{-1}\nu}^{\l}=N_{\mu\nu}^\l\,$, we get
\begin{equation}
     \module_{\mathcal{A}0} \otimes \module_\mu= \module_{0} \otimes \module_{\mathcal{A}\mu} = \module_{\mathcal{A}\mu}~.
 \end{equation}
Therefore $\module_{\cA0}$ generates the outer automorphism action. From now on we denote the representation $\module_{\cA0}$ by $\module_{\cA}$, i.e.\ $\module_{\cA\mu}=\module_{\cA} \otimes \module_{\mu}$.

The invertible element $\module_{\mathcal{A}}$ of the fusion algebra leads to the zero-form symmetry generated by the invertible line $L_{\mathcal{A}}$, and also to the one-form symmetry generated by the local operator $\mathcal{O}_{\mathcal{A}}$.
This one-form symmetry acts on the line operators (see \eqref{ops.on.lines}) and measures their $N$-alities, i.e.
\begin{equation}
    \frac{S_{\mathcal{A},\l}^\ast\,S_{0,0}}{S_{\mathcal{A},0}\,S_{\l,0}} = e^{ \frac{2\pi i}{N} (\l_1+2\l_2\dots+(N-1)\l_{N-1})}~,
\end{equation}
where $\l=(\l_1,\dots,\l_{N-1})$.
While  $\mathcal{O}_{\mathcal{A}}$ acts as the outer automorphism action on primary operators, it acts diagonally \eqref{ops.on.vacua} on the normalized vacua $\ket{\mathsf{v_\lambda}}$ (boundary conditions) with eigenvalue equal to the $N$-ality of $\lambda$.
This is because a local operator cannot cause a transition between the vacua and thus the topological local operators should be diagonalized on a vacuum.

Hence, we conclude that the one-form symmetry of the $\SU(N)_N/\SU(N)_N$ model acts on the primary operators as the $\mathbb{Z}_N$ outer automorphism of the $\hat{\mathfrak{su}}(N)_N$ algebra. This has a rather intuitive interpretation in terms of Chern-Simons theory $\SU(N)_N$. Such Chern-Simons theory has a $\mathbb{Z}_N$ one-form symmetry corresponding to an Abelian subalgebra of lines. These lines act on all the other lines of the theory. Since  $\SU(N)_N/\SU(N)_N$ TQFT is obtained by a circle compactification of $\SU(N)_N$ Chern-Simons theory the results above follow.
This statement generalizes to any $G_k/H_{\tilde k}$ WZW coset model. Namely, the $G_k/H_{\tilde k}$ model has a $Z(H)$ center one-form symmetry that acts as the outer automorphism of the affine Lie algebra of $H_{\tilde k}$ WZW.

\paragraph{G/H TQFT} 
For the case of $\Spin(N^2-1)_1/\SU(N)_N$ TQFT, the $\mathbb{Z}_N^{[1]}$ one-form symmetry acts on the local operators as the outer automorphism of $\hat{\mathfrak{su}}(N)_N$. More precisely, the local operators $\cO^m_{a,\mu}$ of the theory corresponds to the branching spaces in \eqref{branching}, and under the outer automorphism action
\begin{equation}
    \cO^m_{a,\mu} \mapsto \cO^{m'}_{\boldsymbol{\cA} a,\cA\mu}~, \label{out.action}
\end{equation}
where $\cA$ and $\boldsymbol{\cA}$ are elements of the outer automorphism of $\hat{\mathfrak{su}}(N)_N$ and $\hat{\mathfrak{so}}(N^2-1)_1$ respectively.\footnote{The branching rules are invariant under the outer automorphism action: $b_{\boldsymbol{\cA} a,\cA\mu}=b_{a,\mu}$.} 
In particular, the local operator $\cO_{\boldsymbol{\cA},\cA}$ generates the $\bZ_N^{[1]}$ symmetry.
Note that when $b_{a,\mu}>1$, the one-form symmetry can act non-trivially on the multiplicity label $m$ and extra care is needed to determine its action.
The outer automorphisms $\cA$ and $\boldsymbol{\cA}$ are related by a map from the center of $\SU(N)$ into the center of $\Spin(N^2-1)$ that is induced by the ``embeding" of $\SU(N)$ into $\Spin(N^2-1)$.
To be precise, the conformal embedding of the Lie algebras does not correspond to a true embedding of the gauge groups. In fact we have the embeddings
\begin{align}
    \SU(N)/\bZ_{N} \subset \Spin(N^2-1) \qquad &\text{for odd }N~,\notag\\
    \SU(N)/\bZ_{N/2} \subset \Spin(N^2-1) \qquad & \text{for even }N~.\label{embeddings}
\end{align}
For odd $N$, the center of $\SU(N)$ maps trivially into $\Spin(N^2-1)$, therefore $\cA$ maps to the trivial (vacuum) representation and $\boldsymbol{\cA}=0$. Whereas for even $N$, $\cA$ maps to the vector representation $\Module_{\Omega_1}$ generating the $\bZ_2^{\mathrm{v}}=Z(\Spin(N^2-1))$ outer automorphism and $\boldsymbol{\cA}=\Omega_1$.

It is precisely the non-triviality of $\boldsymbol{\cA}$ that is responsible for the mixed anomaly between the $\bZ_N^{[1]}$ one-form symmetry and the $\bZ_2^\chi$ chiral symmetry of the adjoint QCD reviewed in section \ref{sec:review.adjQCD}. This is because
\begin{equation}
    \cO_{\boldsymbol{\cA},\cA} \, \boldsymbol{L}_a = \frac{S_{\boldsymbol{\cA},a}^\ast S_{0,0}}{S_{\boldsymbol{\cA},0}S_{a,0}} \, \boldsymbol{L}_a \, \cO_{\boldsymbol{\cA},\cA}~,
\end{equation}
where $\boldsymbol{L}_a$ is a Verlinde line of the $\Spin(N^2-1)_1$ WZW model that survives the $\SU(N)_N$ gauging -- see section \ref{adjQCD.lines.bos} for a discussion of the topological lines of the coset. In particular, for even $N$ we get $\cO_{\boldsymbol{\cA},\cA}  \boldsymbol{L}_{\Omega_\mathrm{s}} = - \boldsymbol{L}_{\Omega_\mathrm{s}} \cO_{\boldsymbol{\cA},\cA}$, where $\Module_{\Omega_\mathrm{s}}$ is the spinor module of $\hat{\mathfrak{so}}(N^2-1)_1$. This translates to a ``mixed anomaly" between the one-form symmetry generated by $\cO_{\boldsymbol{\cA},\cA}$, and the non-invertible line $\boldsymbol{L}_{\Omega_\mathrm{s}}$. After fermionizing the theory, the Verlinde line $\boldsymbol{L}_{\Omega_\mathrm{s}}$ becomes the chiral $\bZ_2^\chi$ symmetry line \cite{Ji:2019ugf}, and the above anomaly becomes the mixed anomaly between $\bZ_N^{[1]}$ and $\bZ_2^{\chi}$.

Therefore, we learned that the one-form symmetry of the coset TQFT is generated by $\cO_{\boldsymbol{\cA},\cA}$ that acts as \eqref{out.action}. Furthermore, in the NS sector of the fermionic theory there are no multiplicities and equation \eqref{out.action} uniquely determines the action of one-form symmetry. Under this outer automorphism action, a local operators $\cO^m_{a,\mu}$ forms an orbit of some length that divides $N$. As in the $G/G$ case, the true vacua of the theory form a basis that diagonalizes the one-form symmetry action. A vacuum with eigenvalue $q$ under the one-form symmetry -- $\langle \cO_{\boldsymbol{\cA},\cA} \rangle = e^{2\pi i q/N}$ -- belongs to the $q$-th universe. These outer automorphism orbits are studied in appendix \ref{app:branching}, where it has been shown that there are $N$ degenerate universes with at least $2^{N-1}/N+O(2^{N/3})$ vacua in each of them.

\subsection{Other Possible IR TQFTs}\label{sec:IR.TQFT}
Now that we have studied the  $\Spin(N^2-1)_1/\SU(N)_N$ IR TQFT candidate in detail, we study other possibilities that are compatible with the topological lines of the (bosonic) theory in the UV.  Indeed, while it is plausible that  $\Spin(N^2-1)_1/\SU(N)_N$ is the correct answer for the IR TQFT, we have not proven that this is the case and it is worth dwelling on some other logical possibilities. 

As discussed in section \ref{adjQCD.lines.bos}, the bosonized theory has a rich set of topological lines denoted by $\aca$. $\cC=\Rep \hat{\mathfrak{su}}(N)_N$ describes the Verlinde lines of $\SU(N)_N$, and $A$ denotes a gauging
of it which identifies the diagonal $\Spin(N^2-1)_1$ WZW as a non-diagonal $\SU(N)_N$ WZW model
\begin{equation}
	\frac{\SU(N)_N \; \mathrm{WZW}}{A} = \Spin(N^2-1)_1  \mathrm{WZW} ~.
\end{equation} 
Physically, $\aca$ describes the topological lines of the diagonal $\Spin(N^2-1)_1$ WZW model that preserve the chiral algebra $\hat{\mathfrak{su}}(N)_N \subset \hat{\mathfrak{so}}(N^2-1)_1$.  
Recall that the bosonized adjoint QCD is dual to the $\Spin(N^2-1)_1/\SU(N)_N$ gauged WZW model plus a kinetic term. Thus $\aca$ describes precisely the lines that survives after gauging $\SU(N)$, and therefore are present in the bosonized adjoint QCD. Now since these lines remain invariant under the RG flow triggered by the kinetic term, they must also be present in the IR TQFT. Therefore possible gapped phases for the bosonic theory correspond to $\aca$-symmetric TQFTs.

Before moving further, we summarize some crucial facts about symmetric TQFTs that were discussed in sections \ref{sec:gauging} and \ref{sec:modular.invariants} (see also tables \ref{tab:correspondences} and \ref{tab:notations}). Let $\mathcal{D}$ describe a collection of -- possibly non-invertible -- topological lines, then
\begin{itemize}
    \item $\mathcal{D}$-symmetric TQFTs $\;\Leftrightarrow\;$ Different choices of gauging a non-anomalous subpart of $\mathcal{D}$.
    \item For the case of Verlinde lines $\cC$, different gaugings of $\mathcal{C}$ correspond to modular invariants of $\SU(N)_N$.
    \item Since $\aca$ is related to $\cC$ by gauging, there is a one-to-one correspondence between $\aca$-symmetric TQFTs and modular invariant RCFTs of $\SU(N)_N$.
\end{itemize}
Therefore, given any modular invariant RCFT of $\SU(N)_N$ with $\mim\left( B \right)$, there exist the $\aca$-symmetric TQFT $T_{\acb}$. In fact this TQFT is a gauging of the $\Spin(N^2-1)_1/\SU(N)_N$ TQFT by a subpart of its $\cC$ symmetry associated with $B$, and can be intuitively regarded as
\begin{equation}
    \frac{\Spin(N^2-1)_1/\SU(N)_N}{B}~. \label{IR.TQFT} 
\end{equation}
Moreover as explained in \ref{sec:modular.invariants}, there is a map from the lines of $\cC$ into $\aca$, that preserves their fusions and crossing relations.
Therefore any $\aca$-symmetric TQFT, is also a $\cC$-symmetric TQFT or equivalently a modular invariant for $\SU(N)_N$.
Thus the TQFT \eqref{IR.TQFT} is $\cC$-symmetric, and moreover corresponds to the modular invariant matrix
\begin{equation}
    \mim\left( A^{\mathrm{op}} \otimes B \right) = \mim\left( A \right)^\mathsf{T} \mim\left(B\right) ~, \label{TQFT.partition.function}
\end{equation}
where $\mim\left( A \right)=\mim\left( \Spin\right)$ is the $\SU(N)_N$ modular invariant matrix corresponding to the embedding into $\Spin(N^2-1)_1$.
By using this formula, we can decompose the TQFT \eqref{IR.TQFT}, that we also denote by $\underline{\Spin} \otimes \underline{B}$, as a direct sum of indecomposable $\cC$-symmetric TQFTs. This is done by decomposing the RHS of \eqref{TQFT.partition.function} into a sum of $\SU(N)_N$ modular invariants. Furthermore, we can count the number of vacua of this TQFT by
\begin{equation}
    \abs{\frac{\Spin(N^2-1)_1/\SU(N)_N}{B}} = \Tr \left[\mim\left(\Spin\right)^\mathsf{T} \mim\left(B\right) \right] ~. \label{trace.formula}
\end{equation}

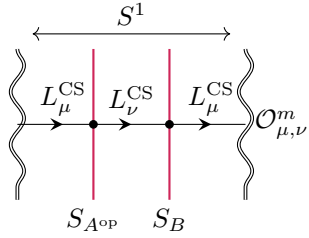
\begin{figure}[t]
    \centering
        \begin{tikzpicture}
        \draw[thick, xred] (ub) ++ (1,0) coordinate(us) -- ++ (0,-2) coordinate (B);
        \draw[thick, xred] (ub) ++ (0,0) -- ++ (0,-2) coordinate (A);
        \node[anchor = north] at (B) {$S_B$};
        \node[anchor = north] at (A) {$S_{A^\mathrm{op}}$};
        \draw (ubb) ++ (-.5,0) coordinate (uls);
        \draw (us) ++ (1,0) coordinate (urs);
        \draw[double, snake=snake, segment length=7mm] (uls) -- ++(0,-2) coordinate (dls);
        \draw[double, snake=snake, segment length=7mm] (urs) -- ++(0,-2) coordinate (drs);
        \draw[<->] (uls) ++ (.2,.2) -- node[midway,anchor = south] {$S^1$} ++ (2.6,0);
        \draw[->-=.5 rotate 0] (uls) ++ (0,-1) coordinate (Ol)  -- ++ (.5,0) coordinate(Ob) --++ (.5,0) coordinate(Obb)  -- node[midway,anchor=south] {$L_\nu^\mathrm{CS}$} ++(1,0)  coordinate(Os) -- ++(1,0) coordinate(Or);
        \draw[->-=.5 rotate 0]  (-0.4,-1); 
        \node[anchor = south] at (-0.4,-1) {$L_\mu^\mathrm{CS}$};
        \draw[->-=.5 rotate 0]  (1.55,-1); 
        \node[anchor = south] at (1.55,-1) {$L_\mu^\mathrm{CS}$};
        \node[anchor = west] at (Or) {$\cO_{\mu,\nu}^m$};
        \draw[fill] (Obb) circle[radius = .05];
        \draw[fill] (Os) circle[radius = .05];
    \end{tikzpicture}
    \caption{The TQFT $T_{\acb}$ is obtained by compactifying the 3d $\SU(N)_N$ CS on $S^1$ with the insertion of surface operators $S_{A^\mathrm{op}}$ and $S_B$ transverse to $S^1$. The left and right wiggly lines are identified in the picture. The topological loop operators $L_\mu^\mathrm{CS}$ and $L_\nu^\mathrm{CS}$ wrapped around $S^1$ correspond to local operators of the 2d TQFT denoted by $\mathcal{O}_{\mu,\nu}^m$ where $m=1,\dots,\mim(A)_{\nu\mu}\mim(B)_{\nu\mu}$ labels the multiplicity.}
    \label{fig:3dtqft}
\end{figure}

The IR TQFT \eqref{IR.TQFT} can also be described from the 3d point of view. The TQFT $T_{\acb}$ can be constructed by compactifying the 3d $\SU(N)_N$ CS theory on $S^1$ with the insertion of two non-intersecting surface operators $S_{A^\mathrm{op}}$ and $S_B$ transverse to $S^1$, see figure \ref{fig:3dtqft}. The surface operator $S_{A^\mathrm{op}}$ is the orientation reversal of $S_A$. In this picture, it is clear that the local operators of the 2d TQFT correspond to topological line operators $L_\mu^\mathrm{CS}$ and $L_\nu^\mathrm{CS}$ wrapped around $S^1$. These topological line operators meet at the surfaces $S_{A^\mathrm{op}}$ and $S_B$, and the dimension of the vector spaces at these intersections is equal to $\mim(A)_{\nu\mu}$ and $\mim(B)_{\nu\mu}$ respectively. Therefore the multiplicity of the corresponding 2d local operator  $\mathcal{O}^m_{\mu,\nu}$ (labeled by $m$), is given by $\mim(A)_{\nu\mu}\mim(B)_{\nu\mu}$. This is consistent with the trace formula \eqref{trace.formula}, counting the number of vacua in this TQFT.

In this 3d picture, the one-form symmetry generator $\cO_\mathcal{A}$ of the 2d TQFT is realized by the topological line $L_\mathcal{A}^\mathrm{CS}$ with highest weight $\mathcal{A}=(0,\dots,0,N)$ wrapped around $S^1$. The multiplicity of this operator is equal to $\mim(A)_{\mathcal{A}\mathcal{A}}\mim(B)_{\mathcal{A}\mathcal{A}}=\mim(B)_{\mathcal{A}\mathcal{A}}$, since $\mim(A)_{\mathcal{A}\mathcal{A}}=1$ from the branching rules given in appendix \ref{app:branching}. Therefore, this shows that the TQFT $T_{\acb}$ admits the symmetries of the bosonic adjoint QCD iff $\mim(B)_{\mathcal{A}\mathcal{A}}\neq0$. Here by symmetries, we mean collectively the topological line operators given by $\aca$ and the topological local operators that generate the $\bZ_N^{[1]}$ one-form symmetry.

Hence, to find the possible IR TQFTs, we have to find all the modular invariants of $\SU(N)_N$ satisfying $\mim(B)_{\mathcal{A}\mathcal{A}}\neq0$. For small values of $N$, namely $N \leq 4$, all modular invariants of $\hat{\mathfrak{su}}(N)_k$ are known \cite{Cappelli:1987xt, DiFrancesco:1989ha, Gannon:1992ty, Ocneanu:2000kj}. But this classification in general is an open problem. In the following we discuss some known modular invariants of $\hat{\mathfrak{su}}(N)_N$ that exist for any $N$, and construct $\aca$-symmetric TQFTs out of them. In appendix~\ref{app:fermionic} we discuss the $N=3,4,5,6$ examples in detail, where for the case of $N=3$ and $N=4$ we provide a complete classification of the possible gapped vacua.

\paragraph{Modular invariants of $\SU(N)_N$:}
As reviewed in \ref{sec:modular.invariants}, modular invariants correspond to left and right extensions of the chiral algebra. There are two practical methods to find such extensions that are useful for $\SU(N)_N$. The first method is by conformal embedding of the affine Lie algebra into a larger affine Lie algebra \cite{Schellekens:1986mb,Bais:1986zs}. The second one, is by gauging a symmetry that is generated by the outer automorphism of the affine Lie algebra \cite{Bernard:1986xy}, and the resulting modular invariant is called a D-type\footnote{This terminology is coming from the ADE classification of $\SU(2)_k$ modular invariants \cite{Cappelli:1987xt}.} invariant. There is also the charge conjugation modular invariant, but since it does not satisfy the requirement $\mim(B)_{\mathcal{A}\mathcal{A}}\neq0$, is not allowed.

The list of all modular invariants of $\SU(N)_N$ that can be obtained by combining these methods is given in appendix~\ref{app:modular.invariants}. Using these modular invariants, we find $\aca$-symmetric TQFTs that can arise as a possible gapped phase of the bosonic theory.

\paragraph{Even $N$}
For even $N$, we find only two types of modular invariants: D-type invariants that we denote by $\underline{{\SU}/{\bZ_k}}$, and the $\underline{\Spin}$ modular invariant given by the conformal embedding $\hat{\mathfrak{su}}(N)_N\subset\hat{\mathfrak{spin}}(N^2-1)_1$. Therefore, as we discussed above, we can build $\aca$-symmetric TQFTs by taking the product of these modular invariant matrices with $\mim(\Spin)$. By decomposing the resulting matrix into a sum of physical\footnote{Note that a physical modular invariant has $\mim_{00}=1$.} modular invariants, and using their explicit formula given in appendix \ref{app:modular.invariants} and equations \eqref{branching1} and \eqref{branching2}, we find
\begin{align}
    \underline{\Spin} \otimes \underline{{\SU}/{\bZ_k}} &=k \,\underline{\Spin}~, \qquad \text{for } \frac{N}{k} \overset{2}{\equiv} 0~,\\
    \underline{\Spin} \otimes \underline{{\SU}/{\bZ_k}} &= \frac{k}{2} \, \underline{\Spin}~, \qquad \text{for }\frac{N}{k} \overset{2}{\equiv} 1 ~,\\
    \underline{\Spin} \otimes \underline{\Spin} &= 2^{N-2} \, \underline{\Spin}~.
\end{align}
Therefore we see that all such $\aca$-symmetric TQFTs, as a $\cC$-symmetric TQFT contain the $\underline{\Spin}$ modular invariant, or equivalently, the $\Spin(N^2-1)_1/\SU(N)_N$ TQFT. Thus the $\Spin(N^2-1)_1/\SU(N)_N$ candidate TQFT, remains the smallest possible IR TQFT with the smallest number of vacua. It is physically reasonable to assume that under the RG flow, some of the vacua get lifted, and the IR TQFT becomes the one with the smallest number of vacua. So the $\Spin(N^2-1)_1/\SU(N)_N$ TQFT seems to be the most natural choice for the IR TQFT of the bosonic theory.

\paragraph{Odd $N$}
When $N$ is odd, apart from the D-series modular invariants $\underline{{\SU}/{\bZ_k}}$ and the $\underline{\Spin}$ invariant, there also exist the $\underline{\SO}$, $\underline{\mathrm{Ss}}$, $\underline{\mathrm{Sc}}$, and $\underline{\PO}^{i=1,2}$ modular invariants obtained by further gaugings of the ${\bZ_2^\tts \times \bZ_2^\ttc}$ center of ${\Spin(N^2-1)_1}$. However some of these theories have the same modular invariant matrices, and thus the same number of vacua -- $\mim({\Spin})=\mim({\SO})$ and $\mim({\mathrm{Ss}})=\mim({\mathrm{Sc}})=\mim(\PO^1)=\mim(\PO^2)$. We can build $\aca$-symmetric TQFTs out of them as
\begin{align}
    \underline{\Spin} \otimes \underline{{\SU}/{\bZ_k}} &=k \,\underline{\Spin} ~,\\
    \underline{\Spin} \otimes \underline{\Spin} &= \big(2^{N-2} - 2^{\frac{N-3}{2}}\big) ~ \underline{\Spin} \oplus 2^{\frac{N-1}{2}} \,\underline{\PO} ~,\\
    \underline{\Spin} \otimes \underline{\PO} &= \begin{cases}
    \big(2^{N-2} + 2^{\frac{N-1}{2}}\big) \, \underline{\PO} & N\overset{8}{\equiv} \pm 1\\
    2^{\frac{N-3}{2}}\,\underline{\Spin} \oplus 2^{N-2}\,\underline{\PO} & N\overset{8}{\equiv} \pm 3
    \end{cases}~,
\end{align}
Using the trace formulae
\begin{align}
    \abs{\underline{\Spin}} &= \Tr\left[ \mim(\Spin) \right] = 3\times2^{N-2} ~,\\
    \abs{\underline{\PO}} &= \Tr\left[ \mim(\PO) \right] = \begin{cases}
    3\times2^{N-3}+2^{\frac{N+1}{2}} & N\overset{8}{\equiv} \pm 1\\
    3\big( 2^{N-3} + 2^{\frac{N-3}{2}} \big) & N\overset{8}{\equiv} \pm 3
    \end{cases}~,
\end{align}
we can count the number of vacua in these TQFTs. As mentioned above there exist four more theories whose number of vacua are given by $\abs{\underline{\Spin} \otimes \underline{\SO}} = \abs{\underline{\Spin} \otimes \underline{\Spin}}$ and $\abs{\underline{\Spin} \otimes \underline{\PO^i}} = \abs{\underline{\Spin} \otimes \underline{\mathrm{Ss}}} =  \abs{\underline{\Spin} \otimes \underline{\mathrm{Sc}}}$.
Although the $\Spin(N^2-1)_1/\SU(N)_N$ TQFT is not contained in all the decompositions above, it still remains the smallest $\aca$-symmetric TQFT among our list, and perhaps the natural expectation for the IR TQFT.
\section*{Acknowledgements}
We would like to thank A. Armoni, A. Cherman, T. Dumitrescu, T. Jacobson, I. Klebanov, G. Moore, M. Ro\v cek, 
B. Ruba, N. Seiberg, M. Shokrian Zini, Y. Tanizaki, M. \"Unsal and Y. Wang for various discussions and comments. ZK, KO and SS are supported in part by the Simons Foundation grant 488657 (Simons Collaboration on the Non-Perturbative Bootstrap) and the BSF grant no. 2018204. KR is supported by the NSF Grant PHY-1915093.

\appendix

\section{Fusion Categories}
\label{app:FusionCat}
\subsection{Axioms \label{app:axioms}}
In this appendix, we review the axioms of fusion categories \cite{etingof2005fusion,etingof2009fusion,etingof2016tensor}; see \cite{Bhardwaj:2017xup} for a physicists’ review, and also \cite{Chang:2018iay} for more physical examples and discussions. A basic and simple example of a fusion category, is the category of representations of some finite group $G$ denoted by $\mathrm{Rep}(G)$. To make the axioms of fusion categories more intuitive, as we define each axiom we also make connection with the simple example of $\mathrm{Rep}(G)$. For more details and pictorial representations of these axioms see \cite{Bhardwaj:2017xup}, which we follow closely.
\begin{enumerate}
	\item \textbf{Lines (Objects):} The objects in a fusion category $\mathcal{C}$ correspond to oriented topological line operators generating the symmetry. More precisely, for any oriented path $C$ and object $a\in \mathcal{C}$, there exist the topological line operator $a(C)$ which can be inserted in the path integral and its dependence on the path $C$ is topological; Objects of $\mathrm{Rep}(G)$ are representations of $G$.
	\item \textbf{Defect Operators (Morphisms):} The morphisms in $\mathcal{C}$, corresponds to local topological defect operators between two oriented lines $a$ and $b$. Such defect operators (morphisms) form a complex vector space denoted by $\mathrm{Hom}(a,b)$; These morphisms in $\mathrm{Rep}(G)$ are the intertwiners between representations $a$ and $b$. Moreover, there is a composition operation between defect operators, which can be thought as a kind of operator product algebra. Namely given defect operators $m\in \mathrm{Hom}(a,b)$ and $n\in \mathrm{Hom}(b,c)$ there exist the defect $n \circ m\in \mathrm{Hom}(a,c)$, which can be thought of as bringing the topological defects $m$ and $n$ close together and viewing the segment starting with $m$ and ending in $n$ as a single defect operators between the lines $a$ and $c$; In $\mathrm{Rep}(G)$, this composition of morphisms is just the compositions of linear maps (intertwiners). An isomorphism is a morphism which has a left and right inverse morphisms; In $\mathrm{Rep}(G)$, isomorphic representations are those which are related by a similarity transformation.
 	\item \textbf{Additive Structure:} Given two lines operators $a$ and $b$, there is a new line operators denoted by $a \oplus b$ which is simply the sum of two such operators such that $\langle \cdots (a \oplus b) (C) \cdots \rangle = \langle \cdots a(C) \cdots \rangle + \langle \cdots b(C) \cdots \rangle$; Additive structure in $\mathrm{Rep}(G)$ is the direct sum of representations.
	\item \textbf{Fusion:} Having two lines $a$ and $b$, one can bring them close and parallel together and consider them as a single line denoted by $a \otimes b$; For $\mathrm{Rep}(G)$, $a \otimes b$ is the tensor product of representation $a$ and $b$. Any fusion category contains the invisible trivial line $\mathbbm{1} \in \mathcal{C}$, which acts as the identity element under fusion, namely $a\otimes \mathbbm{1} = \mathbbm{1} \otimes a =a$. In $\mathrm{Rep}(G)$, the identity object is the trivial one-dimensional representation.
	\item \textbf{Simplicity, Semisimplicity, and Finiteness:} Simple lines such as $a\in\mathcal{C}$ are defined as lines whose defect spaces $\mathrm{Hom}(a,a)$ is one-dimensional and thus isomorphic to $\mathbb{C}$; Simple objects of $\mathrm{Rep}(G)$, are the irreducible representations of $G$. Fusion categories are finite in the sense that the number of isomorphism classes of simple lines is finite; Clearly $\mathrm{Rep}(G)$ has only finitely many non-equivalent irreducible representations for finite $G$.  Moreover, fusion categories are semisimple in the sense that any line is isomorphic to a direct sum of simple lines; In $\mathrm{Rep}(G)$, this is the statement that any representations have a unique decomposition into irreducible representations. Furthermore, in fusion categories as opposed to multifusion categories, one assumes that the identity line $\mathbbm{1}$ is simple, i.e.\ there are no topological local operators; In $\mathrm{Rep}(G)$, this is the condition that the trivial representation is irreducible. Therefore, multifusion category generalizes the notion of both finite 0-form and 1-form symmetries and their interactions.
	
	\item \textbf{Associativity Structure:} The fusion operation is associative $(a\otimes b) \otimes c \simeq a \otimes (b \otimes c)$. Note that, this condition is different from strict associativity where the isomorphism sign $\simeq$ is replaced by the equal sign. In a category, for two isomorphic objects, generally there does not exist a canonical isomorphism between them and an isomorphism between isomorphic objects is a data which one should keep track of. In particular, the associativity structure for lines $a$, $b$, and $c$ is a particular isomorphism
	\begin{equation}
		\alpha_{a,b,c} \in \mathrm{Hom}((a\otimes b) \otimes c, a \otimes (b \otimes c))~,
	\end{equation}
	which is called the \textit{associator}; Wigner's $6j$ symbols (Racah coefficients) in the study of representations of $\SU(2)$ (or Moore-Seiberg fusion $F$ matrices \cite{Moore:1988qv}) capture the data of associators if one fixes a basis for the set of simple lines in the fusion category.
	
	By composing different associator isomorphisms, one arrives at a consistency conditions of such associators known as the pentagon identity, see \cite{Bhardwaj:2017xup, Chang:2018iay}. It is worth mentioning that in practice, the data of a fusion category is captured by the \textit{fusion ring} of the category and a solution of the pentagon identity for that fusion ring, where usually there exist only a handful of such solutions. The fusion ring, is the data of fusion for the isomorphism classes of simple objects, captured by the structure constants of the ring. Solutions to the pentagon identity is a solution for the associators written in terms of a fixed basis of simple lines. For instance, for the Ising fusion ring there exist two solutions to the pentagon identity known as $\bZ_2$ Tambara-Yamagami fusion categories ($\mathrm{TY}_\pm$)~\cite{TAMBARA1998692}. $\mathrm{TY}_+$ and $\mathrm{TY}_-$ are realized by the topological lines of the Ising CFT and the Verlinde lines of the $\SU(2)_2$ WZW model respectively; For the case of $\mathrm{Rep}(G)$, denote $D_8$ and $Q_8$ as the dihedral group and the quaternion group of order 8. The fusion rings of the categories $\mathrm{Rep}(D_8)$ and $\mathrm{Rep}(Q_8)$ are isomorphic while they have different associators and realize different solutions of the pentagon identity, thus as categories $\mathrm{Rep}(D_8)$ and $\mathrm{Rep}(Q_8)$ are not equivalent \cite{TAMBARA1998692}.
	\item \textbf{Dual Structure (Folding):} For any line $a$, there exist the dual line $a^\ast$ which is equivalent to the orientation reversal of $a$, such that $(a^\ast)^\ast \simeq a$ and $(a\otimes b)^\ast = b^\ast \otimes a^\ast$; In $\mathrm{Rep}(G)$, the dual object is the complex conjugate representation. Moreover, a line $a$ can be folded to form the line $a^\ast\otimes a$, and this data is captured by particular defect operators at the end of the line $a^\ast\otimes a$ which are known as the evaluation and co-evaluation morphisms $\epsilon_a \in \mathrm{Hom}(a^\ast\otimes a,\mathbbm{1})$ and $\epsilon^a \in \mathrm{Hom}(\mathbbm{1},a^\ast\otimes a)$; In $\mathrm{Rep}(G)$, the existence of the evaluation morphism follows from the fact that the trivial representation is always contained in $a^\ast \otimes a$ in a canonical way for any representation $a$. Using the folding (evaluation morphisms), one can calculate the expectation value of a loop of line $a$ by
	\begin{equation}
		\begin{tikzpicture}[scale = .7, baseline = -2.5]
			\draw[ ->- = .07 rotate -25] (0,0) arc (0:360:.5);
			\node[anchor = west] at (0,0) {$a$};
			\node[anchor = south] at (-0.5,0.5) {$\epsilon_a$};
			\draw[fill] (-0.5,0.5) circle [radius = .05];
			\node[anchor = south] at (-0.5,-0.5) {$\epsilon^a$};
			\draw[fill] (-0.5,-0.5) circle [radius = .05];
		\end{tikzpicture} = 
		\epsilon_a \circ \epsilon^a = (\mathrm{dim}\,a) \mathbb{I} \in \mathrm{Hom}(\mathbbm{1},\mathbbm{1})\,,
	\end{equation}
	where $\mathrm{dim}\,a$ is known as the quantum dimension of $a$; For $\mathrm{Rep}(G)$, the quantum dimension of a representation is the same as the dimension of that representation.
	\item \textbf{Unitary Structure:} For any defect operator $m \in \mathrm{Hom}(a,b)$ from the line $a$ to $b$, there exist the Hermitian conjugate defect operator $m^\dagger \in \mathrm{Hom}(b,a)$ form $b$ to $a$; For $\mathrm{Rep}(G)$, this is just the Hermitian conjugate of linear maps (the intertwiners) between complex vector spaces. We also require that $\epsilon^{a} = (\epsilon_{a^\ast})^\dagger$.
\end{enumerate}

\subsection{Gauging and Frobenius Algebras \label{app:gauging}}
In this appendix we discuss the relation between gauging and Frobenius algebras in category symmetries. First, we state it for the case of group-like symmetries in a language that can be generalized later on to category symmetries. For a 2d theory $T$ with a discrete group symmetry $G$, a non-anomalous subgroup of it such as $H\subset G$ can be gauged. As stated in \ref{sec:gauging}, the gauging operation is done by summing over the insertions of different topological lines in $H$, or equivalently summing over $H$-gauge fields $a \in H^1(M,H)$. However, this operation is not unique since there is a freedom to insert $a$-dependent phases to get
\begin{equation}
    Z_{T/(H,\omega)}[M] = \# \sum_{a\in H^1(M,H)} e^{2\pi i \omega(a)} Z_T[M,a]~,
\end{equation}
for $\omega \in H^2(H,\mathbb{R}/\bZ)$ known as the discrete torsion \cite{Vafa:1986wx, Brunner:2014lua}. 

Alternatively, one could consider the line
\begin{equation}
    A = \bigoplus_{h\in H} h~,
\end{equation}
given by the sum of the topological lines in $H$. Then summing over all the insertions is equivalent to inserting a fine-enough trivalent mesh of $A$ into the path integral \cite{Carqueville:2012dk,Brunner:2013xna,Bhardwaj:2017xup}. Note that, the choice of  $\omega$ is captured by defect operators in the three-way junctions of $A$ in the mesh. Thus the gauging is determined by a choice of $(H,\omega)$, or equivalently a mesh of $A$ including defect operators $\mu\in \mathrm{Hom}(A\otimes A,A)$ and $u \in \mathrm{Hom}(1,A)$ which are known as the multiplication and unit morphisms respectively. These morphisms satisfy the associativity and unit axioms, and along with some other details stated in \cite{Bhardwaj:2017xup}, form a \emph{symmetric Frobenius algebra} algebra in $\mathrm{Vec}_G$ (see footnote \ref{VecG}). This notion of gauging by algebras can be generalized to any category symmetries.

For a theory $T$ with a category symmetry $\cC$, gauging a non-anomalous subpart of $\cC$ correspond to inserting a fine-enough mesh of a symmetric Frobenius algebra object $A$ in $\cC$ \cite{Fuchs:2002cm, Carqueville:2012dk, Bhardwaj:2017xup}. We denote the gauged theory by $T/A$. However, note that different algebra objects can lead to the same gauged theory and such algebras are called \emph{Morita equivalent} algebras. The Morita equivalence class of an algebra $A\in \cC$, is determined uniquely by the category of $A$-modules in $\cC$ which we denote by $\cC_A$. As we will explained below $\cC_A$ admits a $\cC$-action, or more precisely form a module category over $\cC$.
\subsection{Symmetric 2d TQFTs and Module Categories}
\label{app:module.categories}
In this appendix, we construct all $\mathcal{C}$-symmetric TQFTs for any fusion category $\mathcal{C}$. More precisely, we construct 2d unitary TQFTs that admit topological line operators whose fusion category is given by $\mathcal{C}$. Such construction for group like symmetries with 't Hooft anomalies in arbitrary dimensions was given in \cite{Witten:2016cio, Wang:2017loc, Tachikawa:2017gyf}. For category symmetries in two dimensions this is already answered in \cite{Thorngren:2019iar}, here we give a different and more straightforward construction. Namely we first study the maximal category of boundary conditions (D-branes) of such TQFTs and from it we determine the symmetric TQFT similar to the construction in \cite{Moore:2006dw}. We begin by summarizing some basic facts about two-dimensional unitary TQFTs.

Unitary closed 2d TQFTs are equivalent to semisimple commutative Frobenius algebras, while open/closed (or extended) theories also come with a category $\mathcal{M}$ of boundary conditions that are equipped with trace pairing \cite{Moore:2006dw,Carqueville:2016nqk}. Such categories equipped with traces are called Calabi-Yau categories in the literature \cite{costello2007topological}. As it was pointed out in \cite{Moore:2006dw}, given a semisimple Calabi-Yau category $\mathcal{M}$, then this is a category of boundary conditions of a canonical 2d TQFT whose commutative Frobenius algebra is the ring of endomorphisms of the identity functor of $\mathcal{M}$. Thus for our purposes, the data of a 2d TQFT is just their (maximal) category of boundary conditions. Furthermore for a category of boundary conditions $\cM$, we define the corresponding TQFT by $T_\cM$.

Now for a $\mathcal{C}$-symmetric 2d TQFT, $\mathcal{C}$ acts on the boundary states, therefore $\mathcal{M}$ should form a \emph{module category} \cite{ostrik2003module,Bhardwaj:2017xup,Thorngren:2019iar} over $\mathcal{C}$ with traces that are compatible with the action of $\mathcal{C}$. It is shown in \cite{Schaumann_2013} that any $\mathcal{C}$-module category has a $\mathcal{C}$-compatible traces making it a Calabi-Yau category, and moreover when the module category is indecomposable the trace is unique up to an overall normalization. Therefore $\mathcal{C}$-symmetric 2d TQFTs are in bijection with $\cC$-module categories.
Moreover, the number of boundary states or the number of vacua on the torus is the same as the number of simple objects in $\mathcal{M}$.
\begin{theorem}
	Unitary 2d TQFTs with category symmetry $\mathcal{C}$ are in one to one correspondence with module categories $\mathcal{M}$ over $\mathcal{C}$. Moreover $\mathcal{M}$ is the maximal category of boundary conditions (D-branes), which determines the TQFT uniquely for indecomposable $\mathcal{M}$. The corresponding TQFT is defined by $T_\cM$.
\end{theorem} 
Module category can be thought as a generalization of ``representation" for category symmetries. Note that boundary conditions form a category, as opposed to bulk states which form a vector space. Group symmetries act on the Hilbert space and this action -- without taking into account the anomaly -- is captured by a representation. Whereas for category symmetries -- taking into account the anomaly -- the topological lines act on the boundary conditions and this action is captured by a module category. This action also leads to a nonnegative integer matrix representation for the fusion ring \cite{Cardy:1989ir} which is usually called a NIM-rep \cite{Behrend:1999bn,Gannon:2001ki,Gaberdiel:2002qa}. 

The relation between $\cC$-symmetric 2d TQFTs and $\cC$-module categories was proved in \cite{Thorngren:2019iar}, by putting the 2d TQFT on the boundary of the 3d Turaev-Viro TQFT \cite{Turaev:1992hq} associated to $\mathcal{C}$. The Turaev-Viro theory is basically the 3d TQFT associated with the modular tensor category $Z(\mathcal{C})$ -- the Drinfeld center of $\mathcal{C}$ (see \cite{Bhardwaj:2016clt} for a review). The construction of $\mathcal{C}$-symmetric theories in \cite{Thorngren:2019iar} was given by compactifying the Turaev-Viro theory on the interval. It is known that the gapped boundary conditions of $\mathcal{C}$-Turaev-Viro TQFT are described by the 2-category of $\mathcal{C}$-module categories \cite{Kapustin:2010if,Kitaev_2012,Fuchs:2012dt,Freed:2020qfy}. Thus viewing $\mathcal{C}$ and $\mathcal{M}$ as $\mathcal{C}$-module categories, we could reduce the Turaev-Viro on the interval with these boundary conditions and get a $\mathcal{C}$-symmetric 2d TQFT. Now we give some simple examples.

\paragraph{The Regular \texorpdfstring{$\mathcal{C}$}{C}-Symmetric TQFT}
Any unitary fusion category $\mathcal{C}$ can be viewed as a module category over itself -- known as the regular $\cC$-module category. Thus there exist the canonical $\mathcal{C}$-symmetric TQFT $T_{\cC}$. The category of boundary condition of this theory is given by $\mathcal{C}$, hence the number of vacua in $T_{\cC}$ is the same as the number of simple objects in $\mathcal{C}$. Below we discuss some simple examples for the case of group-theoretical category symmetries such as $\mathrm{Vec}_G$ and $\mathrm{Rep}(G)$, and modular tensor categories.

\paragraph{$\mathrm{Vec}_G$:}
Following the above procedure, namely finding the ring of endomorphisms of the identity functor in $\cC$, we study the regular TQFT $T_\cC$ for the case of $\cC=\mathrm{Vec}_G$. One can identify the commutative Frobenius algebra of $T_\cC$ with functions $f:G \to \mathbb{C}$, with the obvious algebra
\begin{align}
	\left(f + h\right)(g) &= f(g) + h(g)~,  \notag\\
	\left(f \cdot h\right)(g) &= f(g) \cdot h(g)~,
\end{align}
with trace $\theta(f) = \sum_{g\in G} f(g)$. The idempotents are just delta functions $f_h(g)=\delta_{g,h}\,$, thus all the boundary state have the same norm and the partition function is just a constant $Z[\Sigma_g]=\abs{G}$.

\paragraph{$\mathrm{Rep}(G)$:} For $\mathcal{C}=\mathrm{Rep}(G)$, $T_\cC$ is just the discrete $G$-gauge theory. The commutative Frobenius algebra in this case is the center of the group ring $\mathbb{C}[G]$ \cite{Moore:2006dw}. The trace is defined by $\theta(\sum_g f_g\, g) = f_1/\abs{G}$, for $\sum_g f_g\, g$ a central element in $\mathbb{C}[G]$, i.e.\ $f_{hgh^-1}=f_g$. The idempotents or boundary states defined in \ref{app:2d.TQFT}, in this case are given by the characters $\chi_\mu$ of irreducible representations $\mu$ of $G$, i.e.\ $f_g =\chi_\mu(g) \chi_\mu(1)$. Therefore, the partition function of this theory is calculated as 
\begin{equation}
    Z[\Sigma_g] = \sum_\mu \theta_\mu^{1-g} =\abs{G}^{1-g} \sum_\mu \frac{1}{{\chi_\mu(1)}^{2g-2}}~,
\end{equation}
which indeed is the partition function of the discrete $G$-gauge theory.

\paragraph{Modular Categories:}
When $\cC$ is a modular tensor category, $T_{\cC}$ can be obtained by compactifying the (Witten-Reshetikhin-Turaev \cite{Witten:1988hf,Reshetikhin:1990pr}) 3d TQFT associated with $\cC$ on the circle, or equivalently by the folding trick, putting the $\cC$ Turaev-Viro TQFT on the interval. The commutative Frobenius algebra in this case is the same as the fusion ring of $\cC$, with the trace given by the modular $S$-matrix of $\cC$. More precisely, the idempotents are labeled by elements of $\mu\in\cC$ with trace $\theta_\mu = (S_{0\mu})^2$. When $\mathcal{C}=\mathrm{Rep}(\hat{\mathfrak{g}}_k)$ the theory is the $G_k/G_k$ theory, or the $G_k$ CS theory reduced on the circle. The Frobenius algebra and the idempotents are calculated explicitly in \ref{app:G/G}.

\paragraph{Left and Right Symmetries and Bimodule Categories}
More generally, a theory can admit commuting left $\cC$ and right $\mathcal{D}$ category symmetries that we call a $(\cC, \mathcal{D})$-symmetric theory. Accordingly, a $(\cC, \mathcal{D})$-bimodule category describes the topological boundary conditions of a theory with a commuting left $\cC$ and right $\mathcal{D}$ symmetries. Equivalently such a $(\cC, \mathcal{D})$-symmetric theory will have a left $\cC \boxtimes \mathcal{D}^\mathrm{rev}$ symmetry. The product $\cC \boxtimes \mathcal{D}$, is called the Deligne's tensor product which is a generalization of the direct product $G_1 \times G_2$ of group symmetries. Moreover the category symmetry $\mathcal{D}^\mathrm{rev}$, is the orientation reversal of $\mathcal{D}$ where the order of tensor product $\otimes$ has been reversed.

\paragraph{Gauging and Symmetric Gapped Phases}
Here we show a correspondence between different ways of gauging a category symmetry, and topological quantum filed theories -- gapped phases -- with the same category symmetry. This can be easily seen by a theorem about fusion categories relating module categories with Frobenius algebras. Namely for any Frobenius algebra $A$ in $\cC$, one could form the category of right $A$-modules in $\cC$, that we denote by $\cC_A$. Then it is known that $\cC_A$ forms a module category over $\cC$, and any $\cC$-module category can arise in this way \cite{ostrik2003module}. Furthermore algebra objects $A$ and $A'$ with the same module categories $\cC_A\cong\cC_{A'}$ are called Morita equivalent and lead to the same gauged theory. Therefore a way of gauging a subpart of $\cC$ by $A$ is uniquely determined by $\cC_A$, or equivalently with module categories over $\mathcal{C}$. Thus $\mathcal{C}$-module categories classify both $\mathcal{C}$-symmetric TQFTs and ways of gauging $\cC$.

In a 2d TQFT the dimension of Hilbert space on the circle is the same as the number of elementary boundary conditions. Therefore for a theory with category symmetry $\mathcal{C}$, its number of vacua on flat spacetime is constrained by the number of simple objects in $\mathcal{C}$-module categories. In particular, a theory with a category symmetry that does not admit a module category with one simple object cannot be trivially gapped with a unique vacuum. We call such a fusion category \textit{anomalous} -- that is there is an obstruction to gauge the whole category symmetry. This because after gauging the whole symmetry, we expect that all the topological lines are identified. Which means that the module category has a single object, or equivalently there is only a single $A$-module in $\cC$. Thus the anomaly matching condition for group symmetries applied to gapped phases can be thought as a special case of this observation.

\subsection{The Gauged Theory \label{app:lines.of.gauged.th}}
In section \ref{sec:gauging}, we described how the topological lines of a theory change after gauging with an algebra. Given the correspondence between algebras and module categories, we can rephrase those results in terms of module categories. For instance, starting with a theory with category symmetry $\mathcal{C}$ and gauging a subpart of it associated with a $\mathcal{C}$-module category $\mathcal{M}=\ca$, the gauged theory will have a new quantum symmetry given by $\mathrm{Fun}_\mathcal{C}(\mathcal{M},\mathcal{M})^\mathrm{rev}\cong\aca$. Where $\mathrm{Fun}_\mathcal{C}(\mathcal{M},\mathcal{N})$ is the category of left $\mathcal{C}$-module functors from $\mathcal{M}$ to $\mathcal{N}$ \cite{etingof2009fusion,etingof2016tensor}. In particular, $\mathrm{Fun}_\mathcal{C}(\mathcal{M},\mathcal{M})$ is a fusion category with tensor product being composition of functors, and $\mathrm{Fun}_\mathcal{C}(\mathcal{M},\mathcal{M})^\mathrm{rev}$ is the category with reversed tensor product. Note that $\cM$ is a module category over $\cC\boxtimes\mathrm{Fun}_\mathcal{C}(\mathcal{M},\mathcal{M})$, or equivalently a $(\cC,\mathrm{Fun}_\mathcal{C}(\mathcal{M},\mathcal{M})^\mathrm{rev})$-bimodule category. Therefore based on this definition, $\aca$ can be understood as the category of topological lines that act on $\ca$ but commute with the action of $\mathcal{C}$ on $\ca$.

Now that we have discussed how the topological lines change after gauging, we can discuss the gauged theory in more detail. Given a way of gauging $\cC$ such as $A$, there exist a $\cC$-symmetric TQFT denoted by $T_{\ca}$ whose category of boundary conditions is $\ca$. Furthermore as we discussed above, $T_{\ca}$ also has a right $\aca$ symmetry since $\ca$ is actually a $(\cC,\aca)$-bimodule category. Now we can try to either gauge its left or right symmetries. Gauging its left $\cC$ symmetry by an algebra $B\in\cC$, the theory becomes
\begin{equation}
    \frac{T_\ca}{B} = T_\bca~,
\end{equation}
where $\bca$ is the category of $(B,A)$-bimodules in $\cC$, which is a $(\bcb,\aca)$-bimodule category \cite{etingof2016tensor}.\footnote{In terms of module categories $\bca\cong \mathrm{Fun}_\mathcal{C}(\cC_B,\ca)$.} Moreover, $T_{\ca}$ itself can be obtained from $T_\cC$ by gauging its right $\cC$ symmetry by $A$. Therefore, $T_\bca$ can be obtained by gauging (left, right) symmetries of $T_\cC$ by $(B,A)$ and has a commuting $(\bcb,\aca)$ symmetries. Schematically
\begin{equation*}
\begin{tikzcd}
T_\cC  \ar[r,"A"] \ar[d,"B"] & T_\ca \ar[d,"B"]  \\
T_{{}_B\cC} \ar[r,"A"]  & T_\bca
\end{tikzcd}
\end{equation*}
where vertical and horizontal arrows denote gauging left and right symmetries respectively.

As we discussed in section \ref{sec:gauging}, gauging is an invertible and associative operation. Category symmetries $\cC$ and $\aca$ that are related by gauging are called categorically Morita equivalent, and their module categories are in one-to-one correspondence. More precisely, given a $\cC$-symmetric TQFT such as $T_{\cC_B}$ one can gauge its left $\cC$ symmetry by $A$ to get the $\aca$-symmetric TQFT $T_\acb$ and vice versa. Therefore, all $\aca$-symmetric TQFTs are of the form $T_\acb$, for $B$ a way of gauging in $\cC$. Below we describe a practical way of analysing the $T_\acb$ TQFT for when $\mathcal{C}$ is a modular tensor category.

\paragraph{Modular Invariants and Modular Categories}

Here we study $\aca$-module categories when $\cC$ is a modular tensor category. As explained above, since $\aca$ and $\cC$ categorically Morita equivalent there is a one-to-one correspondence between their module categories. More precisely, given a left $\cC$-module category such as $\cC_B$, one can gauge its left $\cC$-action to get the $(\aca,{}_B\cC_B)$-bimodule category ${}_A\cC_B$, and vice versa. Because $\mathcal{C}$ is braided one can turn left $A$-modules in $\cC$, into right $A$-modules. Therefore, one can turn the $(\aca,{}_B\cC_B)$-bimodule category ${}_A\cC_B$ into a left $\mathcal{C}$-module category denoted by $\cC_{A^{\mathrm{op}}\otimes B}$ \cite{Fuchs:2002cm}. The operation $\otimes$ here defines a fusion on algebras in $\cC$.

Using the correspondence between modular invariants and algebras \cite{Kirillov:2001ti,Fuchs:2002cm} reviewed in \ref{sec:modular.invariants}, we denote the modular invariant matrix associated with an algebra $A$ by $\mim(A)$. Then the fusion on algebras reduces to simply taking the product of the corresponding modular invariant matrices \cite{Fuchs:2002cm}
\begin{align}
	\mim\left(A \otimes B\right) &= \mim\left(A\right) \mim\left(B\right) ~,\\
	\mim\left(A \oplus B\right) &= \mim\left(A\right)+ \mim\left(B\right)~,\\
	\mim\left(A^\mathrm{op}\right) &= \mim\left(A\right)^\mathsf{T}~.
\end{align}
With these formulae we can count the number of elementary boundary conditions in $\cC_A$ and simple lines in $\aca$ as
\begin{align}
	\text{Number of simple objects in $\cC_A$} &= \Tr \left[ \mim\left(A\right) \right]~, 
	\label{dim of ca}\\
	\text{Number of simple objects in $\aca$} &= \Tr \left[ \mim\left(A\right) \mim\left(A\right)^\mathsf{T} \right]~. 	\label{dim of aca}\\
	\text{Number of simple objects in ${}_A\cC_B$} &= \Tr \left[ \mim\left(B\right) \mim\left(A\right)^\mathsf{T} \right]~. 	\label{dim of acb}
\end{align}

\begin{example}[$\hat{\mathfrak{su}}(2)_k$ Modular Invariants]
	Modular invariants of $\hat{\mathfrak{su}}(2)_k$ have the ADE classification \cite{Cappelli:1987xt}. Working out the fusions of $\mathrm{Rep}\,\hat{\mathfrak{su}}(2)_k$-module categories one finds \cite{Fuchs:2002cm}
	\begin{align*}
		k\in 4\mathbb{Z}: \quad & \mathrm{D}_{2l+2} \otimes \mathrm{D}_{2l+2} = 2 \mathrm{D}_{2l+2}~,\\
		k\in 4\mathbb{Z}+2: \quad & \mathrm{D}_{2l+1} \otimes \mathrm{D}_{2l+1} = \mathrm{A}_{4l-1}~,\\
		k=10: \quad & \mathrm{D}_7 \otimes \mathrm{E}_6 = \mathrm{E}_6, \;\; \ \quad \mathrm{E}_6 \otimes \mathrm{E}_6 = 2 \mathrm{E}_6~,\\
		k=16:  \quad & \mathrm{D}_{10} \otimes \mathrm{E}_7 = 2\mathrm{E}_7, \quad \mathrm{E}_7 \otimes \mathrm{E}_7 =\mathrm{D}_{10} \oplus \mathrm{E}_7~,\\
		k=28:  \quad & \mathrm{D}_{16} \otimes \mathrm{E}_8 = 2\mathrm{E}_8, \quad \mathrm{E}_8 \otimes \mathrm{E}_8 = 4\mathrm{E}_8~.
	\end{align*}
\end{example}

\section{Action of Topological Lines}
\label{app:ActionOfLines}
\subsection{Ocneanu/Quantum Graphs}
\label{sec:ocv graphs}

Starting with a modular category $\cC$, we saw in section \ref{adjQCD.lines.bos} that the bosonic adjoint QCD is described by two categories, the $\cC$-module category of boundary conditions $\cC_A$ and the fusion category of topological lines $\aca$.  In this subsection we study the action of $\cC$ on $\cC_A$ and $\aca$. The section is mostly a review of known facts in the RCFT literature. Furthermore, we also discuss the action of topological lines on local operators which is important for analyzing the IR phases of adjoint QCD.

The relevance of how the topological lines act on local operators is twofold. On the one hand, it places restrictions on the RG flow. An operators which is not invariant under all topological lines present in the UV cannot be generated by radiative corrections. On the other hand, turning on a non-invariant operators in the UV will explicitly break some of the lines, in the sense that they will not be topological anymore. As a consequence, the number of topological lines present in the IR  will change, which might alter the vacuum structure of the theory. In particular some of the universes might get lifted which may lead to confinement of some Wilson lines. 

In the following we focus on an RCFT based on the Lie group $G$ at level $k$ and we collect several facts that can be found in the literature. For reviews see \cite{Zuber:2000ia,Petkova:2001zn,Coquereaux:2002iw}. Denote the elements of $\cC$ by $U_\m$ and their fusion by
\begin{equation}
U_\m U_\n= N^\r_{\m\n} U_\r~, 
\end{equation}  
where the fusion coefficients $N^\r_{\m\n}$ are known symmetric $N^\r_{\m\n}=N^\r_{\n\m}$ nonnegative integer valued set of matrices. We denote the actions of $\cC$ on $\cC_A$ and $\aca$ by
\begin{align}
U_\m \ket{\mathsf{v}_a} & = (n_\m)_a\!^b \ket{\mathsf{v}_b} \quad  \quad \text{for} \quad \ket{\mathsf{v}_a} \in \cC_A ~, \label{n matrices}\\
U_\m L_\a U_\n &= (V_{\m\n})_\a\!^\b L_\b \quad  \, \text{for} \quad L_\a \in \aca ~,\label{double fusion matrices}
\end{align}
The $n_\mu$ and $V_{\m\n}$ matrices are called NIM-rep and \emph{double fusion matrices} respectively. The above structure constants give the partition function on the cylinder with boundary conditions $a$ and $b$
\begin{equation}
Z_{a|b}=\sum_\m (n_\m)_{ab} \,\x_\m ~,
\end{equation}
and the partition function in the presence of two topological lines $L_\a$ and $L_\b$
\begin{equation}
Z_{\a|\b}=\sum_{\m\n} (V_{\m \n^*})_{\a\b} \,\x_\m \bar{\x}_\n~.
\end{equation}
From \eqref{partition function} we see that $\mim_{\m\n}=(V_{\m\n^*})_{00}$. The structures $n$ and $V$ determine the RCFT completely. They further satisfy the algebras \cite{Petkova:2000ip}
\begin{align}
n_\m n_\n & = \sum_\l N_{\m\n}^\l n_\l ~, \label{fusion.of.n}\\
V_{\m\n} V_{\k\l} &=\sum_{\r \s} N_{\m\k}^\r N_{\n\l}^\s V_{\r\s}~. \label{double fusion algebra}
\end{align}
These algebras can be used as recursion relations to determine all of the $n_\mu$ and $V_{\mu\nu}$ matrices started from some generators, say $n_{i}$ for the former and $V_{i0}$ along with $V_{0i}$ for the latter. Once these matrices are given the RCFT is determined. Since these matrices have non-integer elements, the generators $n_{f_i}$ and $V_{f_i}$ can be encoded as adjacency matrices of graphs. Historically, the classification of RCFT was translated to a classification problem of graphs. The graph that determines the generators for $n_{\m}$ is typically called ADE or Mckay graph, and the graph that determines the generators for $V_{\m\n}$ is called Ocneanu graph. Typically an Ocneanu graph consists of several connected components of the corresponding ADE graphs. For the $\SU(3)_3$ and $\SU(4)_4$ cases studied in \ref{sec:ocn graphs for 3 and 4}, the matrices $n$ are actually equal to the first block of the $V_{\m 0}$ matrices. The summary is that every such pair of graphs determines a modular invariant RCFT. 

The relevant quantities for us are constructed as follows. The spectrum of a non-diagonal RCFT will consist of primary operators $\cO_{\m,\bar{\m}}^m$, where $m$ denotes the multiplicity of operators with the same representation. To determine how line operators act on local operators, we need to diagonalize the commutative sets of matrices $n_\m$ and $V_{\m\n}$ \cite{Behrend:1999bn,Petkova:2000ip}:
\begin{align}
(n_\m)_a\!^b & =\sum_{\n,m} \f^{(\n,m)}_a  \frac{S_{\m \n}}{S_{0\n}}\f^{(\n,m)*}_b ~,\label{nimrep.formula}\\
(V_{\m\bar{\m}})_\a\!^\b &=\sum_{\r,\bar{\r},m, m'} \F^{(\r,\bar{\r};m,m')}_\a  \frac{S_{\m \r}S_{\bar{\m} \bar{\r}}}{S_{0\r}S_{0\bar{\r}}}\F^{(\r,\bar{\r};m,m')\ast}_\b ~.\label{V matrix}
\end{align}

All of the RCFT structures can be expressed in terms of $\f^{(\m,m)}_a $ and $\F^{(\m,\bar{\m};m,m')}_\a$. These are unitary matrices 
\begin{align}
    \sum_{\a} \F^{(\m,\bar{\m};m,m')}_\a  \F^{(\n,\bar{\n};n,{n'})\ast}_\a &= \d^{\m \n}\d^{\bar{\m} \bar{\n}}\d^{m n}\d^{{m'}{n'}}~, \\
    \sum_{\m, \bar{\m},m,{m'}} \F^{(\m,\bar{\m};m,m')}_\a  \F^{(\m,\bar{\m};m,m')*}_\b & = \d_{\a \b }~,
\end{align}
and similarly for $\f^\m_a$. The action of lines on bulk local operators is given by 
\begin{equation}
 L_\a  \cdot \cO_{\m,\bar{\m}}^m = \sum_{m'} \frac{\F^{(\m,\bar{\m};m,m')}_\a}{\sqrt{S_{0\m} S_{0 \bar{\m}}}}  \, \cO_{\m,\bar{\m}}^{m'}  ~, 
\end{equation}
while their fusion algebra
\begin{equation}
L_\a \otimes L_\b= \hat{N}_{\a \b }^\g L_\g~.
\end{equation}
is determined by
\begin{equation}
\hat{N}_{\a\b}^\g= \sum_{\m,\bar{\m},m,m',m''} \frac{  \F^{(\m,\bar{\m};m,m')}_\a \F^{(\m,\bar{\m};m',m'')}_\b \F^{(\m,\bar{\m};m,m'')\ast}_\g}{\sqrt{S_{0\m}S_{0\bar{\m}}}} ~.
\end{equation}
This algebra is not commutative in general because of the existence of multiplicities as can be seen from this formula.
In addition, one can even construct the algebra of bulk operators (Pasquier algebra \cite{Pasquier:1987xj}) from a Verlinde-like formula involving $\F^{(\m,\bar{\m};m,m')}_\a$ which for simplicity we do not reproduce here.

Having computed all the matrices $V_{\m\n}$ one can invert \eqref{V matrix} 
\begin{equation}
\sum_{m, m'} \F^{(\m,\bar{\m};m,m')}_\a \F^{(\m,\bar{\m};m,m')\ast}_\b= S_{0 \m}S_{0 \bar{\m}} \sum_{\r,\bar{\r}} S^*_{\m \r}S^*_{\bar{\m} \bar{\r}} (V_{\r \bar{\r}})_\a\!^\b ~.
\end{equation}
Requiring that the action of the identity line is trivial  $\F^{(\m,\bar{\m};m,m')}_0 = \sqrt{S_{0\m} S_{0 \bar{\m}}} \; \d^{m m'} $. For non-degenerate primary operators there is no summation on the right-hand side of the above equation and we arrive at
\begin{equation}
 \F^{ \m, \bar{\m}}_\a =\sqrt{S_{0\m} S_{0 \bar{\m}}}\sum_{\r,\bar{\r}} S^*_{\m \r}S^*_{\bar{\m} \bar{\r}} (V_{\m \bar{\m}})_\a\!^0 ~.
\end{equation}

So far we have reviewed how to determine a modular invariant RCFT based on $G_k$ given the Mckay and the Ocneanu graphs as an input. Given a partition function the process can be reversed and one can calculate the double fusion matrices through the so called modular splitting equations \cite{Ocneanu:2000kj,Coquereaux:2005hu,Isasi:2006ty}. However, there might be several solutions which correspond to different RCFTs with the same partition function but different set of boundary condition $\cC_A$ and topological lines $\aca$.  In the following, from the known Ocneanu graphs we list the generators of the double fusion algebra for $\Spin(8)_1/\SU(3)_3$ and $\Spin(15)_1/\SU(4)_4$.
\subsection{Ocneanu Graphs for SU(3) \& SU(4)}
\label{sec:ocn graphs for 3 and 4}
From the known Ocneanu graphs, in this appendix we have included the generators of the double fusion matrices \eqref{double fusion matrices}. Using \eqref{dim.of.aca} and the partition function of the conformal embedding $\Spin(8)_1/\SU(3)_3$ \eqref{spin8/su3 embed}, the number of topological lines is $18$. Given the matrices $V_{(1,0)(0,0)}$ and $V_{(0,0)(1,0)}$ all of the double fusion matrices can be recursively computed from \eqref{double fusion algebra} using that $V_{(0,0)(0,0)}=\mathbbm{1}$ and $V_{\m \n }^T=V_{\m^* \n^* }$. The Ocneanu graph for this case can be found in \cite{Bockenhauer:1999wt,Coquereaux:2006gv} and leads to 
\begin{equation}
V_{(1,0)(0,0)}=
\begin{pmatrix}  
G_{10} & 0 & 0 \\
0 & G_{10} & 0 \\
0 & 0 & G_{10} 
\end{pmatrix}~,
\quad
V_{(0,0)(1,0)}=
\begin{pmatrix}  
0 & G_{10} & 0 \\
0 & 0 & G_{10} \\
G_{10} & 0 & 0 
\end{pmatrix}~,
\end{equation}
with 
\begin{equation}
    G_{10}=\left(
\begin{array}{cccccc}
 0 & 0 & 0 & 0 & 1 & 0 \\
 0 & 0 & 0 & 0 & 1 & 0 \\
 0 & 0 & 0 & 0 & 1 & 0 \\
 0 & 0 & 0 & 0 & 1 & 0 \\
 0 & 0 & 0 & 0 & 0 & 2 \\
 1 & 1 & 1 & 1 & 0 & 0 \\
\end{array}
\right)~.\\ \label{su3.nimrep}
\end{equation}
The generator for the NIM-rep matrices $n_\m$ is 
\begin{equation}
    n_{(1,0)}=G_{10}~,
\end{equation}
and the rest of them can be computed recursively from \eqref{fusion.of.n}. However in this case, the unitary matrix $\phi_a^{(v,m)}$ is known and can be used to calculate all the NIM-rep matrices. It is given in equation (6.17) of \cite{Ishikawa:2005ea}
\begin{equation}
    \phi = \frac{1}{2\sqrt{3}} \left(
\begin{array}{cccccc}
 1 & 1 & 1 & \sqrt{3} & \sqrt{3} & \sqrt{3} \\
 1 & 1 & 1 & \sqrt{3} & -\sqrt{3} & -\sqrt{3} \\
 1 & 1 & 1 & -\sqrt{3} & \sqrt{3} & -\sqrt{3} \\
 1 & 1 & 1 & -\sqrt{3} & -\sqrt{3} & \sqrt{3} \\
 2 & 2 e^{\frac{2 i \pi }{3}} & 2 e^{\frac{4 i \pi }{3}} & 0 & 0 & 0 \\
 2 & 2 e^{\frac{4 i \pi }{3}} & 2 e^{\frac{2 i \pi }{3}} & 0 & 0 & 0 \\
\end{array}
\right)~,\label{su3.phi.matrix}
\end{equation}
where the columns are are ordered as $(0,0),(3,0),(0,3),((1,1),\mathrm{v}),((1,1),\mathrm{s}),((1,1),\mathrm{c})$.

Similarly the Ocneanu graph for the conformal embedding $\Spin(15)_1/\SU(4)_4$ can be found in \cite{Coquereaux:2007rw,Coquereaux_2009}. Using \eqref{dim.of.aca} and the partition function of the conformal embedding $\Spin(15)_1/\SU(4)_4$ \eqref{spin16/su4 embed}, the number of topological lines is $48$. In this case, the double fusion algebra has four generators 
\begin{align}
V_{(1,0,0)(0,0,0)}= &
\begin{pmatrix}  
G_{100} & 0 & 0 & 0 \\
0 & G_{100} & 0 & 0 \\
0 & 0 & G_{100} & 0 \\
0 & 0 & 0 & G_{100}
\end{pmatrix} 
~,\\
V_{(0,1,0)(0,0,0)}= &
\begin{pmatrix}  
G_{010} & 0 & 0 & 0 \\
0 & G_{010} & 0 & 0 \\
0 & 0 & G_{010} & 0 \\
0 & 0 & 0 & G_{010}
\end{pmatrix}~,
\end{align}
\begin{align}
V_{(0,0,0)(1,0,0)} &= P \; V_{(1,0,0)(0,0,0)} \; P ~, \\
V_{(0,0,0)(0,1,0)} &= P \; V_{(0,1,0)(0,0,0)} \; P ~,
\end{align}
where 
\begin{equation}
G_{100}= 
\left(
\begin{array}{cccccccccccc}
 0 & 0 & 0 & 0 & 1 & 0 & 0 & 0 & 0 & 0 & 0 & 0 \\
 0 & 0 & 0 & 0 & 1 & 0 & 0 & 0 & 0 & 0 & 0 & 0 \\
 0 & 0 & 0 & 0 & 1 & 1 & 1 & 0 & 0 & 0 & 0 & 0 \\
 0 & 0 & 0 & 0 & 1 & 1 & 1 & 0 & 0 & 0 & 0 & 0 \\
 0 & 0 & 0 & 0 & 0 & 0 & 0 & 2 & 0 & 0 & 0 & 0 \\
 0 & 0 & 0 & 0 & 0 & 0 & 0 & 1 & 1 & 0 & 0 & 0 \\
 0 & 0 & 0 & 0 & 0 & 0 & 0 & 1 & 1 & 0 & 0 & 0 \\
 0 & 0 & 0 & 0 & 0 & 0 & 0 & 0 & 0 & 2 & 1 & 1 \\
 0 & 0 & 0 & 0 & 0 & 0 & 0 & 0 & 0 & 0 & 1 & 1 \\
 1 & 1 & 1 & 1 & 0 & 0 & 0 & 0 & 0 & 0 & 0 & 0 \\
 0 & 0 & 1 & 1 & 0 & 0 & 0 & 0 & 0 & 0 & 0 & 0 \\
 0 & 0 & 1 & 1 & 0 & 0 & 0 & 0 & 0 & 0 & 0 & 0 \\
\end{array}
\right)~,\label{su4.nimrep}
\end{equation}
\begin{equation}
G_{010}= 
\left(
\begin{array}{cccccccccccc}
 0 & 0 & 0 & 0 & 0 & 0 & 0 & 1 & 0 & 0 & 0 & 0 \\
 0 & 0 & 0 & 0 & 0 & 0 & 0 & 1 & 0 & 0 & 0 & 0 \\
 0 & 0 & 0 & 0 & 0 & 0 & 0 & 2 & 1 & 0 & 0 & 0 \\
 0 & 0 & 0 & 0 & 0 & 0 & 0 & 2 & 1 & 0 & 0 & 0 \\
 0 & 0 & 0 & 0 & 0 & 0 & 0 & 0 & 0 & 2 & 1 & 1 \\
 0 & 0 & 0 & 0 & 0 & 0 & 0 & 0 & 0 & 1 & 1 & 1 \\
 0 & 0 & 0 & 0 & 0 & 0 & 0 & 0 & 0 & 1 & 1 & 1 \\
 1 & 1 & 2 & 2 & 0 & 0 & 0 & 0 & 0 & 0 & 0 & 0 \\
 0 & 0 & 1 & 1 & 0 & 0 & 0 & 0 & 0 & 0 & 0 & 0 \\
 0 & 0 & 0 & 0 & 2 & 1 & 1 & 0 & 0 & 0 & 0 & 0 \\
 0 & 0 & 0 & 0 & 1 & 1 & 1 & 0 & 0 & 0 & 0 & 0 \\
 0 & 0 & 0 & 0 & 1 & 1 & 1 & 0 & 0 & 0 & 0 & 0 \\
\end{array}
\right)~,
\end{equation}
and $P$ a permutation matrix that permutes the following pairs (the elements $P_{ij}=P_{ji}$ are $1$ only for these pairs and zero otherwise)
\begin{equation}
\scriptsize
\begin{array}{c|cccccccccccccccccc}
m   & 3  & 4  &  5 &  6 &  7 &  8 & 10 & 11 & 12 & 17 & 18 & 19 & 22 & 23 & 24  & 30 & 31 & 34\\
\hline
m^* & 13 & 14 & 33 & 37 & 38 & 21 & 45 & 25 & 26 & 32 & 39 & 40 & 44 & 27 & 28  & 47 & 48 & 41
\end{array}
\nonumber~.
\end{equation}
Finally, the generators for the $n_\m$ matrices are
\begin{equation}
    n_{(1,0,0)}=G_{100}, \quad n_{(0,1,0)}=G_{010}~.
\end{equation}
\subsection{NIM-reps for SU(5) \label{app:su5.nimrep}}
In this appendix we find the NIM-rep matrices for the module category $\ca$ corresponding to the $\Spin(24)_1$ modular invariant of $\SU(5)_5$. This NIM-rep has not appeared in the literature to our knowledge. We use the $\alpha$-induction tensor functor $\alpha:\cC \to \ca$ to find the action of $\cC$ on $\ca$ from the fusion algebra of $\cC$ and the modular invariant matrix of $\Spin(24)_1$ \cite{Bockenhauer:1998in,Kirillov:2001ti}.

For $\mu\in\cC$ we define its image under the $\alpha$-induction by $\alpha_\mu$. Defining $\langle \alpha_\mu, \alpha_\nu \rangle := \text{DimHom}\left( \a_\mu, \a_\nu \right)$, the starting point of our computation is the equation \cite{Bockenhauer:1998in}
\begin{equation}
    {\langle \alpha_\mu, \alpha_\nu \rangle}_{\ca} = {\langle A \otimes \mu, \nu \rangle}_{\cC} = N_{\mu \rho}^\nu \mim_{\rho 0}~,
\end{equation}
where $\mim_{\mu\nu}$ is the modular invariant matrix associated to $\Spin(24)_1$ chiral algebra. Using the above equation, after some trial and error we find the following decompositions
\begin{equation}
\begin{array}{llll}
    \a_{0000} = \mathsf{v}_1\,, &\a_{0020} = \mathsf{v}_8+\mathsf{v}_9\,,& \frac12(\a_{1020}-\a_{0011}) = \mathsf{v}_{19}\,, \\
    \a_{1000} = \mathsf{v}_2\,, & \a_{1100}-\a_{0010} = \mathsf{v}_{10}+\mathsf{v}_{11}\,, &  \frac12(\a_{0201}-\a_{1100}) = \mathsf{v}_{20}\,, \\
    \a_{0100} = \mathsf{v}_3\,, &
    \a_{1001}=\mathsf{v}_{12}+\mathsf{v}_{13}+\mathsf{v}_{14}\,, & \frac12(\a_{1110}-\a_{1000})-\a_{0020} = \mathsf{v}_{21}\,, \\
    \a_{0010} = \mathsf{v}_4\,, & \a_{0110}=\mathsf{v}_{12}+\mathsf{v}_{13}+\mathsf{v}_{15}+\mathsf{v}_{16}\,, & \frac12(\a_{0111}-\a_{0001})-\a_{0200} = \mathsf{v}_{22}\,, \\
    \a_{0001} = \mathsf{v}_5\,, & \a_{0011}-\a_{0100} = \mathsf{v}_{17}+\mathsf{v}_{18}\,, & \frac12(\a_{1111}-\a_{1001}-\a_{0110}) = \mathsf{v}_{15}+\mathsf{v}_{16}+\mathsf{v}_{23}+\mathsf{v}_{24}\,.
    \\ \a_{0200} = \mathsf{v}_6+\mathsf{v}_7\,,
\end{array} \label{su5.alpha.induction}
\end{equation}
where $\{\mathsf{v}_1,\dots,\mathsf{v}_{24}\}$ are the \emph{simple} objects of $\ca$. Note that the $\a$-induction preserves the fusion algebra, and we have $\a_\mu \cdot \a_\nu = \sum_\rho N_{\mu\nu}^\rho \a_\rho$. Using this equation we can determine the action of $\mu\in\cC$, or equivalently $\a_\mu$, on the $\a$-inducted elements of $\ca$. In particular, the above decompositions determine the first row of the NIM-rep matrices as
\begin{equation}
    \a_\mu = \sum_{a=1}^{24} (n_\m)_1^a \mathsf{v}_a~. 
\end{equation}

Having found the first row of the NIM-reps, there is very little freedom to fix the action of $\cC$ on all the elements of $\ca$. To check that the NIM-rep solutions are compatible with the modular invariant matrix $\mim$, one has to match their ``exponents". In other words, the eigenvalues of $n_\mu$ must be given by ${S_{\mu\nu}}/{S_{0\nu}}$ with multiplicity $\mim_{\nu\nu}$ -- see equation \eqref{nimrep.formula}. With a bit of guesswork we find the NIM-rep that solves the fusion equations \eqref{fusion.of.n} and is compatible with the modular invariant matrix of $\Spin(24)_1$. Here we list some of these NIM-rep matrices. However in this case, knowing $n_{(1,0,0,0)}$ and using \eqref{nimrep.formula} one can finds all the other NIM-rep matrices.
\begin{equation}
    n_{(1,0,0,0)}=\left(
\begin{array}{cccccccccccccccccccccccc}
 0 & 1 & 0 & 0 & 0 & 0 & 0 & 0 & 0 & 0 & 0 & 0 & 0 & 0 & 0 & 0 & 0 & 0 & 0 & 0 & 0 & 0 & 0 & 0 \\
 0 & 0 & 2 & 0 & 0 & 0 & 0 & 0 & 0 & 0 & 0 & 0 & 0 & 0 & 0 & 0 & 0 & 0 & 0 & 0 & 0 & 0 & 0 & 0 \\
 0 & 0 & 0 & 2 & 0 & 0 & 0 & 0 & 0 & 1 & 1 & 0 & 0 & 0 & 0 & 0 & 0 & 0 & 0 & 0 & 0 & 0 & 0 & 0 \\
 0 & 0 & 0 & 0 & 2 & 1 & 1 & 0 & 0 & 0 & 0 & 0 & 0 & 0 & 0 & 0 & 0 & 0 & 0 & 0 & 0 & 0 & 0 & 0 \\
 1 & 0 & 0 & 0 & 0 & 0 & 0 & 0 & 0 & 0 & 0 & 1 & 1 & 1 & 0 & 0 & 0 & 0 & 0 & 0 & 0 & 0 & 0 & 0 \\
 0 & 0 & 0 & 0 & 0 & 0 & 0 & 0 & 0 & 0 & 0 & 1 & 1 & 0 & 1 & 1 & 0 & 0 & 0 & 0 & 0 & 0 & 0 & 0 \\
 0 & 0 & 0 & 0 & 0 & 0 & 0 & 0 & 0 & 0 & 0 & 1 & 1 & 0 & 1 & 1 & 0 & 0 & 0 & 0 & 0 & 0 & 0 & 0 \\
 0 & 0 & 1 & 0 & 0 & 0 & 0 & 0 & 0 & 0 & 0 & 0 & 0 & 0 & 0 & 0 & 1 & 1 & 1 & 0 & 0 & 0 & 0 & 0 \\
 0 & 0 & 1 & 0 & 0 & 0 & 0 & 0 & 0 & 0 & 0 & 0 & 0 & 0 & 0 & 0 & 1 & 1 & 1 & 0 & 0 & 0 & 0 & 0 \\
 0 & 0 & 0 & 0 & 0 & 1 & 1 & 0 & 0 & 0 & 0 & 0 & 0 & 0 & 0 & 0 & 0 & 0 & 0 & 0 & 0 & 0 & 0 & 0 \\
 0 & 0 & 0 & 0 & 0 & 1 & 1 & 0 & 0 & 0 & 0 & 0 & 0 & 0 & 0 & 0 & 0 & 0 & 0 & 0 & 0 & 0 & 0 & 0 \\
 0 & 1 & 0 & 0 & 0 & 0 & 0 & 1 & 1 & 0 & 0 & 0 & 0 & 0 & 0 & 0 & 0 & 0 & 0 & 0 & 0 & 0 & 0 & 0 \\
 0 & 1 & 0 & 0 & 0 & 0 & 0 & 1 & 1 & 0 & 0 & 0 & 0 & 0 & 0 & 0 & 0 & 0 & 0 & 0 & 0 & 0 & 0 & 0 \\
 0 & 1 & 0 & 0 & 0 & 0 & 0 & 0 & 0 & 0 & 0 & 0 & 0 & 0 & 0 & 0 & 0 & 0 & 0 & 0 & 0 & 0 & 0 & 0 \\
 0 & 0 & 0 & 0 & 0 & 0 & 0 & 1 & 1 & 0 & 0 & 0 & 0 & 0 & 0 & 0 & 0 & 0 & 0 & 0 & 1 & 0 & 0 & 0 \\
 0 & 0 & 0 & 0 & 0 & 0 & 0 & 1 & 1 & 0 & 0 & 0 & 0 & 0 & 0 & 0 & 0 & 0 & 0 & 0 & 1 & 0 & 0 & 0 \\
 0 & 0 & 0 & 1 & 0 & 0 & 0 & 0 & 0 & 0 & 0 & 0 & 0 & 0 & 0 & 0 & 0 & 0 & 0 & 1 & 0 & 0 & 0 & 0 \\
 0 & 0 & 0 & 1 & 0 & 0 & 0 & 0 & 0 & 0 & 0 & 0 & 0 & 0 & 0 & 0 & 0 & 0 & 0 & 1 & 0 & 0 & 0 & 0 \\
 0 & 0 & 0 & 0 & 0 & 0 & 0 & 0 & 0 & 1 & 1 & 0 & 0 & 0 & 0 & 0 & 0 & 0 & 0 & 2 & 0 & 0 & 0 & 0 \\
 0 & 0 & 0 & 0 & 0 & 1 & 1 & 0 & 0 & 0 & 0 & 0 & 0 & 0 & 0 & 0 & 0 & 0 & 0 & 0 & 0 & 2 & 0 & 0 \\
 0 & 0 & 0 & 0 & 0 & 0 & 0 & 0 & 0 & 0 & 0 & 0 & 0 & 0 & 0 & 0 & 0 & 0 & 2 & 0 & 0 & 0 & 0 & 0 \\
 0 & 0 & 0 & 0 & 0 & 0 & 0 & 0 & 0 & 0 & 0 & 0 & 0 & 0 & 1 & 1 & 0 & 0 & 0 & 0 & 0 & 0 & 1 & 1 \\
 0 & 0 & 0 & 0 & 0 & 0 & 0 & 0 & 0 & 0 & 0 & 0 & 0 & 0 & 0 & 0 & 0 & 0 & 0 & 0 & 1 & 0 & 0 & 0 \\
 0 & 0 & 0 & 0 & 0 & 0 & 0 & 0 & 0 & 0 & 0 & 0 & 0 & 0 & 0 & 0 & 0 & 0 & 0 & 0 & 1 & 0 & 0 & 0 \\
\end{array}
\right)~,\label{su5.nimrep}
\end{equation}
and
\begin{equation*}
    n_{(0,1,0,0)}=\left(
\begin{array}{cccccccccccccccccccccccc}
 0 & 0 & 1 & 0 & 0 & 0 & 0 & 0 & 0 & 0 & 0 & 0 & 0 & 0 & 0 & 0 & 0 & 0 & 0 & 0 & 0 & 0 & 0 & 0 \\
 0 & 0 & 0 & 2 & 0 & 0 & 0 & 0 & 0 & 1 & 1 & 0 & 0 & 0 & 0 & 0 & 0 & 0 & 0 & 0 & 0 & 0 & 0 & 0 \\
 0 & 0 & 0 & 0 & 2 & 2 & 2 & 0 & 0 & 0 & 0 & 0 & 0 & 0 & 0 & 0 & 0 & 0 & 0 & 0 & 0 & 0 & 0 & 0 \\
 1 & 0 & 0 & 0 & 0 & 0 & 0 & 0 & 0 & 0 & 0 & 2 & 2 & 1 & 1 & 1 & 0 & 0 & 0 & 0 & 0 & 0 & 0 & 0 \\
 0 & 2 & 0 & 0 & 0 & 0 & 0 & 1 & 1 & 0 & 0 & 0 & 0 & 0 & 0 & 0 & 0 & 0 & 0 & 0 & 0 & 0 & 0 & 0 \\
 0 & 1 & 0 & 0 & 0 & 0 & 0 & 2 & 2 & 0 & 0 & 0 & 0 & 0 & 0 & 0 & 0 & 0 & 0 & 0 & 1 & 0 & 0 & 0 \\
 0 & 1 & 0 & 0 & 0 & 0 & 0 & 2 & 2 & 0 & 0 & 0 & 0 & 0 & 0 & 0 & 0 & 0 & 0 & 0 & 1 & 0 & 0 & 0 \\
 0 & 0 & 0 & 2 & 0 & 0 & 0 & 0 & 0 & 1 & 1 & 0 & 0 & 0 & 0 & 0 & 0 & 0 & 0 & 2 & 0 & 0 & 0 & 0 \\
 0 & 0 & 0 & 2 & 0 & 0 & 0 & 0 & 0 & 1 & 1 & 0 & 0 & 0 & 0 & 0 & 0 & 0 & 0 & 2 & 0 & 0 & 0 & 0 \\
 0 & 0 & 0 & 0 & 0 & 0 & 0 & 0 & 0 & 0 & 0 & 1 & 1 & 0 & 1 & 1 & 0 & 0 & 0 & 0 & 0 & 0 & 0 & 0 \\
 0 & 0 & 0 & 0 & 0 & 0 & 0 & 0 & 0 & 0 & 0 & 1 & 1 & 0 & 1 & 1 & 0 & 0 & 0 & 0 & 0 & 0 & 0 & 0 \\
 0 & 0 & 2 & 0 & 0 & 0 & 0 & 0 & 0 & 0 & 0 & 0 & 0 & 0 & 0 & 0 & 1 & 1 & 1 & 0 & 0 & 0 & 0 & 0 \\
 0 & 0 & 2 & 0 & 0 & 0 & 0 & 0 & 0 & 0 & 0 & 0 & 0 & 0 & 0 & 0 & 1 & 1 & 1 & 0 & 0 & 0 & 0 & 0 \\
 0 & 0 & 1 & 0 & 0 & 0 & 0 & 0 & 0 & 0 & 0 & 0 & 0 & 0 & 0 & 0 & 0 & 0 & 0 & 0 & 0 & 0 & 0 & 0 \\
 0 & 0 & 1 & 0 & 0 & 0 & 0 & 0 & 0 & 0 & 0 & 0 & 0 & 0 & 0 & 0 & 1 & 1 & 2 & 0 & 0 & 0 & 0 & 0 \\
 0 & 0 & 1 & 0 & 0 & 0 & 0 & 0 & 0 & 0 & 0 & 0 & 0 & 0 & 0 & 0 & 1 & 1 & 2 & 0 & 0 & 0 & 0 & 0 \\
 0 & 0 & 0 & 0 & 1 & 1 & 1 & 0 & 0 & 0 & 0 & 0 & 0 & 0 & 0 & 0 & 0 & 0 & 0 & 0 & 0 & 1 & 0 & 0 \\
 0 & 0 & 0 & 0 & 1 & 1 & 1 & 0 & 0 & 0 & 0 & 0 & 0 & 0 & 0 & 0 & 0 & 0 & 0 & 0 & 0 & 1 & 0 & 0 \\
 0 & 0 & 0 & 0 & 0 & 2 & 2 & 0 & 0 & 0 & 0 & 0 & 0 & 0 & 0 & 0 & 0 & 0 & 0 & 0 & 0 & 2 & 0 & 0 \\
 0 & 0 & 0 & 0 & 0 & 0 & 0 & 0 & 0 & 0 & 0 & 1 & 1 & 0 & 2 & 2 & 0 & 0 & 0 & 0 & 0 & 0 & 1 & 1 \\
 0 & 0 & 0 & 0 & 0 & 0 & 0 & 0 & 0 & 1 & 1 & 0 & 0 & 0 & 0 & 0 & 0 & 0 & 0 & 2 & 0 & 0 & 0 & 0 \\
 0 & 0 & 0 & 0 & 0 & 0 & 0 & 1 & 1 & 0 & 0 & 0 & 0 & 0 & 0 & 0 & 0 & 0 & 0 & 0 & 2 & 0 & 0 & 0 \\
 0 & 0 & 0 & 0 & 0 & 0 & 0 & 0 & 0 & 0 & 0 & 0 & 0 & 0 & 0 & 0 & 0 & 0 & 1 & 0 & 0 & 0 & 0 & 0 \\
 0 & 0 & 0 & 0 & 0 & 0 & 0 & 0 & 0 & 0 & 0 & 0 & 0 & 0 & 0 & 0 & 0 & 0 & 1 & 0 & 0 & 0 & 0 & 0 \\
\end{array}
\right)~.
\end{equation*}

\section{G/G TQFT and 3d CS theory}
\label{app:TQFT}
\subsection{2d TQFTs}\label{app:2d.TQFT}
Here we summarize the basic facts about 2-dimensional unitary TQFT.
The local operators of a 2d TQFT forms a commutative Frobenius algebra $F$ \cite{Fukuma:1993hy,Karimipour:1995fb,Moore:2006dw}.
That is, $F$ is an commutative algebra over $\mathbb{C}$ (with unit $\mathbbm{1}$) equipped with a linear trace map $\theta:F\to \mathbb{C}$, where $g_{ij}=\langle \mathcal{O}_i,\mathcal{O}_j\rangle := \theta(\mathcal{O}_i\mathcal{O}_j)$ for $\mathcal{O}_{i,j}\in F$ defines a non-degenerate symmetric bilinear form in $F$.
The value $\theta(\mathcal{O})$ is identified with the one-point function of $\mathcal{O}$ on the sphere $S^2$.
We define the Handle-attaching operator $\mathsf{H}:F\to F$ by
\begin{equation}
    \mathsf{H}^i_j = C^{ikl}C_{klj}~,
\end{equation}
where $C_{ijk}$ denotes the three point function $\theta(\mathcal{O}_i\mathcal{O}_j\mathcal{O}_k)$, and the indices are raised by the inverse $g^{ij}$ of the inner product matrix $g_{ij}$.
Then, the partition function on genus $g$ surface $\Sigma_g$ can be presented as
\begin{equation}
    Z[\Sigma_g] = \theta(\mathsf{H}^{g}\mathbbm{1})~.
\end{equation}

For a unitary (to be precise, reflection-positive) TQFT, it is always possible to find an idempotent basis $\varepsilon_i$ satisfying~\cite{Durhuus:1993cq}
\begin{equation}
    \varepsilon_i\varepsilon_j = \delta_{ij}\varepsilon_i~.
\end{equation}
This property for Frobenius algebra is called semisimplicity. Furthermore, unitarity requires that the one-point function $\theta_i := \theta(\varepsilon_i)$ to be a positive real number.
Noting $\mathbbm{1} = \sum_i \varepsilon_i$, the partition function can be explicitly written downs as
\begin{equation}
    Z[\Sigma_g] = \sum_i \theta_i^{1-g}~.
\end{equation}
This formula should be interpreted as the direct some of the invertible TQFTs each of which has the action $S_i[\Sigma_g]=-\frac12\chi[\Sigma_g]\log \theta_i$ where $\chi[\Sigma_g] = 2-2g$ is the Euler number.\footnote{The Euler number counter term defines an invertible TQFT. However, as its coefficient can be deformed continuously, it belongs to the same deformation class of the trivial theory. Therefore, we do not have corresponding anomaly in one lower dimensions.} More physically, the TQFT describes the long range limit of a gapped system with $n=Z[T^2]$ vacua, each of which comes with induced Euler term $S_i$ in the deep IR.
The normalized state of each vacua corresponds to the operator
\footnote{As noted in \cite{Moore:2006dw}, there's a subtlety in the choice of the branch of the square root. In an unitary theory, as $\theta_i$ is real and positive, it is natural to take the positive square root.}
\begin{equation}
    \mathsf{v}_i = \frac{1}{\sqrt{\theta_i}}\varepsilon_i~,
\end{equation}
where $n_i$ are nonnegative integers.
The justification of the interpretation is the theorem by \cite{Moore:2006dw} whose statement is that the boundary  state $\ket{B}$ of a boundary condition of a 2d TQFT with a semisimple Frobenius algebra must have the form of \begin{equation}
    \ket{B} = \sum_i n_i \ket{\mathsf{v}_i}~.
\end{equation}
Therefore, each superselection sectors on $\mathbb{R}^2$ are generated by boundary conditions $\ket{\mathsf{v}_i}$ imposed at the infinity of $\mathbb{R}^2$.

\subsection{3d Chern-Simons Theory}
\label{app:CS}
Because 2d $G_k/G_k$ TQFT is the $S^1$ reduction of 3d the $G_k$ CS theory, it is convenient to summarize the facts on the CS theory -- for more details see \cite{Benini:2018reh} and references therein. The $G_k$ CS theory has topological Wilson lines $L_{\mu}$ labeled by representations of $G$ whose highest weight $\mu$ satisfies $(\mu,\theta_G)\le k$ where $\theta_G$ is the heighest root of $G$.
The fusion coefficient $N^\rho_{\mu\nu}$ characterizes the algebra of these topological lines:
\begin{equation}
    L_\m L_\n = \sum_\r N^\r_{\m\n} L_\r ~.
\end{equation}
Each component of $N^\r_{\m\n}$ is a positive integer and symmetric with respect to the lower indices: $N^\r_{\m\n}=N^\r_{\n\m}.$

The unknot correlation function of a line $L_\m$ in a 3-manifold $M_3$ (divided by the $M_3$ partition function without insertion) is called the quantum dimension $d_\m$:
\begin{equation}
    d_\m = \ 
    \begin{tikzpicture}[scale = .7, baseline = -2.5]
        \draw[thick, ->- = .07 rotate -25] (0,0) arc (0:360:.5);
        \node[anchor = west] at (0,0) {$\m$};
    \end{tikzpicture}~.
\end{equation}
The modular $S$-matrix is defined by the Hopf link observable:
\begin{equation}
    S_{\m\n} = \frac{1}{\mathcal{D}}\,
    \begin{tikzpicture}[scale = .7, baseline =-2.5]
        \draw[thick, ->- = .37 rotate -10] (0,0) ++ (70:.5) arc (70:400:.5);
        \draw[thick, ->- = .6 rotate 160] (0,0) ++ (-60:.5) ++(-.07,.1) arc (223:-107:.5);
        \node[anchor=east] at (-.5,0) {$\m$};
        \node[anchor=west] at (1.1,0) {$\n$};
    \end{tikzpicture}~,
\end{equation}
where $\mathcal{D}$ is the total quantum dimension $\mathcal{D} = \sqrt{\sum_\m d_\m^2}$.
We let the line $L_0$ with the index 0 denotes the trivial line. Then, by definition, we have $\mathcal{D}=S_{00}^{-1}$ and $d_\m=\frac{S_{0\m}}{S_{00}}$. The modular $S$-matrix (for a unitary 3d TQFT) satisfies $S_{\m^* \n}=(S_{\m\n})^* = (S^{-1})_{\n\m}$. Contracting a loop of $L_\mu$ around the line $L_\nu$ we get
\begin{equation}
\begin{tikzpicture}[baseline =-2.5]
\draw[thick,->- = .65 rotate 4] (0,0) ++ (85: 1 and .3) arc (440:100:.5 and .3);
\draw[thick] (0,-.8) to (0,-.4); 
\draw[thick,->- = .5 rotate 0] (0,-.2) to (0,.6) ;
\node at (-.7,0) {$\m$};
\node[anchor = north] at (0,-1) {$\n$};
\end{tikzpicture}
= \frac{S_{\m\n}}{S_{0\n}} \,
\begin{tikzpicture}[ baseline =-2.5]
\draw[thick,->- = 0.7 rotate 0] (0,-.8) to (0,.6) ;
\node[anchor = north] at (0,-1) {$\n$};
\end{tikzpicture}~.
\end{equation}
The Verlinde formula relates the modular $S$-matrix to the fusion coefficients:
\begin{equation}
    N_{\m\n}^\r = \sum_\l \frac{S_{\m \l}S_{\n \l}(S_{\r \l})^*}{S_{0\l}}~.
\end{equation}
\subsection{G/G TQFT \label{app:G/G}}
As \cite{Blau:1993tv} have shown, the $G_k/G_k$ TQFT for a simply connected Lie group $G$ is the $S^1$ reduction of the 3d $G_k$ Chern-Simons theory. Each Wilson line $L_\mu$ in the 3d Chern-Simons theory provides both a line $L_\m$ and a local operator $\mathcal{O}_\n$ for the 2d theory.
The latter form the basis of the Hilbert space of the $G/G$ TQFT on $S^1$.
Both the line and local operators inherit the fusion rule of the 3d Wilson lines:
\begin{align}
    L_\m  L_\n &= \sum_\r N_{\m\n}^\r \, L_\r~,\\
    \mathcal{O}_\m \mathcal{O}_\n &= \sum_\r N_{\m\n}^\r \, \mathcal{O}_\r ~,
\end{align}
Where $N_{\m\n}^\r$ are the fusion coefficients of the $\hat{\mathfrak{g}}_k$ affine Lie algebra. When $\nu$ corresponds to an Abelian anyon in the 3d CS theory, or equivalently an invertible element of the fusion algebra, then these two kinds operators commute with each other up to the monodromy phase
\begin{equation}
    \mathcal{O}_\n L_\m = \frac{S_{\m\n}^\ast S_{00}}{S_{0\m}S_{0\n}} \, L_\m \mathcal{O}_\n~.
    \label{eq:WO}
\end{equation}

In the 3d CS theory, the non-trivial Wilson line has no (topological) endpoint operator (otherwise we can break the line up and trivialize it). Which means that correlation of $L_\m$ inserted along the $S^1$ direction of $S^1\times S^2$ vanishes. In $G/G$ language, this means
\begin{equation}
    \theta(\mathcal{O}_\n) = \delta_{\n0}~.
\end{equation}
Here, the coefficient is because $Z[S^1\times S^2] = 1$ in the 3d Chern-Simons theory. From the Verlinde formula, the idempotent basis can be found as
\begin{equation}
    \varepsilon_{\m} = S_{0\m}\sum_\n S_{\m \n}\mathcal{O}_\n~.
\end{equation}
The corresponding Euler counter term is
\begin{equation}
    \theta_{\m} = \theta(\varepsilon_{\m}) = (S_{0\m})^2~,
\end{equation}
as $S_{0\m}$ is real. 
The normalised vacua are
\begin{equation}
    \mathsf{v}_{\m} = \sum_\n S_{\m\n} \mathcal{O}_\n~.
\end{equation}

\section{Branching Rules and Outer-Automorphism Orbits \label{app:branching}}
The branching rules for the conformal pair $({\hat{\mathfrak{so}}}(\mathfrak{g}),{\mathfrak{g}})$, where $\mathfrak{g} \subset \hat{\mathfrak{so}}(\mathfrak{g})$ via the adjoint representation, is given by Kac-Wakimoto (Remark 4.2.2. in \cite{Kac:1988tf}). Here we are only interested in the branching rules of the embedding $\hat{\mathfrak{su}}(N)_N \subset \hat{\mathfrak{so}}(N^2-1)_1$. Thus we state Kac-Wakimoto result for this case and analyze them. The affine algebra $\hat{\mathfrak{so}}(N^2-1)_1$, has only the vacuum, vector and spinor(s) representations at level one. As derived in \cite{Kac:1988tf}, the spinor representation(s) decompose into $2^{\lfloor N/2 \rfloor-1}$-multiple of $\module_{\rho}$, where ${\rho}=[1,1,\dots,1]$ is the affine Weyl vector of $\hat{\mathfrak{su}}(N)$. The remaining level one representations decompose as
\begin{equation}
    \Module_{{\Omega}_0} = \bigoplus_{\lambda \in P^N_+ \cap R_+} \module_\lambda~, \quad\quad \Module_{{\Omega}_1} = \bigoplus_{\lambda \in P^N_+ \cap R_-}\module_\lambda~, 
    \label{branching.rules}
\end{equation}
where ${\Omega}_0=[1,0,\dots,0]$ and ${\Omega}_1=[0,1,0,\dots,0]$ correspond to the vacuum and vector module of the orthogonal affine algebra respectively, and
\begin{equation}
    R_\pm=\{N{\omega}_0+w\cdot\rho-\rho \mid w \in \mathcal{W},\, \det w = \pm 1 \} \,.
\end{equation}
Here $\Omega_0,\dots,\Omega_{\lfloor(N^2-1)/2\rfloor}$ and $\omega_0,\dots,\omega_{N-1}$ are the fundamental weights of $\hat{\mathfrak{so}}(N^2-1)$ and $\hat{\mathfrak{su}}(N)$ affine algebras respectively, $P^N_+$ denotes the dominant weights at level $N$, and $\mathcal{W}$ the affine Weyl group of $\hat{\mathfrak{su}}(N)_N$. To find the weights that appear in \eqref{branching.rules}, we should find all the dominant weights $\lambda=[\lambda_0,\lambda_1,\dots,\lambda_{N-1}]$ such that $\lambda + \rho = N{\omega}_0+w\cdot\rho$, or equivalently
\begin{equation}
    [\lambda_0+1,\l_1+1,\dots,\lambda_{N-1}+1] = [N,0,\dots,0]+w\cdot[1,1,\dots,1]~.
\end{equation}
One can check the Dynkin labels of $\lambda$ have to satisfy the conditions
\begin{align}
    \sum_{i=0}^{N-1} (\lambda_i+1) = 2N \quad &\mathrm{and} \quad (\lambda_i+1)\geq1~, \nonumber\\
    \sum_{i\leq k \leq j}(\lambda_k+1) \neq N \quad &\mathrm{for} \quad 0\leq i < j \leq N-1~, \label{conditions.app}
\end{align}
and all such weights appear exactly once in \eqref{branching.rules}.

Here we argue that the number of modules in the RHS of the branching rules \eqref{branching.rules} is $2^{N-1}$. Basically we want to count the number of weights $[\lambda_0,\lambda_1,\dots,\lambda_{N-1}]$ that satisfy the conditions \eqref{conditions.app}. Define the partial sums $f_i = \sum_{k=1}^i (\lambda_i+1)$ for $i=1,\dots,N-1$. Then the conditions above translate into
\begin{enumerate}
    \item $0<f_1<\dots<f_{N-1}<2N$ ~,
    \item $f_i \neq N$ ~,
    \item $f_i - f_j \neq N$ ~.
\end{enumerate}
Therefore we have to choose $N-1$ distinct integer numbers from the set $\{1,\dots,N-1,N+1,\dots,2N-1\}$ such that the difference of any two of them is not equal to $N$. Therefore we have to choose exactly one number from each of the sets below
\begin{equation}
    \{1,N+1\},\{2,N+2\},\dots,\{N-1,2N-1\}~.
\end{equation}
Thus there are exactly $2^{N-1}$ choices and therefore $2^{N-1}$ vacua for the fermionized $\Spin(N^2-1)_1/\SU(N)_N$ model or equivalently 2d adjoint QCD at infinite coupling. Moreover, one can also count the number of primaries in the branching of the vacuum and vector representations individually and get
\begin{align}
    \abs{\Module_{{\Omega}_0}} = \begin{cases}
    2^{N-2} & \mbox{even }N \\
    2^{N-2}+2^{\frac{N-3}{2}} & \mbox{odd } N
    \end{cases} \,,\notag\\
    \abs{\Module_{{\Omega}_1}} = \begin{cases}
    2^{N-2} & \mbox{even }N  \\
    2^{N-2}-2^{\frac{N-3}{2}} & \mbox{odd } N
    \end{cases}\,. \label{branching1}
\end{align}
Furthermore, 
\begin{align}
    \module_\rho &\notin \Module_{{\Omega}_0} \oplus \Module_{{\Omega}_1} \quad N \equiv 0 \pmod{2}\,,\notag\\
    \module_\rho &\in \Module_{{\Omega}_0} \qquad\qquad N \equiv \pm1 \pmod{8} \,,\notag\\
    \module_\rho &\in \Module_{{\Omega}_1} \qquad\qquad N \equiv \pm3 \pmod{8} \,. \label{branching2}
\end{align}
This information will be needed later to calculate the product of the $\hat{\mathfrak{su}}(N)_N$ modular invariant matrices.

\subsection*{Outer Automorphism Orbits}
Now by studying the action of the one-form symmetry on these vacua we calculate the number of vacua in each universes in the NS sector of the fermionic theory. The action of the one-form symmetry is the same as the action of the $\mathbb{Z}_N$ outer automorphism of the $\hat{\mathfrak{su}}(N)$ affine Lie algebra which amounts to shifting the $\lambda_i$ Dynkin labels cyclically. For each affine weight $\lambda$ appearing in the branching rules \eqref{branching.rules}, we define its order to be the length of its orbit under the action of the $\mathbb{Z}_N$ outer automorphism. We calculate the number of weights of given order recursively.

Denote $A_d^{(N)}$ to be the number of weights of order $d$ that satisfy \eqref{conditions.app}. First, we note that $A_d^{(N)}$ is nonzero only if $N/d$ is an odd integer. The fact that $N/d$ must be integer is obvious, since for a weight of order $d$, the Dynkin labels have to have preiodicity $d$, i.e.\ decompose into $N/d$ identical blocks of length $d$. Now since the sum of all Dynkin labels is equal to $2N$, the sum of the Dynkin labels in each block will be $2d$. Therefore if $N/d$ were an even integer then the sum of the Dynkin labels in the first $N/2d$ blocks would have been equal to $N$ contradicting the last condition in \eqref{conditions.app}, thus $N/d$ must be an odd integer. Now we prove a key property that
\begin{equation}
    A_d^{(N)} = A_d^{(d)} \quad \text{if} \quad \frac{N}{d}=1\pmod{2}~.
\end{equation}
To prove this, we have to give a one-to-one correspondence between the weights of $\hat{\mathfrak{su}}(d)_d$ and $\hat{\mathfrak{su}}(N)_N$ that have order $d$ and satisfy the conditions \eqref{conditions.app}. We show that the following naive guess works
\begin{equation}
    \lambda=[\lambda_0,\lambda_1,\dots,\lambda_{d-1}] ~\leftrightarrow~ \lambda^{N/d}=[\lambda_0,\lambda_1,\dots,\lambda_{d-1},\dots,\lambda_0,\dots,\lambda_{d-1}]~,
\end{equation}
where the Dynkin labels on the RHS are just $N/d$ copies of the Dynkin labels on the LHS. Therefore we have to show $\lambda$ satisfy the conditions \eqref{conditions.app} if and only if $\lambda^{N/d}$ satisfies them. Let's first assume $\lambda^{N/d}$ satisfies \eqref{conditions.app}, then the only non-trivial condition that we have to check for $\lambda$ is the last condition, that is the consecutive partial sum of the Dynkin labels in each block cannot be equal to $d$. We prove this by contradiction. First note that because of the periodicity property of the Dynkin labels the sum of each $kd$ consecutive Dynkin labels has to be equal to $2kd$ for any integer $k$. Now, assuming that there are some consecutive Dynkin labels in the first block that their sum is equal to $d$, then the sum of those Dynkin labels plus the sum of the next $d\frac{N/d-1}{2}$ labels would be $d+d(N/d-1)=N$, contradicting \eqref{conditions.app}. Now let's prove the opposite, that is $\lambda$ satisfy the conditions, then we must show the sum of any consecutive Dynkin labels of $\lambda^{N/d}$ cannot be equal to $N$. Assume the contrary that there are some consecutive such Dynkin labels with sum $N$, note again that the sum of any $d\frac{N/d-1}{2}$ consecutive labels is equal to $N-d$, thus there has to be some consecutive Dynkin labels whose sum are exactly $d$ contradicting the fact that $\lambda$ satisfies the conditions.

All in all, we have
\begin{equation}
    A_d^{(N)} = \begin{cases} A_d^{(d)} \quad &\text{if $\frac{N}{d}$ is an odd integer}~, \\
    0 \quad &\text{otherwise}~.
    \end{cases}
\end{equation}
Note that $2^{N-1}=\sum_d A_d^{(N)}$, thus keeping only the non-zero terms gives
\begin{equation}
    2^{N-1}=\sum_{2 \nmid d \mid N} A_{\frac{N}{d}}~,
\end{equation}
where we have dropped the upper index on $A$, and the summation is taken over the odd positive divisors of $N$. Using the Möbius inversion formula, we can invert the formula above to get
\begin{equation}
    A_{N} = \sum_{2\nmid d\mid N} \mu(d)\, 2^{\frac{N}{d}-1}~,
\end{equation}
where $\mu$ is the Möbius function. Diagonalizing the action of the outer automorphism on the orbit of an order $d$ weight, gives $d$ universes with eigenvalues $0,\frac{N}{d},2\frac{N}{d},\dots,N \pmod{N}$. Thus if we define $C_m^N$ to be the number of vacua in universe $m$, then
\begin{equation}
    C_m^{N} = \sum_{2 \nmid d \mid \gcd(m,N)} \frac{A_{\frac{N}{d}}}{\frac{N}{d}} = \sum_{\substack{2 \nmid d \mid \gcd(m,N) \\ 2 \nmid d' \mid \frac{N}{d}}}\mu(d')\,\frac{2^{\frac{N}{dd'}-1}}{\frac{N}{dd'}}~.
\end{equation}
In particular, the number of vacua in universe with eigenvalue $1 \pmod{N}$ is $C_1^{N} = \frac{A_{N}}{N}$. Also there are at least
\begin{align}
    C_1^{N} &= \frac{1}{2N} \left(2^N - \sum_{i} 2^{\frac{N}{p_i}} + \sum_{i\neq j} 2^{\frac{N}{p_ip_j}} - \sum_{i<j<k} 2^{\frac{N}{p_ip_jp_k}} + \dots \right) ~,\\
    &> \frac{2^{N-1}}{N} - \frac{2^{\frac{N}{p_1}}}{N} = \frac{2^{N-1}}{N}+O(2^{N/3})~,
\end{align}
vacua in each universes, where $p_1<p_2<\dots$ are the odd prime factors of $N$.

\section{\texorpdfstring{$\SU(N)_N$}{SU(N)} Modular Invariants \label{app:modular.invariants}}
In this appendix we list all the modular invariants of $\SU(N)_N$ that can be obtained by conformal embedding and the method of outer automorphism of the Kac-Moody algebras.

\paragraph{D Series:}
The D-type modular invariants correspond to the usual gauging of the $\SU(N)_N$ WZW by a subgroup of its $\bZ_N$ center symmetry \cite{Bernard:1986xy,Schellekens:1989am, DiFrancesco:1997nk}. As discussed in section \ref{sec:coset.TQFT}, the $\bZ_N$ center of $\SU(N)$ leads to invertible Verlinde lines that generate a $\bZ_N$ zero-form symmetry. It turns out that this group-like symmetry, for the case of $\SU(N)_N$, is non-anomalous. Thus any $\bZ_k$ subgroup of this $\bZ_N$ symmetry can be gauged which leads to a new modular invariant. We denote these modular invariants by $\underline{{\SU}/{\bZ_k}}$ for $k$ a divisor of $N$.

\paragraph{Conformal Embedding:} All the conformal embeddings of affine Lie algebras have been classified in \cite{Schellekens:1986mb,Bais:1986zs}. For the case of $\SU(N)_N$, there are only two such conformal embeddings: $\SU(N)_N \subset \Spin(N^2-1)_1$ where we denote the corresponding modular invariant by $\underline{\Spin}$, and $\SU(6)_6 \subset \Sp(20)_1$ that we will discuss in \ref{example.su6}.

\paragraph{Even N}
Using these methods we find the following modular invariant torus partition functions for even $N$
\begin{align}
    Z\left( \SU \right) &= \sum_{\l} \chi_{\l} \, \bar{\chi}_\l~, \quad Z\left( \SU^\ast \right) = \sum_{\l} \chi_{\l} \, \bar{\chi}_{\l^\ast} ~,\notag\\
    Z\left( {\SU}/{\bZ_k} \right) &= \sum_{p=0}^{k-1}\sum_{i\l_i\overset{k}{\equiv}0} \chi_\l \, \bar{\chi}_{\cA^{pk'}\l} ~, \quad \text{for }k'=\frac{N}{k} \overset{2}{\equiv}0 ~,\notag\\
    Z\left( {\SU}/{\bZ_k} \right) &= \sum_{p=0}^{k/2-1} \left( \sum_{i\l_i\overset{k}{\equiv}0} \chi_\l \, \bar{\chi}_{\cA^{2pk'}\l} +\sum_{i\l_i\overset{k}{\equiv}k/2} \chi_\l \, \bar{\chi}_{\cA^{(2p+1)k'}\l}\right), \quad \text{for }k'\overset{2}{\equiv}1 ~,\notag\\
    Z\left( \Spin \right) &= \abs{\boldsymbol{\chi}_{\Omega_0}}^2 + \abs{\boldsymbol{\chi}_{\Omega_1}}^2 + \abs{2^{\frac{N-2}{2}} \chi_{[1,\dots,1]} }^2 ~,
\end{align}
where the summations are taken over affine weights at level $N$, i.e.\ $\sum_{i=0}^N\l_i=N$, and
\begin{equation}
    \boldsymbol{\chi}_{\Omega_0} = \sum_{\module_\l \in \Module_{\Omega_0}} \chi_\l ~, \qquad \boldsymbol{\chi}_{\Omega_1} = \sum_{\module_\l \in \Module_{\Omega_1}} \chi_\l ~.
\end{equation}
Moreover, $\underline{\SU}$ and $\underline{\SU}^\ast$ denote the diagonal and charge conjugation modular invariants respectively, and $\l^\ast$ is the complex conjugate of $\l$, i.e.\ $\l^\ast_i = \l_{N-i}$. As in \ref{sec:coset.TQFT}, $\cA$ denotes the outer automorphism action, i.e.\ $(\cA\l)_i = \l_{i-1}$.

\paragraph{Odd N}
When $N$ is odd, the chiral algebra of $\Spin(N^2-1)_1$ can be further extended. These extensions are given by the modular invariants of $\Spin(N^2-1)_1$, which have been classified in \cite{Gannon:1992nq}. It turns out that all such modular invariants are D-type that are associated with gauging subgroups of the $\bZ_2^\tts\times\bZ_2^\ttc$ center of $\Spin(N^2-1)$. There are six such modular invariants associated to the five subgroups $1$, $\bZ_2^\tts$, $\bZ_2^\ttc$, $\bZ_2^{\mathrm{v}}$, and $\bZ_2^\tts\times\bZ_2^\ttc$ of the center. The sixth modular invariant arises because there are two ways of gauging the whole center. Different ways of gauging a discrete symmetry group $G$ in 2d are classified by the second group cohomology $H^2(G,\mathrm{U}(1))$, which is known as the discrete torsion \cite{Vafa:1986wx, Brunner:2014lua}. For $G=\bZ_k$ this group is trivial, while $H^2(\bZ_2\times\bZ_2,\mathrm{U}(1))\cong \bZ_2$.

The exact form of these six modular invariants for $\Spin(N^2-1)_1$ depends on whether $N\equiv \pm1 $ or $N \equiv \pm3 \pmod{8}$. We denote the non-diagonal modular invariants by $\underline{\SO}={\Spin}/{\bZ_2^\mathrm{v}}$, $\underline{\mathrm{Ss}}={\Spin}/{\bZ_2^\tts}$, $\underline{\mathrm{Sc}}={\Spin}/{\bZ_2^\ttc}$, and $\underline{\PO}={\Spin}/({\bZ_2^\tts\times\bZ_2^\ttc})$, where there are two $\underline{\PO}$ modular invariants ($\underline{\PO^1}$ and $\underline{\PO^2}$). Note that these six $\Spin(N^2-1)_1$ modular invariants leads to only two different $\SU(N)_N$ modular invariant matrices. More precisely, $\mim({\Spin})=\mim({\SO})$ and $\mim({\mathrm{Ss}})=\mim({\mathrm{Sc}})=\mim(\PO^1)=\mim(\PO^2)$. However, as we will see for the case of $N=3$ in \ref{app:su3}, having the same modular invariant matrix does not mean that these theories are necessarily the same and they can correspond to different boundary conditions (module categories) $\cC_A$. All in all we find the following modular invariants
\begin{align}
    Z\left( \SU \right) &= \sum_{\l} \chi_{\l} \, \bar{\chi}_\l~, \quad Z\left( \SU^\ast \right) = \sum_{\l} \chi_{\l} \, \bar{\chi}_{\l^\ast} ~,\notag\\
    Z\left( {\SU}/{\bZ_k} \right) &= \sum_{p=0}^{k-1}\sum_{i\l_i\overset{k}{\equiv}0} \chi_\l \, \bar{\chi}_{\cA^{pk'}\l}~, \quad \text{for }k'=\frac{N}{k} ~,\notag\\
    Z\left( \Spin \right) &= \abs{\boldsymbol{\chi}_{\Omega_0}}^2 + \abs{\boldsymbol{\chi}_{\Omega_1}}^2 + 2\abs{2^{\frac{N-3}{2}} \chi_{\rho}}^2~,
\end{align}
and
\begin{equation}
    Z\left( \PO \right) = \begin{cases}
    \abs{\boldsymbol{\chi}_{\Omega_0}+2^{\frac{N-3}{2}} \chi_{\rho}}^2 & N \overset{8}{\equiv} \pm 1 \\
    \abs{\boldsymbol{\chi}_{\Omega_0}}^2 + 2^{\frac{N-3}{2}} \left(\boldsymbol{\chi}_{\Omega_1}\bar{\chi}_{\rho} +\chi_{\rho}\bar{\boldsymbol{\chi}}_{\Omega_1} \right)+2^{N-3}\abs{\chi_{\rho}}^2  & N \overset{8}{\equiv} \pm 3
    \end{cases}~,
\end{equation}
where $\rho=[1,\dots,1]$ is the affine Weyl vector of $\hat{\mathfrak{su}}(N)$.

Having discussed the modular invariants of $\SU(N)_N$ for general $N$, in the following we study the $N=3,4,5,6$ examples. For the case of $N=3,4$ the list of the modular invariants that we provide is claimed to be a complete list  \cite{Gannon:1992ty, Ocneanu:2000kj}.
\subsection{\texorpdfstring{$N=3$}{N=3}}
For $N=3$ the only non-diagonal modular invariants are the charge conjugation and the $\mathrm{PSU}(3)_3$ modular invariants \cite{Gannon:1992ty}
\begin{align}
    Z\left( \SU \right) &= \sum_{\l} \chi_{\l} \, \bar{\chi}_\l\,, \quad Z\left( \SU^\ast \right) = \sum_{\l} \chi_{\l} \, \bar{\chi}_{\l^\ast} \,, \\
    Z\left( \Spin \right) = Z\left( \mathrm{PSU} \right) &= \abs{\chi_{00}+\chi_{30}+\chi_{03}}^2 + 3 \abs{\chi_{11}}^2 \,,
\end{align}
where from now on to avoid a cluttered notation we use $\chi_{\l_1\l_2\dots\l_{N-1}}$ instead of $\chi_{[\l_0,\l_1,\l_2,\dots,\l_{N-1}]}$, and suppress the zeroth affine weight $\l_0$. Defining the corresponding modular invariant \emph{matrices} by $\mim$ one gets
\begin{equation}
    \mim(\Spin)\,\mim(\Spin) = 3 \,\mim(\Spin)~.
\end{equation}
\subsection{\texorpdfstring{$N=4$}{N=4}}
For $N=4$, there are still no other $\SU(4)_4$ modular invariant beyond the D-series and the $\underline{\Spin}$ modular invariant \cite{Ocneanu:2000kj}. There are precisely seven modular invariants given by
\begin{align}
    Z\left( \SU \right) =&~ \sum_{\l} \chi_{\l} \, \bar{\chi}_\l~, \quad Z\left( \SU^\ast \right) = \sum_{\l} \chi_{\l} \, \bar{\chi}_{\l^\ast} ~,\notag\\
    Z\left( {\SU}/{\bZ_2} \right) =&~ \sum_{i\l_i\overset{2}{\equiv}0} \chi_\l \left( \bar{\chi}_{\l} + \bar{\chi}_{\cA^{2}\l} \right) ~,\notag\\
    Z\left( {\SU}/{\bZ_4} \right) =&~ \sum_{i\l_i\overset{4}{\equiv}0} \chi_\l \left( \bar{\chi}_{\l} + \bar{\chi}_{\cA^{2}\l} \right) +\sum_{i\l_i\overset{4}{\equiv}2} \chi_\l \left( \bar{\chi}_{\cA\l} + \bar{\chi}_{\cA^{3}\l} \right) ~,\notag\\
     Z\left( \Spin \right) =&~ \abs{\chi_{000}+\chi_{040}+\chi_{210}+\chi_{012}}^2 + \abs{\chi_{400}+\chi_{004}+\chi_{101}+\chi_{121}}^2 +\abs{2 \chi_{111} }^2\,,
\end{align}
along with the charge conjugated modular invariants ${\underline{{\SU}/{\bZ_2}}}^\ast$ and ${\underline{{\SU}/{\bZ_4}}}^\ast$. Furthermore, the product of the corresponding modular invariant matrices are
\begin{align}
    \mim\left(\Spin\right) \, \mim\left(\SU^\ast\right) &= \mim\left(\Spin\right)~,\\
    \mim\left(\Spin\right) \, \mim\left({\SU}/{\bZ_2}\right) &=2 \,\mim\left(\Spin\right)~,\\
    \mim\left(\Spin\right) \, \mim\left({\SU}/{\bZ_4}\right) &= 2 \, \mim\left(\Spin\right) ~,\\
    \mim\left(\Spin\right) \, \mim\left(\Spin\right) &= 4 \, \mim\left(\Spin\right)~.
\end{align}
\subsection{\texorpdfstring{$N=5$}{N=5}}
For the $N=5$ case, interestingly there exist a modular invariant found in \cite{Schellekens:1989uf}, that cannot be obtained by the method of conformal embedding or outer automorphism. We denote this modular invariant by $\underline{\mathrm{SY}}$. In total we find six modular invariants
\begin{align}
    Z\left( \SU \right) =&~ \sum_{\l} \chi_{\l} \, \bar{\chi}_\l~, \quad Z\left( \SU^\ast \right) = \sum_{\l} \chi_{\l} \, \bar{\chi}_{\l^\ast} ~,\notag\\
    Z\left( {\SU}/{\bZ_5} \right) =&~ \abs{\chi_1}^2 + \abs{\chi_2}^2 + \abs{\chi_3}^2 + \abs{\chi_4}^2 + \abs{\chi_5}^2 + 5\abs{\chi_6}^2 ~,\notag\\
    Z\left( \mathrm{SY} \right) =&~ \abs{\chi_1}^2 + \abs{\chi_2}^2 + \abs{\chi_4}^2 + \abs{\chi_5}^2 + \left( \chi_3 \bar{\chi}_6 + \chi_6 \bar{\chi}_3  \right) + 4\abs{\chi_6}^2 ~,\notag\\
    Z\left( \Spin \right) =&~ \abs{\chi_1+\chi_2}^2 + \abs{\chi_3+\chi_6}^2 + 2\abs{2\chi_6}^2 ~,\notag\\
    Z\left( \PO \right) =&~ \abs{\chi_1+\chi_2}^2 + 2 \left( \chi_3 \bar{\chi}_6 + \chi_6 \bar{\chi}_3  \right) + 8\abs{\chi_6}^2 ~, \label{su5.invariants}
\end{align}
written in terms the $\bZ_5$ (outer automorphism) invariant characters
\begin{align*}
    \chi_1 &= \chi_{0000} + \chi_{5000} + \chi_{0500} + \chi_{0050} + \chi_{0005} \,,\\
    \chi_2 &= \chi_{0102} + \chi_{2010} + \chi_{2201} + \chi_{0220} + \chi_{1022} \,,\\
    \chi_3 &= \chi_{0013} + \chi_{1001} + \chi_{3100} + \chi_{1310} + \chi_{0131} \,,\\
    \chi_4 &= \chi_{0021} + \chi_{2002} + \chi_{1200} + \chi_{2120} + \chi_{0121} \,,\\
    \chi_5 &= \chi_{0110} + \chi_{3011} + \chi_{0301} + \chi_{1030} + \chi_{1103} \,,\\
    \chi_6 &= \chi_{1111} \,.
\end{align*}
The product of some of these modular invariant matrices is given by
\begin{align}
    \mim\left(\Spin\right) \mim\left(\SU^\ast\right) &= \mim\left(\Spin\right)~,\notag\\
    \mim\left(\Spin\right) \mim\left({\SU}/{\bZ_5}\right) &=5 \,\mim\left(\Spin\right)~,\notag\\
    \mim\left(\Spin\right) \mim\left(\mathrm{SY}\right) &= \mim\left(\Spin\right) + 4\, \mim\left(\PO\right) ~,\notag\\
    \mim\left(\Spin\right) \mim\left(\PO\right) &= 2 \, \mim\left(\Spin\right) + 8 \,\mim\left(\PO\right)~,\notag\\
    \mim\left(\Spin\right) \mim\left(\Spin\right) &= 6 \, \mim\left(\Spin\right) + 4 \,\mim\left(\PO\right)~.
\end{align}
In particular, these modular invariants form a closed algebra, and no new modular invariants can be constructed by their fusions.
\subsection{\texorpdfstring{$N=6$}{N=6} \label{example.su6}}
For $N=6$, we have an extra conformal embedding $\SU(6)_6 \subset \Sp(20)_1$ \cite{Schellekens:1986mb,Bais:1986zs} that we denote by $\underline{\mathrm{Sp}}$. Furthermore, we will find a new modular invariant by taking the product of $\underline{\Spin}$ with $\underline{\mathrm{Sp}}$. In total we find 13 distinct invariants. The automorphism invariants are
\begin{align}
    Z\left( \SU \right) &= \sum_{\l} \chi_{\l} \, \bar{\chi}_\l~, \quad Z\left( \SU^\ast \right) = \sum_{\l} \chi_{\l} \, \bar{\chi}_{\l^\ast} ~,\notag\\
    Z\left( {\SU}/{\bZ_3} \right) &= \sum_{i\l_i\overset{3}{\equiv}0} \chi_\l \left( \bar{\chi}_{\l} + \bar{\chi}_{\cA^{2}\l} + \bar{\chi}_{\cA^{4}\l} \right) ~,\notag\\
    Z\left( {\SU}/{\bZ_2} \right) &= \sum_{i\l_i\overset{2}{\equiv}0} \abs{\chi_\l}^2 +\sum_{i\l_i\overset{2}{\equiv}1} \chi_\l \, \bar{\chi}_{\cA^{3}\l} ~,\notag\\
    Z\left( {\SU}/{\bZ_6} \right) &= \sum_{i\l_i\overset{6}{\equiv}0} \chi_\l \left( \bar{\chi}_{\l} + \bar{\chi}_{\cA^{2}\l} + \bar{\chi}_{\cA^{4}\l} \right) +\sum_{i\l_i\overset{6}{\equiv}3} \chi_\l \left( \bar{\chi}_{\cA\l} + \bar{\chi}_{\cA^{3}\l} + \bar{\chi}_{\cA^{5}\l} \right) ~,
\end{align}
along with their charge conjugated modular invariants ${\underline{{\SU}/{\bZ_3}}}^\ast$, ${\underline{{\SU}/{\bZ_2}}}^\ast$ and ${\underline{{\SU}/{\bZ_6}}}^\ast$. By using the branching rules of the conformal embedding $\SU(6)_6 \subset \Sp(20)_1$ given in \cite{Aldazabal:1991cj}, we find
\begin{align}
     Z\left( \Spin \right) &= \abs{\chi_{00000}^{(3)} + \chi_{20010}^{(3)} + \chi_{01002}^{(3)} + \chi_{01410}^{(3)} + \chi_{11211}^{(3)} + \chi_{02020}}^2 \notag\\
     & + \abs{\chi_{60000}^{(3)} + \chi_{32001}^{(3)} + \chi_{30100}^{(3)} + \chi_{00141}^{(3)} + \chi_{01121}^{(3)} + \chi_{20202} }^2 +\abs{4 \chi_{11111} }^2~,\notag\\
     Z\left( \mathrm{Sp} \right) &= \abs{\chi_{00000}^{(3)} + \chi_{00200}^{(3)} + \chi_{02020} }^2 + \abs{\chi_{02001}^{(3)} + \chi_{10020}^{(3)} + \chi_{11111} }^2 + \abs{\chi_{30003}^{(3)} + \chi_{11211}^{(3)} + \chi_{02020} }^2 \notag\\
     &+ \abs{\chi_{60000}^{(3)} + \chi_{40020}^{(3)} + \chi_{20202} }^2 + \abs{\chi_{30200}^{(3)} + \chi_{31002}^{(3)} + \chi_{11111} }^2 + \abs{\chi_{00030}^{(3)} + \chi_{11011}^{(3)} + \chi_{20202} }^2\notag\\
     &+ \abs{\chi_{00100}^{(3)} + \chi_{01110}^{(3)} }^2 + \abs{\chi_{01010}^{(3)} + \chi_{12021}^{(3)} }^2 \notag\\
     &+ \abs{\chi_{50010}^{(3)} + \chi_{30111}^{(3)} }^2 + \abs{\chi_{40101}^{(3)} + \chi_{01202}^{(3)} }^2 + \abs{\chi_{20102}^{(6)} }^2\,, \label{su5.embedding.invs}
\end{align}
where
\begin{align}
    \chi_{\l_1\l_2\l_3\l_4\l_5}^{(3)} = \chi_{\l_1\l_2\l_3\l_4\l_5} + \chi_{\l_5\l_0\l_1\l_2\l_3} + \chi_{\l_3\l_4\l_5\l_0\l_1} \,,\\
    \chi_{\l}^{(6)} = \chi_{\l} + \chi_{\cA\l} + \chi_{\cA^2\l} + \chi_{\cA^3\l} + \chi_{\cA^4\l} + \chi_{\cA^5\l} \,,  
\end{align}
denote the $\bZ_3$ and $\bZ_6$ (outer automorphism) invariant characters. By taking the product the above invariants, we find three more invariants
\begin{align}
     Z\left( \mathrm{X|Spin} \right) &= \left( \chi_{00000}^{(3)} + \chi_{00200}^{(3)} + \chi_{30003}^{(3)} + \chi_{11211}^{(3)} + 2 \chi_{02020} \right) \left( \bar{\chi}_{00000}^{(3)} + \bar{\chi}_{20010}^{(3)} + \bar{\chi}_{01002}^{(3)} + \bar{\chi}_{01410}^{(3)} + \bar{\chi}_{11211}^{(3)} + \bar{\chi}_{02020} \right) \notag\\
     & + \left( \chi_{60000}^{(3)} + \chi_{40020}^{(3)} + \chi_{03000}^{(3)} + \chi_{01121}^{(3)} + 2 \chi_{20202} \right) \left( \bar{\chi}_{60000}^{(3)} + \bar{\chi}_{32001}^{(3)} + \bar{\chi}_{30100}^{(3)} + \bar{\chi}_{00141}^{(3)} + \bar{\chi}_{01121}^{(3)} + \bar{\chi}_{20202} \right) \notag\\ & + \left( \chi_{02001}^{(6)} + \chi_{10020}^{(6)} + 2\chi_{11111} \right)4\bar{\chi}_{11111} ~,\notag\\
     Z\left( \mathrm{X} \right) &= \abs{\chi_{00000}^{(3)} + \chi_{00200}^{(3)} + \chi_{30003}^{(3)} + \chi_{11211}^{(3)} + 2 \chi_{02020} }^2 \notag\\
     & +\abs{\chi_{60000}^{(3)} + \chi_{40020}^{(3)} + \chi_{03000}^{(3)} + \chi_{01121}^{(3)} + 2 \chi_{20202} }^2 \notag\\ & + \abs{\chi_{02001}^{(6)} + \chi_{10020}^{(6)} + 2\chi_{11111} }^2 ~,
\end{align}
where the third one is $\mim\left( \mathrm{Spin|X} \right) = \mim\left(\mathrm{X|Spin}\right)^\mathsf{T}$.
They are related to the previous invariants by
\begin{align}
    \mim\left(\Spin\right)  \mim\left( \mathrm{Sp} \right) &= 4 \,  \mim\left( \mathrm{Spin|X} \right)~,\notag\\
    \mim\left(\mathrm{Sp}\right) \mim\left(\Spin\right) &= 4 \, \mim\left(\mathrm{X|Spin}\right)~,\notag\\
    \mim\left(\mathrm{Sp}\right)  \mim\left(\Spin\right)  \mim\left(\mathrm{Sp}\right) &= 16 \,\mim\left(\mathrm{X}\right)~.
\end{align}
In total we have found 13 modular invariants, and taking the product of these invariants with $\underline{\Spin}$, we get
\begin{align}
    \mim\left(\Spin\right) \mim\left(\SU^\ast\right) &= \mim\left(\Spin\right)\,,\notag\\
    \mim\left(\Spin\right) \mim\left({\SU}/{\bZ_3}\right) &=3 \,\mim\left(\Spin\right)\,,\notag\\
    \mim\left(\Spin\right) \mim\left({\SU}/{\bZ_2}\right) &= \mim\left(\Spin\right)\,,\notag\\
    \mim\left(\Spin\right) \mim\left({\SU}/{\bZ_6}\right) &=3 \,\mim\left(\Spin\right)\,,\notag\\
    \mim\left(\Spin\right) \mim\left(\mathrm{Sp}\right) &= 4\, \mim\left(\mathrm{Spin|X}\right) \,,\notag\\
    \mim\left(\Spin\right) \mim\left(\mathrm{X}\right) &= 8\, \mim\left(\mathrm{Spin|X}\right)\,,\notag\\
    \mim\left(\Spin\right) \mim\left(\mathrm{Spin|X}\right) &= 16 \,\mim\left(\mathrm{\Spin|X}\right)\,,\notag\\
    \mim\left(\Spin\right) \mim\left(\mathrm{X|Spin}\right) &= 8 \,\mim\left(\mathrm{Spin}\right)\,,\notag\\
    \mim\left(\Spin\right) \mim\left(\Spin\right) &= 16 \, \mim\left(\Spin\right) \,,
\end{align}
where
\begin{align}
    \Tr\left[ \mim\left(\Spin\right) \right] &= 48\,,\\
    \Tr\left[ \mim\left(\mathrm{Spin|X}\right) \right] &= 24 \,.
\end{align}

\section{Fermionic Symmetric TQFTs}
\label{app:fermionic}
In the bulk of the paper we have discussed the bosonic version of adjoint QCD, where one sums over all spin structures. This changes the symmetries of the theory a little bit, the number of vacua changes etc. It however does not affect the question of confinement vs deconfinement. For the sake of completeness in this appendix we discuss the symmetries and  vacua of the fermionic theory. 
In order to return to the fermionic theory from the bosonic one, we can ``gauge back" the $(-1)^F$ symmetry in \eqref{eq:NABos} by gauging a $\mathbb{Z}_2$ symmetry with the twist defined by the Arf invariant \cite{AlvarezGaume:1987vm, Gaiotto:2015zta, Kapustin:2017jrc,Thorngren:2018bhj,Karch:2019lnn,Gaiotto:2020iye}.
Applying this operation to the bosonization \eqref{eq:NABos}, we get the duality
\begin{equation}
    \text{$n$ Majorana fermions}
    \leftrightarrow
    \text{$\mathrm{Spin}(n)_1$ WZW}/_{\mathrm{Arf}}\mathbb{Z}_2~.
    \label{eq:NAFer}
\end{equation}
Here, the $\mathbb{Z}_2$ symmetry in $\mathrm{Spin}(n)_1$ WZW model is a subgroup of the center of the chiral $\mathrm{Spin}(n)$ symmetry. The notation $/_{\mathrm{Arf}}\mathbb{Z}_2$ represents the gauging of the $\mathbb{Z}_2$ with the $\mathrm{Arf}$ twist.
By taking $n=N^2-1$ and gauging $\SU(N)$, we arrive at the following duality for the adjoint QCD with infinite coupling:
\begin{equation}
    \text{($\SU(N)$ adj. QCD with $g_\mathrm{YM}\to\infty$)}
    \leftrightarrow
    \text{$\mathrm{Spin}(N^2-1)_1/\SU(N)_N/_{\mathrm{Arf}}\mathbb{Z}_2$ TQFT}~.
    \label{eq:QcdDualFer}
\end{equation}

Having analyzed the vacua of the $\Spin(N^2-1)_1/\SU(N)_N$ theory, we can now gauge the $\bZ_2^\mathrm{v}$ center of $\Spin(N^2-1)$ with an Arf twist to analyze the vacua of adjoint QCD. The question is then which vacua survive the gauging and whether more appear. In the following we show that this gauging amounts to dropping the spinor representations of $\Spin(N^2-1)_1$. Therefore, the vacua of adjoint QCD in the NS sector are in one-to-one correspondence with the branching coefficients in the decomposition of the trivial and vector representations of $\Spin(N^2-1)_1$.

Consider the theory on a torus. To gauge a discrete subgroup we sum over insertions of topological lines generating the action of the symmetry \cite{Bhardwaj:2017xup}. We count the number of vacua in the NS sector of the fermionic theory by the NSNS partition function which is a sum of terms 
\begin{equation}
Z_\mathrm{NSNS}= 
    \frac12 \left(
    \begin{tikzpicture}[baseline = 13pt]
        \draw (0,0) -- (0,1);
        \draw (0,1) -- (1,1);
        \draw (1,0) -- (1,1);
        \draw (0,0) -- (1,0);
    \end{tikzpicture}
    +
    \begin{tikzpicture}[baseline = 13pt]
        \draw (0,0) -- (0,1);
        \draw (0,1) -- (1,1);
        \draw (1,0) -- (1,1);
        \draw (0,0) -- (1,0);
        \draw[blue] (0,.5) -- (1,.5);
    \end{tikzpicture}
    +
    \begin{tikzpicture}[baseline = 13pt]
        \draw (0,0) -- (0,1);
        \draw (0,1) -- (1,1);
        \draw (1,0) -- (1,1);
        \draw (0,0) -- (1,0);
        \draw[blue] (.5,0) -- (.5,1);
    \end{tikzpicture}
    \pm
    \begin{tikzpicture}[baseline = 13pt]
        \draw (0,0) -- (0,1);
        \draw (0,1) -- (1,1);
        \draw (1,0) -- (1,1);
        \draw (0,0) -- (1,0);
        \draw[blue] (.5,0) -- (.5,1);
        \draw[blue] (0,.5) -- (1,.5);
    \end{tikzpicture}
    \right)~, \label{tqft.pf}
\end{equation}
where the blue line is the $\bZ_2^\mathrm{v}$ symmetry line. The minus sign in front of the last term corresponds to the insertion of the term $i \pi \mathrm{Arf}[\r s]$ where $s$ is the gauge field of the $\bZ_2^\mathrm{v}$. To see this notice that the boundary conditions along the two circles are periodic and therefore $i \pi \int \mathrm{Arf}[\r s]=i \pi$. 
Moreover the last two terms are related by the action of modular $T$ transformation
\begin{equation}
    \begin{tikzpicture}[baseline = 13pt]
        \draw (0,0) -- (0,1);
        \draw (0,1) -- (1,1);
        \draw (1,0) -- (1,1);
        \draw (0,0) -- (1,0);
        \draw[blue] (.5,0) -- (.5,1);
        \draw[blue] (0,.5) -- (1,.5);
    \end{tikzpicture}
    =
    \begin{tikzpicture}[baseline = 13pt]
        \draw (0,0) -- (0,1);
        \draw (0,1) -- (1,1);
        \draw (1,0) -- (1,1);
        \draw (0,0) -- (1,0);
        \draw[blue] (.5,0) -- (1,.5);
        \draw[blue] (0,.5) -- (.5,1);
    \end{tikzpicture}
    =
    T\left(
    \begin{tikzpicture}[baseline = 13pt]
        \draw (0,0) -- (0,1);
        \draw (0,1) -- (1,1);
        \draw (1,0) -- (1,1);
        \draw (0,0) -- (1,0);
        \draw[blue] (.5,0) -- (.5,1);
    \end{tikzpicture}
    \right)~,
\end{equation}
and thus, they contribute the same to the partition function. This is because in a TQFT, the partition function is independent of $\tau$ and therefore each term in \eqref{tqft.pf} is invariant under modular transformations. Also, we have used the fact that the $\bZ_2^\ttv$ symmetry is non-anomalous and thus the four-way junction can be resolved unambiguously. Without the Arf term (upper sign) the theory has two invariant sectors, the untwisted one (first two boxes) and a twisted one (last two boxes). However, gauging $\bZ_2^\mathrm{v}$ with the Arf term amounts to dropping the odd states without introduction a twisted sector for this symmetry. Going back to the case of interest, out of the three primaries of $\Spin(N^2-1)_1$, only the spinor is charged under the $\bZ_2^\mathrm{v}$ center. In conclusion, gauging the $\bZ_2^\ttv$ in the $\Spin(N^2-1)_1/\SU(N)_N$ gauged WZW model with an Arf twist and going to the NS sector of the fermionic theory, is equivalent to dropping the branching coefficients appearing in the decomposition of the spinor representations.

Having discussed all the essential ingredients, now we discuss the $N=3,4,5,6$ examples.

\subsection{SU(2)}
We begin by analyzing the simplest case of $N=2$. In this case the $\Spin(3)_1$ WZW model is actually the same as the $\SU(2)_2$ WZW and the IR TQFT candidate is the $\SU(2)_2/\SU(2)_2$ coset. The branching rules in this case are trivial since every module decomposes into exactly one module of the same $\SU(2)_2$ representation. The primaries of $\SU(2)_2$ are $\mathcal{O}_{0}$, $\mathcal{O}_{\mathrm{fund}}$ and $\mathcal{O}_{\mathrm{adj}}$ with affine weights $\l=[2,0]$, $[1,1]$ and $[0,2]$ respectively. Hence, before gauging the $\bZ_2^\mathrm{v}$ center of $\SU(2)$ with Arf, there are three vacua in the bosonic theory. After gauging with Arf twist, we are instructed to drop the fundamental representation and we end up with two vacua which are permuted under the action of the one-form symmetry. Thus the $\Spin/\SU$ IR TQFT candidate predicts two universes and two vacua -- one vacuum in each universe -- for the $\SU(2)$ adjoint QCD in its NS sector. This is in agreement with the scenarios proposed by both \cite{Gross:1995bp} and \cite{Cherman:2019hbq}.

Furthermore, since $\SU(2)_2$ does not admit any non-diagonal modular invariant, there exist no other possible IR TQFTs, based on the existence of topological lines, beside a direct sum of decoupled $\SU(2)_2/\SU(2)_2$ TQFTs.

\subsection{SU(3) \label{app:su3}}
To study the $N=3$ case, we begin by the explicit branching rules
\begin{align}
    \Module_{[1,0,0,0,0]}&=\module_{[3,0,0]}\oplus \module_{[0,3,0]}\oplus \module_{[0,0,3]}~, \nonumber \\
    \Module_{[0,1,0,0,0]}&=\module_{[1,1,1]}~,\nonumber\\
    \Module_{[0,0,0,1,0]}&=\module_{[1,1,1]}~,\nonumber\\
    \Module_{[0,0,0,0,1]}&=\module_{[1,1,1]}~. \label{spin8/su3 embed}
\end{align}
After fermionization we drop the two spinor representations and we are left with two vacua in universe $0$ and one in each of universes with $1$ and $2$. Therefore, the $\Spin(8)_1/\SU(3)_3$ TQFT predicts four vacua (in the NS sector) and deconfinement for adjoint QCD.

There exist only two allowed non-diagonal modular invariants for $\SU(3)_3$, namely $\underline{\Spin}$ and $\underline{\SO}$ modular invariant theories. Note that the other invariants such as the $\underline{\mathrm{Ss}}$, $\underline{\mathrm{Sc}}$, and $\underline{\PO}^{1,2}$ invariants do not lead to new theories because of the triality of $\Spin(8)_1$, see also \cite{Tong:2019bbk}. The possible IR TQFTs compatible with the existence of the topological lines of $\aca$, are in one-to-one correspondence with the $\SU(3)_3$ modular invariant matrices
\begin{align}
    \underline{\Spin} \otimes \underline{\SU} &= \,\underline{\Spin} ~,\\
    \underline{\Spin} \otimes \underline{\Spin} &= 3\, \underline{\Spin} ~,\\
    \underline{\Spin} \otimes \underline{\SO} &= 3\, \underline{\SO} ~,
\end{align}
which shows that $\Spin(8)_1/\SU(3)_3$ TQFT is the smallest possible IR TQFT and probably the most reasonable candidate.

Note that although the $\underline{\Spin}$ and $\underline{\SO}$ have the same modular invariant matrices, they correspond to different theories. The $\SO(8)_1$ WZW model is obtained by gauging the $\bZ_2^\ttv$ symmetry of the $\Spin(8)_1$ WZW model, and these two theories have different boundary conditions $\cC_A$. These different boundary conditions correspond to different $\SU(3)_3$ Ocneanu graphs denoted by $A_3/3$ and $3A_3^c$ in \cite{Ocneanu:2000kj}.

\subsection{SU(4)}
The branching rules for $N=4$ are \cite{Abolhassani:1993iz,Coquereaux:2007rw}
\begin{align}
    \Module_{[1,0,0,0,0,0,0,0]}&=\module_{[4,0,0,0]}\oplus \module_{[0,0,4,0]}\oplus \module_{[1,2,1,0]}\oplus \module_{[1,0,1,2]}~, \nonumber \\
    \Module_{[0,1,0,0,0,0,0,0]}&=\module_{[0,4,0,0]}\oplus \module_{[0,0,0,4]}\oplus \module_{[2,1,0,1]}\oplus \module_{[0,1,2,1]}~,\nonumber \\
    \Module_{[0,0,0,0,0,0,0,1]}&=2\module_{[1,1,1,1]}~.
     \label{spin16/su4 embed}
\end{align}
Similarly here after dropping the spinor representation and going to the fermionic theory, there are $8$ vacua that organize themselves into two orbits of the one-form symmetry, each one of length $4$. Hence we conclude that the $\Spin(15)_1/\SU(4)_4$ TQFT candidate predicts four universes, each with two vacua, for the $\SU(4)$ adjoint QCD. Looking for other possible IR TQFTs, note that for $\SU(4)_4$ there are no other modular invariants beside D-series and the $\underline{\Spin}$ invariant that we discussed for general values of $N$. Thus we find possible TQFTs
\begin{align}
    \underline{\Spin} \otimes \underline{\SU} &= \,\underline{\Spin} ~,\\
    \underline{\Spin} \otimes \underline{\SU/\bZ_2} &= 2\, \underline{\Spin} ~,\\
    \underline{\Spin} \otimes \underline{\SU/\bZ_4} &= 2 \,\underline{\Spin} ~,\\
    \underline{\Spin} \otimes \underline{\SU/\bZ_2}^\ast &= 2\, \underline{\Spin} ~,\\
    \underline{\Spin} \otimes \underline{\SU/\bZ_4}^\ast &= 2 \,\underline{\Spin} ~,\\
    \underline{\Spin} \otimes \underline{\Spin} &= 4\, \underline{\Spin} ~.
\end{align}
Again, the $\Spin(15)_1/\SU(4)_4$ TQFT has the smallest number of vacua among the possible IR TQFTs.

\subsection{SU(5)}
By studying the branching rules for $N=5$, using the modular invariants given in \eqref{su5.invariants}, we find 16 vacua for the $\Spin(24)_1/\SU(5)_5$ TQFT after fermionizing the theory. Where there are 4 vacua in universe 0, and 3 vacua in the other four universes. Furthermore, five modular invariants satisfying $\mim(B)_{\mathcal{A}\mathcal{A}}\neq0$ were found in appendix \ref{app:modular.invariants}, which lead to possible IR TQFTs
\begin{align}
    \underline{\Spin} \otimes \underline{\SU} &= \,\underline{\Spin} ~,\\
    \underline{\Spin} \otimes \underline{{\SU}/{\bZ_5}} &= 5\, \underline{\Spin} ~,\\
    \underline{\Spin} \otimes \underline{\mathrm{SY}} &=  \underline{\Spin} \oplus 4\, \underline{\PO}~,\\
    \underline{\Spin} \otimes \underline{\PO} &= 2 \,\underline{\Spin} \oplus 8\, \underline{\PO}~,\\
    \underline{\Spin} \otimes \underline{\Spin} &= 6\, \underline{\Spin} \oplus 4\, \underline{\PO} ~.
\end{align}
There are four more theories whose number of vacua are given by $\abs{\underline{\Spin} \otimes \underline{\SO}} = \abs{\underline{\Spin} \otimes \underline{\Spin}}$ and $\abs{\underline{\Spin} \otimes \underline{\PO^{1,2}}} = \abs{\underline{\Spin} \otimes \underline{\mathrm{Ss}}} =  \abs{\underline{\Spin} \otimes \underline{\mathrm{Sc}}}$. Again, the $\Spin(24)_1/\SU(5)_5$ TQFT has the smallest number of vacua among our list.

\subsection{SU(6)}
For $N=6$, by dropping the spinor representation, the $\Spin(35)_1/\SU(6)_6$ TQFT has $32$ vacua after fermionization. There are six vacua in universes 0 and 3, and five vacua in other universes. In appendix \ref{app:modular.invariants}, we found 12 allowed modular invariants, that lead to other possible IR TQFTs
\begin{align}
    \underline{\Spin} \otimes \underline{\SU} &= \,\underline{\Spin} ~,\notag\\
    \underline{\Spin} \otimes \underline{{\SU}/{\bZ_2}} &= \underline{\Spin} ~,\notag\\
    \underline{\Spin} \otimes \underline{{\SU}/{\bZ_3}} &= 3\, \underline{\Spin} ~,\notag\\
    \underline{\Spin} \otimes \underline{{\SU}/{\bZ_6}} &= 3\, \underline{\Spin} ~,\notag\\
     \underline{\Spin} \otimes \underline{{\SU}/{\bZ_2}}^\ast &= \underline{\Spin} ~,\notag\\
    \underline{\Spin} \otimes \underline{{\SU}/{\bZ_3}}^\ast &= 3\, \underline{\Spin} ~,\notag\\
    \underline{\Spin} \otimes \underline{{\SU}/{\bZ_6}}^\ast &= 3\, \underline{\Spin} ~,\notag\\
    \underline{\Spin} \otimes \underline{\mathrm{Sp}} &=  4\, \underline{\mathrm{Spin|X}}~,\notag\\
    \underline{\Spin} \otimes \underline{\mathrm{X}} &=  8\, \underline{\mathrm{Spin|X}}~,\notag\\
    \underline{\Spin} \otimes \underline{\mathrm{Spin|X}} &=  16\, \underline{\mathrm{Spin|X}}~,\notag\\
    \underline{\Spin} \otimes \underline{\mathrm{X|Spin}} &=  8\, \underline{\Spin}~,\notag\\
    \underline{\Spin} \otimes \underline{\Spin} &= 6\, \underline{\Spin} ~.
\end{align}
By using the trace formula we get
\begin{align}
    \abs{\underline{\Spin}} &= \Tr\left[ \mim(\Spin) \right] = 48~,\\
    \abs{\underline{\PO}} &= \Tr\left[ \mim(\PO) \right] = 24~.
\end{align}
The $\Spin(35)_1/\SU(6)_6$ TQFT is the one with the smallest number of vacua, namely 48 vacua for the bosonic theory. The slightly larger TQFT is the $\underline{\Spin} \otimes \underline{\mathrm{Sp}}$ TQFT with $96$ vacua, which is twice as big as the $\Spin(35)_1/\SU(6)_6$ TQFT.

\section{Topological Lines in Fermionic Theory}
\label{app:fermioniclines}
In most of this paper we have worked with the topological lines in the bosonized theory.
This is to avoid technical complications, and also because of lack of literature completing the theory of lines in a fermionic theory, though there are remarkable papers \cite{Gaiotto:2015zta,Novak:2015ela,Bhardwaj:2016clt,brundan2017monoidal,bruillard2017fermionic,Creutzig:categorysuper201705,usher2018fermionic,Aasen:2017ubm,Runkel:2020zgg,Lou:2020gfq}.\footnote{In particular, most of the literature talks about super-commutative Frobenius algebra object $A$ and the corresponding module category $\mathcal{C}_A$, but not about the bimodule category ${}_A\mathcal{C}_B$ in detail. In \cite{Lou:2020gfq} constructed the bimoduel category based on \cite{Fuchs:2002cm}.}
Based on this literature, here we give a brief outlook of the topological lines in fermionic theories, and give an explicit example of topological lines in the \emph{fermionic} $\SU(3)$ adjoint QCD.

\subsection{Topological lines in fermionic theories}
The topological lines in a 1+1d \emph{fermionic} QFT is described by a \emph{super}-fusion category.
A difference between a fusion category and its super-version is that in the latter the space of morphisms is a $\mathbb{Z}_2$ graded $\mathbb{C}$-vector space (super-vector space). The grading denotes whether the defect operator is bosonic or fermionic.\footnote{The dimension of a $\mathbb{Z}_2$-graded vector space is denoted as $p|q$, where $p$ is the dimension of the degree-even (bosonic) subspace and $q$ is the dimension of the degree-odd (fermionic) subspace.}
Correspondingly, there are two types of simple object in the category distinguished by their endmorphism algebra:\footnote{The m-type and q-type are referred to as bosonic and Majorana lines respectively in \cite{Gaiotto:2015zta,Bhardwaj:2016clt}.}
\begin{itemize}
    \item m-type simple object: the endmorphism algebra is $\mathbb{C}$, whose dimension is $1|0$. An object of this type shares many properties with an object in a (bosonic) fusion category.
    \item q-type simple object: the endmorphism algebra is the Clifford algebra $\mathrm{Cl}_1$ with a single odd generator, whose dimension is $1|1$. The odd isomorphism $f$ in $\mathrm{Cl}_1$ squares to 1.
\end{itemize}
Physically, an m-type line is a defect on which an even number of 1d Majorana fermion reside, while on a q-type line an odd number of 1d Majorana fermion reside.
In addition, if $\ket{0}$ is a gapped vacuum and $L$ is a topological line of q-type, the vacuum $L\ket{0}$ should differ from $\ket{0}$ by the $\mathrm{Arf}$ invertible phase, to match with the anomaly possessed by the 1d Majorana fermion on the line. In particular, $\ket{0} \neq L\ket{0}$.
An example of a q-type operator is the $\mathbb{Z}_2$ symmetry operator when the $\mathbb{Z}_2$ symmetry has an odd element of the $\mathbb{Z}_8$ classification of the $\mathbb{Z}_2$ 't Hooft anomaly.
Note that the quantum dimension of the anomalous $\mathbb{Z}_2$ line is $\sqrt{2}$, even though it is invertible. 
Indeed the $\sqrt{2}$ is a natural contribution from a 1d Majorana fermion living on the line.

A super-fusion category $\cC$ should contain the ``transparent fermion" line, which is denoted by $\Pi$. This line satisfy
\begin{equation}
    \mathrm{Dim}\;\mathrm{Hom}(\Pi,\mathbbm{1}) = 0|1, \quad \mathrm{Dim}\;\mathrm{Hom}(\Pi,L) = 0|0 \qquad \text{for $L\neq\mathbbm{1}$}.
\end{equation}
In other words $\Pi$ is odd-isomorphic to the trivial line $\mathbbm{1}$.
For a q-type line $L$, the line $\Pi L := \Pi\otimes L$ is identified with $L$ itself by the odd-isomorphism in $\mathrm{Hom}(L,L)\simeq \mathrm{Cl}_1$.
The simplest super-fusion category is the cateogry $\mathrm{SVect}$ of super-vector spaces, whose simple objects are even and odd one-dimensional vector spaces.
Given a bosonic fusion category $\cC_\text{bos}$, one can obtain a super-fusion category $\cC_\text{bos}\boxtimes \mathrm{SVect}$ by including the transparent fermion line. These are the topological lines in a bosonic system, when it is declared to be put on spin manifolds without actually be coupled with the spin structure.

A general discrete gauging procedure in a fermionic theory is characterised by a ``super-commutative" Frobenius algebra object, which generalises a symmetric Frobenius algebra object in the bosonic case.
The ``fermionization" gauging often considered in the main text is the special case. There, the starting theory is bosonic whose symmetry category $\cC_\text{bos}$ contains an anomaly-free $\mathbb{Z}_2$ line $a \in \cC_\text{bos}$. The Frobenius algebra object for this special case is $\mathbbm{1} \oplus \Pi a \in \cC_\text{bos}\boxtimes \mathrm{SVect}$.
For a line $L\in \cC_\text{bos}$ to survive the fermionization, $L$ should commute with $a$ with respect to the fusion $\otimes$.
If $L\otimes a \simeq a \otimes L\simeq L$,
after the fermionization $L$ becomes a q-type line because $a$ is identified with $\Pi$ after the gauging and the above fusion provides the odd isomorphism to itself.
One the other hand, if $L_1\otimes a \simeq a \otimes L_1 \simeq L_2$ with $\mathrm{Hom}_{\cC_\text{bos}}(L_1,L_2)\simeq\{0\}$, $L_1$ becomes an m-type line after the gauging and $L_2$ is identified with $\Pi L_1$.
Further, there should be a quantum $\mathbb{Z}_2$ symmetry, which is the fermion parity $(-1)^F$.

For a general super-commutative Frobenius algebra object $A\in\cC$, we expect the category symmetry of the $A$-gauged theory to be the bimodule category ${}_A\cC_A$ defined in a suitable sense. Here we assume that $\cC = \cC_\text{bos}\boxtimes\mathrm{SVect}$ with a bosonic braided fusion category $\cC_\text{bos}$.
We further expect that the 3d picture involving surface operator $S_A$, explained in Section~\ref{sec:3d} still holds, by allowing $S_A$ to couple to the spin structure.\footnote{The categorical condensation construction of $S_A$ in \cite{Gaiotto:2019xmp,Johnson-Freyd:2020usu} is applicable to fermionic cases.}
Namely, we expect that there are tensor functors $\alpha^\pm : \cC \to {}_A\mathcal{C}_A$, and the analog of \eqref{eq:alphainner}:
\begin{equation}
    \mathrm{Dim}\;\mathrm{Hom}(\alpha^+(L_i)\otimes\alpha^-(L_{\bar i}),\alpha^+(L_j)\otimes\alpha^-(L_{\bar j})) = \sum_{k,\bar{k}}N_{ik}^j N_{\bar j\bar k}^{\bar i}\mim^\mathrm{NS}_{k\bar{k}},
    \label{eq:alphainnerfer}
\end{equation}
where $\mim^\mathrm{NS}_{k\bar{k}}$ defines the $T^2$ partition function with $\mathrm{NS}$ spin structure along the space direction.\footnote{$\mim^\mathrm{NS}_{k\bar k}$ is understood as a pair of integers $p|q$ where $p$ counts the bosonic operators while $q$ counts the fermionic operators.}

\subsection{Lines in Fermionic SU(3) adjoint QCD}
The lines in the bosonized adjoint QCD, equivalently that of the $\Spin(8)_1/\SU(3)_3$ gauged WZW model, are enumerated in \eqref{a ind in su3}.
By analysing the fusion between the $\mathbb{Z}_2^\mathrm{v}$ line $a$ and other lines, we can determine the fate of the lines in \eqref{a ind in su3} after the fermionization.
One finds that $L_i$, $i=1,2,\cdots,6$ are fixed by the fusion with $a$ and thus survive after the gauging as q-type lines. The $\mathbb{Z}_2^\mathrm{s}$ line will become the chiral $\mathbb{Z}_2^\chi$ line which is anomaly-free in 8 fermions. Actually there are left chiral symmetry $\chi_L$ and the right chiral symmetry $\chi_R$, which differ from each other by the quantum symmetry $(-1)^F$.
The other invertible lines $s_i$ does not commute with $a$. However, if we let $s$ denote the $\mathbb{Z}_3$ generator with multiplication $a s = s a b$, the non-simple lines $s+ s b$, $s a + s a b$, $s^2 + s a b$, $s^2 a+ s^2 b$ commutes with $a$ and these give 2 m-type simple lines $L_7,L_8$ and the lines odd-isomorphic to them: $\Pi L_7, \Pi L_8$ after the gauging.
Therefore we expect that the lines are generated by the q-type lines $L_i$, $i=1,2,\cdots 6$ and the m-type lines $\Pi,L_7,L_8,\chi_L,(-1)^F$ over the fusion $\otimes$.

The fact that $L_1,L_2$ are q-type has significance in physics, because it indicates the universes 0 and 1,2 have different invertible phase regarding the $\mathrm{Arf}$ invariant. This in turn reproduces the expectation made in section~\ref{sec:deconf} that a phase transition is expected at a finite mass in the string universes.

One can further try to fit these lines into the image of the functors $\alpha^\pm:\cC \to {}_A\cC_A$, where $A$ is now taken to be the super-commutative Frobenius algebra in $\cC=\Rep\,\hat{\mathfrak{su}}(3)_3\boxtimes\mathrm{SVect}$ defining the 8 fermions CFT. 
The left-chiral symmetry line $\chi_L$ belongs to $\alpha^+_{(1,1)}$, since the symmetry comes from the invertible line $b$ belongs to $\alpha_{(1,1)}^\pm$ in the bosonized theory, and $\chi_L$ acts only on the left-moving sector. Then the right-chiral symmetry $\chi_R$ should belongs to $\alpha^-_{(1,1)}$.
We also assume that the relation between $\alpha^\pm$ lines and $L_i$ lines with $i=1,2,\cdots 6$ remains the same as \eqref{a ind in su3} and that $L_7, L_8$ belong to $\alpha^+_{(1,0)}\alpha^-_{(0,1)}$ and $\alpha^+_{(0,1)}\alpha^-_{(1,0)}$, respectively.
The rest of the data can be fixed by demanding the equation \eqref{eq:alphainnerfer} and the result is 

    \begin{align}
        &\a^+_{(0,0)}=\mathbbm{1}~,
        \qquad&&\a^+_{(0,1)}\a^-_{(1,0)}=L_8 + \Pi L_8 ~, \notag\\
        &\a^+_{(1,0)}=L_1~,
        &&\a^+_{(1,1)}=\mathbbm{1}+\chi_L + \Pi\chi_L~, \notag\\
        &\a^+_{(0,1)}=L_2~,
        &&\a^-_{(1,1)}=\mathbbm{1}+\chi_R + \Pi\chi_R~, \notag\\
        &\a^-_{(1,0)}=L_3~,
        &&\a^+_{(1,0)}\a^-_{(1,1)}=L_1 + 2 \chi_R L_1~, \notag\\
        &\a^-_{(0,1)}=L_4~,
        &&\a^+_{(0,1)}\a^-_{(1,1)}=L_2 + 2 \chi_R L_2~,\\
        &\a^+_{(0,1)}\a^-_{(0,1)}=2L_5 ~,
        &&\a^+_{(1,1)}\a^-_{(1,0)}=L_3 + 2 \chi_L L_3~, \notag\\
        &\a^+_{(1,0)}\a^-_{(1,0)}=2L_6 ~,
        &&\a^+_{(1,1)}\a^-_{(0,1)}=L_4 + 2 \chi_L L_4~, \notag\\
        &\a^+_{(1,0)}\a^-_{(0,1)}=L_7 + \Pi L_7,
        &&\a^+_{(1,1)}\a^-_{(1,1)}=\mathbbm{1}+\chi_L+\Pi \chi_L + \chi_R + \Pi \chi_R + 2(-1)^F + 2\Pi(-1)^F~. \notag
    \end{align}
In summary, there are 10 q-type simple lines:
\begin{equation}
    L_1,L_2,L_3,L_4,L_5,L_6,\chi_R L_1,\chi_R L_2,\chi_L L_3, \chi_L L_4
\end{equation}
and 6 m-type lines up to $\Pi$ action:
\begin{equation}
    \mathbbm{1}, \chi_L, \chi_R, (-1)^F, L_8, L_9~.
\end{equation}
The quantum dimensions are 1 for the invertible lines $\mathbbm{1},\chi_L,\chi_R,(-1)^F$ and 2 for all of the others.

The analysis regarding deformations can also be done. The mass operator $\cO_\mathrm{m}$ breaks all but $(-1)^F$, and the double trace quartic coupling preserves the m-type lines while breaking the q-type lines.
\bibliographystyle{ytphys}
\bibliography{ref}

\end{document}